\documentclass{aa} 
\usepackage{graphicx}
\usepackage{times}
\usepackage{latexsym}
\usepackage{epsfig}
\usepackage{txfonts}
\usepackage{natbib}
\usepackage{supertabular,multirow,longtable,rotating}
\usepackage{amssymb}
\usepackage{color,afterpage}
\usepackage{longtable,lscape}

\def\Ebv{\ensuremath{{\mathrm E}_{\mathrm (B-V)}}}

\bibpunct{(}{)}{;}{a}{}{,} 

\begin{document}

\title{Diffuse interstellar bands in Upper Scorpius: Probing variations in the DIB spectrum due to changing environmental
  conditions\thanks{Based on observations collected at the European Southern Observatory, Paranal, Chile (ESO program 63.H-0456)}}
\author{D.A.I. Vos~\inst{1} \and N.L.J. Cox~\inst{2} \and L. Kaper~\inst{3} \and M. Spaans~\inst{4} \and P. Ehrenfreund~\inst{5}}

\institute{%%
	Radboud University Nijmegen, Toernooiveld 1, Postbus 9010, 6500 GL, Nijmegen, The Netherlands
	\and
	Instituut voor Sterrenkunde, K.U.Leuven, Celestijnenlaan 200D, bus 2401, 3001, Leuven, Belgium
        \and
        Astronomical Institute ''Anton Pannekoek'', Universiteit van Amsterdam, Postbus 94249, 1090~GE~~Amsterdam, The Netherlands
	\and	
	Kapteyn Astronomical Institute, Rijksuniversiteit Groningen, Postbus 800, 9700~AV~~Groningen, The Netherlands
	\and
	Astrobiology Group, Leiden Institute of Chemistry, Leiden University, Einsteinweg 55, 2300 RA, Leiden, The Netherlands}

   \offprints{Nick Cox, \email{nick.cox@ster.kuleuven.be}}
   \date{Received 7 March 2008; Accepted 28 July 2011}

\abstract{%%
}{%%
We study the effects of local environmental conditions affecting the diffuse interstellar band (DIB) 
carriers within the Upper Scorpius subgroup of the Sco~OB2 association. 
The aim is to reveal how the still unidentified DIB carriers respond to different physical conditions prevailing 
in interstellar clouds, in order to shed light on the origin of the DIB carriers.  
}{%%
We obtained optical spectra with FEROS on the ESO 1.52m telescope at La Silla, Chile, and measured the equivalent widths of five DIBs 
(at 5780, 5797, 6196, 6379, and 6613~\AA) as well as those of absorption lines of di-atomic molecules (CH, CH$^+$, CN) and atoms (\ion{K}{i}, \ion{Ca}{i})
towards 89\ targets in the direction of Upper Scorpius. We construct a simple radiative transfer and chemical network model of the diffuse 
interstellar medium (ISM) sheet in front of Upp Sco\ to infer the effective radiation field.
}{%%
By measuring the DIB and molecular spectrum of diffuse clouds towards 89\ sightlines in the Upper Scorpius region, 
we have obtained a valuable statistical dataset that provides information on the physical conditions that influence the band strengths of the DIBs. 
Both the interstellar radiation field strength, $I_{\rm UV}$, and the molecular hydrogen fraction, $f_{\mathrm{H}_2}$, have been derived for 
55 sightlines probing the Upp Sco\ ISM.
We discuss the relations between DIB strengths, CH and CH$^+$ line strengths, \Ebv, $I_{\rm UV}$, and $f_{\mathrm{H}_2}$.
The ratio between the 5780 and 5797~\AA\ DIBs reveals a (spatial) dependence on the local environment in terms of cloud density and 
exposure to the interstellar radiation field, reflecting the molecular nature of these DIB carriers.
}{%%
}


\titlerunning{Diffuse interstellar bands in Upper Scorpius}
\authorrunning{Vos, Cox, Kaper, Spaans \& Ehrenfreund}
\maketitle

\section{Introduction}\label{sec:intro}

The diffuse interstellar medium contains compounds of unidentified origin that absorb in the UV-visual to
near-infrared spectral range. More than 300 different diffuse interstellar bands (DIBs) are currently
identified (\citealt{1995ARA&A..33...19H}; \citealt{2008ApJ...680.1256H}). Many possible carriers have been
proposed,  ranging from grain impurities and exotic molecules to H$_2$.   In the past two decades the field
has converged towards larger carbonaceous molecules,  like the fullerenes and polycyclic aromatic
hydrocarbons (PAHs), which have electronic transitions in the optical (see for example
\citealt{1999ApJ...526..265S}, \citealt{2005A&A...432..515R}; \citealt{2006JPCA..110.6173K};
\citealt{2006ApJ...638L.105Z} and \citealt{2011ApJ...728..154S}). New diffuse bands have been detected in
one line-of-sight which appear to match with naphthalene and anthracene cations 
(\citealt{2008ApJ...685L..55I,2010arXiv1005.4388I}) and the weak 5450~\AA\ DIB is found to match with an
absorption band arising  from a hydrocarbon plasma created in the laboratory
(\citealt{2010A&A...511L...3L}).  Linear-C$_3$H$_2$ has been put forward as a carrier of the 5450 and
4881~\AA\ DIBs by \citet{2011ApJ...726...41M}. These assignments are, however, tentative and disputed
(\citealt{2011MNRAS.412.1259G}).

In order to understand the chemical and physical properties of the DIB carrier(s) it is important to study
their behaviour in different interstellar environments, both in our own galaxy and beyond.  Studies of DIBs
in the Magellanic Clouds (\citealt{2002ApJ...576L.117E}; \citealt{2006A&A...447..991C,2007A&A...470..941C};
\citealt{2006ApJS..165..138W}), \object{M31} (\citealt{2008A&A...480L..13C,2008A&A...492L...5C}) and beyond
(\emph{e.g.}  \citealt{2000ApJ...537..690H}; \citealt{2006ApJ...647L..29Y}; \citealt{2005A&A...429..559S};
\citealt{2008A&A...485L...9C})  illustrate that DIB carrier abundances (per amount of dust and gas) can be
similar to galactic values. However, these studies have also revealed systematic differences in these 
extragalactic environments.

\begin{figure*}[t!]
\centering
\includegraphics[angle=-90,width=0.7\textwidth]{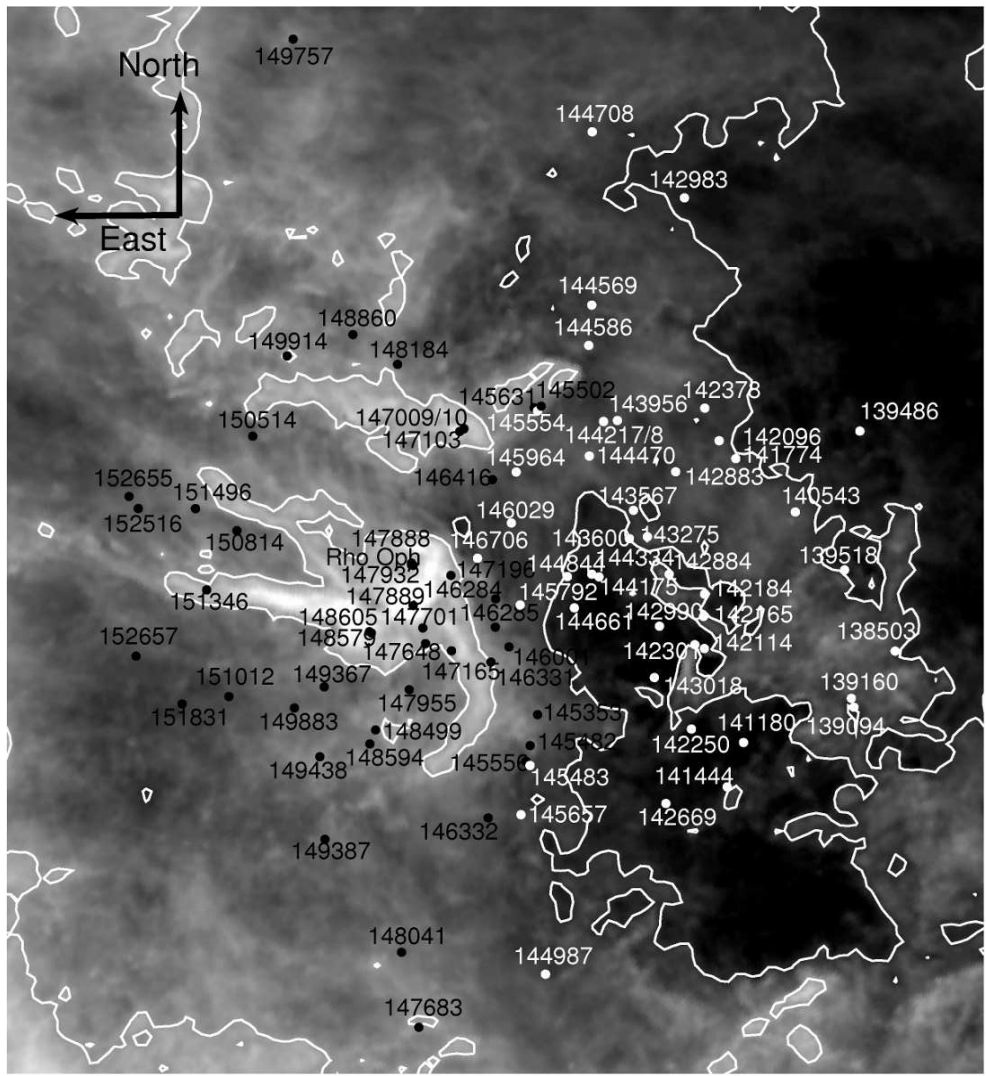}
\caption{The positions (black and white dots) of the 89\ mainly B-type members of
Upp~Sco are shown (with HD numbers) on top of a 100~$\mu$m infrared dust map of this region 
(\citealt{1998ApJ...500..525S}). The north-east arrows are 3\degr\ in length.
The $\rho$\,Oph cloud can be identified by the
bright filamentary emission located just left to the center of the figure.
The dust emission (on a logarithmic grey scale) is proportional to the
reddening \Ebv\ of sightlines penetrating these clouds; the lowest intensities
correspond to \Ebv\ $\sim$0.02 mag (black) and highest intensities to $\geq$~2~mag (white). 
The well-known targets $\sigma$~Sco (\object{HD 147165}, in the $\rho$\,Oph cloud) and $\zeta$~Oph 
(\object{HD 149757}, top left) are included in our study.} 
\label{fig:targets}
\end{figure*}

A large amount of published information is available regarding DIBs in many sightlines probing the Galaxy
(\emph{e.g.} \citealt{1993ApJ...407..142H,1986ApJ...305..455C,1999A&A...347..235K, 
2003ApJ...584..339T,2004MNRAS.355..169G,2004A&A...414..949W,2008A&A...484..381W,2011ApJ...727...33F}),
yielding relations  of DIB properties with respect to each other and to other diffuse ISM gas \& dust
tracers.  Most of these studies focused on DIBs probing various galactic environments, and provided average 
results for the Milky Way. Studies dealing with a particular region usually only include a very limited
number of sightlines. An exception is the study of the Orion region by \citet{1994A&A...281..517J}  which
entails 22 lines of sight. 
Another multi-object study, by \citet{2009MNRAS.399..195V}, used the distant globular cluster $\omega$~Cen 
to probe fluctuations of \ion{Ca}{ii}, \ion{Na}{i} and the $\lambda\lambda$5780 and 5797 DIBs in the diffuse - low reddening - foreground ISM. This study revealed small-scale structure - on parsec scales - in the warm neutral and weakly ionised medium of the Disc-Halo interface. The observed low 5797/5780 DIB ratio was found to be consistent with the relative high UV radiation levels typically inferred for the extra-planar warm medium.

Nearby OB associations host many bright early-type stars confined in a relatively small area of the sky. 
These stars have only few stellar lines in the optical spectrum contaminating the interstellar spectrum.
Thus, these associations provide a setup that is perfectly suited to study the effect of varying local
conditions on the DIB spectrum.   One of these associations, Scorpius OB2, is a young ($5-20$ Myr),
low-density ($\approx 0.1~$M$_{\odot}$~pc$^{-3}$) grouping of stars divided in three subgroups
(\citealt{1999AJ....117..354D}; \citealt{2005A&A...430..137K}). %(\citealt{2004RMxAC..21..139K}).  The Upper
Scorpius (Upp Sco) region is the subgroup near the Ophiuchus star forming region and the $\rho$ Oph cloud at
a distance of 145~$\pm$~2~pc \citep{1999AJ....117..354D}.  Combining 2MASS extinction maps with Hipparcos
and Tycho parallaxes, \citet{2008A&A...489..143L} found a distance of 119$\pm$6 pc for the $\rho$ Ophiuchi
cloud (with the core at 128$\pm$8~pc). \citet{2008AN....329...10M} suggested a mean distance of 139$\pm$6 pc
for the distance of the Ophiuchus molecular cloud, which they placed within 11~pc of the centroid of the
Upper Scorpius subgroup.

Filamentary - interstellar - material connected to the $\rho$\,Ophiuchus cloud complex is observed towards
Upper Scorpius (\citealt{1992A&A...262..258D}). The densest part of this complex is the $\rho$\,Oph dark
cloud, a site of ongoing low-mass star formation (\citealt{1973ApJ...184L..53G};
\citealt{1992ApJ...395..516G}; \citealt{1997PASP..109..549W}; \citealt{2001ASPC..243..791P}) that is exposed
to the radiation fields and stellar winds produced by nearby early-type stars.  A detailed review on the
stellar population and star formation history of the Sco OB2 association is given by
\citet{2008hsf2.book..235P} and \citet{2008hsf2.book..351W}.

The advantages of studying the properties of DIBs in the Upper Scorpius region are numerous. It is in close
vicinity and it has been extensively studied in the past. Detailed information is available on both the
stellar content (spectral types, photometry, distances, kinematics, etc.) and the conditions of the
surrounding interstellar medium (dust emission and absorption, IR-to-far-UV extinction curves, UV emission,
molecular content, etc.). It exhibits a significant variation in local environmental conditions which should
translate into changing properties of the DIBs (if they depend on these conditions) when probing different
parts of the Upp Sco\ region.

Previous studies of interstellar gas and dust in the Upp Sco\ region focused on the $\rho$ Oph cloud and a
few other nearby bright B stars. \citet{2008ApJ...679..512S} give a concise summary of different studies of
the Upp Sco\ region covering a range of topics  including UV extinction, atomic and molecular hydrogen,
atomic and molecular gas, astrochemistry, and DIBs. For example, $H_2$ observations show that sightlines in
this region have both low ($\leq$0.1) and high ($\sim$0.3--0.6)  molecular fractions $f_{\mathrm{H}_2}$
(\emph{e.g.} \citealt{1977ApJ...216..291S}).

In this paper we investigate the behaviour of five well-known DIBs (at 5780, 5797, 6196, 6379, and 6613~\AA)
and the molecular  lines of CH, CH$^+$, and CN in the sightlines towards 89\ B-type stars in the
direction of Upp~Sco (Fig.~\ref{fig:targets}).  These targets, within a field of 20\degr$\times$20\degr,
provide a unique and detailed view of the gas and dust in this nearby association. In
Sect.~\ref{sec:observations} we introduce our sample and provide information on the reduction of the
obtained spectra.  Sect.~\ref{sec:reddening} briefly discusses line-of-sight reddening and dust towards
Upp Sco. In Sect.~\ref{sec:results} we present the observational details of atomic and molecular lines as
well as diffuse bands. We explore the results in Sect.~\ref{sec:discussion}, where we discuss first the
relation between DIB strength, the dust tracer \Ebv,  and the molecular content. Then, we demonstrate that
the DIB ratio 5797/5780 may be useful to distinguish between  lines-of-sight probing diffuse cloud edges and
those penetrating denser cloud cores.  The observed differences in physical properties of both types of
sightlines are often attributed to the {\it skin effect}, the increase in effective shielding of molecules
from UV radiation as one moves deeper into an interstellar cloud (\citealt{1988A&A...190..339K};
\citealt{1995ARA&A..33...19H}; \citealt{1997A&A...326..822C}).  Furthermore, we have studied the effect of
local environmental conditions, such as density and UV field strength on DIB strengths and ratios.  The line
strengths of CH, CH$^+$, and CN can be used to characterise the physical and chemical conditions in the
respective sightlines. We have constructed a simple dust cloud model to derive the intensity of the
interstellar radiation field (ISRF) from the observed CH and CN line strengths.  The paper concludes with a
summary of the main results (Sect.~\ref{sec:conclusion}).

\begin{figure}[t!]
\centering 
   \includegraphics[angle=-90,width=\columnwidth]{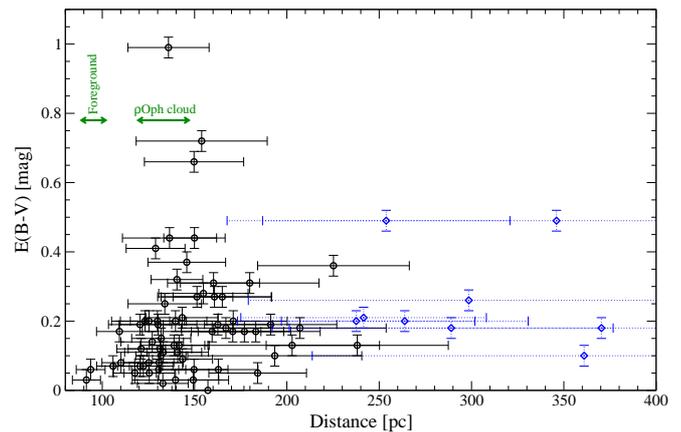}
   \caption{The colour excess \Ebv\ (mag) is 
   plotted against the distance of the observed targets (\citealt{1997A&A...323L..49P}).
   Stars with distance errors larger than 50~pc are shown as squares in grey/blue.  
   The low \Ebv\ below 120~pc indicates that there is little foreground
   material in front of Upper Scorpius.  The increased scatter at
   140~$\pm$~20~pc reflects the density variation associated primarily
   with the $\rho$\,Oph cloud complex.  Beyond that distance, no
   substantial increase of reddening is observed up to 400~pc.
   A similar figure for the $\rho$\,Oph cloud region was shown in \citet{2008A&A...489..143L} 
   (also based on optical photometry, with a partial overlap in the selected sky region).}
   \label{fig:ebv_vs_distance}
\end{figure}

\section{Optical spectra of B-type stars in Upp Sco}\label{sec:observations}

The observed targets cover a region in the sky of approximately 20\degr$\times$20\degr\  (\emph{i.e.}
$\sim$$50 \times 50$~pc at a distance of 145~pc). Within this relatively small region 89\ sightlines are
measured and analysed.  The positions of the observed targets in the Upper Scorpius region are shown on a
100~$\mu$m image (far-infrared dust reddening map;  Fig.~\ref{fig:targets}), a reprocessed composite of the
COBE/DIRBE and IRAS/ISSA maps with the zodiacal foreground  and confirmed point sources removed
(\citealt{1998ApJ...500..525S}). 

Seven out of eight stars that generate 90\% of the local interstellar radiation field (ISRF) are located in
this region. The dust in these lines-of-sight imposes \Ebv\ values from $\sim$0.02 up to $\sim$0.99~mag (see
Sect.~\ref{sec:reddening}),  implying local variations in the cloud - column or volume - density and
structure, and subsequently the attenuation of the ISRF. Therefore this large dataset is extremely valuable
to investigate the effects of environmental conditions on the DIB carriers on a local scale.

Echelle spectra were obtained with the FEROS instrument on the ESO 1.52m telescope at La Silla from  26--30
April 1999. The spectra were taken at a resolving power of $R\approx 48,000$ covering a spectral range of
3800 to 8500~\AA. The data were reduced using the FEROS context within the ESO-MIDAS data reduction
package.  Data reduction was performed in a standard fashion, the CCD images were first bias subtracted and
subsequently the Echelle orders were straightened, extracted, unblazed (flat fielded), rebinned (wavelength 
calibrated) and finally merged. We extracted and normalised the spectral ranges of interest. Final spectra
have signal-to-noise ratios between 100 and 400 in the wavelength regions of the measured lines. Typically,
the S/N values are lower for the blue region ($\sim$3900-4500~\AA) with respect to the red part of the
spectrum. Furthermore, exact values differ for each line-of-sight (due to differences in exposure time,
visual magnitude of the star and the weather conditions). The S/N is reflected in the reported equivalent
width uncertainties.

Table~\ref{tb:basic-data} (Online) summarises the basic data for the observed targets:  Henry-Draper (HD)
and Hipparcos (HIP) number, spectral type, right ascension \& declination,  colour $B-V$, intrinsic colour
$(B-V)_0$, reddening \Ebv\ (see Sect.~\ref{sec:results}),  total-to-selective visual extinction $R_V$, and
Hipparcos distance (pc).

\onltab{1}{
\begin{table*}[t]
\begin{center}
\caption{%\scriptsize 
List of the observed targets. 
Cols.~1 to 8 give the Henry-Draper (HD) number, Hipparcos (Hip) number, 
spectral type, right ascension \& declination, colour $(B-V)$, intrinsic stellar colour $(B-V)_0$, and reddening \Ebv.
Spectral types are from the Michigan Spectral Catalog of HD stars (\citealt{1982MSS...C03....0H,1988MSS...C04....0H}). %III/80/vol3 and III/133/vol4
Intrinsic colours, $(B-V)_0$, are taken from \citealt{1970A&A.....4..234F}.
$B$ and $V$ photometric magnitudes are taken from Tycho-2 catalog and converted to the Johnson colour index $B-V$ 
(using Bessell 2000 and Mamajek et al. 2002); subsequently E$_{B-V} = (B-V) - (B-V)_0$. 
Typical error on \Ebv\ is $\sim$0.03~mag (see Sect.~\ref{sec:reddening}).
Col.~9 lists the total-to-selective visual extinction ratio R$_{v}$  from \citet{2007ApJ...663..320F} [f], \citet{2004ApJ...616..912V} [v], 
\citet{2002BaltA..11....1W} [w], \citet{1988AJ.....96..695C,1989ApJ...345..245C} [c], \citet{2005ApJ...619..357L} [l], and \citet{2005ApJ...623..897L} [l2]
Col.~10 gives the parallax distance $d$ (pc) from Hipparcos (\citealt{1997A&A...323L..49P}).
}
\label{tb:basic-data}
\resizebox{0.7\textheight}{!}{%%
\begin{tabular}{llllllrrrp{3cm}l}\hline\hline
Nr.   &HD      		    & Hip    & Spectral & Right        & Declination      &  \multicolumn{1}{c}{$(B-V)$}  & \multicolumn{1}{c}{$(B-V)_{0}$}  & \multicolumn{1}{c}{E$_{B-V}$}   & $R_V$		     & $d^5$   \\
      &        		    &        & type     & Ascension    &      	          &  \multicolumn{1}{c}{(mag)}    & \multicolumn{1}{c}{(mag)}	  & \multicolumn{1}{c}{(mag)}	    &	   		     & (pc)    \\\hline									     
1     & \object{HD 138503}  & 76161  & B2/B3IV  & 15:33:18.91  & $-$25:01:37.0    &  $-$0.048	   & $-$0.22	  &  0.17  &		      & 	\\
2     & \object{HD 139094}  & 76473  & B8IV	& 15:37:06.89  & $-$26:29:32.6    &	0.082	   & $-$0.10	  &  0.18  &			      & 289.0	\\
3     & \object{HD 139160}  & 76503  & B9IV	& 15:37:28.50  & $-$26:16:47.5    &  $-$0.025	   & $-$0.07	  &  0.05  &			      & 184.2	\\
4     & \object{HD 139486}  & 76633  & B9V	& 15:39:00.06  & $-$19:43:57.2    &	0.033	   & $-$0.07	  &  0.10  &			      & 193.4	\\
5     & \object{HD 139518}  & 76666  & B9.5V	& 15:39:21.37  & $-$23:09:00.8    &	0.021	   & $-$0.04	  &  0.06  &			      & 93.7	\\
6     & \object{HD 140543}  & 77131  & B1Iab/Ib & 15:44:56.66  & $-$21:48:53.9    &  $-$0.013	   & $-$0.19	  &  0.18  &   3.16$^w$ 	      & 735.3	\\
7     & \object{HD 141180}  & 77449  & B9III	& 15:48:42.90  & $-$27:34:56.3    &	0.002	   & $-$0.08	  &  0.08  &			      & 442.5	\\
8     & \object{HD 141444}  & 77569  & B9.5V	& 15:50:11.47  & $-$28:42:15.6    &	0.12	   & $-$0.04	  &  0.16  &			      & 	\\
9     & \object{HD 141774}  & 77677  & B9V	& 15:51:29.84  & $-$20:35:14.5    &	0.063	   & $-$0.07	  &  0.13  &			      & 202.8	\\
10    & \object{HD 142096}  & 77811  & B3V	& 15:53:20.06  & $-$20:10:01.3    &  $-$0.032	   & $-$0.20	  &  0.17  &			      & 109.3	\\
11    & \object{HD 142114}  & 77840  & B2.5Vn	& 15:53:36.72  & $-$25:19:37.7    &  $-$0.105	   & $-$0.22	  &  0.11  &			      & 133.0	\\
12    & \object{HD 142165}  & 77858  & B5V	& 15:53:53.91  & $-$24:31:59.4    &  $-$0.023	   & $-$0.16	  &  0.14  &			      & 127.1	\\
13    & \object{HD 142184}  & 77859  & B2V	& 15:53:55.86  & $-$23:58:41.1    &  $-$0.052	   & $-$0.24	  &  0.19  &			      & 120.5	\\
14    & \object{HD 142250}  & 77900  & B7V	& 15:54:30.11  & $-$27:20:19.1    &  $-$0.070	   & $-$0.13	  &  0.06  &			      & 162.9	\\
15    & \object{HD 142301}  & 77909  & B8III/IV & 15:54:39.53  & $-$25:14:37.5    &  $-$0.074	   & $-$0.10	  &  0.03  &   3.39$^w$ 	      & 139.7	\\
16    & \object{HD 142378}  & 77939  & B2/B3V	& 15:55:00.36  & $-$19:22:58.5    &  $-$0.030	   & $-$0.22	  &  0.19  &			      & 191.2	\\
17    & \object{HD 142669}  & 78104  & B2IV-V	& 15:56:53.08  & $-$29:12:50.7    &  $-$0.192	   & $-$0.24	  &  0.05  &			      & 125.5	\\
18    & \object{HD 142883}  & 78168  & B3V	& 15:57:40.46  & $-$20:58:59.1    &	0.002	   & $-$0.20	  &  0.20  &			      & 139.7	\\
19    & \object{HD 142884}  & 78183  & B8/B9III & 15:57:48.80  & $-$23:31:38.3    &  $-$0.018	   & $-$0.09	  &  0.07  &			      & 122.2	\\
20    & \object{HD 142983}  & 78207  & B8Ia/Iab & 15:58:11.37  & $-$14:16:45.7    &  $-$0.089	   & $-$0.02	  &  $-$0.07&		      & 157.2	\\
21    & \object{HD 142990}  & 78246  & B5V	& 15:58:34.87  & $-$24:49:53.4    &  $-$0.100	   & $-$0.16	  &  0.06  &   3.22$^w$ 	      & 149.7	\\
22    & \object{HD 143018}  & 78265  & B1V	& 15:58:51.11  & $-$26:06:50.8    &  $-$0.154	   & $-$0.26	  &  0.11  &			      & 140.8	\\
23    & \object{HD 143275}  & 78401  & B0.3IV	& 16:00:20.01  & $-$22:37:18.2    &  $-$0.083	   & $-$0.28	  &  0.20  &   3.60$^c$ 3.09$^w$  {\bf 2.80$\pm$0.12}$^l$     & 123.2	\\
24    & \object{HD 143567}  & 78530  & B9V	& 16:01:55.46  & $-$21:58:49.4    &	0.078	   & $-$0.07	  &  0.15  &			      & 131.9	\\
25    & \object{HD 143600}  & 78549  & B9.5V	& 16:02:13.56  & $-$22:41:15.2    &	0.085	   & $-$0.04	  &  0.12  &			      & 121.1	\\
26    & \object{HD 143956}  & 78702  & B9V	& 16:04:00.24  & $-$19:46:02.9    &	0.140	   & $-$0.07	  &  0.21  &			      & 241.5	\\
27    & \object{HD 144175}  & 78809  & B9V	& 16:05:19.15  & $-$23:40:08.8    &	0.053	   & $-$0.07	  &  0.12  &			      & 131.2	\\
28    & \object{HD 144217}  & 78820  & B0.5V	& 16:05:26.23  & $-$19:48:19.6    &  $-$0.070	   & $-$0.26	  &  0.19  &   4.00$^c$ 2.72$^w$ {\bf 3.90$\pm$0.27}$^l$      & 162.6	\\
29    & \object{HD 144218}  & 78821  & B2V	& 16:05:26.55  & $-$19:48:06.7    &  $-$0.007	   & $-$0.24	  &  0.23  &			      & 	\\
30    & \object{HD 144334}  & 78877  & B8V	& 16:06:06.38  & $-$23:36:22.7    &  $-$0.077	   & $-$0.11	  &  0.03  &			      & 149.3	\\
31    & \object{HD 144470}  & 78933  & B1V	& 16:06:48.43  & $-$20:40:09.1    &  $-$0.057	   & $-$0.26	  &  0.20  &   3.75$^c$ 3.35$^w$ 3.37$\pm$0.29$^v$ {\bf 3.71$\pm$0.3}$^f$ 3.20$\pm$0.13$^l$ & 129.9   \\
32    & \object{HD 144569}  & 78956  & B9.5V	& 16:07:04.67  & $-$16:56:35.8    &	0.160	   & $-$0.04	  &  0.20  &			      & 170.9	\\
33    & \object{HD 144586}  & 78968  & B9V	& 16:07:14.93  & $-$17:56:09.7    &	0.102	   & $-$0.07	  &  0.17  &			      & 159.5	\\
34    & \object{HD 144661}  & 79031  & B8IV/V	& 16:07:51.89  & $-$24:27:44.5    &  $-$0.052	   & $-$0.10	  &  0.05  &			      & 117.6	\\
35    & \object{HD 144708}  & 79005  & B9V	& 16:07:36.42  & $-$12:44:43.5    &  $-$0.006	   & $-$0.07	  &  0.06  &			      & 130.5	\\
36    & \object{HD 144844}  & 79098  & B9V	& 16:08:43.72  & $-$23:41:07.5    &	0.012	   & $-$0.07	  &  0.08  &   4.04$^w$ 	      & 130.7	\\
37    & \object{HD 144987}  & 79199  & B8V	& 16:09:52.59  & $-$33:32:44.9    &  $-$0.086	   & $-$0.11	  &  0.02  &			      & 132.8	\\
38    & \object{HD 145353}  & 79343  & B9V	& 16:11:33.52  & $-$27:09:03.1    &	0.012	   & $-$0.07	  &  0.17  &			      & 177.0	\\
39    & \object{HD 145482}  & 79404  & B2V	& 16:12:18.20  & $-$27:55:34.9    &  $-$0.154	   & $-$0.24	  &  0.09  &			      & 143.5	\\
40    & \object{HD 145483}  & 79399  & B9Vvar	& 16:12:16.04  & $-$28:25:02.3    &  $-$0.044	   & $-$0.07	  &  0.03  &			      & 91.4	\\
41    & \object{HD 145502}  & 79374  & B2IV	& 16:11:59.73  & $-$19:27:38.6    &	0.009	   & $-$0.24	  &  0.25  &   4.10$^c$ 3.48$^w$ {\bf 3.57$\pm$0.15}$^l$ & 133.9   \\
42    & \object{HD 145554}  & 79410  & B9V	& 16:12:21.83  & $-$19:34:44.6    &	0.125	   & $-$0.07	  &  0.19  &   {\bf 3.65$\pm$0.43}$^f$        & 130.2	\\
43    & \object{HD 145556}  & 79437  & B4II/III & 16:12:43.56  & $-$28:19:18.1    &	0.015	   & $-$0.15	  &  0.16  &			      & 	\\
\hline									     				    
\end{tabular} 
}												     
\end{center}
\end{table*}
}

\onltab{1}{
\begin{table*}[t]
\begin{center}
\caption{continued. }\label{tb:basic-data2}
%\resizebox{\textwidth}{!}{%%
\resizebox{0.7\textheight}{!}{%%
\begin{tabular}{llllllrrrp{3cm}l}\hline\hline
Nr.   & HD	& Hip	 & Spectral & Right	& Declination  &  \multicolumn{1}{c}{$(B-V)$}  & \multicolumn{1}{c}{$(B-V)_{0}$}  & \multicolumn{1}{c}{E$_{B-V}$}   & $R_V$		     & $d^5$   \\
      & 	&	 & type     & Ascension &	       &  \multicolumn{1}{c}{(mag)}    & \multicolumn{1}{c}{(mag)}	  & \multicolumn{1}{c}{(mag)}	    &			     & (pc)    \\ \hline  								     
44    & \object{HD 145631}  & 79439  & B9V	& 16:12:44.10  & $-$19:30:10.4    &	 0.133     &   $-$0.07     &  0.20  &	{\bf 3.82$\pm$0.53}$^{l2}$   & 125.3   \\
45    & \object{HD 145657}  & 79473  & B9.5V	& 16:13:09.51  & $-$29:38:50.1    &	 0.143     &   $-$0.04     &  0.18  &			   & 370.4   \\
46    & \object{HD 145792}  & 79530  & B6IV	& 16:13:45.50  & $-$24:25:19.5    &   $-$0.005     &   $-$0.14     &  0.13  &  {\bf 3.72$\pm$0.75}$^v$     & 138.9	 \\
47    & \object{HD 145964}  & 79599  & B9V	& 16:14:28.88  & $-$21:06:27.5    &	 0.001     &   $-$0.07     &  0.07  &		      & 105.8	 \\
48    & \object{HD 146001}  & 79622  & B8V	& 16:14:53.43  & $-$25:28:37.1    &	 0.022     &   $-$0.11     &  0.13  &  3.14$^w$ 	      & 141.6	 \\   
49    & \object{HD 146029}  & 79621  & B9V	& 16:14:53.33  & $-$22:22:49.1    &	 0.060     &   $-$0.07     &  0.13  &			      & 238.1	\\
50    & \object{HD 146284}  & 79740  & B9III/IV & 16:16:26.69  & $-$24:16:55.0    &	 0.125     &   $-$0.08     &  0.20  &  {\bf 3.10}$^f$	      & 263.9	\\
51    & \object{HD 146285}  & 79739  & B8V	& 16:16:25.17  & $-$24:59:19.5    &	 0.201     &   $-$0.11     &  0.31  &  {\bf 3.83$\pm$0.18}$^f$      & 179.9   \\
52    & \object{HD 146331}  & 79771  & B9V	& 16:16:50.63  & $-$25:51:46.7    &	 0.301     &   $-$0.07     &  0.37  &			      & 145.8	\\
53    & \object{HD 146332}  & 79775  & B3III	& 16:16:52.55  & $-$29:44:37.4    &	 0.132     &   $-$0.20     &  0.33  &			      & 	\\
54    & \object{HD 146416}  & 79785  & B9V	& 16:16:58.77  & $-$21:18:14.9    &	 0.006     &   $-$0.07     &  0.08  &			      & 125.2	\\
55    & \object{HD 146706}  & 79897  & B9V	& 16:18:28.26  & $-$23:16:27.5    &	 0.103     &   $-$0.07     &  0.17  &  3.36$^w$ 	      & 170.6	\\
56    & \object{HD 147009}  & 80019  & RN	& 16:20:03.98  & $-$20:02:41.5    &	 0.256     &   $-$0.01     &  0.27  &			      & 160.8	\\
57    & \object{HD 147010}  & 80024  & B9II/III & 16:20:05.49  & $-$20:03:23.0    &	 0.126     &   $-$0.08     &  0.21  &  {\bf 2.81$\pm$0.45}$^{l2}$   & 143.3   \\
58    & \object{HD 147103}  & 80063  & B9/A0V	& 16:20:30.26  & $-$20:07:03.9    &	 0.338     &   $-$0.04     &  0.38  &			      & 	\\
59    & \object{HD 147165}  & 80112  & B1III	& 16:21:11.32  & $-$25:35:34.1    &	 0.097     &   $-$0.26     &  0.36  &  3.80$^c$ 3.22$^w$ 3.86$\pm$0.52$^v$ {\bf 3.60$\pm$0.14}$^f$ 4.22$\pm$0.20$^l$	      & 225.2	\\
60    & \object{HD 147196}  & 80126  & B5/8V	& 16:21:19.19  & $-$23:42:28.7    &	 0.127     &   $-$0.14     &  0.27    &  4.03$^w$ {\bf 3.10}$^f$      & 151.3	\\
61    & \object{HD 147648}  & 80338  & B8II	& 16:24:02.89  & $-$25:24:54.0    &	 0.620     &   $-$0.10     &  0.72  &  3.49$^w$ 	      & 153.8	\\
62    & \object{HD 147683}  & 80405  & B4V	& 16:24:43.72  & $-$34:53:37.5    &	 0.119     &   $-$0.18     &  0.30  &			      & 591.7	\\
63    & \object{HD 147701}  & 80371  & B5III	& 16:24:21.32  & $-$25:01:31.4    &	 0.500     &   $-$0.16     &  0.66  &  4.05$^c$ 3.60$^w$ 3.86$\pm$0.24$^v$ {\bf 4.03$\pm$0.09}$^f$  & 149.7   \\
64    & \object{HD 147888}  & 80461  & B3/B4V	& 16:25:24.28  & $-$23:27:36.8    &	 0.253     &   $-$0.19     &  0.44  &  4.03$^c$ 3.89$\pm$0.20$^v$ {\bf 4.08$\pm$0.13}$^f$		      & 136.4	\\
65    & \object{HD 147889}  & 80462  & B2III/IV & 16:25:24.32  & $-$24:27:56.6    &	 0.750     &   $-$0.24     &  0.99  &  4.20$^c$ 3.68$^w$ 3.95$\pm$0.13$^v$ {\bf 4.19$\pm$0.073}$^f$ & 135.9   \\
66    & \object{HD 147932}  & 80474  & B5V	& 16:25:35.08  & $-$23:24:18.8    &	 0.254     &   $-$0.16     &  0.41  &  4.58 {\bf 4.58}$^c$	      & 128.9	\\
67    & \object{HD 147933}  & 80473  & BIV	& 16:25:35.12  & $-$23:26:49.8    &	 0.170     &   $-$0.2	   &  0.4   &			      & 	\\     
68    & \object{HD 147955}  & 80493  & B9.5V	& 16:25:50.66  & $-$26:34:06.3    &	 0.206     &   $-$0.07     &  0.28  &			      & 154.8	\\
69    & \object{HD 148041}  & 80552  & B9IV	& 16:26:46.59  & $-$33:04:23.1    &	 0.316     &   $-$0.07     &  0.39  &			      & 442.5	\\
70    & \object{HD 148184}  & 80569  & B2Vne	& 16:27:01.43  & $-$18:27:22.5    &	 0.203     &   $-$0.24     &  0.44  &  {\bf 4.15}$^c$	      & 149.9	\\
71    & \object{HD 148499}  & 80778  & B9III	& 16:29:36.73  & $-$27:34:13.7    &	 0.414     &   $-$0.08     &  0.49  &			      & 346.0	\\
72    & \object{HD 148579}  & 80804  & B9V	& 16:29:59.20  & $-$25:08:52.0    &	 0.245     &   $-$0.07     &  0.31  &  {\bf 4.01$\pm$0.19}$^f$      & 160.3   \\
73    & \object{HD 148594}  & 80819  & B8V	& 16:30:15.68  & $-$27:54:58.5    &	 0.074     &   $-$0.11     &  0.18  &  {\bf 3.10}$^f$	      & 166.9	\\
74    & \object{HD 148605}  & 80815  & B3V	& 16:30:12.48  & $-$25:06:54.8    &   $-$0.132     &   $-$0.20     &  0.07  &  3.18$\pm$0.52$^w$      & 120.5	\\
75    & \object{HD 148860}  & 80940  & B9.5III  & 16:31:38.71  & $-$17:42:48.3    &	 0.132     &   $-$0.05     &  0.18  &			      & 207.0	\\
%%    & \object{HD 149363}  & 81153  & B0.5III  & 16:34:28.28  & $-$06:08:10.3    &	-0.02	   &		   &  0.26    &  2.90$\pm$0.16$^w$	&	\\
76    & \object{HD 149367}  & 81211  & B8/9IV/V & 16:35:15.21  & $-$26:28:42.1    &	 0.105     &   $-$0.07     &  0.18  &  3.49$\pm$0.10$^w$      & 	\\
77    & \object{HD 149387}  & 81236  & B7II/III & 16:35:29.22  & $-$30:16:03.5    &	 0.141     &   $-$0.12     &  0.26  &			      & 	\\
78    & \object{HD 149438}  & 81266  & B0V	& 16:35:52.95  & $-$28:12:57.7    &   $-$0.198     &   $-$0.30     &  0.10  &			      & 131.8	\\
79    & \object{HD 149757}  & 81377  & O9.5Vn	& 16:37:09.54  & $-$10:34:01.5    &	 0.020     &   $-$0.30     &  0.32  &  3.09$^c$ 2.55$\pm$0.24$^v$ {\bf 3.08$\pm$0.19}$^f$   & 140.4   \\
80    & \object{HD 149883}  & 81487  & B9V	& 16:38:36.29  & $-$26:59:21.6    &	 0.128     &   $-$0.07     &  0.20  &			      & 237.5	\\
81    & \object{HD 149914}  & 81474  & B9.5IV	& 16:38:28.65  & $-$18:13:13.7    &	 0.231     &   $-$0.04     &  0.27  &			      & 165.0	\\
82    & \object{HD 150514}  & 81785  & B8III	& 16:42:17.41  & $-$20:11:13.3    &	 0.103     &   $-$0.10     &  0.20  &			      & 	\\
83    & \object{HD 150814}  & 81941  & B9.5V	& 16:44:17.66  & $-$22:31:24.9    &	 0.129     &   $-$0.04     &  0.17  &			      & 183.2	\\
84    & \object{HD 151012}  & 82069  & B9.5V	& 16:45:48.46  & $-$26:38:57.9    &	 0.044     &   $-$0.04     &  0.08  &			      & 110.1	\\
85    & \object{HD 151346}  & 82217  & B8II	& 16:47:46.53  & $-$23:58:27.5    &	 0.386     &   $-$0.10     &  0.49  &  3.70$^w$ {\bf 3.81$\pm$0.10}$^f$       & 253.8	\\
86    & \object{HD 151496}  & 82271  & B9V	& 16:48:36.25  & $-$21:56:28.1    &	 0.191     &   $-$0.07     &  0.26  &			      & 298.5	\\
%%    & \object{HD 151831}  & 82451  & B8/B9Ib  & 16:50:59.07  & $-$26:46:51.5    &   $-$0.043     &   $-$0.02     & -0.02  &			      & 139.7	\\
87    & \object{HD 152516}  & 82734  & B2III	& 16:54:38.40  & $-$21:52:46.6    &	 0.061     &   $-$0.24     &  0.30  &			      & 	\\
88    & \object{HD 152655}  & 82822  & B9III	& 16:55:32.54  & $-$21:34:10.2    &	 0.100     &   $-$0.08     &  0.18  &			      & 425.5	\\
89    & \object{HD 152657}  & 82839  & B8II	& 16:55:44.38  & $-$25:32:02.1    &	 0.004     &   $-$0.10     &  0.10  &			      & 361.0	\\	     
%%    & \object{153919}     & 83499  & O6...	& 17:03:56.77  & $-$37:50:38.9    &	  0.27   &	  ?	   &  0.59  &  3.59,3.87$\pm$0.22$^v$,3.58$^w$        & 	\\     
\hline									     			      					 
\end{tabular}	
}															    
\end{center}
\end{table*}
}

\begin{figure}[t!]
\centering
\includegraphics[angle=-90,width=.9\columnwidth]{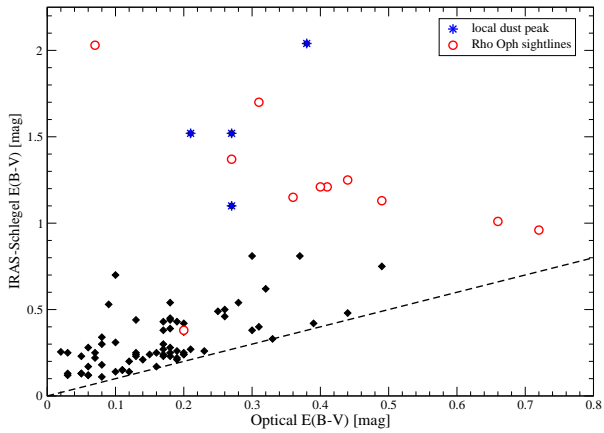}
\caption{Comparison of the amount of dust, as indicated by \Ebv, inferred from
optical photometry observations and stellar spectral classification with those inferred from infrared dust 
emission (IRAS 100~$\mu$m map; \citealt{1998ApJ...500..525S}).
$\rho$ Oph sightlines are indicated separately as well as 4 sightlines connected to an infrared 
dust emission structure north of $\rho$ Oph. 
The dashed line indicates the 1-to-1 relation between the two measurements.
} 
\label{fig:iras_ebv}
\end{figure}

\section{Line-of-sight reddening and dust}\label{sec:reddening}

For each target we derive the reddening \Ebv\ from the $B$ and $V$ photometry (taken from Tycho-2 and
converted to Johnson system) after assigning the  {\it intrinsic} colour $(B-V)_0$ (from
\citealt{1970A&A.....4..234F}) according to the spectral type of the target (as provided by the Michigan
Spectral Catalog of HD stars; \citealt{1982MSS...C03....0H,1988MSS...C04....0H}).  Visual inspection of the
spectral range from 4000 to 5000~\AA, used for the classification of OB-type stars
(\citealt{1990PASP..102..379W}), gives results in good agreement with the spectral types listed in the
Michigan Spectral Catalog.  The adopted magnitudes and results for \Ebv\ are listed in
Table~\ref{tb:basic-data}. The total error for \Ebv\ is $\sim$0.03~mag, which is derived from the error of
the Tycho-2 $B$  and $V$ photometry ($\sim$0.02~mag), the assumed uncertainty ($\sim$0.01~mag) in the
transformation  to the Johnson system, the colour range of spectral sub-types, and the uncertainty in the
spectral  classification (both $\sim$0.01~mag for our B stars).

Two structures of interstellar medium are observed towards the Upp Sco\ complex.  A recent study of the
distribution and motions of the interstellar gas in the $\rho$ Oph region provides evidence for  a low
density/extinction ISM component, at a distance of 50--80~pc, located in front of the $\rho$ Oph complex
(\citealt{2008ApJ...679..512S}). This nearest sheet-like structure was also observed at a distance of
$\sim$60~pc towards the Sco-Cen region by \citet{2004MNRAS.347.1065C}. This structure has a very low column
density and an almost negligible effect on the observed reddening.

The second structure is located at a distance of $\sim$110 -- 150~pc, consisting of diffuse extended
portions of the dense $\rho$\,Oph cloud at 122$\pm$8~pc (\citealt{2008ApJ...679..512S}).  This is consistent
with a mean thickness of $\sim$30~pc found by \citet{2008A&A...489..143L}. Combined with the gas densities
measured by \citet{2003ApJ...589..319Z}, which suggest a cloud thicknesses between 1 and 15~pc, this implies
that these clouds are not spread homogeneously throughout the Upp Sco\ region but form a patchy complex of
scattered  and loosely connected clouds.

Note that the column density of this dust sheet (N(H) $\sim$3.2 -- 50 $\times 10^{20}$~cm$^{-2}$;
\citealt{1978ApJ...224..132B}; \citealt{1994ApJ...427..274D}) located at approximately 125~pc is an order of
magnitude higher than  that of the nearer sheet. For additional information on interstellar material
observed towards the $\rho$\,Oph molecular cloud complex see also \citet{1998A&A...336..150M}.

The above is supported by the measured colour excess \Ebv\ as a function of target distance
(Fig.~\ref{fig:ebv_vs_distance}).  Six stars are probably in front of Upp Sco\, while about 10 to 15 of
these are background stars. The strong increase of the reddening around a distance of 140~pc suggests that
most material contributing to the extinction is  associated with the Upp Sco\ complex, with the observed
scatter resulting from variations in the - column or volume -  density within this region.   The reddening
values extracted directly from the dust reddening map (\citealt{1998ApJ...500..525S}) are compared with the 
\Ebv\ values obtained from optical photometry and stellar classification for the individual sightlines
(Fig.~\ref{fig:iras_ebv}). We note, however, that extinction maps based on infrared emission or
optical/near-infrared star counts show systematic offsets with respect to each other and are unreliable at
small scales ($\leq$ 5\arcmin), with typical 1$\sigma$ uncertainties of  1.2~mag in $A_V$ (\emph{e.g.} 
\citealt{2005ApJ...634..442S}). For all sightlines the reddening inferred from far-infrared emission is
higher than that derived from optical photometry and spectroscopy, which suggests that, for most sightlines,
the infrared emission also traces dust that is located behind the observed star. Some caution is required
comparing these two results as some variation would be expected due to calibration/systematic  and
statistical errors on both values. Nevertheless, most sightlines with \Ebv$_{\rm optical} < 0.3$~mag have
\Ebv$_{\rm infrared} < 0.6$~mag, which is fully consistent with a dust sheet associated to the Upp Sco\ OB
association, with both stars and dust inter-dispersed with each other. For sightlines in the direction of
the dense $\rho$ Oph cloud much higher values for \Ebv\ ($>$1~mag) are inferred from the dust map with
respect to the optical  photometric data. Logically, stars visible in this direction are likely situated at
the front side of this dense cloud (the stars at the back will be much fainter / invisible due to higher
extinction).

The contribution from foreground material to the observed total reddening is very small
(\Ebv~$\la$~0.02~mag). Therefore, we conclude that the dust distribution inferred from both the 100~$\mu$m
infrared emission  and the line-of-sight reddening is predominantly due to the Upper Scorpius complex. In
other words, the low-density foreground dust sheet contributes very little to the total observed values for
the infrared emission and the reddening.

\section{Interstellar absorption lines}\label{sec:results} % In this section we present the properties of
\begin{figure}[h!]
\centering
\includegraphics[bb=105 40 570 300,clip,width=10.5cm,height=8cm,angle=-90]{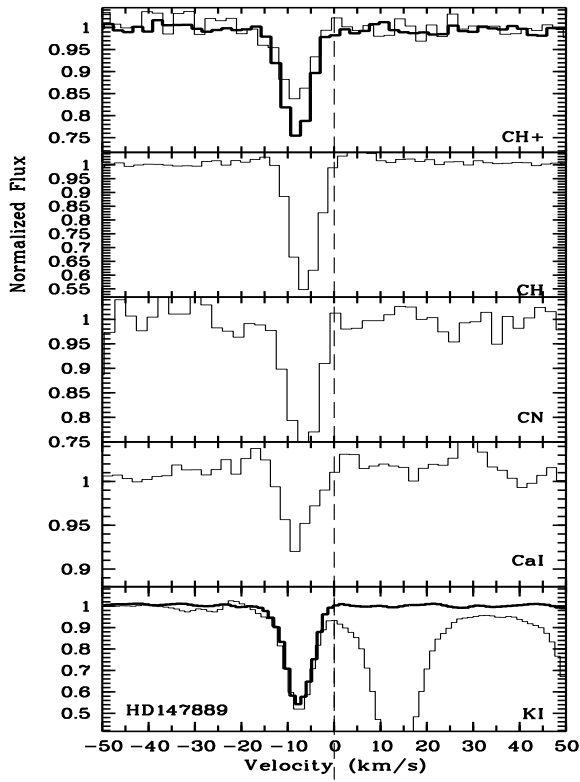}
\caption{%%
 Example of the velocity profiles of interstellar absorption lines. The line-of-sight shown is that towards
 HD\,147889. From top to bottom the CH$^+$ (both at 3957.7 and 4232.5~\AA, latter as thick line), CH, CN
 R(0), \ion{Ca}{i}, and \ion{K}{i} line (both 7665 and 7699, latter as thick solid line). The strong feature
 at $\sim$15~km~s$^{-1}$\ in the bottom panel is a telluric absorption line. Similarly, velocity profiles for all
 sightlines are included in Appendix~\ref{sec:profiles} (Online).
 }
\label{fig:ex_velprof}
\end{figure}

the interstellar absorption lines observed  towards the 89\ Upper Scorpius targets. We determined
equivalent widths for the five strong DIBs at  5780, 5797, 6196, 6379, and 6613~\AA, for the di-atomic lines
of CH, CH$^+$, and CN  as well as for the \ion{K}{i} and \ion{Ca}{i} lines (Table~\ref{tb:DIB-data}). The
\ion{Na}{i}~D doublet is omitted because it is saturated for the majority of sightlines. Line profiles and
central heliocentric velocities for the atomic and di-atomic lines are given in  Fig.~\ref{fig:lineprofiles}
and Table~\ref{tb:velocities} (Online), respectively.  To illustrate, the velocity absorption profiles of
the atomic and molecular absorption lines towards HD\,147889  are shown in Fig.~\ref{fig:ex_velprof}.

\onltab{2}{
\begin{table*}[h!]
\caption{%%
DIB and atomic line equivalent widths, $W$ (m\AA), with 1$\sigma$ errors (upper limits are given as
2$\sigma$), inferred  molecular hydrogen fraction $f_{\mathrm{H}_2}$ (= 2 N(H$_2$)/[N(\ion{H}{i}) + 2
N($H_2$)]; $f_{\mathrm{H}_2}$ from Friedman et al. 2011 are given in brackets for 8 sources present in both
studies), interstellar radiation field $I_{\rm UV}$\ (in units of the Draine field), 5797/5780 ratio, 
$\zeta/\sigma$ classification and spatial subgroup association (low $\rho$ for the low density region and
$\rho$\,Oph for the high density region; see Sect.~\ref{subsec:DIBratios}). The error bars for the DIBs and
atomic lines are calculated as described in Appendix A. Systematic errors on CH, CH$^+$, CN, and \ion{Ca}{i}
equivalent widths are adopted to be 10\%. The error on the CH/CH$^+$ ratio is typically 20\%. Column
densities for CH, CH$^+$, CN and \ion{Ca}{i} can be obtained adopting the relation for the linear part of
the curve-of-growth: $N({\rm cm}^{-2}) = 1.13 10^{17} W(m\AA) / (f \lambda^2)$. Adopted values for $f$ and
$\lambda$ (\AA) are given in Sect.~\ref{sec:molecules}.
}
\label{tb:DIB-data}
\label{tb:ratio}
\label{tb:sigma-zeta}
\centering
\rotatebox{90}{%%
\resizebox{1.3\textwidth}{!}{%%
\begin{tabular}{lrrrrrrrrrrr|rrr|rrll}\hline\hline
HD       & \multicolumn{11}{c|}{Equivalent Width $W$ (m\AA)}							     												 				 & \multicolumn{3}{c|}{Line strength ratios}  			      & $f_{\mathrm{H}_2}$ & $I_{\rm UV}$ & $\zeta/\sigma$ & subgroup  \\ \hline
         & $\lambda$5780      & $\lambda$5797	    & $\lambda$6196 	  & $\lambda$6379	& $\lambda$6613	     &CH(4300)		     &CH+(4232) 	   & CN(3875)	     & Ca{\sc i}(4227)& K{\sc i}(7699) &K{\sc i}(7665)     &$\lambda$5797/$\lambda$5780 & $\lambda$6196/$\lambda$6613& CH/CH$^+$ &     &	    &		 &	  \\
\hline
138503   &  98.0   $\pm$ 4.2  &  24.4	$\pm$ 2.2   &  15.2   $\pm$ 2.4   &  15.9   $\pm$ 2.1	& 35.9   $\pm$ 2.7    & 0.2  		     & 10.6                &-		  & -        & 48.4   $\pm$ 3.8   &  63.6  $\pm$ 3.6 &     0.25  $\pm$ 0.02    &    0.42  $\pm$  0.07  & 0.02          &  0.02         &  (50)        & int	   &	  \\
139094   &  46.0   $\pm$ 3.0  &  19.7	$\pm$ 2.2   &  13.0   $\pm$ 1.9   &   9.5   $\pm$ 1.9   & 17.3   $\pm$ 1.7    & 5.7                  & 4.1                 & 5.0          & -	     & 121.7  $\pm$ 3.9   &  162.1 $\pm$ 3.4 &     0.43  $\pm$ 0.06    &    0.75  $\pm$  0.13  & 1.4           &  0.53	       &  3.6	      & $\zeta$    &	 \\
139160   &            -       &  10.8	$\pm$ 2.9   &	      $<$   6.9   &   9.3   $\pm$ 3.1   & 21.2   $\pm$ 2.9    & 1.5                  & 7.1		   &-		  & -	     & 25.6   $\pm$ 5.8   &  39.5  $\pm$ 5.9 &  	 -	       &	   $<$   0.33  & 0.21	       &  -    	       &  3.5	      & 	   &		   \\
139486   &  38.5   $\pm$ 6.9  & 	$<$   11.8  &	      $<$   11.8  & 	    $<$   19.6	&  	 $<$   15.3   & -		     & -		   &-		  & -	     & 25.9   $\pm$ 9.5   &  18.5  $\pm$ 8.3 &     $<$   0.32	       &	  -	       & -	       &  -	       &  -	      & 	   &		   \\
139518   &         $<$   17.6 & 	$<$   6     &	      $<$   6	  &	    $<$   5.1	&  	 $<$   6      & -                    & -		   &-		  & -	     &        $<$3.3      &	       -     &  	 -	       &	  -	       & -	       &  -	       &  -	      & $\sigma$   &		  \\
140543   & 123.0   $\pm$ 5.1  &  14.7	$\pm$ 4.0   &  23.7   $\pm$ 3.7   &  25.0   $\pm$ 3.4   & 51.6	 $\pm$ 4.4    & -		     & -                   &-		  & -	     & 43.6   $\pm$ 6.8   &  50.0  $\pm$ 8.0 &     0.12  $\pm$ 0.03    &    0.46  $\pm$  0.08  & -	       &  -	       &  -	      & $\sigma$   &	       \\
141180   &  39.3   $\pm$ 4.6  &         $<$   10.3  &  11.7   $\pm$ 2.2   & 	    $<$   12.6	&  9.5   $\pm$ 2.6    & -		     & -		   &-		  & -	     & 6.3    $\pm$ 5.1   &	       -     &     $<$   0.26	       &    1.23  $\pm$  0.40  & -	       &  -	       &  -	      & $\sigma$   &	       \\
141444   &  43.8   $\pm$ 6.3  &  12.7	$\pm$ 3.7   &	9.5   $\pm$ 3.5   & 	    $<$   22.2	& 14.7   $\pm$ 3.3    & -		     & -		   &-		  & -	     & 49.4   $\pm$ 9.0   &  51.8  $\pm$ 4.8 &     0.29  $\pm$ 0.09    &    0.64  $\pm$  0.28  & -	       &  -	       &  -	      & $\zeta$    &	       \\
141774   &  82.7   $\pm$ 4.1  &  15.6	$\pm$ 1.9   &  10.2   $\pm$ 1.6   &   9.7   $\pm$ 1.8	& 28.8   $\pm$ 2.3    & 1.2                  & 2.5 		   &-		  & -	     & 24.4   $\pm$ 3.1   &  22.7  $\pm$ 3.8 &     0.19  $\pm$ 0.02    &    0.35  $\pm$  0.06  & 0.48          &  0.11         &  10.3        & $\sigma$   &	       \\
142096   &  89.5   $\pm$ 3.1  &  13.7	$\pm$ 1.6   &	5.6   $\pm$ 1.2   & 	    $<$   8.4	& 22.9   $\pm$ 1.7    & -                    & 0.45		   &-		  & 5.8      & 23.0   $\pm$ 3.1   &  19.0  $\pm$ 2.9 &     0.15  $\pm$ 0.02    &    0.25  $\pm$  0.05  & -	       &               &  (9.0)       & $\sigma$   & Low $\rho$  \\
142114   &  65.5   $\pm$ 2.8  &  12.5	$\pm$ 2.4   &	8.1   $\pm$ 1.2   & 	    $<$   5.5	& 20.9   $\pm$ 1.4    & 1.3                  & -                   &-		  & -	     & 22.4   $\pm$ 2.6   &  20.1  $\pm$ 4.0 &     0.19  $\pm$ 0.04    &    0.39  $\pm$  0.06  & -             &  0.15	       &  7.5	      & $\sigma$   & Low $\rho$  \\
142165   &  89.5   $\pm$ 3.4  &  11.3	$\pm$ 1.6   &	4.0   $\pm$ 1.2   &	    $<$   9.7 	& 14.3   $\pm$ 1.4    & -		     & -		   &-		  & -	     & 18.8   $\pm$ 4.0   &  15.7  $\pm$ 3.4 &     0.13  $\pm$ 0.02    &    0.28  $\pm$  0.09  & -             &  -	       &  -	      & $\sigma$   & Low $\rho$  \\
142184   &  74.5   $\pm$ 2.7  &  15.5	$\pm$ 1.5   &	7.3   $\pm$ 1.2   &  10.9   $\pm$ 1.8   & 19.9	 $\pm$ 1.8    & 2.8                  & 2.0                 &-		  & -	     & 32.2   $\pm$ 3.2   &  48.2  $\pm$ 4.4 &     0.21  $\pm$ 0.02    &    0.36  $\pm$  0.07  & 1.4           &  0.25	       &  6.4	      & $\sigma$   & Low $\rho$  \\
142250   &  34.5   $\pm$ 4.5  &      	$<$   8.5   & 	      $<$   10.3  &	    $<$   11 	&  	 $<$   13.1   & -                    & -		   &-		  & -	     & 8.7    $\pm$ 4.2   &	       -     &     $<$   0.26	       &	  -	       & -	       &  -	       &  -	      & 	   &	       \\
142301   &  61.0   $\pm$ 3.7  &  14.1	$\pm$ 2.8   &  22.1   $\pm$ 2.4   &  10.1   $\pm$ 2.2   & 10.6	 $\pm$ 2.6    & -		     & -		   &-		  & -	     & 10.5   $\pm$ 2.4   &	       -     &     0.23  $\pm$ 0.05    &    2.09  $\pm$  0.56  & -	       &  -	       &  -	      & $\sigma$   & Low $\rho$  \\
142378   &  87.4   $\pm$ 2.8  &  14.7	$\pm$ 2.2   &  10.7   $\pm$ 1.1   &   6.9   $\pm$ 1.8	& 17.4   $\pm$ 1.6    & 0.9                  & 4.2                 &-		  & -	     & 22.6   $\pm$ 2.1   &  23.0  $\pm$ 3.7 &     0.17  $\pm$ 0.03    &    0.61  $\pm$  0.09  & 0.21          &  0.08         &  20.0        & $\sigma$   & Low $\rho$  \\
142669   &  	   $<$   11.4 &      	$<$   5.5   &	      $<$   2.0	  &   	    $<$   4.2	&  	 $<$   3.1    & -                    & -                   &-		  & -	     & 4.5    $\pm$ 2.3   &	       -     &  	 -	       &	  -	       & -	       &  -	       &  -	      & 	   &	       \\
142883   &  95.1   $\pm$ 2.2  &  11.5	$\pm$ 1.4   &	8.1   $\pm$ 1.1   &  	    $<$   7.8	& 24.4   $\pm$ 1.7    & 4.9                  & 10.9                &-		  & -	     & 65.9   $\pm$ 2.1   &  84.1  $\pm$ 3.8 &     0.12  $\pm$ 0.02    &    0.33  $\pm$  0.05  & 0.45          &  0.32	       &  3.7	      & $\sigma$   & Low $\rho$  \\
142884   & 107.7   $\pm$ 3.1  &  13.9   $\pm$ 1.8   &  11.5   $\pm$ 1.3   &  15.3   $\pm$ 1.2   & 28.6   $\pm$ 2.2    & 1.7                  & 5.8                 &-		  & -	     & 35.3   $\pm$ 2.3   &  44.8  $\pm$ 5.7 &     0.13  $\pm$ 0.02    &    0.40  $\pm$  0.06  & 0.29          &  0.13	       &  2.8	      & $\sigma$   & Low $\rho$  \\
142983   &  81.8   $\pm$ 2.3  &   9.7   $\pm$ 1.5   &   8.4   $\pm$ 1.1   &	   	-       & 27.0   $\pm$ 1.5    & -                    & -		   &-		  & -	     & 7.2    $\pm$ 2.6   &  6.8   $\pm$ 3.3 &     0.12  $\pm$ 0.02    &    0.31  $\pm$  0.04  & -	       &  -	       &  -	      & $\sigma$   &		\\
142990   &  51.5   $\pm$ 2.0  & 	$<$   10.1  &  12.9   $\pm$ 1.6   &         $<$   6.9   & 10.1   $\pm$ 1.5    & 0.9                  & 4.6                 &-		  & -	     & 11.3   $\pm$ 2.0   &  6.8   $\pm$ 2.9 &     $<$   0.20	       &    1.28  $\pm$  0.25  & 0.20          &  0.13	       &  5.1	      & $\sigma$   & Low $\rho$  \\
143018   &  28.8   $\pm$ 3.0  & 	$<$   8.2   &         $<$   7.8   & 	    $<$   11.5  & 10.5   $\pm$ 2.0    & -                    & -                   &-		  & -	     & 4.7    $\pm$ 2.9   &  2.6   $\pm$ 2.4 &     $<$   0.28	       &    0.74  $\pm$  0.21  & -	       &  [0.08]       &  -	      & 	   &		\\
143275   &  68.2   $\pm$ 3.1  &  20.9   $\pm$ 2.2   &   7.4   $\pm$ 1.5   &  12.3   $\pm$ 1.5   & 20.3   $\pm$ 1.2    & 1.3                  & 2.1                 &-		  & -	     & 30.9   $\pm$ 2.2   &  41.3  $\pm$ 1.8 &     0.31  $\pm$ 0.04    &    0.36  $\pm$  0.08  & 0.62          &  0.14 [0.11]  &  11.4        & $\zeta$    & Low $\rho$  \\
143567   & 162.5   $\pm$ 3.6  &  19.8   $\pm$ 2.5   &  12.1   $\pm$ 2.3   &  13.2   $\pm$ 2.0   & 40.1   $\pm$ 2.7    & -                    & -                   &-		  & -	     & 21.0   $\pm$ 3.1   &  19.0  $\pm$ 3.6 &     0.12  $\pm$ 0.02    &    0.30  $\pm$  0.06  & -             &  -	       &  -	      & $\sigma$   & Low $\rho$  \\
143600   &  81.3   $\pm$ 5.3  &  17.7   $\pm$ 2.4   &   8.6   $\pm$ 1.6   &  13.8   $\pm$ 2.5   & 19.2   $\pm$ 3.0    & -                    & -                   &-		  & -	     & 30.0   $\pm$ 4.0   &  46.2  $\pm$ 2.8 &     0.22  $\pm$ 0.03    &    0.45  $\pm$  0.11  & -	       &  -	       &  -	      & $\sigma$   & Low $\rho$  \\
143956   & 178.0   $\pm$ 4.5  &  19.4   $\pm$ 2.8   &  17.2   $\pm$ 2.0   &  17.8   $\pm$ 2.4   & 49.9   $\pm$ 2.2    & -                    & -                   &-		  & -	     & 23.1   $\pm$ 5.8   &  29.5  $\pm$ 5.0 &     0.11  $\pm$ 0.02    &    0.34  $\pm$  0.04  & -             &  -	       &  -	      & $\sigma$   & Low $\rho$  \\
144175   &  35.8   $\pm$ 3.6  &      	$<$   2.7   &  11.9   $\pm$ 3.2   &  11.4   $\pm$ 2.1   &      	 $<$   5.7    & -                    & -		   &-		  & -	     & 5.4    $\pm$ 2.1   &	       -     &     $<$   0.08	       &	  $>$	 2.1   & -	       &  -	       &  -	      & 	   & Low $\rho$  \\
144217   & 152.3   $\pm$ 2.1  &  21.6   $\pm$ 1.7   &  12.7   $\pm$ 0.9   &  15.6   $\pm$ 1.2   & 44.4   $\pm$ 1.4    & 1.4                  & 4.8                 &-		  & 1.2      & 23.1   $\pm$ 2.0   &  32.3  $\pm$ 2.2 &     0.14  $\pm$ 0.01    &    0.29  $\pm$  0.02  & 0.29          &  0.08         &  13.3        & $\sigma$   & Low $\rho$  \\
144218   & 158.8   $\pm$ 2.3  &  17.6   $\pm$ 1.8   &  14.0   $\pm$ 1.0   &  19.2   $\pm$ 1.4   & 42.9   $\pm$ 1.1    & 0.8                  & 4.7                 &-		  & 1.5      & 28.0   $\pm$ 2.0   &  33.8  $\pm$ 1.5 &     0.11  $\pm$ 0.01    &    0.33  $\pm$  0.02  & 0.17	       &  0.04         &  20.0        & $\sigma$   & Low $\rho$  \\
144334   &  52.9   $\pm$ 1.5  &  12.0   $\pm$ 1.5   &  18.5   $\pm$ 1.2   &   8.2   $\pm$ 1.6   & 10.3   $\pm$ 1.7    & -                    & -		   &-		  & -	     & 4.5    $\pm$ 1.7   &	       -     &     0.23  $\pm$ 0.03    &    1.79  $\pm$  0.32  & -	       &  -	       &  -	      & $\sigma$   & Low $\rho$  \\
144470   & 165.8   $\pm$ 2.1  &  28.0   $\pm$ 1.6   &  19.0   $\pm$ 1.0   &  26.2   $\pm$ 1.2   & 58.2   $\pm$ 1.5    & 3.0                  & 5.9                 &-		  & 1.5      & 29.1   $\pm$ 1.4   &  34.8  $\pm$ 2.0 &     0.17  $\pm$ 0.01    &    0.33  $\pm$  0.02  & 0.51          &  0.14 [0.13]  &  7.0	      & $\sigma$   & Low $\rho$  \\
144569   &  86.7   $\pm$ 4.8  &  11.5   $\pm$ 2.7   &  11.3   $\pm$ 2.2   &  18.4   $\pm$ 2.7   & 25.5   $\pm$ 2.9    & -                    & -                   &-		  & -	     & 29.7   $\pm$ 5.2   &  30.0  $\pm$ 4.5 &     0.13  $\pm$ 0.03    &    0.44  $\pm$  0.10  & -	       &  -	       &  -	      & $\sigma$   &	     \\
144586   &  81.7   $\pm$ 3.4  &  15.0   $\pm$ 2.2   &  10.0   $\pm$ 1.8   &  11.1   $\pm$ 2.0   & 30.8   $\pm$ 2.5    & -                    & -                   &-		  & -	     & 27.2   $\pm$ 2.7   &  39.7  $\pm$ 2.0 &     0.18  $\pm$ 0.03    &    0.32  $\pm$  0.06  & -	       &  -	       &  -	      & $\sigma$   &	     \\
144661   &  57.4   $\pm$ 1.6  &   9.0   $\pm$ 1.4   & 	      $<$   5.6   & 	    $<$   5.8   &  6.9   $\pm$ 1.6    & -                    & -		   &-		  & -	     & 4.1    $\pm$ 2.4   &  2.4   $\pm$ 2.2 &     0.16  $\pm$ 0.02    &	  $<$	 0.81  & -	       &  -	       &  -	      & $\sigma$      & Low $\rho$  \\
144708   &  61.0   $\pm$ 2.6  &  11.8   $\pm$ 2.1   &   4.8   $\pm$ 1.2   &   8.5   $\pm$ 1.8   & 17.3   $\pm$ 2.0    & -                    & -                   &-		  & -	     & 13.1   $\pm$ 2.3   &  17.5  $\pm$ 3.1 &     0.19  $\pm$ 0.04    &    0.28  $\pm$  0.08  & -	       &  -	       &  -	      & $\sigma$   &	     \\
144844   &          -         & 	$<$   8.0   &  	      $<$   5.7   & 	    $<$   8.6   & 21.3   $\pm$ 1.3    & -                    & -		   &-		  & -	     & 9.0    $\pm$ 1.8   &  9.5   $\pm$ 3.0 &  	 -	       &	     $<$ 0.27  & -	       &  -	       &  -	      & 	       & Low $\rho$  \\
144987   & 	   $<$   7.0  & 	$<$   7.8   & 	      $<$   4.6   &         $<$   3.4   &  7.6   $\pm$ 1.4    & -             	     & -                   &-		  & -	     & 4.7    $\pm$ 1.5   &	       -     &  	  -	       &    0.61  $\pm$  0.13  & -	       &  -	       &  -	      & 	       &	    \\
145353   &  95.5   $\pm$ 2.9  &  32.7   $\pm$ 1.7   &   9.0   $\pm$ 1.4   &  11.5   $\pm$ 1.7   & 31.8   $\pm$ 1.9    & 4.1                  & 7.1                 &-		  & 1.8      & 48.9   $\pm$ 2.9   &  57.0  $\pm$ 3.0 &     0.34  $\pm$ 0.02    &    0.28  $\pm$  0.05  & 0.58          &  0.28	       &  5.1	      & $\zeta$    &	       \\
145482   & 	   $<$   10.3 &      	$<$   3     &         $<$   1.4   &         $<$   4.4   &  4.6   $\pm$ 1.6    & -                    & -                   &-		  & -	     & 3.0    $\pm$ 1.8   &	       -     &  	  -	       &	  $<$	 0.3   & -	       &  -	       &  -	      & 	&	     \\
145483   & 	   $<$   13.2 & 	$<$   12.2  &         $<$   2.3   & 	    $<$   4.7   &  4.9   $\pm$ 1.4    & -                    & -                   &-		  & -	     & 7.5    $\pm$ 2.1   &	       -     &  	 -	       &	  $<$	 0.47  & -	       &  -	       &  -	      & 	&	     \\
145502   & 166.2   $\pm$ 2.0  &  35.0   $\pm$ 1.5   &  15.8   $\pm$ 1.1   &  33.6   $\pm$ 1.2   & 57.0   $\pm$ 1.7    & 2.0                  & 4.1                 &-		  & 3.8      & 38.9   $\pm$ 1.1   &  55.5  $\pm$ 2.0 &   0.21  $\pm$ 0.01      &    0.28  $\pm$  0.02  & 0.49          &  0.10 [0.09]  &  8.2	      & $\sigma$   & Low $\rho$  \\
145554   & 143.0   $\pm$ 3.8  &  25.2   $\pm$ 2.0   &  16.0   $\pm$ 2.0   &  24.5   $\pm$ 1.8   & 48.3   $\pm$ 2.0    & 1.1                  & 2.7                 &-		  & 1.7      & 39.2   $\pm$ 3.5   &  47.9  $\pm$ 4.1 &   0.18  $\pm$ 0.01      &    0.33  $\pm$  0.04  & 0.41          &  0.07         &  17.4        & $\sigma$   & Low $\rho$  \\
145556   & 108.5   $\pm$ 5.5  &  24.0   $\pm$ 3.1   &  14.9   $\pm$ 2.3   &  11.3   $\pm$ 2.1   & 36.9   $\pm$ 3.6    & 1.8                  & 2.6                 &-		  & -	     & 27.9   $\pm$ 3.8   &  17.6  $\pm$ 4.0 &   0.22  $\pm$ 0.03      &    0.40  $\pm$  0.07  & 0.69          &  0.13	       &  6.5	      & $\sigma$   &		     \\%
\hline
\end{tabular}
}}
\end{table*}
}

\onltab{2}{
\begin{table*}[h!]
\begin{center}
\caption{continued.}
\label{tb:DIB-data2}
\label{tb:ratio2}
\label{tb:sigma-zeta2}
\rotatebox{90}{
\resizebox{1.3\textwidth}{!}{%%
\begin{tabular}{lrrrrrrrrrrr|rrr|rrll}\hline\hline
HD       & \multicolumn{11}{c|}{Equivalent Width $W$ (m\AA)}							     												 				 & \multicolumn{3}{c|}{Line strength ratios}  			      & $f_{\mathrm{H}_2}$ & $I_{\rm UV}$ & $\zeta/\sigma$ & subgroup  \\ \hline
         & $\lambda$5780      & $\lambda$5797	    & $\lambda$6196 	  & $\lambda$6379	& $\lambda$6613	     &CH(4300)		     &CH+(4232) 	   & CN(3875)	     & Ca{\sc i}(4227)& K{\sc i}(7699) &K{\sc i}(7665)     &$\lambda$5797/$\lambda$5780 & $\lambda$6196/$\lambda$6613& CH/CH$^+$ &     &	    &		 &	  \\
\hline
145631   & 129.9   $\pm$ 3.7  &  27.1   $\pm$ 2.1   &  14.9   $\pm$ 1.7   &  19.0   $\pm$ 2.0   & 36.5   $\pm$ 2.3   & 3.8                   & 3.8                 &-		     & 2.9	  & 47.9   $\pm$ 2.3   &  61.0  $\pm$  4.0 &   0.21  $\pm$ 0.02  &    0.41  $\pm$  0.05  & 1.0  	&  0.21        &  5.4	    & $\sigma$   & Low $\rho$  \\
145657   &  97.2   $\pm$ 5.8  &  34.7   $\pm$ 4.3   &  10.6   $\pm$ 2.8   &  29.4   $\pm$ 2.7   & 57.0   $\pm$ 3.3   & 5.4                   & 14.1                &-		     & 5.7	  & 78.0   $\pm$ 4.1   &  93.7  $\pm$  6.9 &   0.36  $\pm$ 0.05  &    0.19  $\pm$  0.05  & 0.38 	&  0.33        &  3.4	    & $\zeta$	 &	      \\
145792   &  66.0   $\pm$ 2.6  &  12.8   $\pm$ 1.9   &   8.9   $\pm$ 1.7   &  14.1   $\pm$ 1.5   & 14.6   $\pm$ 1.4   & 1.5                   & 0.2 		   &-		     & -	  & 14.4   $\pm$ 1.8   &  12.9  $\pm$  2.3 &   0.19  $\pm$ 0.03  &    0.61  $\pm$  0.13  & 7.5  	&  0.17        &  7.7	    & $\sigma$   & Low $\rho$  \\
145964   &  43.1   $\pm$ 2.3  &   8.1   $\pm$ 1.4   &   3.3   $\pm$ 1.1   &   7.6   $\pm$ 1.3   & 16.2   $\pm$ 1.9   & -                     & -                   &-		     & -	  & 4.6    $\pm$ 1.7   &  -		   &   0.19  $\pm$ 0.03  &    0.21  $\pm$  0.07  & -		&  -	       &  -	    & $\sigma$   & Low $\rho$  \\
146001   &  68.8   $\pm$ 5.4  &  10.1   $\pm$ 2.5   &  11.7   $\pm$ 3.4   &         $<$   10.3  & 17.7   $\pm$ 3.3   & -                     & -                   &-		     & -	  & 7.8    $\pm$ 3.5   &  -		   &   0.15  $\pm$ 0.04  &    0.66  $\pm$  0.23  & -		&  -	       &  -	    & $\sigma$   & Low $\rho$  \\
146029   &  96.8   $\pm$ 3.3  &  19.9   $\pm$ 2.4   &   6.5   $\pm$ 1.5   &  	    $<$   11.3  & 22.5   $\pm$ 1.8   & -                     & -                   &-		     & -	  & 13.1   $\pm$ 3.0   &  10.5  $\pm$  3.5 &   0.21  $\pm$ 0.03  &    0.29  $\pm$  0.07  & -		&  -	       &  -	    & $\sigma$   & Low $\rho$  \\
146284   &  99.7   $\pm$ 2.8  &  22.8   $\pm$ 1.7   &   9.6   $\pm$ 1.0   &  	    $<$   7.8   & 28.8   $\pm$ 1.5   & 5.8                   & 3.9                 &-		     & 1.4	  & 56.3   $\pm$ 2.7   &  58.2  $\pm$  2.6 &   0.23  $\pm$ 0.02  &    0.33  $\pm$  0.04  & 1.5  	&  0.34        &  -	    & $\sigma$   & Low $\rho$  \\
146285   & 158.5   $\pm$ 3.9  &  23.1   $\pm$ 2.3   &   9.9   $\pm$ 1.4   &  13.2   $\pm$ 1.6   & 43.1   $\pm$ 2.0   & 5.7                   & 9.1                 &-		     & -	  & 31.9   $\pm$ 2.7   &  34.1  $\pm$  2.9 &   0.15  $\pm$ 0.01  &    0.23  $\pm$  0.04  & 0.63 	&  0.25        &  3.3	    & $\sigma$   & Low $\rho$  \\
146331   & 189.8   $\pm$ 3.4  &  46.1   $\pm$ 2.5   &  17.8   $\pm$ 1.5   &  19.6   $\pm$ 1.3   & 70.2   $\pm$ 3.3   & 6.4                   & 3.0                 &-		     & 3.9	  & 57.0   $\pm$ 2.7   &  87.0  $\pm$  3.0 &   0.24  $\pm$ 0.01  &    0.25  $\pm$  0.02  & 2.1  	&  0.24        &  6.6	    & $\sigma$   &	      \\
146332   & 103.0   $\pm$ 2.9  &  65.3   $\pm$ 2.4   &  15.3   $\pm$ 2.2   &  37.1   $\pm$ 2.0   & 69.2   $\pm$ 2.1   & 15.1                  & 3.8                 & 9.3	     & 3.0	  & 106.5  $\pm$ 4.2   &  107.5 $\pm$  2.4 &   0.63  $\pm$ 0.03  &    0.22  $\pm$  0.03  & 4.0  	&  0.57        &  5.5	    & $\zeta$	 &	      \\
146416   &  63.9   $\pm$ 2.7  &  10.4   $\pm$ 1.2   &   4.9   $\pm$ 1.2   &         $<$   3.4   & 19.7   $\pm$ 1.8   & -                     & -                   &-		     & -	  & 6.5    $\pm$ 2.1   &  3.8	$\pm$  2.5 &   0.16  $\pm$ 0.02  &    0.25  $\pm$  0.06  & -		&	       &  2.5	    & $\sigma$   & Low $\rho$  \\
146706   &  45.2   $\pm$ 4.0  &  27.7   $\pm$ 2.6   &   4.0   $\pm$ 2.3   &         $<$   4.2   & 17.2   $\pm$ 2.7   & 7.2                   & 5.0                 &-		     & 1.4	  & 53.9   $\pm$ 5.2   &  49.1  $\pm$  5.0 &   0.61  $\pm$ 0.08  &    0.23  $\pm$  0.14  & 1.4  	&  0.58        &  2.4	    & $\zeta$	 &	      \\
147009   & 123.7   $\pm$ 4.3  &  31.7   $\pm$ 3.5   &  12.9   $\pm$ 2.5   &  23.5   $\pm$ 2.3   & 51.7   $\pm$ 3.5   & 8.0                   & 18.3                &-		     & -	  & 47.6   $\pm$ 3.3   &  52.4  $\pm$  6.9 &   0.26  $\pm$ 0.03  &    0.25  $\pm$  0.05  & 0.44 	&  0.37        &  4.1	    & int	 &	      \\
147010   &            -       &        	  -         &       	-	  &  12.1   $\pm$ 2.0   &	  -	     & 7.0         	     & 14.1                & 1.9	     & -	  & 50.5   $\pm$ 3.1   &  55.8  $\pm$  5.2 &	      - 	 &	      - 	 & 0.50 	&	       &  3.4	    &		 &	      \\
147103   & 131.0   $\pm$ 3.4  &  71.3   $\pm$ 2.8   &  16.1   $\pm$ 1.5   &  28.8   $\pm$ 1.6   & 56.7   $\pm$ 2.3   & 23.8                  & 7.4                 & 8.6	     & -	  & 75.0   $\pm$ 1.8   &  78.4  $\pm$  3.7 &   0.54  $\pm$ 0.03  &    0.28  $\pm$  0.03  & 3.2  	&  0.63        &  1.6	    & $\zeta$	 &	      \\
147165   & 234.5   $\pm$ 3.9  &  55.2   $\pm$ 2.2   &  18.6   $\pm$ 1.8   &  25.7   $\pm$ 2.1   &	  -	     & 3.8                   & 4.8                 &-		     & -	  & 27.1   $\pm$ 2.9   &  28.1  $\pm$  2.7 &   0.24  $\pm$ 0.01  &	      - 	 & 0.79 	&  0.13 [0.05] &  9.1	    & $\sigma$   & $\rho$ Oph  \\
147196   &  31.0   $\pm$ 3.0  &  31.3   $\pm$ 2.4   &   4.9   $\pm$ 1.7   &  	    $<$   9.9   & 25.6   $\pm$ 2.0   & 15.0                  & 4.9                 & 2.4	     & 2.1	  & 61.6   $\pm$ 3.7   &  58.9  $\pm$  3.6 &   1.01  $\pm$ 0.12  &    0.19  $\pm$  0.07  & 3.1  	&  0.80        &  2.4	    & $\zeta$	 & $\rho$ Oph  \\
147648   & 253.9   $\pm$ 4.8  & 113.8   $\pm$ 3.4   &  33.2   $\pm$ 2.2   &  72.0   $\pm$ 2.2   &142.6   $\pm$ 3.2   & 21.4                  & 7.4                 &-		     & -	  & 89.3   $\pm$ 2.3   &  90.3  $\pm$  6.1 &   0.45  $\pm$ 0.02  &    0.23  $\pm$  0.02  & 2.9  	&  0.45        &  3.2	    & $\zeta$	 & $\rho$ Oph  \\
147683   &  96.7   $\pm$ 3.2  &  33.9   $\pm$ 1.8   &  11.2   $\pm$ 1.7   &  15.0   $\pm$ 1.6   & 45.2   $\pm$ 2.1   & 20.0                  & 18.0                & 17.4	     & 2.4	  & 100.4  $\pm$ 2.2   &  84.3  $\pm$  4.7 &   0.35  $\pm$ 0.02  &    0.25  $\pm$  0.04  & 1.1  	&  0.65        &  1.6	    & $\zeta$	 &	     \\
147701   & 256.4   $\pm$ 4.7  &  95.6   $\pm$ 2.3   &  21.3   $\pm$ 2.0   &  51.3   $\pm$ 1.7   & 65.6   $\pm$ 2.2   & 27.4                  & 17.7                & 19.3	     & 4.1	  & 102.1  $\pm$ 3.1   &  97.1  $\pm$  2.9 &   0.37  $\pm$ 0.01  &    0.33  $\pm$  0.03  & 1.6  	&  0.51        &  2.4	    & $\zeta$	 & $\rho$ Oph  \\
147888   & 196.9   $\pm$ 6.6  &  49.7   $\pm$ 4.4   &  19.2   $\pm$ 4.0   &  32.3   $\pm$ 2.8   & 62.0   $\pm$ 4.3   & 14.8                  & 6.9                 & 3.6	     & 6.1	  & 90.6   $\pm$ 7.0   &  110.1 $\pm$  9.9 &   0.25  $\pm$ 0.02  &    0.31  $\pm$  0.07  & 2.1  	&  0.41 [0.18] &  1.9	    & int	 & $\rho$ Oph  \\
147889   & 343.7   $\pm$ 4.1  & 144.7   $\pm$ 2.9   &  39.9   $\pm$ 2.8   &  83.5   $\pm$ 2.1   &165.4   $\pm$ 2.4   & 45.5                  & 24.7                & 25.6	     & 7.8	  & 83.0   $\pm$ 2.1   &  80.6  $\pm$  3.0 &   0.42  $\pm$ 0.01  &    0.24  $\pm$  0.02  & 1.8  	&  0.56        &  1.5	    & $\zeta$	 & $\rho$ Oph  \\
147932   & 207.3   $\pm$ 4.1  &  60.3   $\pm$ 2.0   &  14.4   $\pm$ 1.5   &  25.8   $\pm$ 1.7   & 57.1   $\pm$ 1.7   & 13.1                  & 4.9                 & 13.0	     & 3.6	  & 87.0   $\pm$ 3.4   &  96.5  $\pm$  2.5 &   0.29  $\pm$ 0.01  &    0.25  $\pm$  0.03  & 2.7  	&  0.37        &  3.1	    & $\zeta$	 & $\rho$ Oph  \\
147933   & 194.5   $\pm$ 2.8  &  58.9   $\pm$ 1.7   &  16.7   $\pm$ 1.3   &  23.5   $\pm$ 1.0   & 62.9   $\pm$ 1.8   & 17.9                  & 12.3                & 5.8	     & 5.7	  & 91.6   $\pm$ 3.2   & 98.3	$\pm$ 4.2  &   0.30  $\pm$ 0.01  &    0.27  $\pm$  0.02  & 1.5  	&  0.46        &  2.4	    & $\zeta$	 & $\rho$ Oph  \\
147955   & 118.2   $\pm$ 3.0  &  62.8   $\pm$ 1.5   &  10.6   $\pm$ 1.3   &  22.6   $\pm$ 1.2   & 35.0   $\pm$ 1.8   & 12.2                  & 27.1                &-		     & -	  & 58.7   $\pm$ 1.7   &  67.8  $\pm$  3.4 &   0.53  $\pm$ 0.02  &    0.30  $\pm$  0.04  & 0.45 	&  0.49        &  4.7	    & $\zeta$	 & $\rho$ Oph  \\
148041   & 148.1   $\pm$ 6.1  &  50.6   $\pm$ 2.8   &  25.9   $\pm$ 3.6   &  48.0   $\pm$ 3.4   & 94.4   $\pm$ 3.1   & 10.5                  & 20.0                &-		     & 5.2	  & 125.9  $\pm$ 6.3   &  118.0 $\pm$  6.4 &   0.34  $\pm$ 0.02  &    0.27  $\pm$  0.04  & 0.53 	&  0.40        &  3.6	    & $\zeta$	 &	      \\
148184   &  81.7   $\pm$ 1.6  &  57.7   $\pm$ 1.2   &  14.2   $\pm$ 0.8   &  19.0   $\pm$ 1.7   & 45.5   $\pm$ 1.3   & 23.6                  & 10.3                & 2.0	     & 1.5	  & 72.9   $\pm$ 1.5   &  85.1  $\pm$  3.2 &   0.71  $\pm$ 0.02  &    0.31  $\pm$  0.02  & 2.3  	&  0.72        &  1.4	    & $\zeta$	 &	      \\
148499   & 105.4   $\pm$ 6.7  &  34.1   $\pm$ 4.0   &  16.5   $\pm$ 4.5   &  24.7   $\pm$ 3.6   & 25.8   $\pm$ 4.9   & 6.4                   & 23.9                &-		     & -	  & 104.0  $\pm$ 6.9   &  110.9 $\pm$  7.7 &   0.32  $\pm$ 0.04  &    0.64  $\pm$  0.21  & 0.27 	&  0.36        &  15.7      & $\zeta$	 &	      \\
148579   & 104.6   $\pm$ 3.2  &  55.0   $\pm$ 1.8   &  13.8   $\pm$ 1.6   &  18.4   $\pm$ 1.6   & 39.9   $\pm$ 1.9   & 11.1                  & 27.2                &-		     & -	  & 54.3   $\pm$ 2.5   &  65.0  $\pm$  2.8 &   0.53  $\pm$ 0.02  &    0.35  $\pm$  0.04  & 0.41 	&  0.49        &  5.8	    & $\zeta$	 & $\rho$ Oph  \\
148594   &  78.8   $\pm$ 3.0  &  30.7   $\pm$ 1.9   &   7.3   $\pm$ 1.1   &   8.3   $\pm$ 1.1   & 21.9   $\pm$ 2.1   & 1.8                   & 7.8                 &-		     & 1.9	  & 39.6   $\pm$ 3.2   &  54.6  $\pm$  3.5 &   0.39  $\pm$ 0.03  &    0.33  $\pm$  0.06  & 0.23 	&  0.17        &  10.2      & $\zeta$	 &	       \\
148605   &  29.0   $\pm$ 1.2  &  12.2   $\pm$ 1.6   &   2.2   $\pm$ 0.7   &         $<$   1     & 8.3    $\pm$ 1.1   & 0.6                   & 1.2                 &-		     & -	  & 6.1    $\pm$ 1.6   &  5.3	$\pm$  1.8 &   0.42  $\pm$ 0.06  &    0.27  $\pm$  0.10  & 0.50 	&  0.15        &  7.4	    & $\zeta$	 & $\rho$ Oph  \\
148860   & 108.9   $\pm$ 3.3  &  28.7   $\pm$ 2.8   &       	-	  &  15.3   $\pm$ 1.9   &	  -	     & 2.0                   & -                   &-		     & 14.0	  & 62.3   $\pm$ 4.0   &  78.4  $\pm$  4.6 &   0.26  $\pm$ 0.03  &	      - 	 & -		&  0.14        &  9.0	    &  int	 &	    \\
%% 149363 & 239.5  $\pm$ 2.0  &  58.9   $\pm$ 1.1   &  24.8   $\pm$ 0.9   &  24.7   $\pm$ 1.0   & 73.6   $\pm$ 1.2   &	         -	     & -		   &-		     & -	  &		       &		   &   0.25  $\pm$ 0.005 &    0.34  $\pm$  0.01  & -		&  -	       &  -	    &  int	 &	    \\
149367   & 108.2   $\pm$ 3.9  &        	   -        &   6.6   $\pm$ 2.2   &   9.7   $\pm$ 2.2   & 33.4   $\pm$ 2.4   & 4.2                   & 5.0                 &-		     & 3.1	  & 62.4   $\pm$ 4.9   &  63.2  $\pm$  9.5 &	      - 	 &    0.20  $\pm$  0.07  & 0.84 	&  0.26        &  4.8	    &		 &	    \\
149387   & 106.1   $\pm$ 6.2  &  32.0   $\pm$ 4.0   &  16.4   $\pm$ 3.1   &  22.0   $\pm$ 3.0   &	  -	     & 11.9                  & 11.2                &-		     & 2.1	  & 90.7   $\pm$ 8.5   &  95.9  $\pm$  8.7 &   0.30  $\pm$ 0.04  &	      - 	 & 1.1  	&  0.50        &  9.8	    &	$\zeta$    &	       \\
149438   &  11.0   $\pm$ 2.0  &  15.5   $\pm$ 2.1   &         $<$   4     &         $<$   5.9   & 	 $<$   5.9   & -                     & -                   &-		     & -	  &	   $ <$  4.2   &  -		   &   1.41  $\pm$ 0.32  &	      - 	 & -		&  -	       &  -	    &	$\zeta$    &	       \\
149757   &  51.0   $\pm$ 1.8  &  41.0   $\pm$ 1.3   &   7.7   $\pm$ 0.8   &  18.9   $\pm$ 1.0   & 45.7   $\pm$ 1.3   & 18.7                  & 23.3                & 6.2	     & -	  & 67.5   $\pm$ 2.1   &  69.8  $\pm$  1.9 &   0.80  $\pm$ 0.04  &    0.17  $\pm$  0.02  & 0.8  	&  0.76        &  1.8	    &	$\zeta$    &	       \\
149883   &  57.5   $\pm$ 3.0  &  49.7   $\pm$ 3.0   &  13.3   $\pm$ 2.3   &  33.6   $\pm$ 2.0   & 33.0   $\pm$ 2.1   & 5.2                   & 10.0                &-		     & 2.4	  & 65.5   $\pm$ 4.8   &  81.5  $\pm$  5.9 &   0.86  $\pm$ 0.07  &    0.40  $\pm$  0.07  & 0.52 	&  0.45        &  5.3	    &	$\zeta$    &	       \\
149914   & 116.0   $\pm$ 2.3  &  53.2   $\pm$ 1.3   &  12.1   $\pm$ 0.9   &  26.9   $\pm$ 1.1   & 57.0   $\pm$ 2.0   & 18.1                  & 6.2                 & 10.2	     & 3.8	  & 109.9  $\pm$ 2.1   &  141.6 $\pm$  2.8 &   0.46  $\pm$ 0.01  &    0.21  $\pm$  0.02  & 2.9  	&  0.59        &  2.5	    &	$\zeta$    &	       \\
150514   & 114.2   $\pm$ 3.6  &  18.1   $\pm$ 2.5   &  13.2   $\pm$ 2.5   &  20.5   $\pm$ 2.3   &	  -	     & -              	     & -		   &-		     & -	  & 45.2   $\pm$ 4.4   &  55.2  $\pm$  4.0 &   0.16  $\pm$ 0.02  &	      - 	 & -		&  -	       &  -	    &	$\sigma$   &	       \\
150814   & 102.6   $\pm$ 4.5  &  20.2   $\pm$ 3.0   &   6.6   $\pm$ 2.0   &  15.1   $\pm$ 2.2   & 27.2   $\pm$ 2.5   & 2.9                   & 4.1                 &-		     & 1.8	  & 40.7   $\pm$ 3.6   &  54.5  $\pm$  6.4 &   0.20  $\pm$ 0.03  &    0.24  $\pm$  0.08  & 0.71 	&  0.20        &  6.7	    &	$\sigma$   &	       \\
151012   &         $<$   2.0  &         $<$   4.7   &         $<$   6.2   &         $<$   4.1   &  	 $<$   11.0  & -                     & -                   &-		     & -	  &	   $<$   5.9   &  -		   &	      - 	 &	      - 	 & -		&  -	       &  -	    &		   &	       \\
151346   & 107.4   $\pm$ 3.3  &  55.8   $\pm$ 3.7   &  12.1   $\pm$ 1.4   &  26.7   $\pm$ 2.0   & 69.8   $\pm$ 2.4   & 26.9                  & 25.6		   & 8.7	     & -	  & 53.6   $\pm$ 2.2   &  58.9  $\pm$  3.5 &   0.52  $\pm$ 0.04  &    0.17  $\pm$  0.02  & 1.1  	&  0.69        &  2.5	    &	$\zeta$    &	       \\
151496   & 148.1   $\pm$ 4.8  &  34.1   $\pm$ 2.7   &  18.6   $\pm$ 2.5   &  32.3   $\pm$ 3.5   &          -         & 3.3                   & 12.0                &-		     & -	  & 57.8   $\pm$ 6.0   &  65.8  $\pm$  7.3 &   0.23  $\pm$ 0.02  &	      - 	 & 0.28 	&  0.17        &  9.9	    &	$\sigma$   &	       \\
%151831  &            -       &        	  -         &       	-	  & 	    $<$   33.9  &	  -	     & -                     & -                   &-		     & -	  & 34.4   $\pm$ 29.2  &  52.7  $\pm$  30.0&	      - 	 &	      - 	 & -		&  -	       &  -	    &		   &	       \\
152516   & 121.7   $\pm$ 4.6  &  26.6   $\pm$ 2.8   &  12.4   $\pm$ 2.1   &  29.2   $\pm$ 2.7   & 41.8   $\pm$ 2.6   & -                     & -                   &-		     & -	  & 40.4   $\pm$ 4.8   &  53.6  $\pm$  7.0 &   0.22  $\pm$ 0.02  &    0.30  $\pm$  0.05  & -		&  -	       &  -	    &	$\sigma$   &	       \\
152655   &  95.2   $\pm$ 4.0  &  34.1   $\pm$ 2.3   &  15.4   $\pm$ 2.6   &  22.6   $\pm$ 2.7   & 44.4   $\pm$ 3.6   & 3.3                   & 14.2                &-		     & 1.7	  & 55.1   $\pm$ 5.5   &  59.4  $\pm$  5.7 &   0.36  $\pm$ 0.03  &    0.35  $\pm$  0.07  & 0.23 	&  0.24        &  7.0	    &	$\zeta$    &	       \\
152657   &  75.2   $\pm$ 3.5  &  29.0   $\pm$ 3.1   &   8.2   $\pm$ 1.4   & 	    $<$   4.9   & 16.6   $\pm$ 2.1   & 2.0                   & -                   &-		     & 2.9	  & 30.5   $\pm$ 3.7   &  -		   &   0.39  $\pm$ 0.04  &    0.49  $\pm$  0.10  & -		&  0.19        &  5.1	    &	$\zeta$    &	       \\
%153919  & 305.1   $\pm$ 1.3  &        	  -  	    &  32.9   $\pm$ 0.9   &  26.9   $\pm$ 2.5   & 107.1  $\pm$ 0.9   & 		    	     &	     		   \\				
\hline
\end{tabular}
}}
\end{center}
\end{table*}
}

\subsection{Diffuse interstellar bands}\label{sec:dibs}

Although more than 300 DIBs are known we focus here on the five strong and narrow bands  at 5780, 5797,
6196, 6379, and 6613~\AA. The strength and width of these features facilitates the measurement of modest
column densities of their carriers in slightly reddened sightlines.  Additionally, the Galactic
relationships between DIB strength and reddening are well  established for these DIBs both in the Galaxy and
beyond. The equivalent width, $W$, is measured via a straight line continuum integration across the
absorption feature (see Appendix.~\ref{sec:appendix-ew}; Online).  For the DIB measurement we do not expect
significant contamination from stellar atmosphere lines (see Appendix.~\ref{sec:stellar}; Online). The
measured equivalent widths, or 2$\sigma$ upper limits, are listed in Table~\ref{tb:DIB-data} (Online)  for
the five DIBs towards the 89\ targets in Upp Sco. This is the first consistently measured data set
containing this many sightlines within one region.

\begin{figure}[t!]
\centering
   \includegraphics[angle=-90,width=\columnwidth,clip]{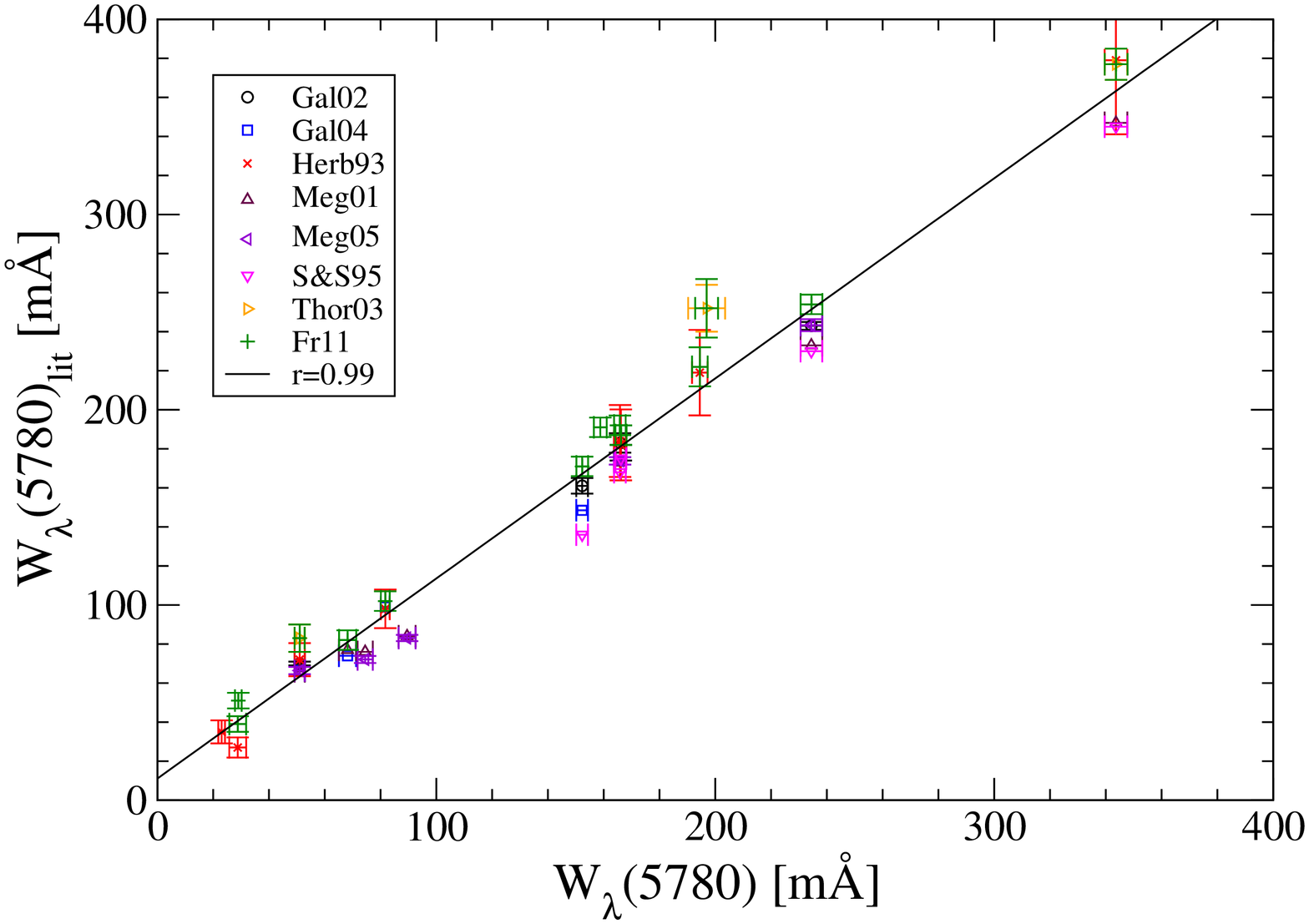}
   \includegraphics[angle=-90,width=\columnwidth,clip]{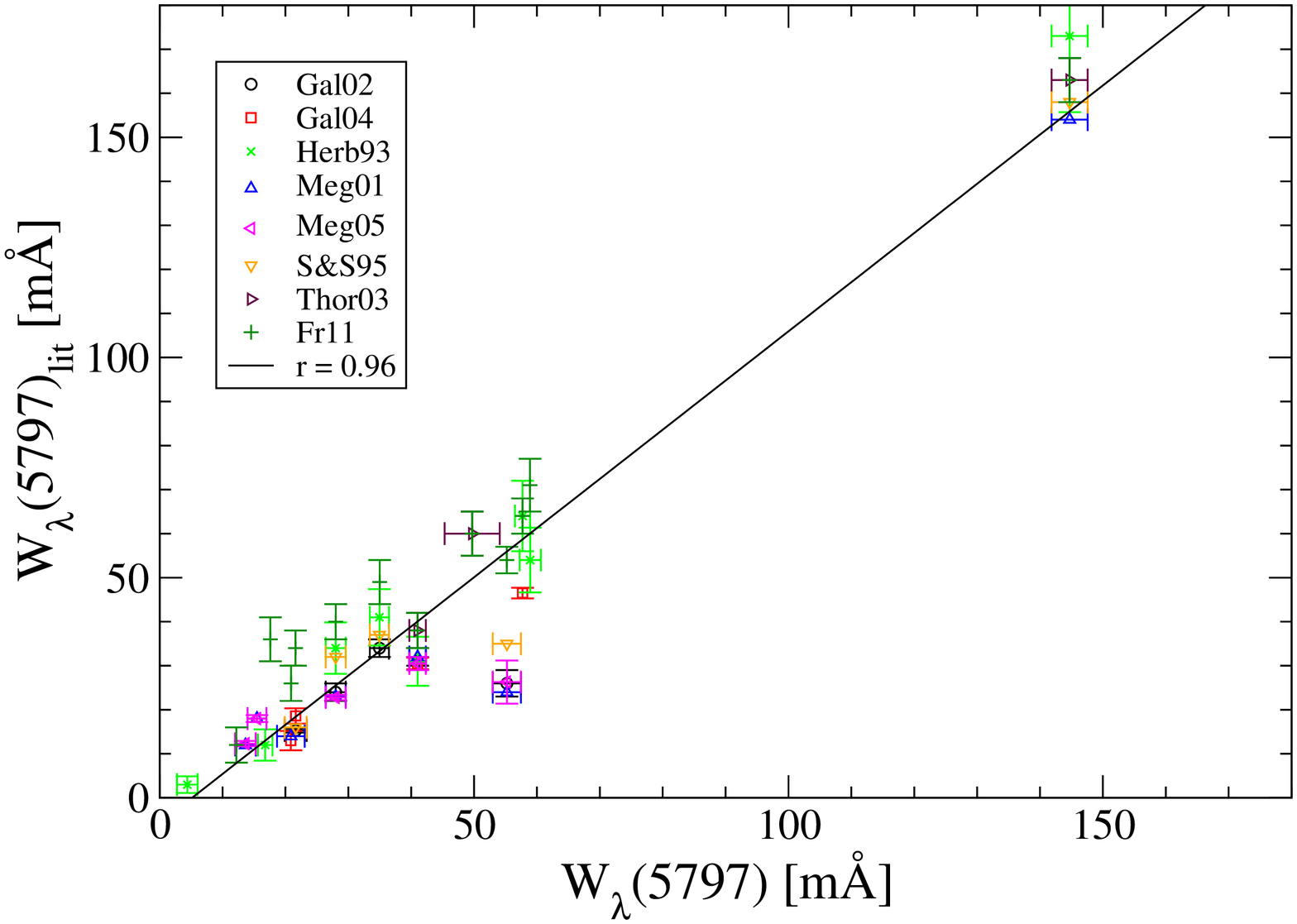}
   \caption{%%						       
   Equivalent widths of the $\lambda\lambda$5780 and 5797 DIBs previously measured for sightlines included in 
   this survey compared to values measured in this work.
   For several sightlines several literature values are available, illustrating the ``intrinsic'' scatter
   in equivalent widths due to measurement methods.
   Literature values are taken from \citet{1993ApJ...407..142H}; \citet{1995ApJ...443..698S}; \citet{2002A&A...384..215G}; 
   \citet{2003ApJ...584..339T}; \citet{2004MNRAS.355..169G}; \citet{2005A&A...429..559S}; \citet{2001MNRAS.326.1095M,2005MNRAS.358..563M};
   and \citet{2011ApJ...727...33F}.
   Linear regressions are shown in each panel. Correlation coefficients $r$ are 0.99 and 0.96 for the 5780 and 5797~\AA\ DIBs, respectively.
   Slopes and intercepts of these regressions are 1.02 and 11~m\AA\ for the 5780~\AA\ DIB and 1.12 and -6~m\AA\ for the 5797~\AA\ DIB, respectively.}
   \label{fig:EWLit_vs_ThisWork}
\end{figure}

\onlfig{6}{
\begin{figure}[t!]
%\centering
   \includegraphics[angle=-90,width=\columnwidth,clip]{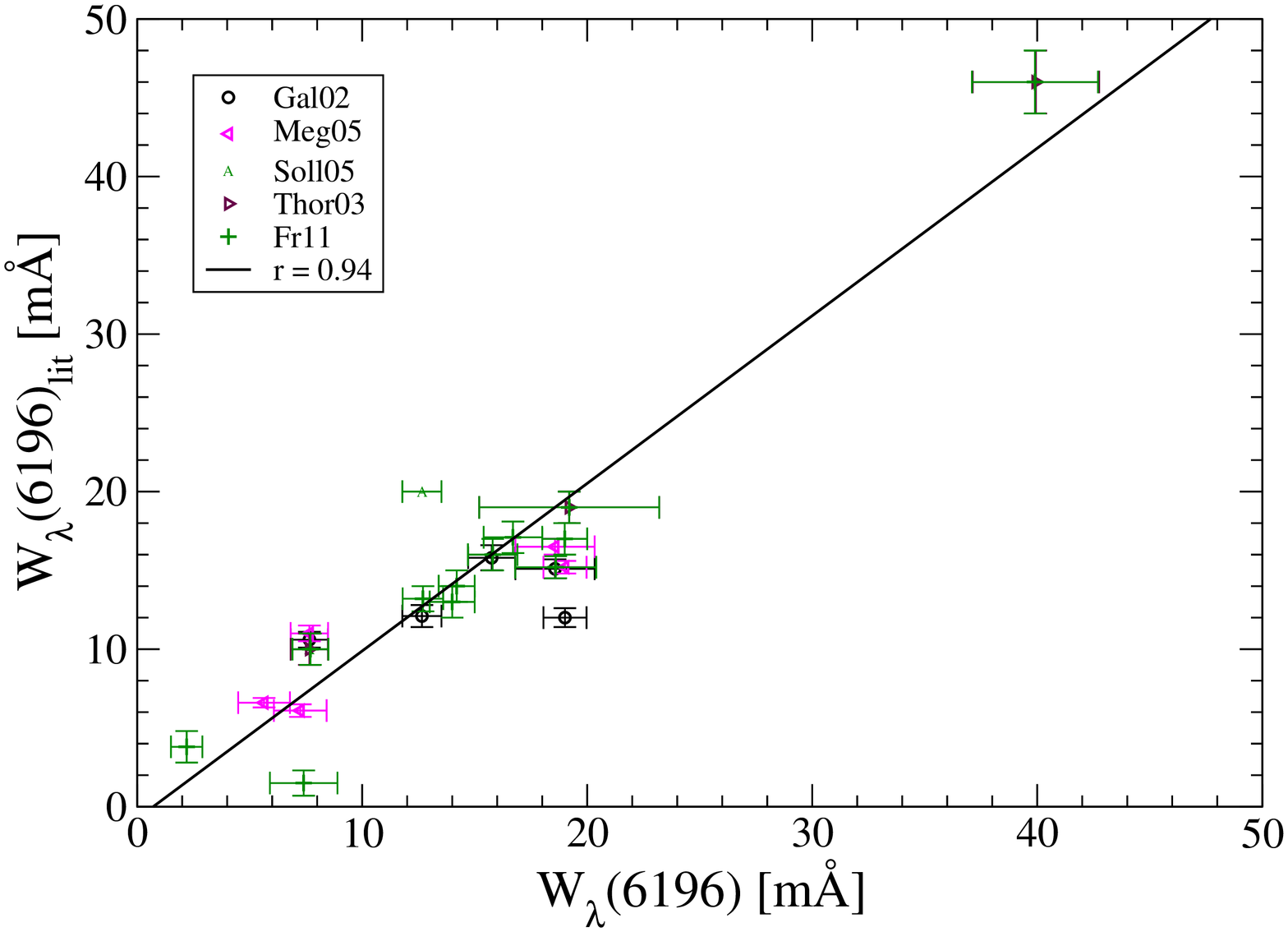}
   \includegraphics[angle=-90,width=0.925\columnwidth]{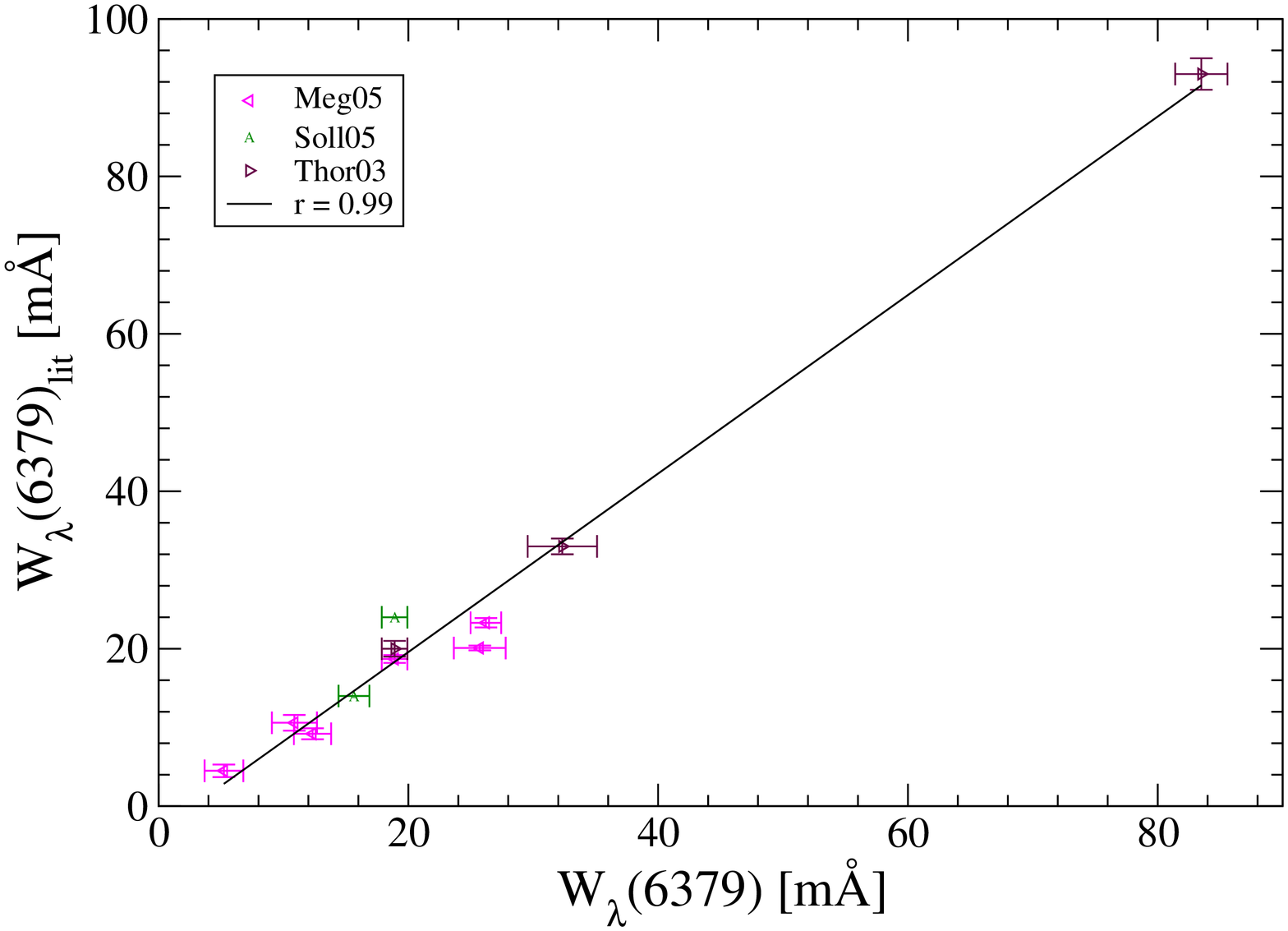} 
   \includegraphics[angle=-90,width=\columnwidth,clip]{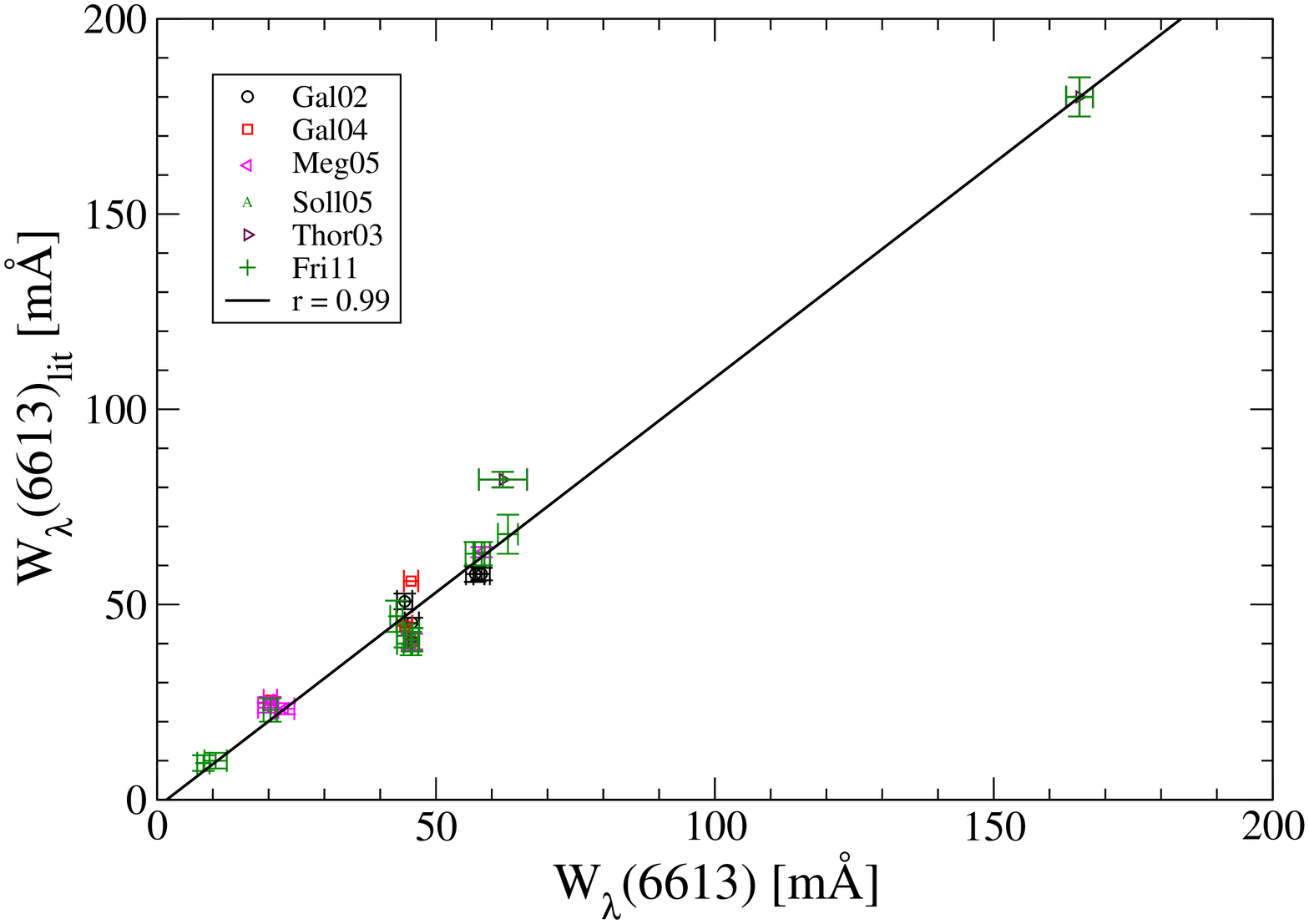}
   \caption{%%						       
   The equivalent widths of the 6196, 6379, and 6613~\AA\ DIBs in lines-of-sight previously measured 
   and included here are compared with the measurements obtained in this work.
   Literature values are taken from \citet{1993ApJ...407..142H}; \citet{1995ApJ...443..698S}; \citet{2002A&A...384..215G}; 
   \citet{2003ApJ...584..339T}; \citet{2004MNRAS.355..169G}; \citet{2005A&A...429..559S}; \citet{2001MNRAS.326.1095M,2005MNRAS.358..563M}.
   Linear regressions with the corresponding correlation coefficients $r$ are shown in each panel.
   Slopes and intercepts of these regressions are 1.06 and -0.8~m\AA, 1.13 and -3.1~m\AA, and 0.995 and -1.9~m\AA, for the 
   6196, 6379, and 6613~\AA\ DIBs, respectively.}
   \label{fig:EWLit_vs_ThisWork2}
\end{figure}
}

To ascertain the accuracy and consistency of our results we compare our measured values with those available
in the literature (\citealt{1993ApJ...407..142H}; \citealt{1995ApJ...443..698S}; 
\citealt{2001MNRAS.326.1095M}; \citealt{2002A&A...384..215G}; \citealt{2003ApJ...584..339T};
\citealt{2004MNRAS.355..169G}; \citealt{2005MNRAS.358..563M}; \citealt{2005A&A...429..559S}; and
\citealt{2011ApJ...727...33F}). The result is shown in Fig.~\ref{fig:EWLit_vs_ThisWork} for the 5780 and
5797 DIBs, and in Fig.~\ref{fig:EWLit_vs_ThisWork2} (Online only) for the 6196, 6379, and 6613~\AA\ DIBs.
The correlation between values from this work and from literature are good (for the 5 DIBs, the correlation
coefficients $r$ range from 0.92 to 0.99).  The linear regressions reveal a small non-zero offset,
indicating that our values are systematically lower by a few percent. We note that for individual cases the
values between different studies vary significantly by as much as 20\%. Small inconsistencies (between all
studies) arise naturally from differences in the data quality (S/N, resolution)  as well as differences in
the adopted methods for equivalent width measurements (adopted stellar continuum,  contamination from nearby
weak features, adopted integration limits, inclusion or removal of underlying broad bands). In conclusion,
the measured equivalent widths are consistent with previous studies but do show a small systematic offset.

\citet{1997A&A...326..822C} found that the $\lambda\lambda$5797, 6379, and 6613 DIBs show a good
correspondence to each other,  with $r \sim 0.8$.            These authors also found that the $\lambda$5780
DIB is moderately correlated with the $\lambda$6613 DIB ($r = 0.65$) and weakly  to the 6379 DIB ($r =
0.47$).  Recently, \citet{2011ApJ...727...33F} found high values for $r$ (ranging from 0.93 to 0.99) for the
Galactic DIB pairs  in Table~\ref{tb:r}. \citet{2010ApJ...708.1628M} reported a nearly perfect correlation
($r=0.99$) between the $\lambda$$\lambda$6196 and 6613 DIBs  toward 114 Galactic diffuse cloud sightlines.
In this work we find $r=0.8$ for $\lambda\lambda$6196-6613 DIB pair,  which is less than for other pairs.
The correlation coefficient between the five DIBs measured in this study are given in Table~\ref{tb:r}.  In
line with previous results, the $\lambda$5797 DIB has a good correlation with both $\lambda$$\lambda$6379
and 6613 DIBs, however, it shows a poor correlation with both $\lambda$$\lambda$5780 and 6196 DIBs. The
$\lambda$$\lambda$6379 and 6613 DIB pair shows the strongest correlation, with $r = 0.92$. In fact, the
$\lambda$6613 DIB correlates well with all four DIBs. The $\lambda$5780 DIB shows a good correlation with
the $\lambda$6613 DIB ($r = 0.85$). Unexpectedly, the other two DIB family members of $\lambda$6613 (i.e.
$\lambda$$\lambda$5797 and 6379 DIBs) have a weaker correlation. Restricting the computation of $r$ to the
13 sightlines present in both \citet{2011ApJ...727...33F} and this work,  increases $r$ for our data (but
lowers $r$ slightly for the Friedman sample). For example, for the $\lambda\lambda$5780-5797 DIB pair
$r=0.75$ (this work)  and $r=0.86$ (Friedman); for the $\lambda\lambda$6196-6613 DIB pair $r=0.97$ (this
work) and $r=0.99$ (Friedman), and for the  $\lambda\lambda$5780-6196 DIB pair both studies give $r=0.93$
(for the complete dataset Friedman report $r=0.97$). The higher Pearson correlation coefficient (independent
of quoted error bars) for the Friedman data suggests that the overall uncertainties on the measurements are
lower than for this work, resulting in an improved correlation. Partly this is due to the fact that our
sample includes a large fraction of sightlines with low values for \Ebv, and thus weak DIBs.  On the other
hand, restricting the comparison to the Upp Sco\ sightlines in common lowers $r$ in both samples  (probably
as there are fewer data points), and also reduces the difference between the two sets.  This could be partly
due to an increased effect of local variations in the DIB spectrum on the correlation coefficient  (such
effects would be averaged out in a larger Galactic survey probing many different regions as opposed to
probing a peculiar region like Upp Sco).

\begin{table}[!ht]
\begin{center}
\caption{Pearson correlation coefficients $r$ between the observed DIBs.}  % add CH and CH+???
\label{tb:r}
\begin{tabular}{lccccc}\hline\hline
DIB	& 5780	& 5797 & 6196 & 6379 & 6613 \\ \hline
5780	& 1     & 0.72 & 0.74 & 0.75 & 0.85 \\
5797	&	&  1   & 0.69 & 0.87 & 0.85 \\
6196	&	&      &   1  & 0.81 & 0.80 \\
6379	&	&      &      &   1  & 0.92 \\
6613	&	&      &      &      &	 1  \\
\hline
\end{tabular} 
\end{center}
\end{table}

\subsection{Molecular lines}\label{sec:molecules}

We have measured equivalent widths and heliocentric radial velocities  for the CH ($\lambda_{\rm rest} =
4300.313$~\AA), CH$^+$ ($\lambda_{\rm rest} = 4232.548$~\AA), and CN R(0) ($\lambda_{\rm rest} =
3874.608$~\AA) lines (Tables~\ref{tb:DIB-data} and~\ref{tb:velocities}, respectively; Online). In a few
(about 5) cases, the CN (3874.608~\AA), CH (3886.410~\AA), and CH$^+$ (3957.70~\AA) lines are tentatively
detected (see \emph{e.g.}  Fig.~\ref{fig:ex_velprof}).  These lines are weak and have large ($>$50\%) uncertainties.
It may be that the strongest CH line is saturated, which can occur for individual components with $W$(CH)
$\geq$~20~m\AA\ \citep{1989ApJ...340..273V}. The CN R(0) transition is also prone to saturation for
(individual) components with $W>6$~m\AA, leading to underestimated column densities, though the corrections
are less than about 20\% up to $W=15$~m\AA\  (\citealt{2008MNRAS.390.1733S}). The CN lines toward
\object{HD 147683}, \object{HD 147701}, \object{HD 147889}, and \object{HD 147932} likely suffer from saturation.  
For the lines-of-sight
including the strongest CN lines in our sample,  \object{HD 147932}, \object{HD 147701}, and \object{HD 147889}, the column
densities would need to be corrected by a factor 1.27, 1.7, and 2.1, respectively (following
\citealt{2008MNRAS.390.1733S}; adopting a value of 1~km~s$^{-1}$\ for the Doppler broadening). Saturation also
occurs for CH$^+$ if $W\geq$20 or $\geq$40~m\AA\ for components with $b=1$ or 2~km~s$^{-1}$, respectively
(\citealt{1994ApJ...424..754A}). Only a few sightlines have measured total $W$ larger than these limits, and
even for these cases the individual (unresolved) velocity  components are not expected to be strongly
saturated as noted above.  For the sightlines towards \object{HD 147683}, \object{HD 147889}, \object{HD 147933}, 
and \object{HD 149757} the
equivalent width ratio between the (tentatively) detected weaker  and stronger lines of both CH and CH$^+$
are close to - within the uncertainties - the expected ratio of $\sim$3.9, and $\sim$1.9, respectively. 
Only for the latter two sightlines are these ratios significantly lower ($\sim$1.5) indicative of some
saturation.

For certain spectral types stellar line contamination can complicate measurements. However, for the majority
of spectra presented here this problem could be well resolved (see also Appendix~\ref{sec:stellar}).
Equivalent widths for CH and CH$^+$ given by \citet{1994ApJ...424..772F}; \citet{2005MNRAS.358..563M},  and
\citet{2008A&A...484..381W} for nine Upp Sco\ sightlines in common with this work are consistent with our
values.  Furthermore, the reported $W$(CN) are consistent with values given by \citet{2008MNRAS.390.1733S}
for six targets in common, although their reported error bars are smaller. The velocity profiles for CH,
CH$^+$, and CN are also shown, for the relevant sightlines, in Figure~\ref{fig:lineprofiles} (Online).

\subsection{Atomic lines}\label{sec:atoms}

Inspection of the \ion{Na}{i} (5889.951 \& 5895.924~\AA; \citealt{2003ApJS..149..205M}) and \ion{K}{i}
(7664.91 \& 7698.974~\AA; \citealt{2003ApJS..149..205M}) doublets shows that most  (75 of $89$)
sightlines are dominated by one strong velocity component. This strong component displays asymmetries and
broadening for a number of sightlines suggesting that in reality multiple unresolved narrow components may
be present (see \emph{e.g.}  Sect.~\ref{sec:velocity} and \citealt{2008ApJ...679..512S}). The obtained spectral
resolution is not sufficient to resolve hyperfine splitting (order of $\sim$1~km~s$^{-1}$). For 10 sightlines (all
with low reddening; \Ebv\ $\lesssim 0.2$) two or three weaker components, clearly separated in velocity
space, could be discerned in the \ion{Na}{i} profiles. For a small number of sightlines we also detect
\ion{Ca}{i} at 4226.73~\AA. Equivalent widths, heliocentric radial velocities, and profiles for \ion{K}{i}
and \ion{Ca}{i} are  included in Tables~\ref{tb:DIB-data} and~\ref{tb:velocities}, and
Figure~\ref{fig:lineprofiles}, respectively (Online). Equivalent widths and profiles are not provided for
the highly saturated \ion{Na}{i} doublet as these preclude any column density measurements. However,
approximate central velocities are included in Table~\ref{tb:velocities} (Online). Similarly, the
\ion{Ca}{ii} line is saturated, but also suffers from stellar contamination and reduced spectral quality in
the blue.

\section{Results \& Discussion}\label{sec:discussion}

In the following we present and discuss the relation between equivalent widths of the observed DIBs,
molecules,  and the line-of-sight reddening.  We investigate whether the correlations we found can be
explained in terms of the {\it skin effect} and explore the spatial variation of DIB strength and strength
ratios. Furthermore, we discuss the velocity structure of the ISM, as well as a model of the dust sheet and
the inferred effective interstellar  radiation field (ISRF).  The ISRF strength, $I_{\rm UV}$, and molecular
hydrogen fraction, $f_{\mathrm{H}_2}$, are both discussed in view of $\sigma$ and $\zeta$-type clouds.

\subsection{DIBs and dust}

\begin{figure}[t!]
\centering
\includegraphics[angle=-90,width=.95\columnwidth,clip]{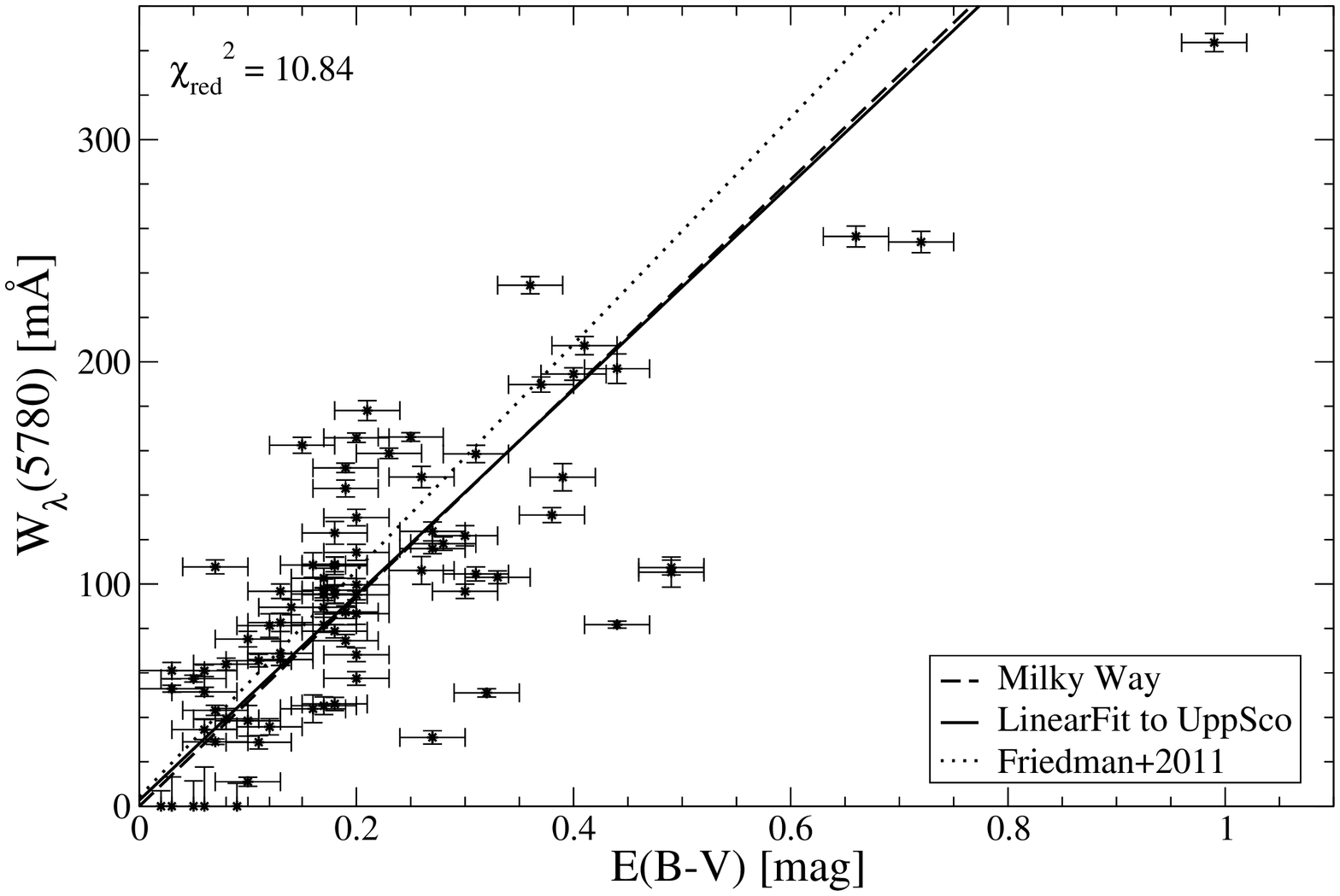}
\includegraphics[angle=-90,width=.95\columnwidth,clip]{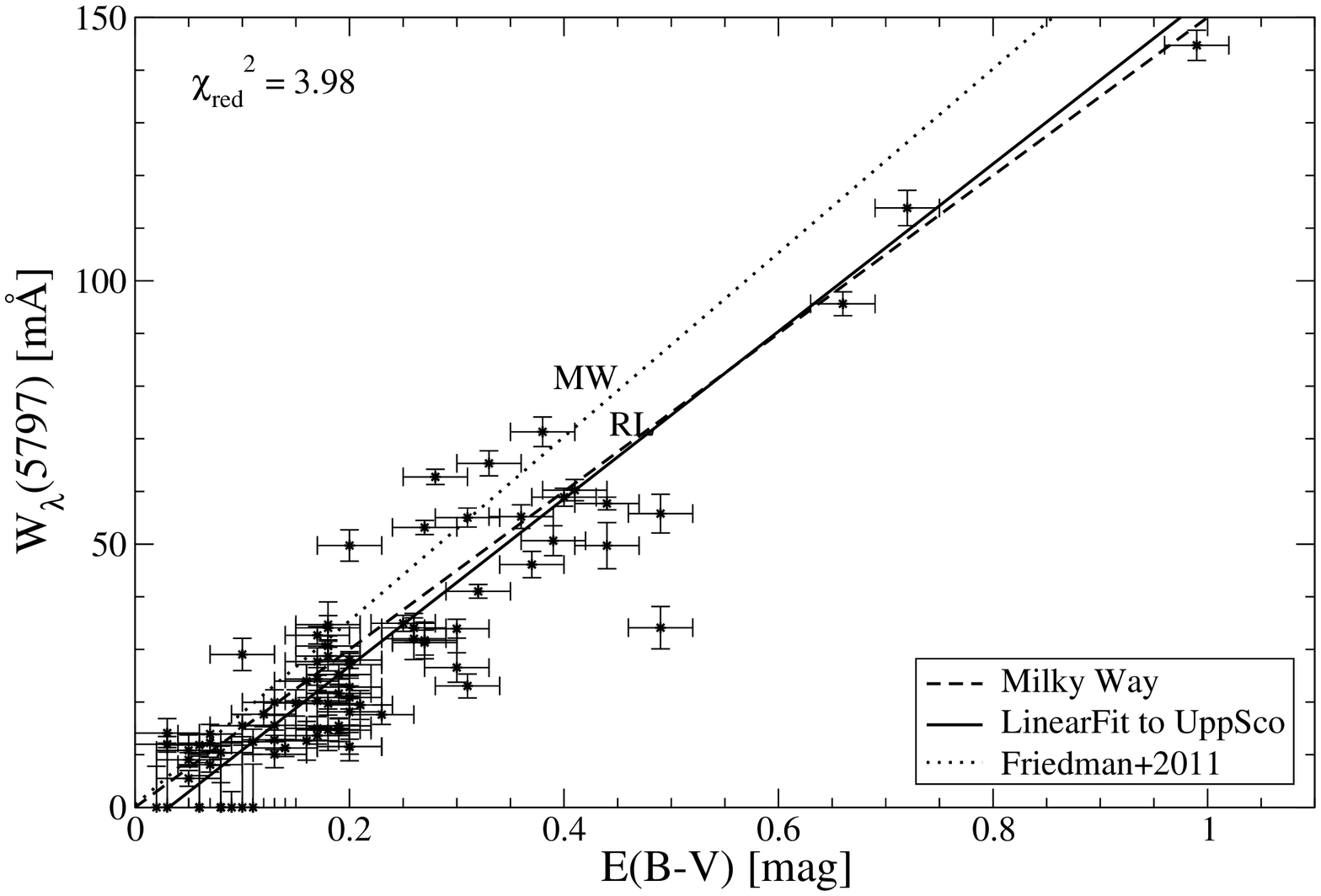}
   \caption{Equivalent width versus \Ebv\ for the 5780 (top) and 5797~\AA\
   (bottom) DIBs. The average Galactic relationships (dashed: \citealt{2005A&A...438..187C},
   and dotted: \citealt{2011ApJ...727...33F}) and 
   the linear least-squares fit for Upp Sco\ (solid; Sect.~\ref{sec:discussion}) are shown. 
   Reduced chi-squared values ($\chi_{\rm red}^{2}$) for the latter are indicated in the respective panels.
   Intercepts and slopes for the linear least-squares fits are given in Table~\ref{tb:linearfit}.
   To avoid biases, the regressions were not forced to go through the origin, and upper 
   limits for the Upp Sco\ data were not taken into account.
   }
   \label{fig:5780-5797-Ebv}
\end{figure}

\begin{figure}[t!]
\centering
   \includegraphics[bb=80 5 570 700, angle=-90,width=.95\columnwidth,clip]{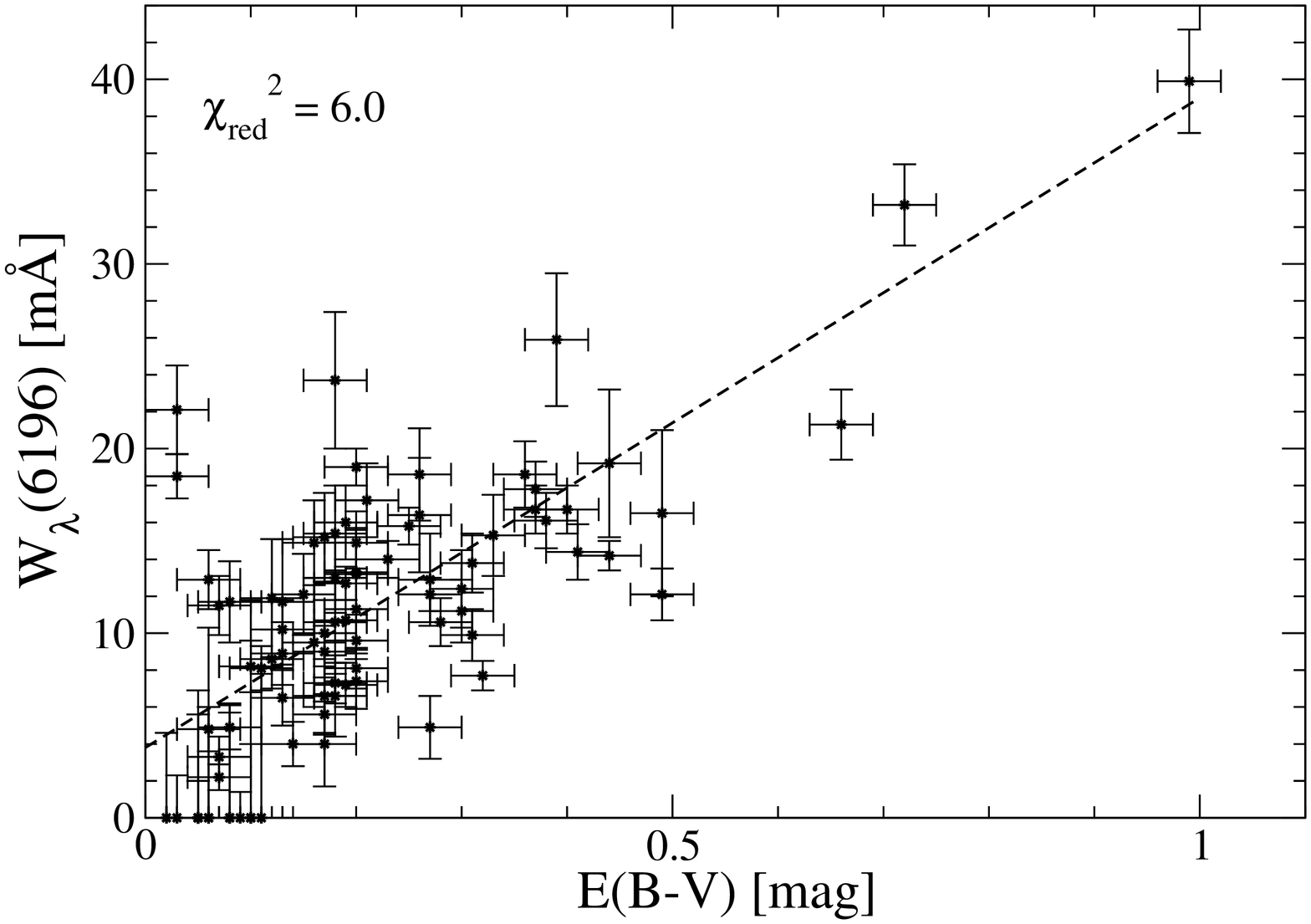}
   \includegraphics[bb=80 5 570 700, angle=-90,width=.95\columnwidth,clip]{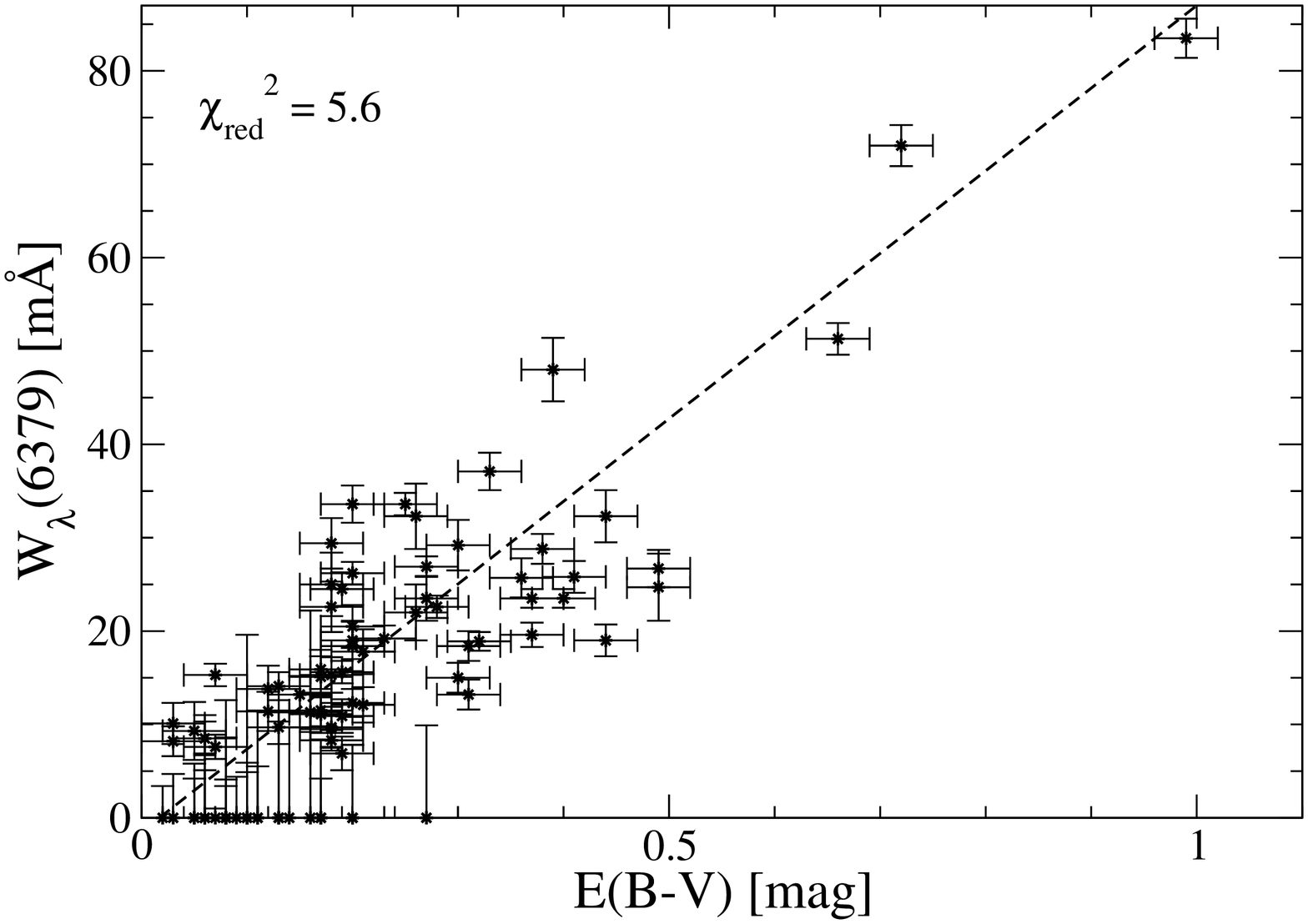}
   \includegraphics[bb=80 5 570 700, angle=-90,width=.95\columnwidth,clip]{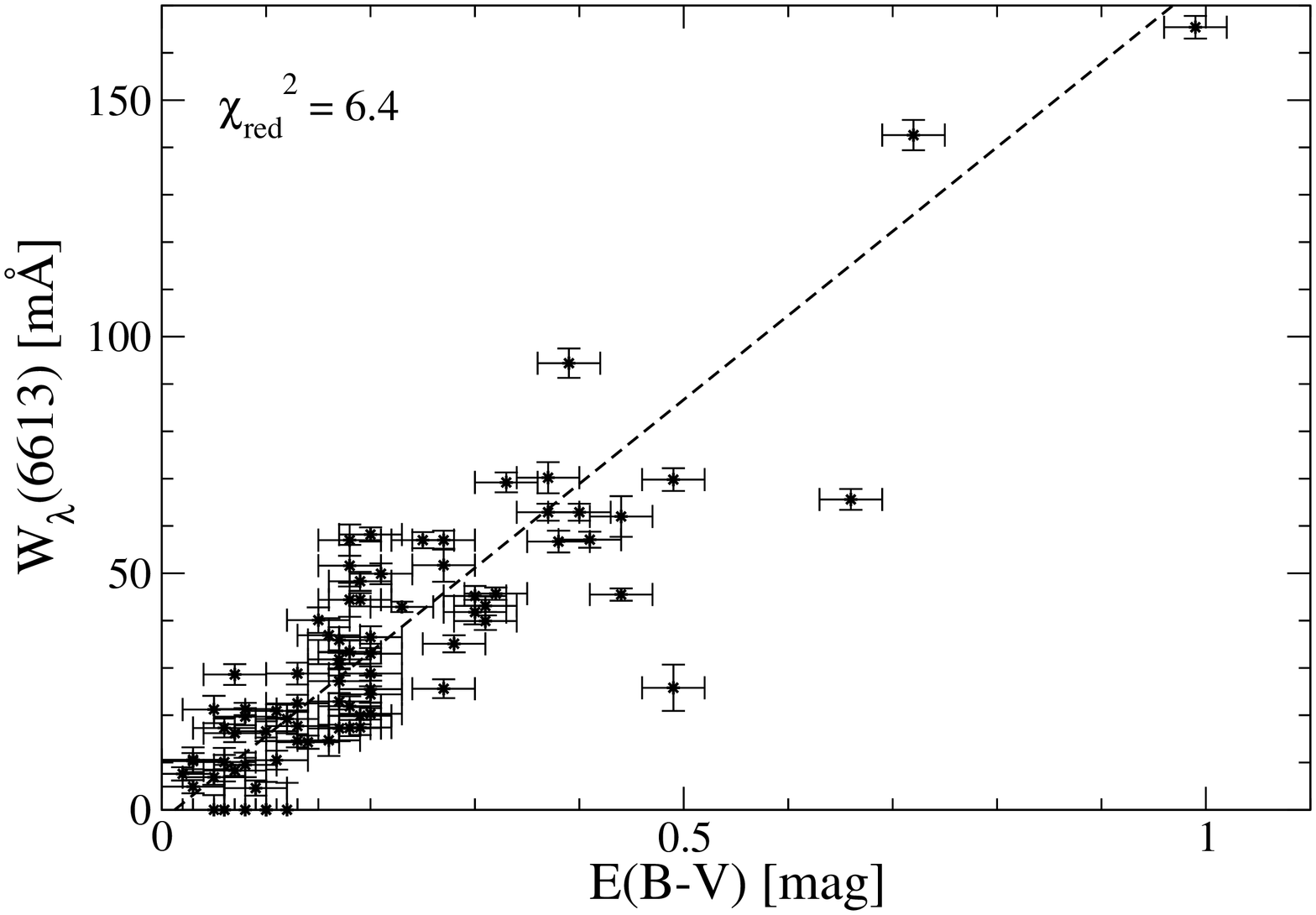}
   \caption{Equivalent width versus \Ebv\ for the 6196, 6379, and 6613~\AA\ DIBs observed towards the Upp Sco\ lines-of-sight.
   The reduced chi-squared ($\chi_{red}^{2}$) for the linear fits (dashed lines) are indicated in the respective panels. 
   The linear fit parameters are given in Table~\ref{tb:linearfit}.}
   \label{fig:EWDIBvsEbv}
\end{figure}

In Fig.~\ref{fig:5780-5797-Ebv} we show $W_\lambda(5780)$ (top) and $W_\lambda(5797)$ (bottom) against \Ebv.
Both the average Galactic and Upp Sco\ relationships are shown. Several conclusions can be drawn immediately
from this initial result:  1) A linear model does not adequately describe the relation between the measured
values ($\chi^2 \gg 1$); 2) The average DIB strength per unit reddening in Upp Sco\ is similar to the
Galactic average; 3) There is a positive trend between the amount of DIB carriers and the amount of dust in
the diffuse ISM; 4) There is a significant scatter from this mean linear relationship (which is also
observed for the Galaxy-wide surveys),  especially for the $\lambda$5780 DIB. This results in a poor
$\chi_{\rm red}^{2}$. 5) In particular, for sightlines with \Ebv $\approx$ 0.2 to 0.3~mag (which would
typically be expected to be single diffuse cloud sightlines) there is marked range in strength of the DIBs
(for both $\lambda\lambda$5780 and 5797~DIBs  the strength can vary by factor of about four to five). The
scatter (standard deviation) around the mean is equally high for higher \Ebv, but for those multiple cloud
components are more likely to contribute and confuse the true variations in individual clouds. The
strength-reddening relations for the $\lambda\lambda$6196, 6379, and 6613 DIBs are similar to that for
$\lambda$5797,  albeit with different slopes and an increased scatter (see Fig.~\ref{fig:EWDIBvsEbv}). The
linear fit method using uncertainties in both parameters is an implementation of the routine {\sc fitexy} 
from Numerical Recipes (\citealt{Press1992}) where $\chi_{red}^{2} = \chi^{2}/(N-2)$, with $N$ the number of
data points. A good fit will have $\chi_{red}^{2} \approx 1$. Despite significant intrinsic scatter in the
DIB versus reddening relations the least-square linear fit results are  given in Table~\ref{tb:linearfit} to
facilitate estimates of interstellar line-of-sight reddening from observed band strengths.

\begin{table}[!ht]
\begin{center}
\caption{%
Slopes and intercepts of the linear least-square fits. Uncertainties in both coordinates are taken into
account. Non-detections and upper limits were excluded from the fit procedure. The fits were not forced to
go through the origin, though it can be noted that in most cases the  derived intercept is within 2$\sigma$
of the origin. These relations can be used to derive estimates for the (interstellar) line-of-sight
reddening from measurements of the diffuse band strengths.
}
\label{tb:linearfit}
\begin{tabular}{llllll}\hline\hline
Correlated parameters  	  &  Intercept        	 & Slope	       & $\chi_{red}^{2}$   & r \\
			  & (m\AA)		 & (m\AA/\Ebv)	       &			\\
\hline
$W(5780)$ -- $E_{B-V}$    &    2.7   $\pm$ 3.3   & 462.0  $\pm$  12.7  & 10.8   & 0.79  	\\
$W(5797)$ -- $E_{B-V}$    &  $-$5.0  $\pm$ 1.1   & 159.0  $\pm$  4.1   &  4.0   & 0.92  	\\
$W(6196)$ -- $E_{B-V}$    &	3.8  $\pm$ 0.4   &  35.2   $\pm$ 1.8   & 6.0    & 0.72      	\\
$W(6379)$ -- $E_{B-V}$    &  $-$1.5  $\pm$ 0.8   &  88.5  $\pm$  0.8   & 5.6    & 0.85      	\\
$W(6613)$ -- $E_{B-V}$    &  $-$2.2  $\pm$ 1.2   & 177.8  $\pm$  4.6   & 6.4    & 0.86      	\\
\hline
$W(5780)_{\sigma}$ -- $E_{B-V}$   &    3.0  $\pm$ 7.8	& 640.2  $\pm$ 43.6   &  3.3  &  \\
$W(5780)_{\zeta}$ -- $E_{B-V}$    & $-$23.8 $\pm$ 5.0	& 419.2  $\pm$ 13.7   &  9.3  &  \\
$W(5797)_\sigma$ -- $E_{B-V}$	  & $-$1.7  $\pm$ 1.8   & 127.0  $\pm$ 9.6    &  2.8  &  \\
$W(5797)_\zeta$  -- $E_{B-V}$	  & $-$0.5  $\pm$ 1.9   & 153.3  $\pm$ 5.3    &  5.1  &  \\
\hline
\end{tabular} 
\end{center}
\end{table}

These deviations could reveal the effects of local conditions on the balance between DIB carrier formation 
and destruction (including changes in \emph{e.g.} ionisation and hydrogenation state), and therefore the
abundance and physical properties of the DIB carrier. The generally positive correlation between DIB
carriers and reddening suggests a link between the presence of dust  grains and the molecules responsible
for the diffuse bands. Figs.~\ref{fig:EWDIBvsEbv2} and~\ref{fig:EWDIBvsEbv3} (Online appendix) illustrate
that there are large variations, particularly at intermediate \Ebv $\approx$ 0.2-0.3~mag, in the DIB
strengths normalised by the amount of dust in the sightline. At lower \Ebv\ the measurements are inaccurate,
and at higher \Ebv\ the presence of multiple clouds in the line-of-sight  appears to reduce the effect of
variations in individual clouds on the composite, total line-of-sight DIB spectrum. The behaviour of the
DIBs in relation to molecular tracers and the local environmental conditions  will be discussed in the next
sections. The different behaviour of the $\lambda\lambda$5780 and 5797 DIBs is used as a tool to study the
deviations of both DIBs from the mean trend with \Ebv.

\subsection{The skin effect}\label{subsec:skineffect}

DIB carriers seem to reflect the evolutionary cycle of molecular carbon species (such as aromatic molecules)
through formation, ionisation, recombination, and destruction (\citealt{1997A&A...326..822C};
\citealt{2005A&A...432..515R}).

Uncharged aromatic molecules exhibit strong absorption bands in the UV and visible (blue) range  while their
cations and  anions show specific transitions in the visible (green-yellow) and near-infrared 
(\citealt{1999ApJ...526..265S}). Each DIB carrier is thus influenced by the interstellar radiation field in
a particular way, since its molecular properties such as ionisation potential and electron affinity are
unique.

Interstellar clouds are exposed to the interstellar radiation field which drives their photochemistry
(\citealt{2006ARA&A..44..367S}). The UV radiation is attenuated (by dust) increasingly from cloud edge to
core, giving  different steady-state solutions for the photochemical reactions (like the ionisation-state) 
in different parts of the diffuse cloud.  Thus interstellar species are subjected to stronger radiation at
the edge than in the centre of the cloud.

Especially the (molecular) DIB carriers are believed to be sensitive to UV radiation. The signatures of more
stable DIB carriers (such as corresponding to the $\lambda$5780 DIB)  show a relative higher intensity in
lower density, higher $I_{\rm UV}$\ regions with respect to less  stable DIB carriers which are more rapidly
destroyed at high $I_{\rm UV}$\  (\emph{e.g.}  $\lambda$5797 DIB). These reach higher intensity only in more UV protected
denser regions (where more stable DIB carriers like the $\lambda$5780 DIB are less efficiently ionized and
thus reduced in strength). For a sightline probing Upper Scorpius a larger amount of dust is expected to
correlate with, on average, higher densities, especially as there often is only one apparent strong
interstellar velocity component.

The effect of shielding (to a certain degree) of molecules from strong UV radiation is often  referred to as
the skin-effect ({\emph{e.g.} \citealt{1988A&A...190..339K}; \citealt{1995ARA&A..33...19H}).  The
skin-effect reflects the life cycle and charge distribution of DIB carriers, which can  can lead to an
interpretation of high DIB carrier concentrations in the outer cloud layers.  However, DIB carriers are also
expected to be present in high concentrations in denser regions although in a different  charge state
(neutral) that can only be observed in the UV.

\begin{figure}[t!]
\centering
   \includegraphics[angle=-90,width=\columnwidth]{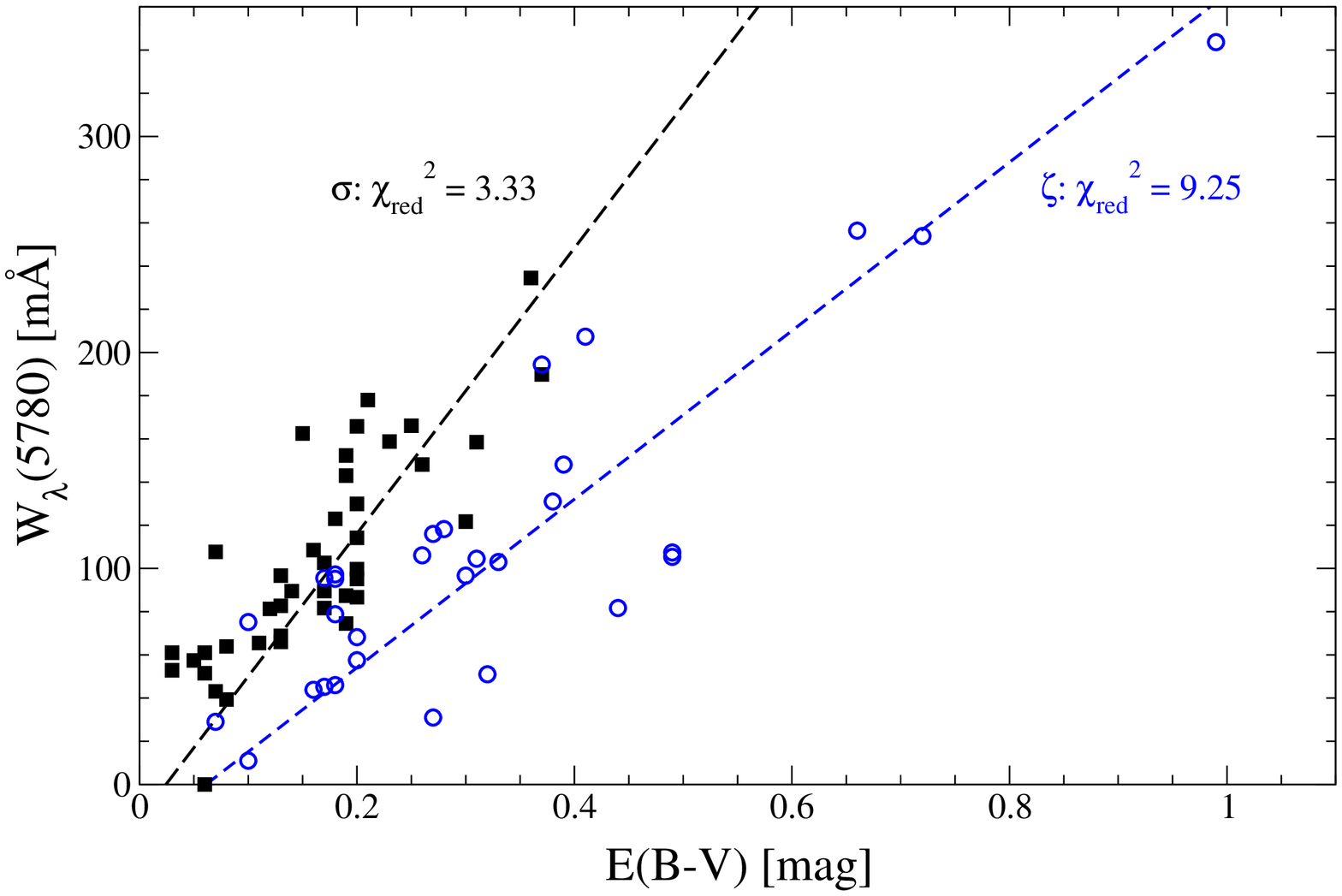}
   \includegraphics[angle=-90,width=\columnwidth]{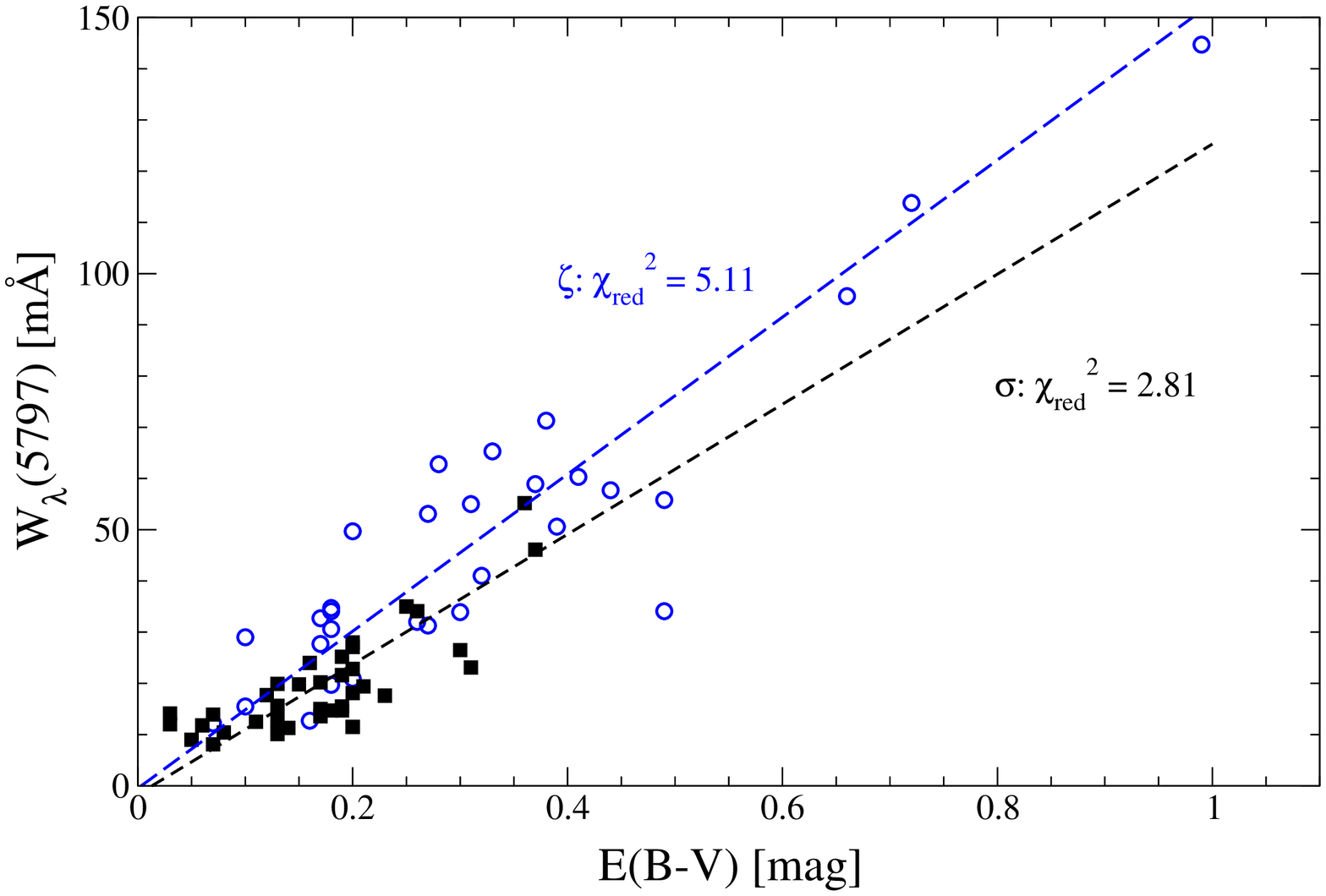}
   \caption{%%
   $W_\lambda(5780)$ (top) and $W_\lambda(5797)$ (bottom) versus \Ebv\ for the $\sigma$ 
   %(\mbox{\scalebox{0.6}{$\blacksquare$}}) 
   and $\zeta$  %(\mbox{\textcolor{blue}{$\bullet$}}) 
   subgroups, respectively.
   Error bars, identical to those in Fig.~\ref{fig:5780-5797-Ebv}, have been omitted for 
   clarity but taken into account for the linear fit.
   Parameters for the linear fit and regression are given in Table~\ref{tb:linearfit}.   }
   \label{fig:57805797-sigmazeta}
\end{figure}

\citet{1997A&A...326..822C} inferred that the $\lambda$5780~DIB carrier reaches its maximum abundance when
exposed to the interstellar UV radiation field (typically near the edge of a cloud), whereas the
$\lambda$5797 DIB carrier is more easily ionised and destroyed.  Even more, at very low \Ebv\  ($<$0.1~mag)
only very few DIB carriers survive due to the high rate of UV photons (\citealt{1994A&A...281..517J}). The
relative abundance between the $\lambda\lambda$5780 and 5797 DIBs reflects an interplay  between neutral,
ionised, and destroyed DIB carriers along the entire line of sight. This balance is affected not only by the
impinging radiation field, but also by the carbon abundance and the dust particle size distribution
(\citealt{2006A&A...451..973C}). A difference in the observed ratio of these two DIBs is thus directly
related to the skin-effect.

\citet{1989IAUS..135...67K,1991PASP..103.1005S,1992MNRAS.258..693K} identified two types of clouds, referred
to as $\sigma$ and $\zeta$-type. $\sigma$-type clouds show atomic lines and DIBs, but the molecular lines
are weak or absent, while $\zeta$-type clouds have strong diatomic lines in addition to DIBs.  The main
difference between both types lies in a combination of density and UV irradiation by the ISRF, with $\sigma$
clouds associated with low density and/or strong exposure to UV radiation, while $\zeta$ clouds are
associated with higher densities and/or more protection from UV radiation. Therefore, differentiation
between $\sigma$ and $\zeta$-type clouds is directly linked to the skin-effect described previously. For
sightlines probing $\zeta$-type clouds the 5797~\AA\ DIB is deeper than the 5780~\AA, while  for
$\sigma$-type clouds the reverse is observed.  Therefore, the $W(5797)/W(5780)$ ratio has been used to
distinguish between UV exposed ($\sigma$) and UV protected ($\zeta$) sightlines. The nomenclature for the
$\sigma$ and $\zeta$ type sightlines is historical and based on the representative lines-of-sight towards
\object{$\sigma$ Sco} and \object{$\zeta$ Oph}, respectively (\citealt{1988A&A...190..339K};
\citealt{1992MNRAS.258..693K}; \citealt{1995dib..book...13K}).  Note that both sightlines are included in
our analysis.

\begin{figure}[t!]
\centering 
   \includegraphics[angle=-90,width=\columnwidth,clip]{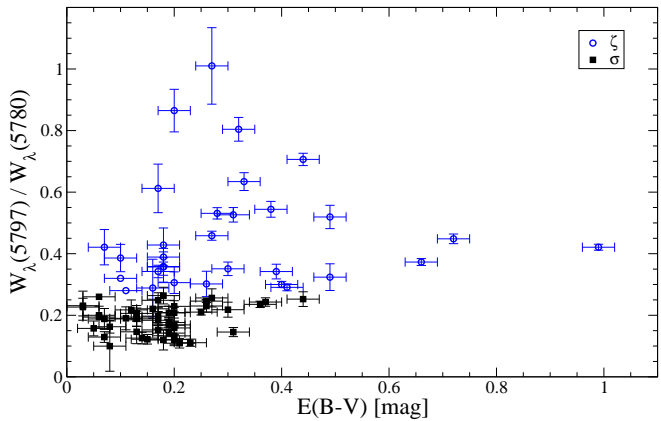}
   \caption{%%
   The $W(5797)/W(5780)$ ratio plotted against \Ebv. 
   The distribution peaks at an \Ebv\ of $\sim$0.25~mag, indicating most optimal conditions for 
   formation of the 5797~\AA\ DIB carrier and the destruction c.q. insufficient excitation of 
   the molecule giving rise to the 5780~\AA\ DIB.
   Nonetheless, the significant scatter suggests that processes additional to dust extinction are important.}
   \label{fig:ratio_vs_ebv}
\end{figure}

In this work we re-establish this classification, assuming a relatively equal distribution of  sightlines
probing dense versus diffuse clouds. Sightlines are classified $\sigma$ when the ratio is lower than the
weighted mean of the ratio minus 1$\sigma$, while ratios higher than the weighted mean plus 1$\sigma$ are
classified as $\zeta$. The remaining sightlines are classified as intermediate. The results of this
selection for individual lines-of-sight are included  in Table~\ref{tb:sigma-zeta} (Online).
Figure~\ref{fig:57805797-sigmazeta} shows that the application of our classification to the data in 
Fig.~\ref{fig:5780-5797-Ebv} improves the relation (reduced scatter) between DIB strength and reddening. 
Indeed, Fig.~\ref{fig:57805797-sigmazeta} shows that $W$(5780)-\Ebv\  has an improved reduced $\chi^2$  for
the  $\sigma$ and $\zeta$ sightlines respectively,  though only a marginal improvement is found for
$W$(5797)-\Ebv\  (where higher $W$(5797) tend to correspond to $\zeta$-type environments)  revealing the
$\lambda$5780 DIB is primarily giving rise to variations in the $W$(5797)/$W$(5780) ratio.

Note that the original classification is based on the central depth, $A$, of the two DIBs. If $A_{5797} >
A_{5780}$ (corresponding to $W(5797)/W(5780) \ga 0.4$)  the line-of-sight is considered as $\zeta$-type. 
Increasing our selection threshold for the DIB ratio to 0.4 would imply that the ``intermediate'' sources
would be included in the $\sigma$-group as well as a few $\zeta$-types. However, sightlines with both low
($f_{\mathrm{H}_2}$$<$0.3) and high ($f_{\mathrm{H}_2}$$>$0.4) molecular content currently classified
$\zeta$ would  also be re-assigned as $\sigma$-type. We note that it is impossible to make a sharp
distinction between $\sigma$ and $\zeta$-type sightlines as there is - as expected - a smooth transition of
physical conditions characterising both types.

Fig.~\ref{fig:ratio_vs_ebv} shows $W(5797)/W(5780)$ as a function of reddening. The distribution peaks at an
\Ebv\ of $\sim$0.25~mag, indicating most optimal conditions for formation  of the 5797~\AA\ DIB carrier
(sufficient shielding), or alternatively less optimal conditions for the carrier associated  to the
5780~\AA\ DIB carrier (insufficient UV photons to transform it into its ionic form and thus not absorbing at
the visible wavelength). For sightlines with \Ebv $>$0.4~mag the conditions for formation of the
$\lambda$5797~DIB are sub-optimal,  but still more  favourable with respect to the $\lambda$5780~DIB carrier
than for $\sigma$ sightlines. At very low \Ebv\  ($<$0.1~mag) the $\lambda$5797~DIB carrier is
under-abundant (due to more efficient destruction  of molecules by the stronger ISRF) with respect to the
$\lambda$5780~DIB carrier. Also, the $W$(5797)/$W$(5780) ratio itself displays a bimodal distribution with a
strong peak at about 0.2$\pm$0.05  (Fig.~\ref{fig:DIBratioHistogram}; Sect.~\ref{subsec:DIBratios}). The
sightlines associated to this peak are predominantly $\sigma$-type (which indeed we may consider to
represent typical diffuse ISM though this should be confirmed by studies of other regions).  There is a
second smaller peak ``bump'' at $\sim$0.45, corresponding to the $\zeta$-type sightlines. Although, like the
$\lambda$5797 DIB the $\lambda\lambda$ 6196, 6379 and 6613 DIBs also show signficant scatter on the
respective $W$-\Ebv\ trends,  the $\lambda$5780 DIB reveals the clearest distinction in behaviour between
$\sigma$ and $\zeta$ sightlines. The link between the $W(5797)/W(5780)$ ratio and the strength of the ISRF
is discussed in more detail in Sect.~\ref{subsec:isrf}.

For comparison we plot also the $W(6196)/W(6613)$ ratio as a function of \Ebv\ in
Fig.~\ref{fig:ratio_vs_ebv2}. This ratio is less sensitive to reddening and therefore is not such a useful
tracer of local conditions  such as density and UV irradiation.  This is indeed expected from the recent
results by \citet{2010ApJ...708.1628M} who found an excellent correlation between  the 6196 and 6613~\AA\
DIB strengths.

\subsection{Diatomic molecules and dust}

Different interstellar species are restricted to different regions (see \emph{e.g.} Fig. 6 in
\citealt{2005ApJ...633..986P}):  CN and CO are present in dense regions, CH and \ion{K}{i} are predominantly
present in moderately high  density regions ($n >$ 30~cm$^{-3}$), and CH$^+$ and \ion{Ca}{i} in intermediate
density regions ($n \sim 10-300$~cm$^{-3}$). 

\begin{figure}[t!]
\centering
\includegraphics[angle=-90,width=\columnwidth,clip]{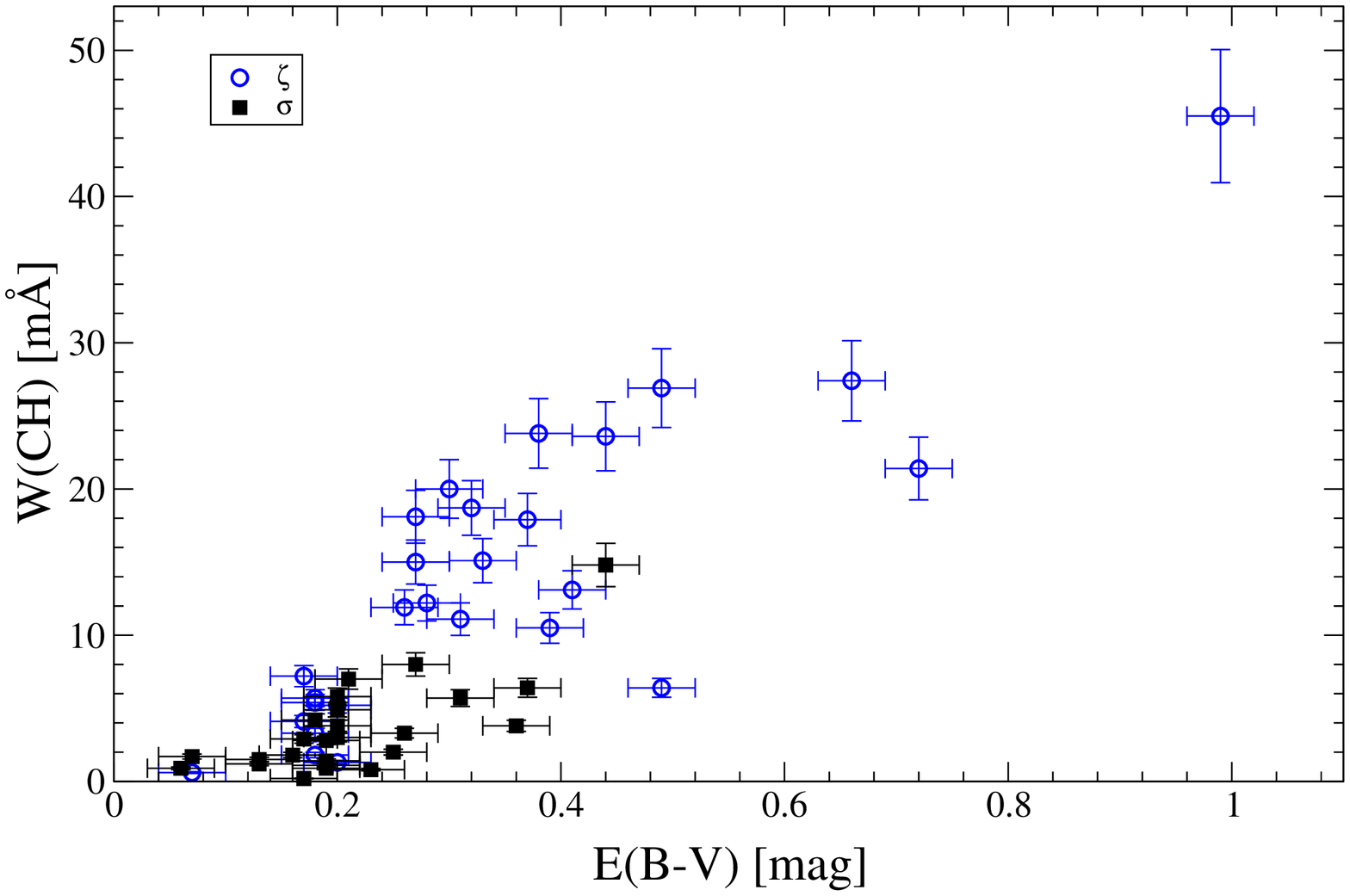}
\includegraphics[angle=-90,width=\columnwidth,clip]{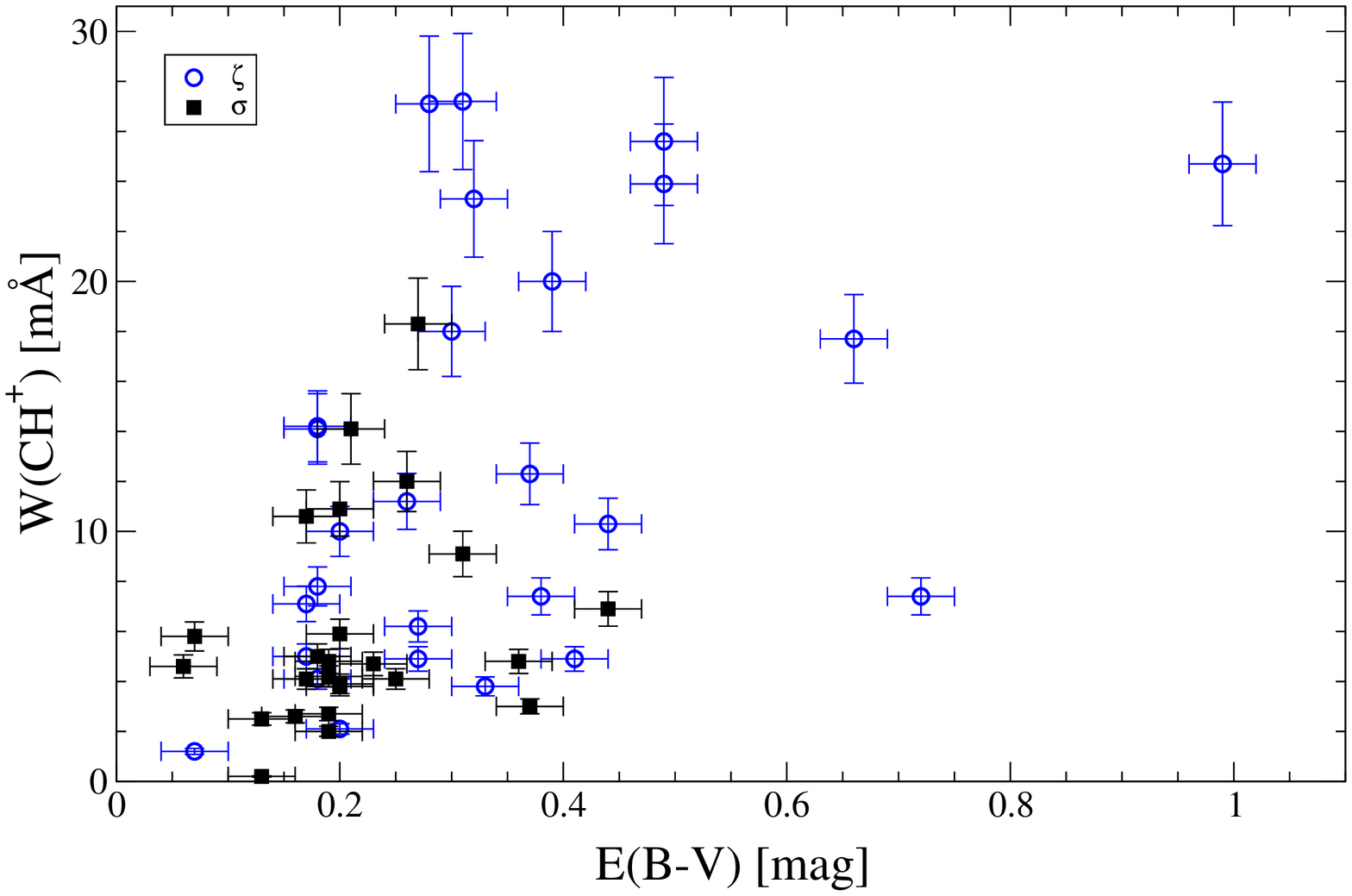}
   \caption{Equivalent width versus \Ebv\ for the CH 4300~\AA\ (top) and CH$^+$ 4232~\AA\ (bottom) transitions.
   For $\sigma$ (black squares) and $\zeta$-type (blue circles) sightlines designations see Sect.~\ref{subsec:skineffect} and Sect.~\ref{subsec:DIBratios}.
   %% For $\sigma$-type sightlines both CH and CH$^+$ hardly change with \Ebv.
   %%Standard correlation coefficients for both $\sigma$ and $\zeta$ sightlines are indicated in the respective panels.
   %%Linear fit parameters are given in Table~\ref{tb:linearfit}. 
   }
   \label{fig:CH-CH+-Ebv}
\end{figure}

In Fig.~\ref{fig:CH-CH+-Ebv} we show $W$(CH) (top) and $W$(CH$^+$) (bottom) versus \Ebv. It can be seen that
CH correlates much better with \Ebv\ than CH$^+$, which is in line with previous observations
(\citealt{1989MNRAS.241..575C},  \citealt{1999A&A...347..235K}). CH traces the dense, molecular gas and its
abundance is directly proportional to N(H$_2$) as N(CH)/N(H$_2$) = 3.5 $\times$ 10$^{-8}$
(\citealt{1982ApJ...257..125F}; \citealt{1986A&A...160..157M}; \citealt{2004A&A...414..949W};
\citealt{2008ApJ...687.1075S}). For 8 lines-of-sight direct measurements of N(H$_2$) (IUE or FUSE; compiled
in \citealt{2011ApJ...727...33F}) can be compared to those derived from N(CH) in this work
(Table~\ref{tb:sigma-zeta}).  The scatter is less than $\sim$0.5~dex, and in good agreement with the results
of \citet{2008ApJ...687.1075S} and references therein.  Theoretically, one can thus infer the molecular
hydrogen fraction $f_{\mathrm{H}_2}$ from N(H$_2$) derived from CH together with N(\ion{H}{i}) derived from
$W$(5780) (log N(\ion{H}{i}) = 19.00 + 0.94 log($W$(5780)); \citealt{2011ApJ...727...33F}). The resulting
values for $f_{\mathrm{H}_2}$ are given in Table~\ref{tb:DIB-data}.  These values are consistent with -
though systematically higher than - the directly measured $f_{\mathrm{H}_2}$  (\emph{e.g.} 
\citealt{2011ApJ...727...33F}; Table~\ref{tb:DIB-data}) for the eight sightlines in common.  Here we have
used the average Galactic relation between $W$(5780) and N(\ion{H}{i}), whereas this relation may actually
be lower for  Upp Sco\  (similar to the lower gas-to-dust ratio in this region;
\citealt{1998ApJ...500..525S}) thus leading to a higher estimate  of $f_{\mathrm{H}_2}$. The strongly
improved regression coefficient between CH and \Ebv\ for the $\zeta$-type sightlines ($r = 0.83$) compared
to its $\sigma$-type equivalent ($r = 0.53$) supports the interpretation that $\zeta$-type lines-of-sight
trace dense gas. It is noteworthy to recall that the significant scatter for the diffuse band strengths at
low \Ebv\  ($\sim$0.2~mag) as illustrated in Figs.~\ref{fig:5780-5797-Ebv} and~\ref{fig:EWDIBvsEbv} is not
observed for CH. CH$^+$, on the other hand, is not a good tracer of H$_2$ (\citealt{2008A&A...479..149W}).
Therefore, the low value of the correlation coefficient for CH$^+$ is not unexpected. Furthermore, note that
significant amounts of CH are needed before CN is produced (\citealt{1984ApJ...287..219F}), with the latter
tracing also relatively dense material (\citealt{1986ApJ...309..771J}).

\begin{figure}[t!]
\centering
   \includegraphics[angle=-90,width=\columnwidth]{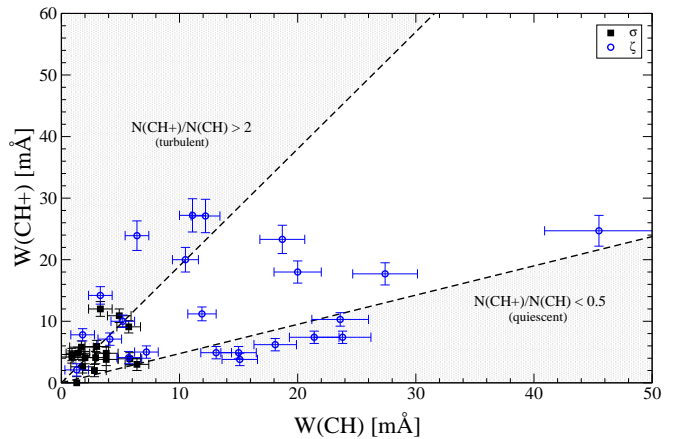}
   \caption{$W$(CH$^+$) versus $W$(CH). Assuming optically thin lines the N(CH$^+$)/N(CH) ratio
   is equal to $0.95 \times$ $W$(CH$^+$)/$W$(CH). Turbulent and quiescent regions are indicated
   by shaded areas. See text for further details.
   For $\sigma$ and $\zeta$-type sightlines designations see Sect.~\ref{subsec:skineffect} and
   Sect.~\ref{subsec:DIBratios}.}
   \label{fig:CH-CH+}
\end{figure}

Work by \citet{1989MNRAS.241..575C} suggests that the ratio of N(CH$^+$) and N(CH) is indicative of the
turbulent or quiescent nature of the interstellar medium in the line-of-sight. For shocked environments an
offset velocity between CH and CH$^+$ or a velocity broadening of CH$^+$ is predicted by models. For the
sightlines in this work we obtain an average $\Delta v$ of 0.3~km~s$^{-1}$, with individual velocity measurements
that have errors of about 1 to 2~km~s$^{-1}$\  (see also Sect.~\ref{sec:velocity}).  Our data support recent
surveys which find no evidence for a velocity difference between CH and CH$^+$
(\citealt{1995ApJS...99..107C};  \citealt{2005ApJ...633..986P}). The data do not allow for an accurate
measurement and comparison of CH and CH$^+$ line widths. The line profiles of atomic and di-atomic species
can be compared in Fig.~\ref{fig:ex_velprof} (and associated Fig.~\ref{fig:lineprofiles} in the Online
material). W(CH$^+$) is plotted against W(CH) in Fig.~\ref{fig:CH-CH+} with the turbulent (N(CH$^+$)/N(CH)
$>$ 2)  and quiescent (N(CH$^+$)/N(CH) $<$ 0.5) regions indicated by the shaded areas. The general
correlation between CH$^+$ and CH (\citealt{2005ApJ...633..986P}) is poor, but it appears that two separate
trends might in fact exist for the quiescent and turbulent regions, respectively, potentially indicative of
different dominant CH$^+$ production mechanisms.} The dense cloud tracer CN is only detected towards
$\zeta$-type lines-of-sight, supporting the interpretation  that the latter probe dense clouds.  The
$\sigma$ and $\zeta$ type sightlines show different trends for W(CH), but not so clearly for W(CH$^+$).

\begin{figure}[t!]
\centering
   \includegraphics[angle=-90,width=\columnwidth]{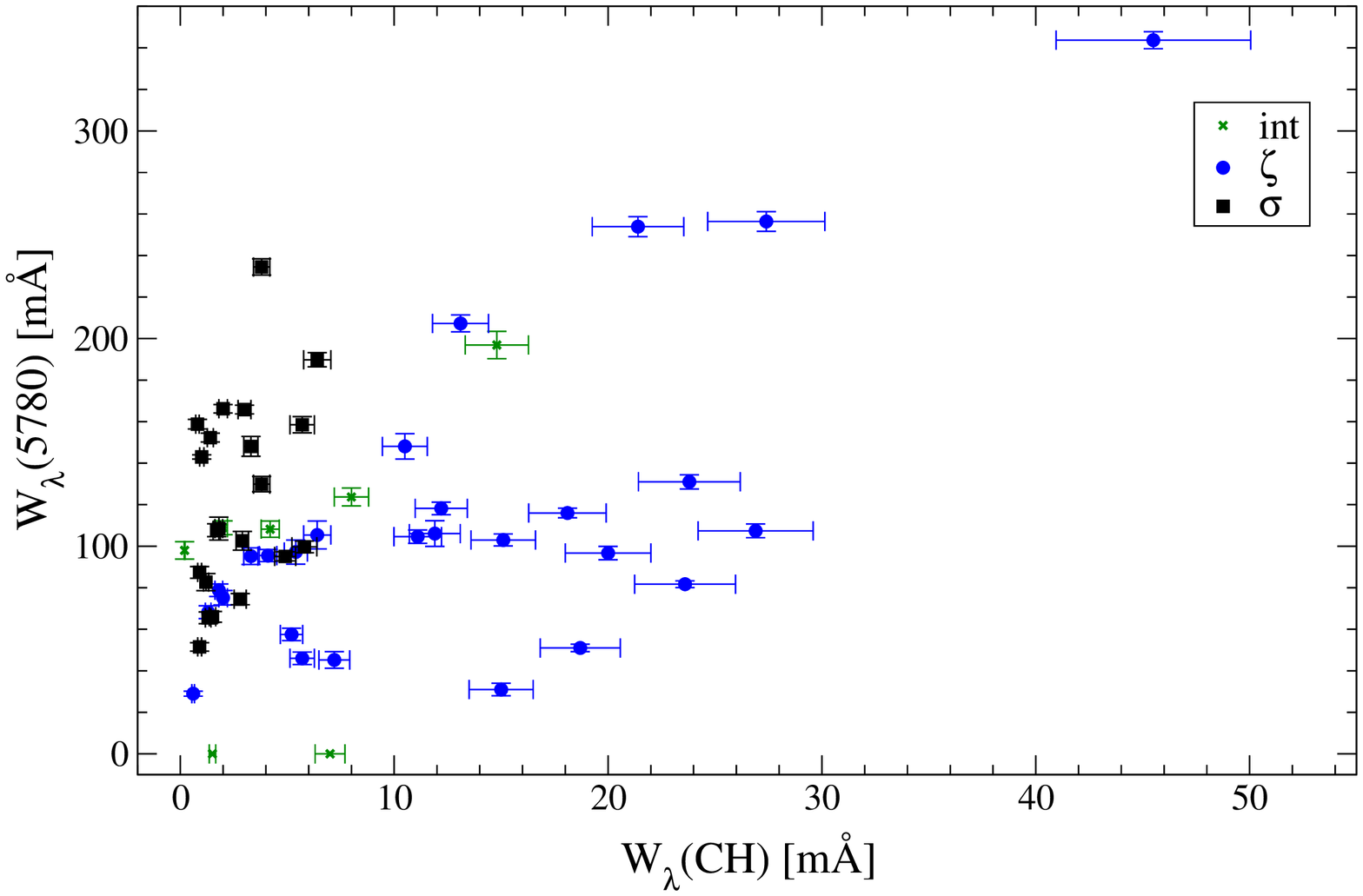}
   \includegraphics[angle=-90,width=\columnwidth]{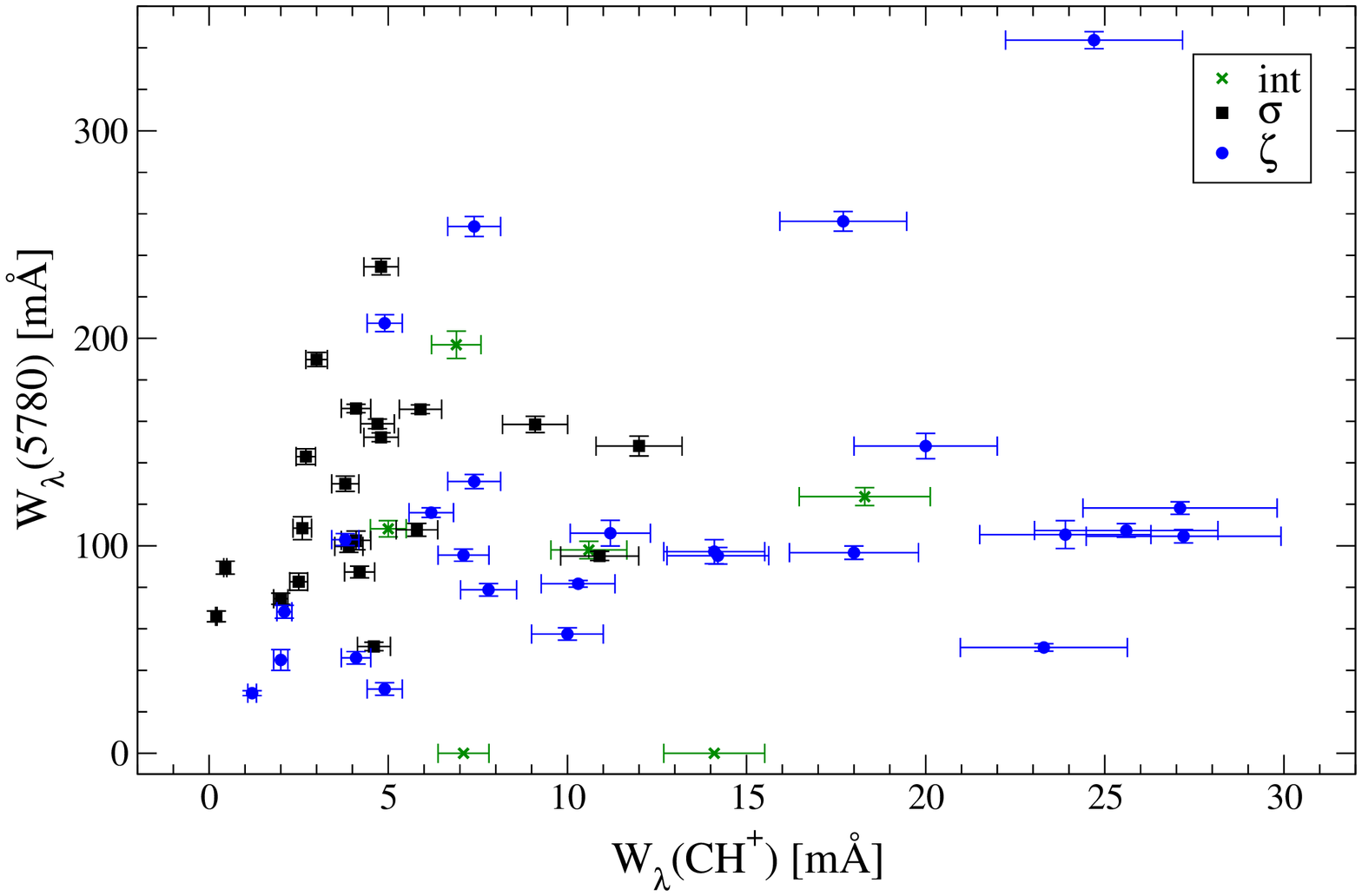}
   \caption{$W$(CH) (top) and $W$(CH$^+$) (bottom) versus $W_\lambda$(5780).
   The $\lambda$5780 DIB shows no direct correlation with either CH or CH$^+$. 
   Looking separately at the $\sigma$ and $\zeta$-type sightlines one can distinguish
   different behaviour between the molecular lines and the $\lambda$5780 DIB for both types.
   Sightlines classified as intermediate are indicated by green crosses.
   DIBs are stronger with respect to CH and CH$^+$ line strengths for $\sigma$-type sightlines.
   In other words, the 5780 DIB carrier abundance is lower for $\zeta$-type clouds which have a 
   higher molecular content.}     
   \label{fig:DIBvsCH2}
\end{figure}

\begin{figure}[t!]
\centering
   \includegraphics[angle=-90,width=\columnwidth]{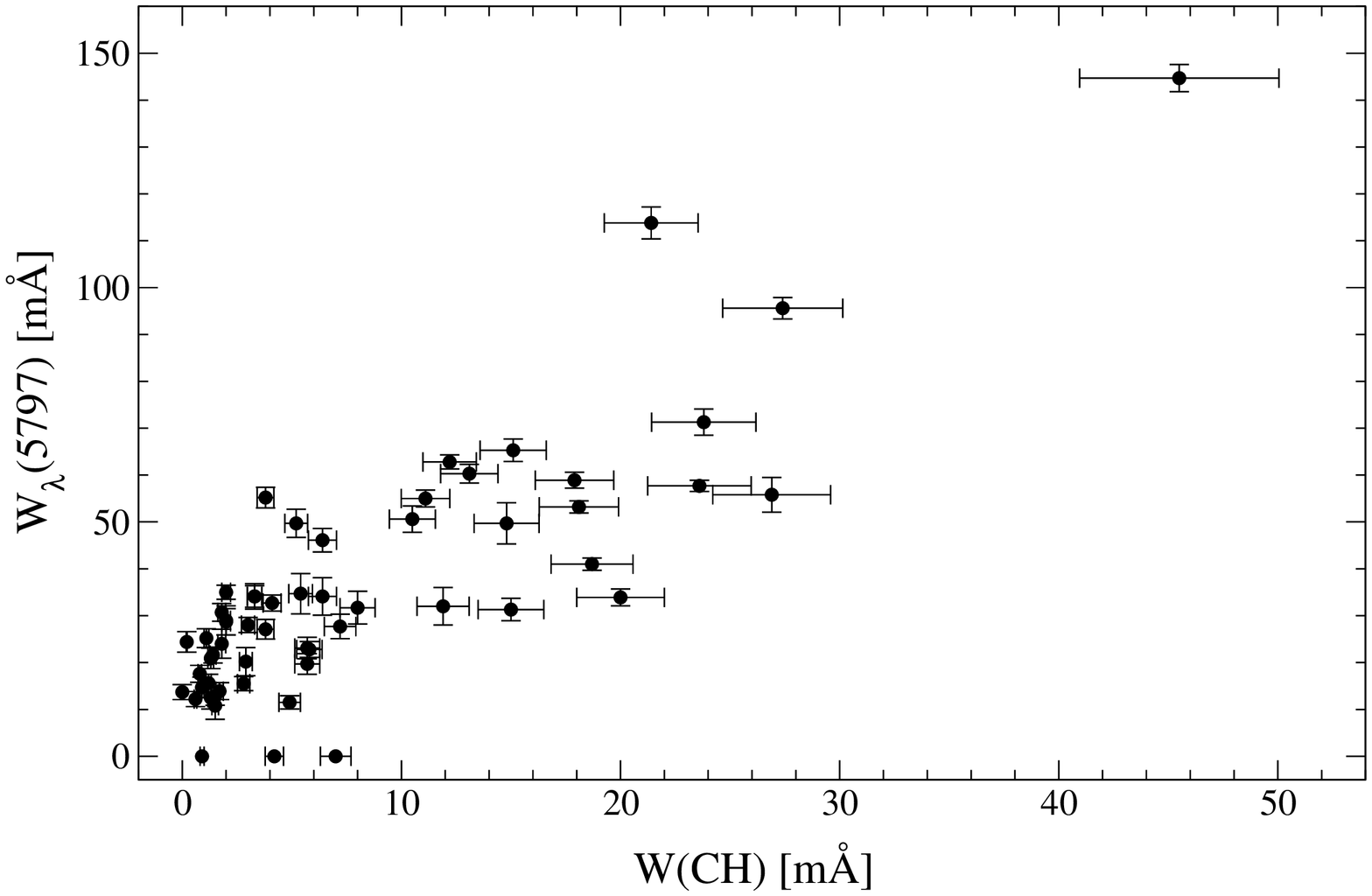}
   \includegraphics[angle=-90,width=\columnwidth]{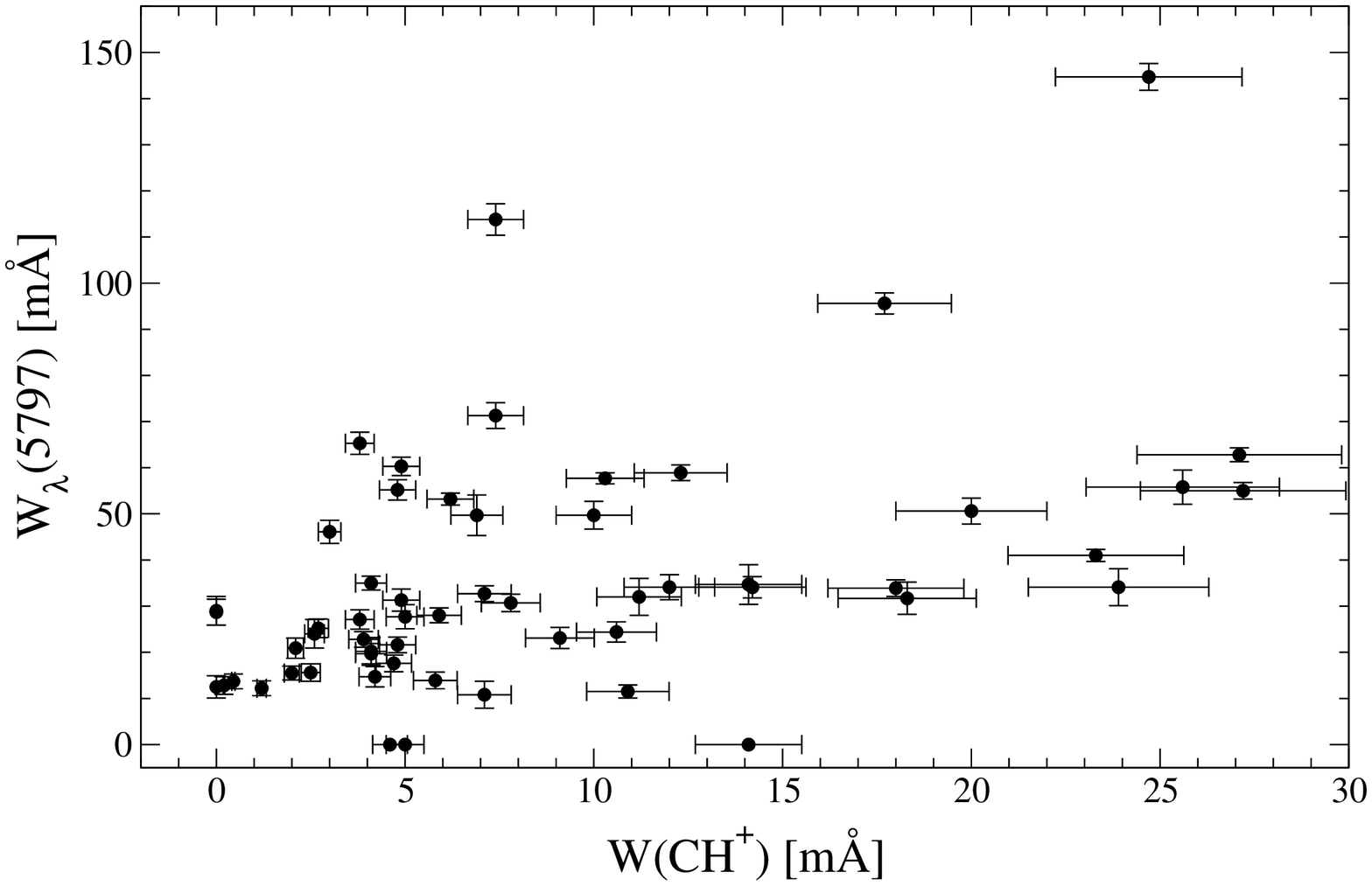}
   \caption{$W$(CH) (top) and $W$(CH$^+$) (bottom) versus $W_\lambda$(5797).
   There is a moderately good correlation ($r=0.84$) between $W$(CH) and $W_\lambda$(5797).
   There is no correlation ($r=0.47$) between CH$^+$ and $W_\lambda$(5797).
   }     
   \label{fig:DIBvsCH1}
\end{figure}

\subsection{DIBs and small molecules}\label{sec:dibs-diatomics}

In this section we discuss the behaviour of the $\lambda\lambda$5797 and 5780 DIBs with respect to CH and
CH$^+$. \citet{1992MNRAS.258..693K} found that CH and CN are only detected if the $\lambda$5797 DIB is {\it
deeper} than the $\lambda$5780 DIB. \citet{2008A&A...484..381W} studied CH, CH$^+$, and CN in relation to
DIBs for a large, inhomogeneous sample of sightlines. These authors found a good correlation between
$W$(5797) and N(CH), but a poor correlation  between $W$(CH)/\Ebv\ or $W$(CN)/\Ebv\ versus
$W$(5797)/$W$(5780).  The correlation of $W$(5797) vs. N(CH) is further improved by excluding sightlines
with overabundant CH. Their conclusion is that the $\lambda$5797 DIB carrier is favoured in environments
with higher molecular gas content. On the other hand CN traces a denser medium where the production of the
$\lambda$5797 DIB is apparently more inefficient.

Figures~\ref{fig:DIBvsCH2} and~\ref{fig:DIBvsCH1} show the relationship between molecular line strengths
($W$(CH) and $W$(CH$^+$)) and diffuse interstellar band strengths ($W_\lambda$(5780) and $W_\lambda$(5797)).
These results are in line with \citet{1993ApJ...407..142H} and \citet{2008A&A...484..381W},  who concluded
that DIB strengths correlate better with \Ebv\ and \ion{H}{i} than with any other feature originating from
the gas phase. These DIBs have a stronger correlation with CH than with CH$^+$ (this work) or CN
(\citealt{2008A&A...484..381W}). The positive correlation with \Ebv\ suggests that even though grains do not
give rise to the diffuse bands they do play an important role in the either the DIB carrier formation - via
\emph{e.g.} grain surface reactions - or destruction - \emph{e.g.} attenuation of UV radiation - processes.
The CH molecule and the $\lambda$5797 DIB correlate tightly, indicating that the $\lambda$5797 DIB carrier
is most abundant in CH / H$_2$ clouds (see also \citealt{2004A&A...414..949W}). Some correlation is expected
since both species correlate with \Ebv. For individual clouds a larger $W$(CH) is indicative for the
formation in denser clouds, which explains the tighter correlation with the $\lambda$5797~DIB compared to
the $\lambda$5780 DIB. Note however, that the strongest molecular features potentially arise from
(unresolved) multiple  components of the ISM which are not necessarily denser (see \emph{e.g.}  velocity profiles
for \ion{K}{i} in Fig.~\ref{fig:lineprofiles}).

\begin{figure}[t!]
\centering
   \includegraphics[angle=-90,width=\columnwidth]{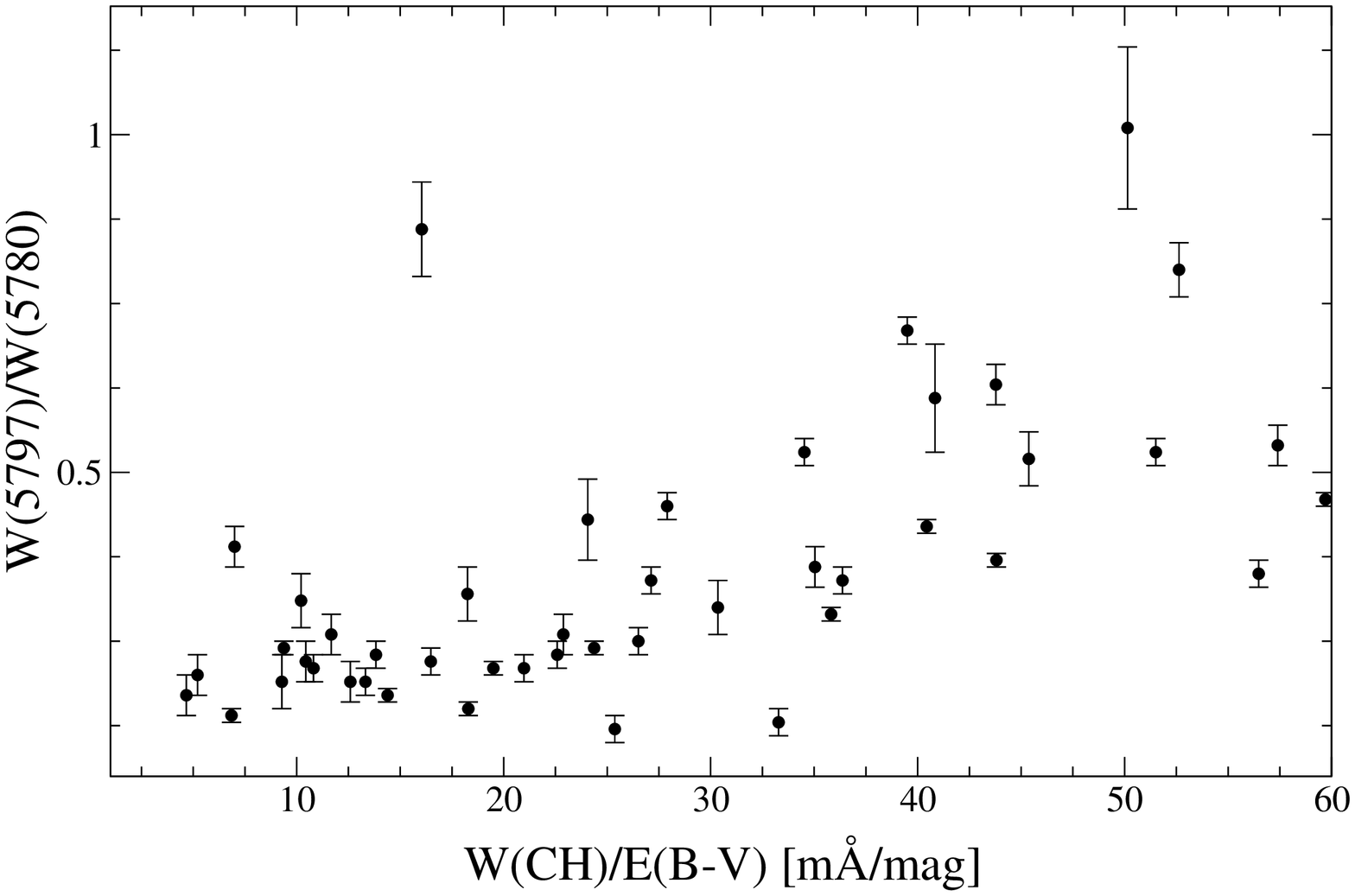}
   \includegraphics[angle=-90,width=\columnwidth]{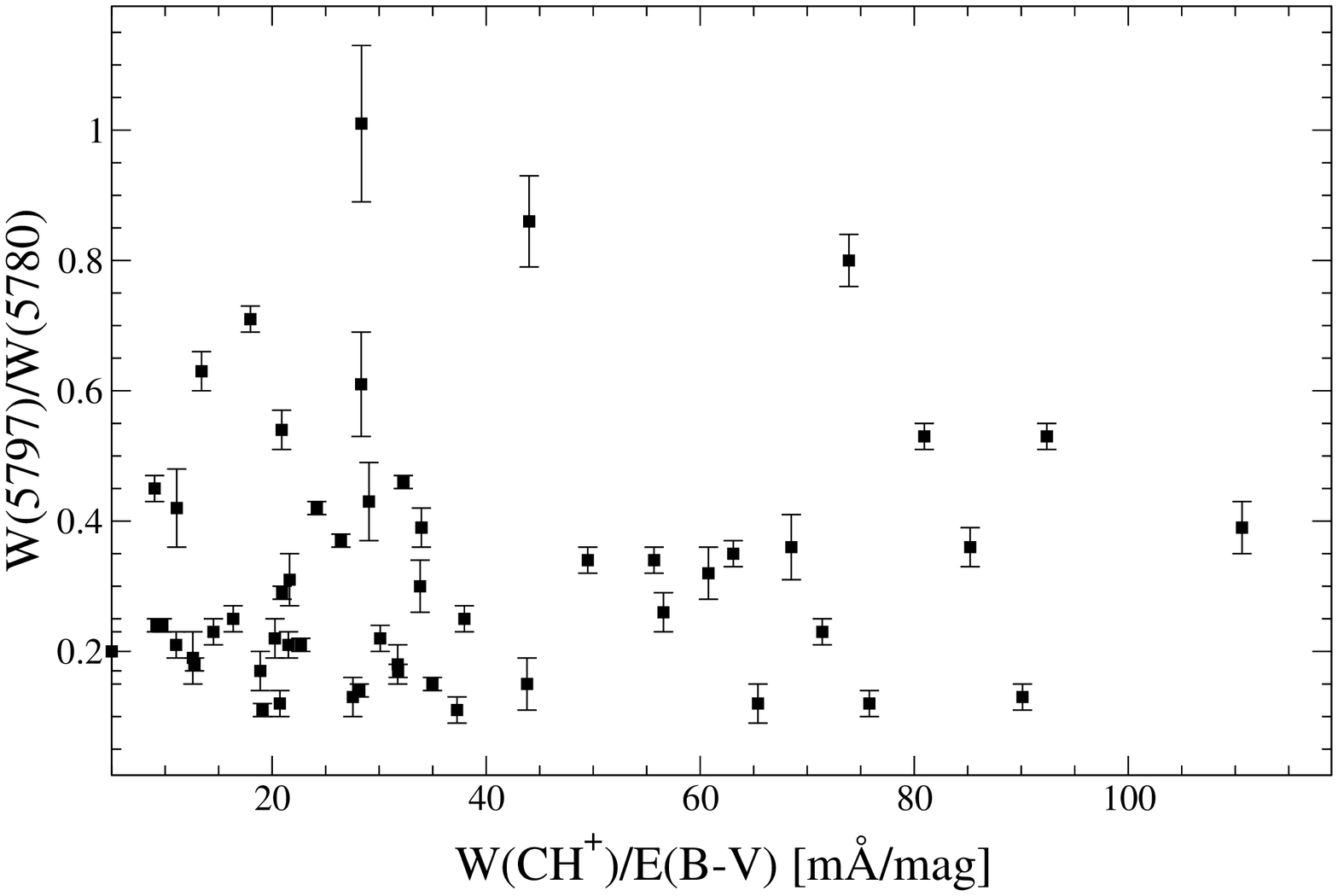}
   \caption{$W_\lambda(5797)/W_\lambda(5780)$ against $W$(CH)/\Ebv\ (top) and $W$(CH$^+$)/\Ebv\ (bottom).
   Tentatively a positive trend can be discerned between the DIB ratio and $W$(CH)/\Ebv\ (top) but not between
   the DIB ratio and $W$(CH$^+$)/\Ebv\ (bottom).}
   \label{fig:DIBratioCH.CHp}
\end{figure}

Figure~\ref{fig:DIBratioCH.CHp} shows the $W_\lambda(5797)/W_\lambda(5780)$ ratio  versus the $W$(CH) (top)
and $W$(CH$^+$) (bottom) normalised to \Ebv. In agreement with \citet{1999A&A...347..235K}, a stronger
correspondence is observed for the $W_\lambda(5797)/W_\lambda(5780)$ ratio versus $W$(CH)/\Ebv\ compared to
that for $W_\lambda(5797)/W_\lambda(5780)$ versus $W$(CH$^+$)/\Ebv. This confirms that $\zeta$-type clouds
(dense, $\lambda$5797 DIB favoured) are connected to a higher molecular content, implying furthermore that
the DIB ratio is related to the abundance of cold cloud molecular species and properties of interstellar
dust as suggested in Sect.~\ref{subsec:skineffect}. The poorer correlation between this DIB ratio and
$W$(CH$^+$)/\Ebv\ then suggests that CH$^+$ forms in regions with different conditions, such as in the
clouds outer edge, where the UV radiation field is much stronger.

\begin{figure}[t!]
\centering
   \includegraphics[angle=-90,width=\columnwidth]{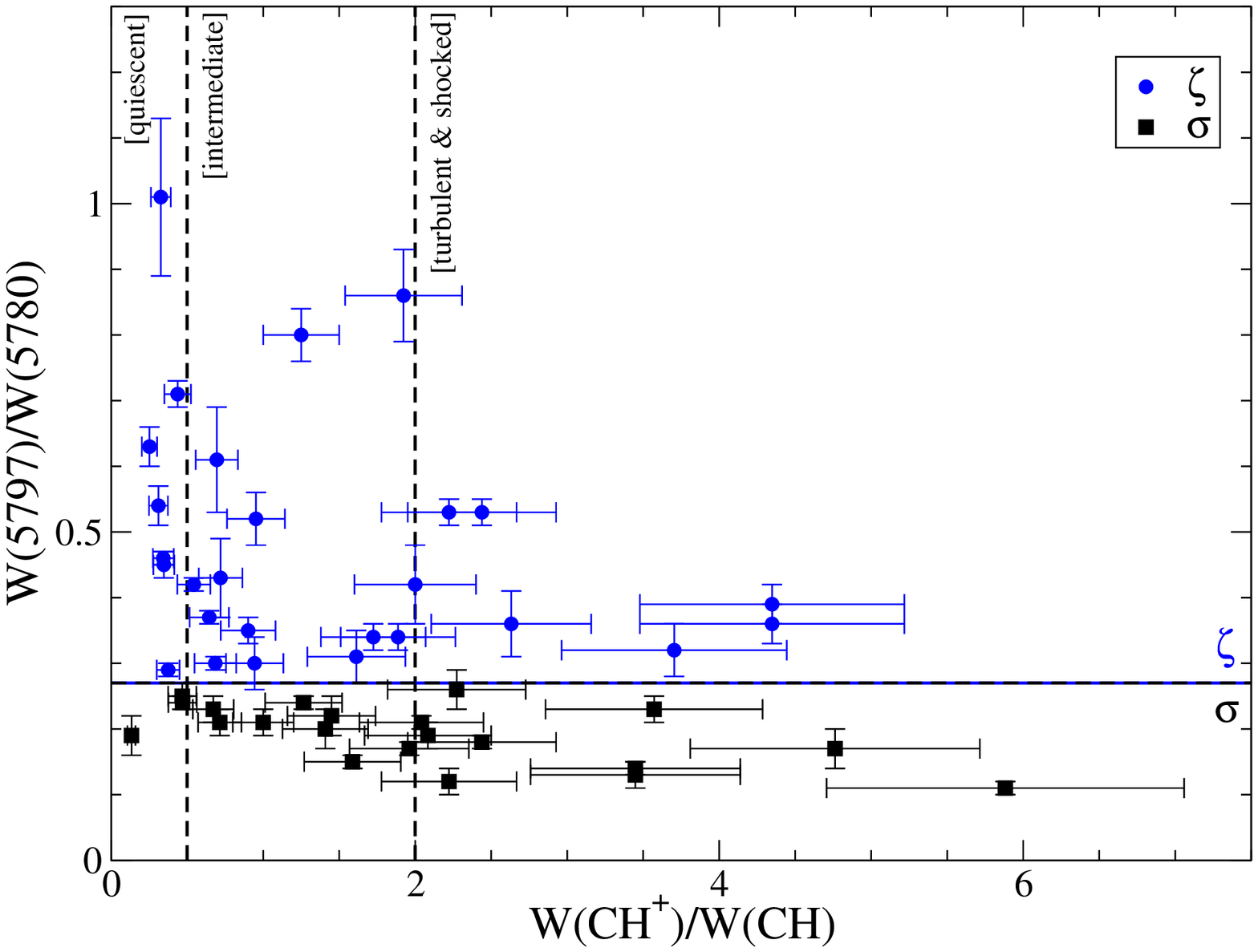}
   \caption{$W_\lambda(5797)/W_\lambda(5780)$ vs. $W$(CH)/$W$(CH$^+$). There is a tentative trend
   for decreasing DIB ratio with increasing CH$^+$/CH ratio.}     
   \label{fig:DIBratio-CHpCHratio}
\end{figure}

In Fig.~\ref{fig:DIBratio-CHpCHratio} the  $W_\lambda(5797)/W_\lambda(5780)$ ratio is plotted against the
$W$(CH$^+$)/$W$(CH) ratio. Again, turbulent and quiescent ISM are indicated. This plot reveals no marked
correlation between these ratios. Tentatively, it shows a high DIB ratio (\emph{i.e.} $\zeta$-type) for
quiescent clouds and a low DIB ratio (\emph{i.e.} $\sigma$-type) for turbulent clouds, which supports the
idea that both CH and the $\lambda$5797 DIB trace moderately dense regions, while CH$^+$ traces the cloud
edges and inter-cloud regions. Although both the $\lambda$5780 DIB and CH$^+$ are related to the outer edges
of diffuse clouds they do not reveal a strong correlation (although there appears to be a positive trend
when considering only $\sigma$-type sightlines) and thus appear to react to changes in  the ISRF
differently. In agreement with \citet{2004A&A...414..949W} and Sect.~\ref{subsec:skineffect} it seems that
the Upp Sco\ region is somewhat turbulent, but is absent of extreme shocks. However, uncertainties in
$W$(CH) and $W$(CH$^+$) are too large to draw firm conclusions.

\begin{figure*}[t!]
\centering
   \includegraphics[width=1.5\columnwidth]{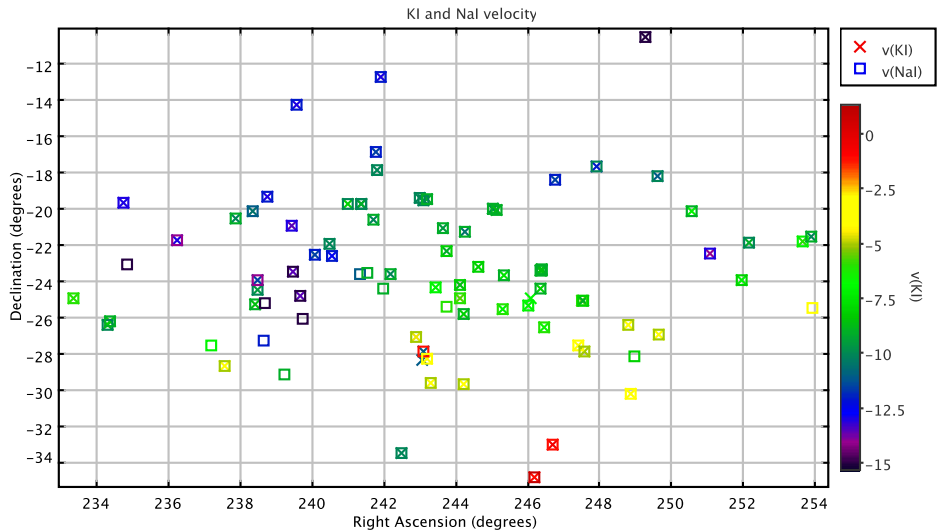}
   \caption{The heliocentric radial peak velocity for \ion{K}{i} and  \ion{Na}{i} are shown in a 
   sky coordinate (right ascension and declination) plot.
   The radial velocity of the gas (in the diffuse ISM) in Upp Sco\ is highest in the upper-left 
   corner, the material is approaching with velocities up to 15~km~s$^{-1}$\ (along the line-of-sight). 
   The gas in the lower-right corner has the lowest velocity (with respect to the Sun). 
   (See on-line electronic version for colour figure).}
   \label{fig:radvel}
\end{figure*}

\subsection{The ISM velocity distribution}\label{sec:velocity}

We measured the heliocentric radial velocities for both atomic and molecular lines towards Upp Sco. The
strongest interstellar lines are observed at a radial velocity of about -9~km~s$^{-1}$, and a weaker absorption
component is detected at about -22~km~s$^{-1}$\  (\emph{e.g.} Fig.~\ref{fig:ex_velprof}). This is fully in-line
with recent results reported by \citet{2008ApJ...679..512S} who studied 16 lines-of-sight towards the
Upp Sco\ region.  The velocity component of -9~km~s$^{-1}$\ corresponds to the patchy dust sheet at a distance of
110 -- 150~pc, which is loosely connected to the $\rho$\,Oph dense/molecular cloud at 122~pc (see also
Sect.~\ref{sec:reddening}). The other, weaker velocity component ($v = -22$~km~s$^{-1}$) is linked to the tenuous
low density dust layer at $\sim$50~pc. The average velocity difference between CH and CH$^+$ is 0.3~km~s$^{-1}$\ 
(for the \ion{K}{i} doublet lines the average velocity difference is 0.07~km~s$^{-1}$). Within the limits of the
observations we confirm that there is no evidence for a CH-CH$^+$ velocity offset in Upp Sco\ which is 
predicted by models for regions with strong shocks.

The relatively broad DIBs preclude a detailed radial velocity determination (for the obtained  S/N,
resolving power, and spectral quality). However, first order estimates (for the sightlines with strong DIBs)
show no  systematic differences between atomic, molecular, and DIB velocities. For the relatively narrow
5797~\AA\ DIB we measure radial velocities roughly between -20 to 0~km~s$^{-1}$.

This large set of radial velocity information allows us to map the velocity of the diffuse ISM clouds in
front of the observed stars, very similar to the work by \citet{2008ApJ...679..512S}. In
Fig.~\ref{fig:radvel} we show the color-coded interstellar radial velocity of \ion{K}{i} and \ion{Na}{i} as
a function of declination and right ascension. Assuming that indeed the observed sightlines probe different
parts of a single dust-sheet, it thus appears that this sheet is moving differentially. The upper-left
corner of the sheet (where most young OB stars are) is moving towards us while the lower-right corner
remains stationary  (ignoring any velocity component perpendicular to the line-of-sight). The
three-dimensional kinematic motions can not be fully reconstructed with these data.

\onltab{5}{
\begin{table*}[bh!]
\begin{center}
\caption{Central radial velocities (in the heliocentric frame) for selected atomic and molecular lines.
With an instrumental profile FWHM of about 6--7~km~s$^{-1}$\ and pixels of 2~km~s$^{-1}$, the accuracy of the
measured radial velocity is 1--2~km~s$^{-1}$.}
\label{tb:velocities}
%\resizebox{!}{0.325\textheight}{%%
\begin{tabular}{lllllllll}\hline\hline
         & \multicolumn{8}{c}{Radial velocities (km~s$^{-1}$)}\\ 
HD	 & $v_{CH^+}$ & $v_{CH}$	& $v_{CN}$   & $v_{Ca\,I}$ & $v_{K\,I}$ &          & \multicolumn{2}{l}{$v_{Na\,I}$}\\\hline
138503   & -6	      &    -6	        &    	     &    	   &	-6	&    -6    &      -6    &		     \\ 
139094   & -8	      &    -9	   	&     -11    &     -9	   &	-9	&    -8    &      -10   &		     \\ 
139160   & -7	      &    -7	        &    	     &       	   &	-8	&    -8    &      -9    &		     \\ 
139486   &   	      &     	        &    	     &       	   &	-13	&    -13   &      -13   &		     \\ 
139518   &  	      &     	        &    	     &       	   &	 	&          &      -15   & 		     \\ %very weak
140543   & -10        &     	        &    	     &       	   &	-13	&    -13   &      -14   &		     \\ %very broad KI!
141180   &  	      &     	        &    	     &       	   &	 	&          &      -7    & -17, +11   	     \\ 
141444   &  	      &     	        &    	     &       	   &	-4	&    -4    &      -5    &		     \\ 
141774   & -10        &    -9	        &    	     &       	   &	-10	&    -10   &      -9    & -29, -18, +16      \\ 
142096   &  	      &     	        &            &    	   &	-11	&    -10   &      -11   &		     \\ 
142114   &  	      &    -8           &    	     &       	   &	-9   	&    -8    &      -9    & -14   	     \\ 
142165   &  	      &     	        &    	     &       	   &	-10	&    -9    &      -10   &		     \\ 
142184   &  	      &    -12          &    	     &       	   &	-11	&    -12   &      -14   & -33		     \\%w/ weaker comp. Between -10 and -20
142250   &  	      &     	        &    	     &       	   &	 	&          &      -12   &		     \\ 
142301   &  	      &     	        &    	     &       	   &	 	&          &      -15   &		     \\%possible weak comp. At -14
142378   & -14        &     	        &    	     &       	   &	-13	&    -13   &      -12   &		     \\ 
142669   &  	      &     	        &    	     &       	   &	 	&          &      -9    & -21   	     \\%NaI: Both -20 and -10 comp. Very weak 
142883   & -15        &    -14          &    	     &       	   &	-14	&    -13   &      -13   &		     \\ 
142884   & -19        &    -18 	        &    	     &       	   &	-14	&    -13   &      -15   &		     \\%KI weak extended component upt to -20km/s
142983   &  	      &     	        &    	     &      	   &	-12	&    -14   &      -12   & -24                \\%NaI feature at -63 km/s???? 
142990   & -17        &    -15          &    	     &       	   &	-13	&    -14   &      -15   &		     \\%higher delta CH+ and other lines 
143018   &  	      &     	        &    	     &       	   &	 	&          &      -15   &		     \\%possible very weak KI at -14
143275   & -13        &      	        &    	     &       	   &	-12	&    -11   &      -12   &		     \\ 
143567   &  	      &     	        &    	     &       	   &	-10	&    -10   &      -10	& -22		     \\ 
143600   &  	      &      	        &    	     &       	   &	-12	&    -13   &      -12   & -23   	     \\ 
143956   &            &                 &    	     &       	   &	-8	&    -8    &      -10   & -22   	     \\ 
144175   &  	      &     	        &    	     &      	   &	 	&          &      -11   & -21   	     \\ 
144217   & -11        &    -10          &    	     &     -10     &	-10	&    -10   &      -10   & -23   	     \\ 
144218   & -11        &                 &    	     &     -10	   &	-10	&    -9    &      -10   & -23   	     \\ 
144334   &  	      &     	        &    	     &       	   &	 	&          &      -9    & -22   	     \\ %NaI at -20 2xstronger than -10 comp.
144470   & -11        &    -10          &    	     &      	   &	-10	&    -10   &      -9    & -23   	     \\ 
144569   & -12        &     	        &    	     &       	   &	-12	&    -11   &      -12   & -27   	     \\ 
144586   & -11        &    -9           &    	     &       	   &	-10	&    -10   &      -10   & -26   	     \\ 
144661   &  	      &     	        &    	     &       	   &	 	&    	   &      -9    & -22   	     \\%NaI at -20 only ½ of -10 comp.
144708   &  	      &     	        &    	     &       	   &	-13	&    -13   &      -12   &		     \\ 
144844   &            &     	        &            &       	   &	-10	&    -10   &      -9    & -23   	     \\
144987   &  	      &     	        &    	     &       	   &	 	&    -10   &      -10   & -22	     	     \\%(very weak KI); hardly no -10 NaI comp.
145353   & -2	      &    -2	        &    	     &     -7	   &	-4	&    -5    &      -5    & -21   	     \\ 
145482   &  	      &     	        &    	     &       	   &	 	&    -11   &      -1    & -20   	     \\ %-10 NaI weak, -20 also weak but 2x stronger still than -10
145483   &  	      &     	        &            &       	   &	 	&    -11   &    	& -25   	     \\ %no NaI at -10, only -25 component!
145502   & -11        &    -9	        &    	     &     -10     &	-10	&    -10   &      -10   & -25   	     \\ 
145554   & -11        &      	        &    	     &     -13     &	-10	&    -10   &      -10   & -25   	     \\ 
145556   & -3	      &    -4	        &            &       	   &	-4	&    -4    &      -4    & -25   	     \\ 
\hline	
\end{tabular}
%}																																						  																																							  
\end{center}																																							  
\end{table*}
}

%\addtocounter{table}{-1}

\onltab{5}{
\begin{table*}[bh!]
\begin{center}
\caption{continued.}
\label{tb:velocities2}
\begin{tabular}{lllllllll}\hline\hline%
         & \multicolumn{8}{c}{Radial heliocentric velocities (km~s$^{-1}$)}\\ %
HD	 & $v_{CH^+}$    & $v_{CH}$	& $v_{CN}$   & $v_{Ca\,I}$ & $v_{K\,I}$ &          & \multicolumn{2}{l}{$v_{Na\,I}$}\\\hline
145631   & -8	 	 & -8	    	& 	      &   -9	    &   -9         &  -9   &     -9  	&    -25   	  \\			   
145657   & -6	 	 & -4	    	&             &   -3	    &   -3         &  -4   &     -5  	&    -19  	  \\  
145792   & -5	 	 &	    	&             &             &   -7         &  -7   &     -7  	&    -22  	  \\  
145964   &  	 	 &	    	&             &             &   -9         &  -10  &     -9  	&    -22  	  \\ %very weak KI; -10&-20 comp.equal strength
146001   & -5	 	 &	    	&             &             &   -8?        &  -9?  &     -6  	&    -21  	  \\ %KI?
146029   &  	 	 &	    	&             &             &   -9         &  -9   &     -8  	&    -23  	  \\  
146284   & -8	 	 & -8	    	&             &   -8	    &   -9         &  -8   &     -8  	&    -24  	  \\  
146285   & -6	 	 & -5	    	&             &             &   -5         &  -6   &     -6  	&    -22   	  \\  
146331   & -7	 	 & -7	    	&             &   -9	    &   -8         &  -8   &     -9  	&   		  \\  
146332   & -1	 	 & -3	    	&   -4        &   -4	    &   -4         &  -3   &     -5  	&    +14  	  \\  
146416   &  	 	 &	    	&             &             &   -11        &  -10  &     -9  	&    -25  	  \\ %very weak KI
146706   & -8	 	 & -7	    	&             &   -7	    &   -8         &  -8   &     -8  	&   		  \\  
147009   & -9	 	 & -9	    	&             &             &   -9         &  -9   &     -9  	&   		  \\  
147010   & -10   	 & -10      	&             &             &   -11        &  -9   &     -9  	&   		  \\  
147103   & -8	 	 & -8	    	&   -8        &             &   -9         &  -8   &     -9  	&    -24   	  \\  
147165   & -7	 	 & -7	    	&             &             &   -7         &  -7   &     -6  	&    		  \\ %NaI slightly broadened  
147196   & -7	 	 & -7	    	&   -7        &             &   -7         &  -8   &     -9  	&   		  \\  
147648   & -7	 	 & -7	    	&             &             &   -7         &  -7   &     -7   	&   		  \\  
147683   & -1	 	 &  0	    	&   -1        &   -1	    &   -1         &  -1   &     +1 	&    -18  	  \\  
147701   & -7	 	 & -6	    	&   -6        &   -6	    &   -7         &  -7   &    	&   		  \\  
147888   & -9	 	 & -8	    	&   -9        &   -9	    &   -9         &  -8   &     -9  	&   		  \\  
147889   & -8	 	 & -7	    	&   -7        &   -8	    &   -8         &  -8   &     -9  	&   		  \\  
147932   & -7	 	 & -7	    	&   -8        &   -8	    &   -9         &  -8   &     -9  	&   		  \\  
147933   & -8     	 & -8		&   -9	      &   -9	    &   -9	   & -8	   &     -9  	&   		  \\
147955   & -6	 	 & -5	    	&             &             &   -7         &  -6   &     -6   	&   		  \\  
148041   & -1	 	 & -1	    	&             &   -4	    &   -1         &   0   &     -1   	&           	  \\ %broad NaI 
148184   & -11   	 & -11      	&   -11       &   -12	    &   -11        &  -12  &     -12	&                 \\ %broad wing in NaI 
148499   & -4	 	 & -4	    	&             &             &   -4         &  -4   &     -4  	&    +12  	  \\  
148579   & -5	 	 & -5	    	&             &             &   -6         &  -5   &     -6  	&   		  \\  
148594   & -5	 	 & -4	    	&             &   -6	    &   -5         &  -5   &     -5  	&   		  \\  
148605   & -8	 	 &	    	&             &   -8	    &   -9         &	   &     -8  	&   		  \\
148860   &	 	 & -12      	&             &   -7	    &   -12        &  -12  &     -10    &    -55  	  \\  
%%149363          									      		   
149367   & -4	 	 & -4     	&             &             &   -4         &  -4   &     -5  	&   		  \\  
149387   & -3	 	 & -3	    	&             &   -2	    &   -4         &  -3   &     -4  	&   		  \\  
149438   &  	 	 &	    	&             &             &              &  -8? ?&     -8  	&   		  \\  
149757   & -15   	 & -15      	&   -15       &             &   -15        &  -15  &     -15    &    -27  	  \\  
149883   & -5	 	 & -4	    	&             &   -6	    &   -4         &  -4   &     -5  	&   		  \\  
149914   & -12   	 & -11      	&   -12       &   -10	    &   -11        &  -11  &     -10 	&   		  \\  
150514   & -8    	 &	    	&             &             &   -7         &  -7   &     -9  	&    -27, +7      \\  
150814   & -12   	 &	    	&             &             &   -14        &  -14  &     -13    &     +2  	  \\  
151012   &  	 	 & -20?     	&             &             &   -19?       &  -22? &       	&   		  \\  
151346   & -7	 	 & -7	    	&   -7        &             &   -7         &  -7   &     -8  	&   		  \\  
151496   & -9	 	 & -8	    	&             &             &   -9         &  -9   &     -10 	&   		  \\  
%%151831  & -4	 	 &	    	&             &             &   -3         &  -3   &     -5  	&    -15  	  \\  
152516   & -7	 	 &   	    	&             &             &   -6         &  -6   &     -7  	&   		  \\  
152655   & -10   	 & -9	    	&             &   -8	    &   -10        &  -10  &     -9  	&   		  \\  
152657   &  	 	 &	    	&             &    	    &	           &	   &     -4  	&   		  \\
%%153919  & 	 	 &	    	&	      & 	    &	           &	   &  	   	&   		  \\
\hline
\end{tabular}
\end{center}
\end{table*}
}

\subsection{The interstellar radiation field strength}\label{subsec:isrf}

In order to estimate the effective interstellar radiation field for each interstellar cloud probed by the 
Upp Sco\ stars we constructed a simplified model of a sheet of dust irradiated by several OB-type stars. 
The thickness of the dust sheet will roughly depend on the volume and column density of \ion{H}{i}.  For
$n_H = 100$~cm$^{-3}$ and N(H) = 5 $\times$ 10$^{21}$~cm$^{-2}$ (\Ebv $\sim$1~mag) the thickness is
$\sim$16~pc. The dust sheet can be represented by a homogeneous thin slab at a distance of 120~pc and a
thickness of 20~pc.  In this way, the distribution of individual clouds can be represented by a single
sheet, which is a valid assumption because  (1) $\tau_{UV} > 1$ so photons are scattered frequently enough
to loose most of their directional memory and (2) the distribution of individual clouds has a surface area
covering factor larger than unity. Property (1) assures that the radiation field strength $I_{\rm UV}$ is
the roughly isotropic flux that impinges on the individual  clouds making up the sheet.  Aspect (2) assures
that each line of sight through the representative sheet has approximately the same {\it total} extinction, 
relevant for the attenuation of $I_{\rm UV}$. This ensures that the radiative transfer problem to be solved is that
for a slab geometry. As eight OB stars contribute over 90\% of the ISRF in this region these are included as
the only source of the ionising radiation (\citealt{2005ApJ...633..257S}). These stars illuminate the
interstellar cloud from behind.  In this particular model one star, \object{HD 143275} (B0.3IV), dominates the
effective ISRF, even while $\zeta$ Oph (\object{HD 149757}) has the earliest spectral type.

The radiative transfer model (\citealt{1995PhDT.......306S}; \citealt{1996A&A...307..271S}) takes into
account both absorption and scattering.  The effective optical depth $\tau_V$ is computed from the observed
\Ebv\ and subsequently $e^{-\tau_{\nu}}$  is multiplied by the  individual stellar fluxes for an appropriate
extinction curve for standard Milky Way dust (with $R_V$ taken either as 3.1 or 4). The latter case is also
considered since the Upp Sco\ region contains sightlines with high $R_V$ values for the dust extinction 
(see Sect.~\ref{sec:Rv}). In addition, this method is also applied to compute the amount of back-scattered
radiation. 

For a sheet geometry, it is possible to express $I_{\rm UV}$\  (in units of the Draine field)  as a function of
$R_V$ and cloud position $R$.

\begin{equation}
I_{\rm UV} = 4.7 [(R+a)/a]^2\ {\rm exp} \left[ -6.9 \left(\frac{\Ebv}{0.20}\right) \left(\frac{R_V}{3.1}\right) (20-R) \cdot 10^{-2} \right] \label{eq:Iuv}
\end{equation}

\noindent In this parametrisation of the radiative transfer grid the parameter $a$ depends on the distance
to \object{HD 143275} ($a=3.4$ for a distance of 123~pc). \Ebv\ is the individual extinction of the cloud. The cloud
position $R$ is set between 0 and 20~pc (0~pc being the sheet edge closest to the observer).  Hence, $I_{\rm UV}$\
$\sim$1 at the shielded edge, close to the mean  Galactic value, and $I_{\rm UV}$\ $\sim$200 at the bright edge. 
This latter value is relatively high and depends on the distance to \object{HD 143275}. Placing this star 1~pc
further away results in $a=5.1$ and $I_{\rm UV}$\ decreases by a factor two. With the nominal values, $I_{\rm UV}$\
$\leq$~20 for about one third of the sheet structure. Eq.~\ref{eq:Iuv} allows a range of impinging field
strengths which has subsequently been used to set up a grid of chemical models (including non-thermal
production of CH; \citealt{1995PhDT.......306S}) for a given measured extinction, to determine which model
clouds yield  the best match  to the available data.  Thus effectively, for a given \Ebv\ we extracted the
$I_{\rm UV}$\ reproducing best the observed CH and CN, where the derived $I_{\rm UV}$\ is  also constrained by the observed
upper limits for CN. In this, CH$^+$ has been excluded because it is well known that canonical chemical
models under produce its abundance by about two orders of magnitude.  Turbulent dissipation and/or shocks
are likely needed in the (endothermic) formation of CH$^+$. To first order, $I_{\rm UV}$/$n_{\rm H}$ is the
controlling parameter for the chemical and thermal balance. So an increase in density by a factor of 2
corresponds to an increase in $I_{\rm UV}$ by a factor of 2. Due to the limited information available for
each line-of-sight we adopted a generic density $n_H$ = 300~cm$^{-3}$. This is representative of a cloud
that is slightly denser than the ambient medium, the dust sheet, in which it is embedded. In other words,
the clouds do not fill the region and the sheet is seen as a patchy complex of individual, but connected, 
clouds scattered in distance. Previous detailed modeling of the Upp Sco\ line-of-sight towards \object{HD 147889}
shows that this is likely a conservative lower limit  for sightlines probing the denser parts of the
$\rho$\,Oph cloud. With detailed modeling, including observational constraints for additional species,
\citet{2005A&A...432..515R} found a density  of 1200~cm$^{-3}$ and an $I_{\rm UV}$\ $\sim$10 for this line-of-sight.
Increasing the input density by a factor of four in the model above for \object{HD 147899} would give a revised
$I_{\rm UV}$\ of 6, already in  better agreement with the detailed analysis. Also, \citet{2003ApJ...589..319Z} found
\ion{C}{i} densities between 100 and 300~cm$^{-3}$ for \object{HD 143275} and \object{HD 147165}, but lower values,
$\sim$50 to $\sim$200, for \object{HD 144470} and \object{HD 144217}. However, as \ion{C}{i} traces the purely atomic phase
of clouds, it is likely that this yields lower densities than for the  molecular/shielded parts as traced by
\emph{e.g.}  C$_2$, CN and CH. Clearly, the simplifications introduced in the model presented in this work do not
fully incorporate all the intricacies of  a full-fledged analysis. However, the strength of this model,
which relies only on the CH and CN abundance, is in  giving statistically relevant predictions of the ISRF
for a larger dataset for which only limited information is available. For accurate equivalent width
measurements of both CH and CN the computed $I_{\rm UV}$ has an uncertainty of approximately 25\%, not
including any unknown systematic effects. Uncertainties in the density, $D = dn/n$, propagate into $I_{\rm UV}$\ as
$D^{1/2}$. If only CH is detected the value of $I_{\rm UV}$ should therefore be considered indicative only
(like a model dependent lower limit). The resulting interstellar radiation field strengths are presented in
Table~\ref{tb:DIB-data}.

For diffuse clouds the ISRF can also be estimated from steady-state gas phase chemistry
(see \emph{e.g.}  \citealt{2006ApJS..165..138W}, \citealt{2006ApJ...649..788R}):
\begin{equation}
I_{\rm UV}/n_H \propto\ \frac{N(CH^+)}{N(CH)}\ f_{\mathrm{H}_2}, \label{eq:Iuv-nH}
\end{equation} %%
which is valid for non-thermal CH production and for small values of $f_{\mathrm{H}_2}$.
On the other hand, rotational excitation modeling of H$_2$ gives (see \emph{e.g.} 
\citet{1975ApJ...197..581J,1987ApJ...322..412B,2007ApJ...655..940L}):
\begin{equation}
\log(n_H / I_{\rm UV}) \propto \log\ f \label{eq:Iuv-nH2} %%- \log(N(H)) 
\end{equation}
which is appropriate for $n$(H$_2$) $<< n(H) \approx n_H$ (but the linearity holds also for higher N(H$_2$)
(\emph{e.g.} \citealt{2007ApJ...655..940L}). Note that Eqs.~\ref{eq:Iuv-nH} and~\ref{eq:Iuv-nH2} show an opposite
dependence of $I_{\rm UV}$/$n_H$ on $f_{\mathrm{H}_2}$.

\begin{figure}[t!]
\centering
   \includegraphics[angle=-90,width=1.\columnwidth,clip]{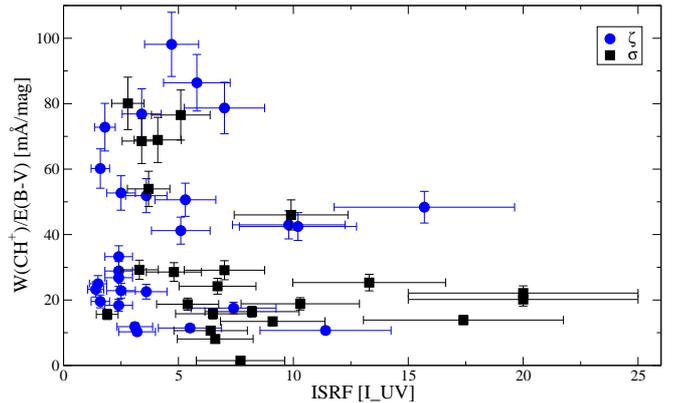}
   \caption{%%
   The CH$^+$ line strength per unit reddening is shown as a function of the ISRF strength, $I_{\rm UV}$.
   Highest values for $I_{\rm UV}$\ are found for low CH$^+$ abundances per unit reddening.
   Because N(CH) $\propto$ N(H$_2$) and \Ebv\ $\propto$ \ion{H}{i} Eq.~\ref{eq:Iuv-nH} gives
   $I_{\rm UV}$/$n_H$ $\propto$ N(CH$^+$)/\Ebv.
   %The observed trend is consistent - though not conclusive - with non-thermal processes 
   %playing a role in the production of CH$^+$. 
   There is some evidence for enhanced CH$^+$ production (UV pumping?) in clouds with moderate $I_{\rm UV}$ $\sim$ 5.}
   \label{fig:iuv_chp}
\end{figure}

\begin{figure}[t!]
\centering
   \includegraphics[angle=-90,width=1.\columnwidth,clip]{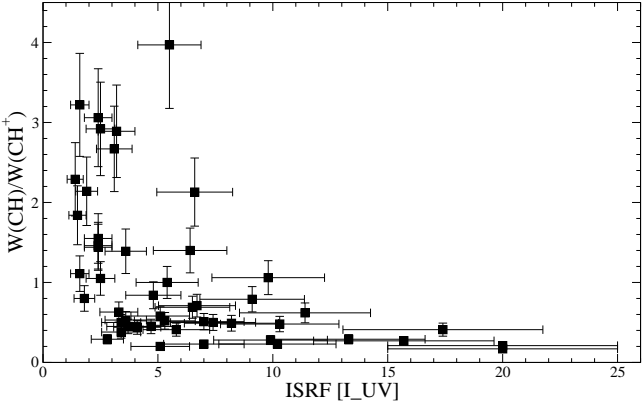}
   \caption{%%
   The CH over CH$^+$ line strength ratio is plotted as a function of the ISRF strength, $I_{\rm UV}$.
   There is a  drop in this ratio (\emph{i.e.} enhanced CH$^+$ or reduced CH production) for stronger 
   radiation fields ($I_{\rm UV}$).
   This trend is consistent with non-thermal production of CH, otherwise no trend would be expected.
   }
   \label{fig:IuvCHCHp}
\end{figure}

In addition, UV pumping can produce excited H$_2^*$  leading to an enhancement in the production of CH$^+$
via C$^+$ + H$_2^*$ $\rightarrow$ CH$^+$ + H.  Therefore, we compare the independently obtained values for
CH$^+$ and $I_{\rm UV}$, as well as CH/CH$^+$ and $I_{\rm UV}$\ to investigate whether this process is important.
Fig.~\ref{fig:iuv_chp} illustrates that sightlines with high CH$^+$ abundances show only moderate values for
$I_{\rm UV}$\ (\emph{i.e.} less than 10),  while sightlines with high $I_{\rm UV}$\  (\emph{i.e.} larger than $\sim$10) all
show low-to-normal CH$^+$ abundances. From this relation it appears that a strong ISRF ($I_{\rm UV}$ $>$ 10) does
not lead to enhanced CH$^+$ production,  possibly because the molecular hydrogen abundance of these
sightlines is too low. However, W(CH$^+$)/\Ebv\ peaks at $I_{\rm UV}$ = 5 which may reveal a delicate balance for
the presence of UV pumping at intermediate $I_{\rm UV}$\ and moderate $f_{\mathrm{H}_2}$. On the other hand,
Fig.~\ref{fig:IuvCHCHp} shows that the CH/CH$^+$ ratio drops rapidly for $I_{\rm UV}$ $>$ 4. Thus despite a lower
total CH$^+$ abundance (per unit reddening) for higher values of $I_{\rm UV}$\ the relative production of CH$^+$
with respect to CH increases. This could be due to more efficient production of CH$^+$ or less efficient
formation of CH in these low density, strongly UV exposed environments.  The latter is indeed expected as
N(CH) correlates with N(H$_2$) whose relative presence reduces also with increasing $I_{\rm UV}$\  (see below). UV
pumping may thus contribute significantly to CH$^+$ formation only in diffuse clouds with sufficient
abundance  of both H$_2$ and UV photons. Other mechanisms, like turbulent dissipation of mechanical energy,
could also be important for CH$^+$ formation in this region. 

\begin{figure}[t!]
\centering
   \includegraphics[angle=-90,width=1.\columnwidth,clip]{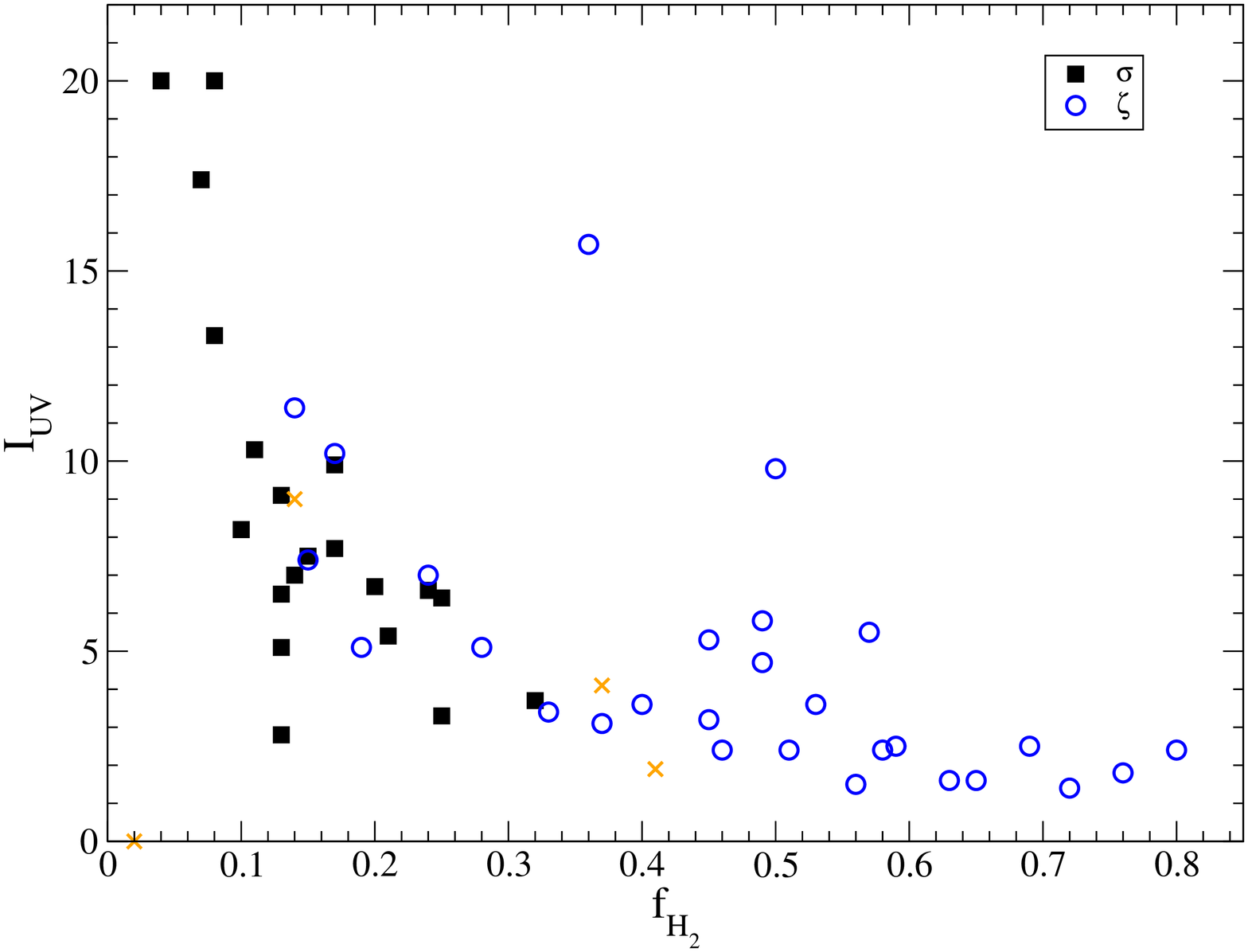}
   \includegraphics[angle=-90,width=1.\columnwidth,clip]{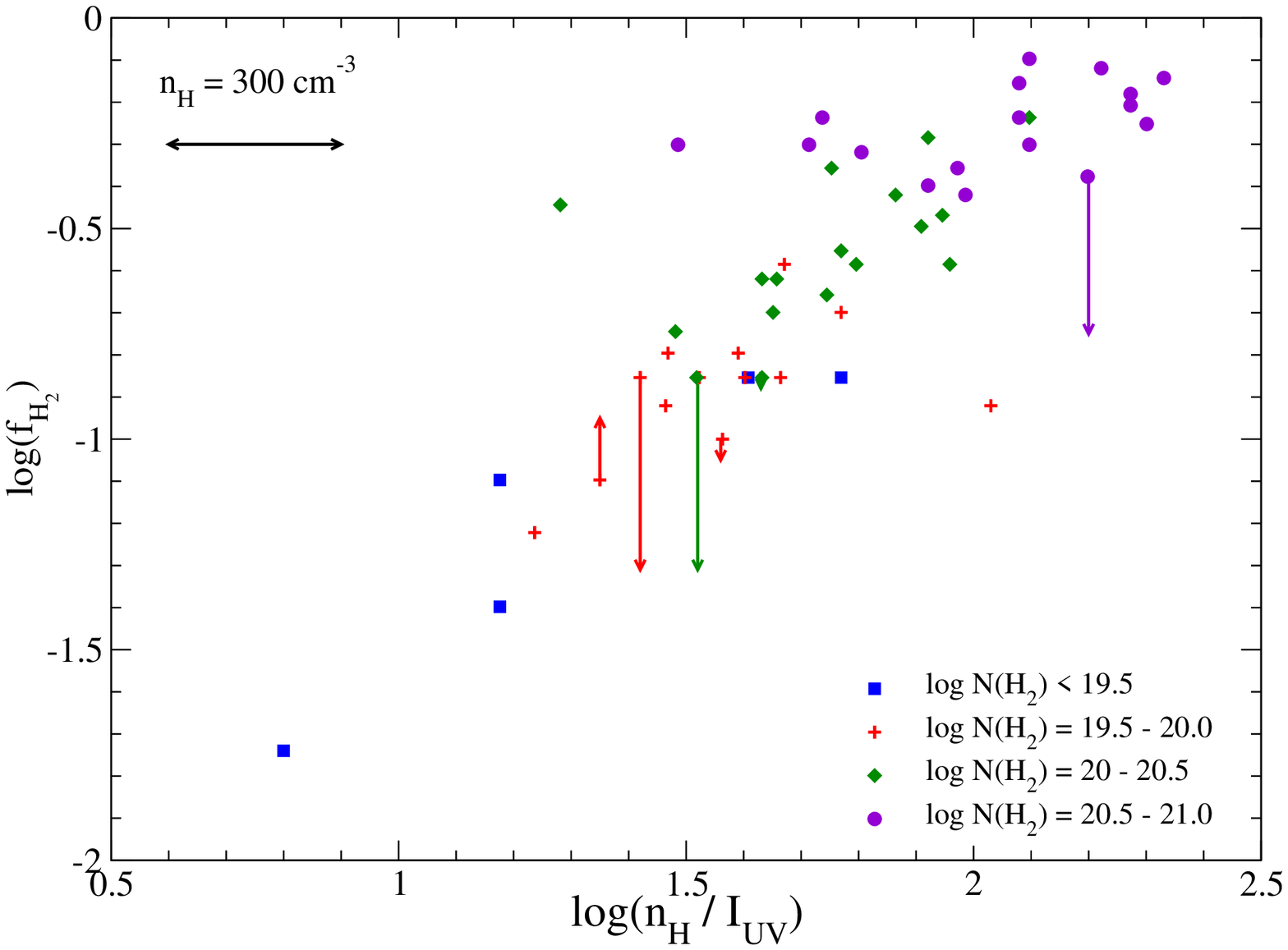}
   \caption{%%
   (top) The model $I_{\rm UV}$\ is plotted as a function of inferred $f_{\mathrm{H}_2}$.  (bottom) Alternatively,
   the molecular hydrogen fraction can be plotted as a function of  the ratio of the hydrogen density of the
   UV radiation field strength, for our general value  of $n_H$ = 300.0~cm$^{-3}$. As expected from
   \citet{1987ApJ...322..412B}, log($f_{\mathrm{H}_2}$) is directly proportional to log($n_H$/$I_{\rm UV}$), where
   the intercept of this relation depends on the total $H_2$ column density (see also Fig. 2 in
   \citet{2007ApJ...655..940L}). Ranges for inferred N($H_2$) are indicated by different symbols. The
   horizontal arrow gives the change in log($n_H$/$I_{\rm UV}$) for an increase or decrease of $n_H$ by a factor of
   2. The vertical arrows on the data points indicate the correction of the inferred fraction to the
   directly observed fraction. Note that the inferred molecular fraction, $f_{\mathrm{H}_2}$, directly
   depends (non-linearly) on the ratio of $W$(CH) over $W$(5780), and the $I_{\rm UV}$\ depends also on CH, as well
   as CN and \Ebv.
   } 
   \label{fig:ratio_iuv}
\end{figure}

Fig.~\ref{fig:ratio_iuv} (top panel) reveals an evident inverse relation between the strength of the ISRF,
$I_{\rm UV}$,  and the molecular hydrogen fraction, $f_{\mathrm{H}_2}$. %% (see Sect.~\ref{sec:dibs-diatomics}).
This effect of lower $I_{\rm UV}$\ for interstellar clouds with higher molecular fractions (and thus more efficient
shielding of the UV radiation)  is expected from Eq.~\ref{eq:Iuv-nH2}. This figure also illustrates the
general trend that the $\sigma$-type clouds have a higher $I_{\rm UV}$\ and a  lower molecular content
$f_{\mathrm{H}_2}$, while $\zeta$-type sightlines have a higher molecular content and are exposed to a
weaker ISRF.  Note that although a few sightlines with low $f_{\mathrm{H}_2}$ and higher $I_{\rm UV}$\ values were
classified as $\zeta$-type based on the observed  $W$(5797)/$W$(5780) ratio, there is a clear separation -
based on physical conditions - between the $\sigma$- and $\zeta$-type sightlines.
\citet{2004A&A...414..949W} also show a similar distinction between $\sigma$ and $\zeta$ at
$f_{\mathrm{H}_2} \sim 0.4$ (although their  $\sigma$-$\zeta$ classification is based on central depth
ratios resulting in a slightly different division between the two types).

The linear relation between log($f_{\mathrm{H}_2}$) and log($n_H$/$I_{\rm UV}$) (\emph{i.e}\ Eq.~\ref{eq:Iuv-nH2}) in
Fig.~\ref{fig:ratio_iuv} (bottom panel)  can be compared directly to Fig.~2 in \citet{2007ApJ...655..940L}
showing indeed a close relation between the molecular fraction and the ratio of hydrogen density over
radiation field strength, $n_H$/$I_{\rm UV}$.  This relation is sensitive to the total $H_2$ column density but does
not depend strongly on the hydrogen particle density, $n_H$.  Nonetheless, knowledge of the latter value
(either estimated or derived from complementary data) is required to derive $I_{\rm UV}$.  Non-thermal H$_2$
excitation due to turbulence can mimic UV pumping and thus alter the relation between the model $I_{\rm UV}$\ and 
the observed $f_{\mathrm{H}_2}$ (\citealt{1995PhDT.......306S}).  Thus, in addition to deriving the
effective ISRF strength in interstellar clouds (averaged along the line-of-sight) in Upp Sco\ with the model
above we can use measurements of the $\lambda$5780 DIB, CH and CN absorption line strengths to estimate
N(H$_2$), N(\ion{H}{i}), and the molecular hydrogen fraction, $f_{\mathrm{H}_2}$.

\begin{figure}[t!]
\centering
   \includegraphics[angle=-90,width=1.\columnwidth,clip]{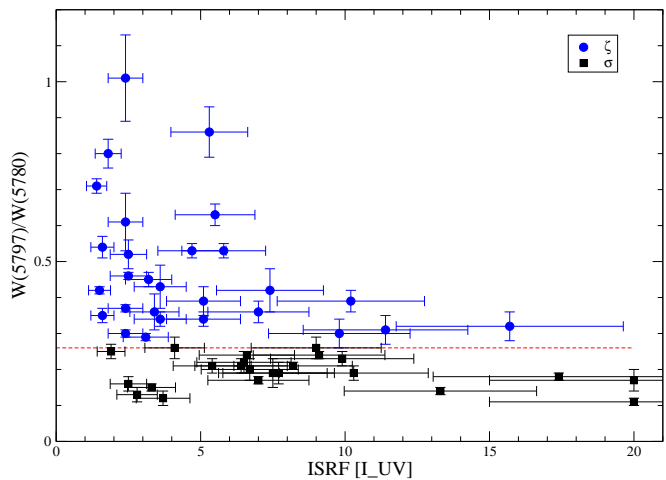}
   \caption{%%
   The $W$(5797)/$W$(5780) DIB ratio is plotted against the ISRF, $I_{\rm UV}$.
   High DIB ratios (\emph{i.e.} $\zeta$) correspond to a lower $I_{\rm UV}$.
   $\sigma$-type clouds show a similar range in $I_{\rm UV}$.}
   \label{fig:DIBratio_Iuv}
\end{figure}

Figure~\ref{fig:DIBratio_Iuv} shows the dependence of the $W$(5797)/$W$(5780) ratio on the ISRF, $I_{\rm UV}$. In
general, sightlines with low $I_{\rm UV}$\ values are $\zeta$-type sightlines for which also CN has been detected.
And lines-of-sight for which we find high values of $I_{\rm UV}$\ have, on average, lower values for the
$W$(5797)/$W$(5780) ratio. The few $\zeta$-type clouds with high $I_{\rm UV}$\ have in fact DIB ratios that are
close to the average ratio used to discriminate between $\sigma$  and $\zeta$-type environments. On the
other end, there are also a few $\sigma$-type clouds associated with a weak ISRF.  We recall that a change
in the density will give an equal change of $I_{\rm UV}$, which could consequently shift individual sightlines to
either lower or higher $I_{\rm UV}$, thus introducing additional scatter.  Fig.~\ref{fig:DIBratio_Iuv} is consistent
with a 5797~\AA\ DIB carrier which requires sufficient protection from UV radiation in order to survive in
the diffuse ISM, while conversely the 5780~\AA\ DIB carrier requires UV photons for excitation (possibly
because the carrier needs to be ionised in order to absorb at 5780~\AA). At this point it is important to
note that the sightlines with higher inferred $I_{\rm UV}$\ all rely on CH measurements only and should  therefore
be considered indicative. Also, the average $I_{\rm UV}$\ values for respectively $\sigma$ and $\zeta$ type
sightlines are within 1$\sigma$ of each other (where the mean of $I_{\rm UV}$$_{\sigma}$ is two times the mean of
$I_{\rm UV}$$_{\zeta}$). Higher sensitivity data of CN transitions in Upp Sco\ are required to accurately probe
$I_{\rm UV}$\ throughout the region. In that case, subsequent comparisons with accurate CH$^+$ line-widths (to
determine the Doppler velocity parameter $b$)  could be used to distinguish between the production of CH$^+$
in shocks (\emph{c.q.} turbulent media) and the  effect of UV pumping on enhanced abundances of CH$^+$ (see
also Sect.~5.3).

\subsection{Spatial distribution of DIBs and DIB ratios}\label{subsec:DIBratios}

\begin{figure}[ht!]
\centering
   \includegraphics[width=.9\columnwidth,clip]{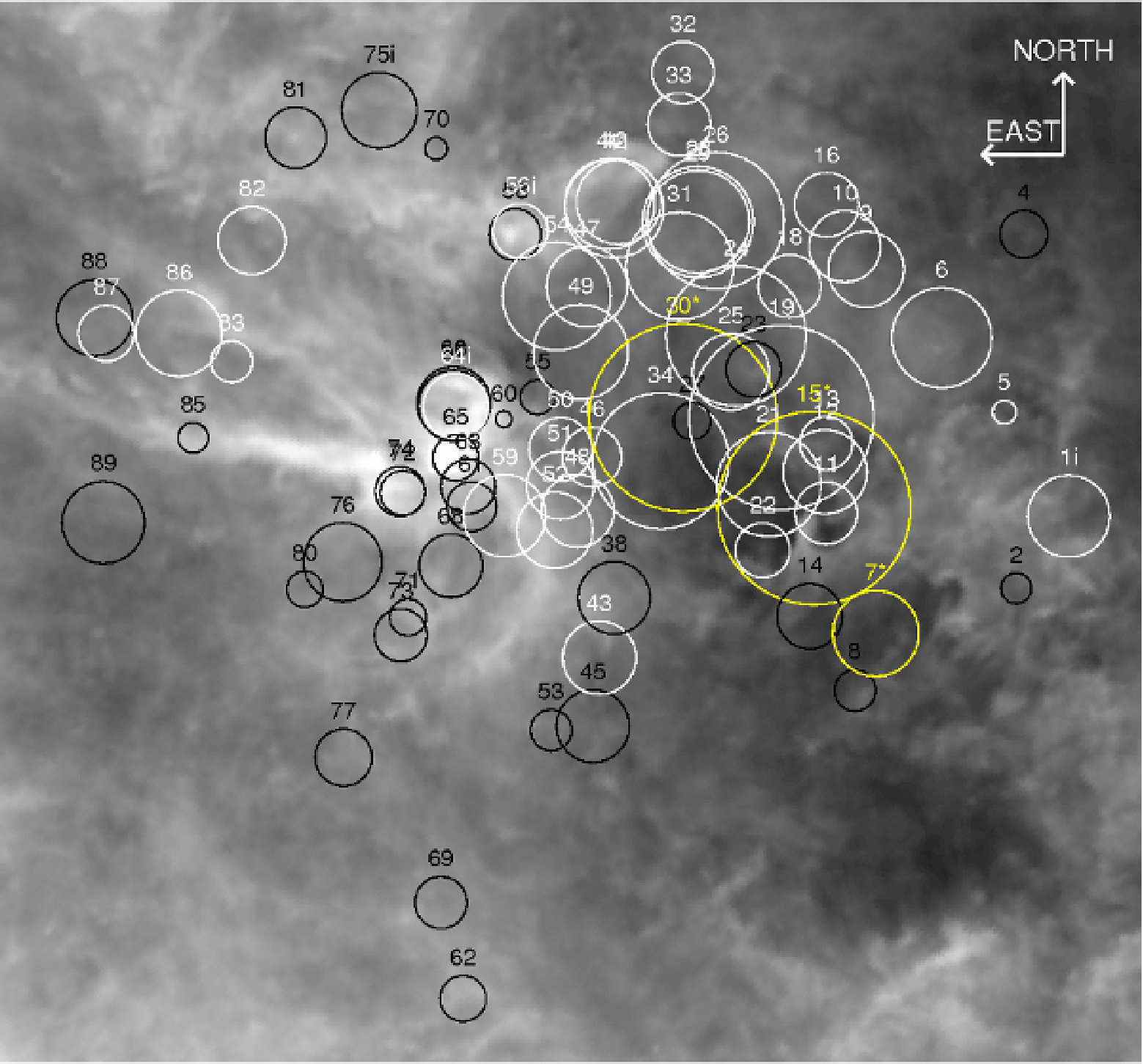}
   \includegraphics[width=.9\columnwidth,clip]{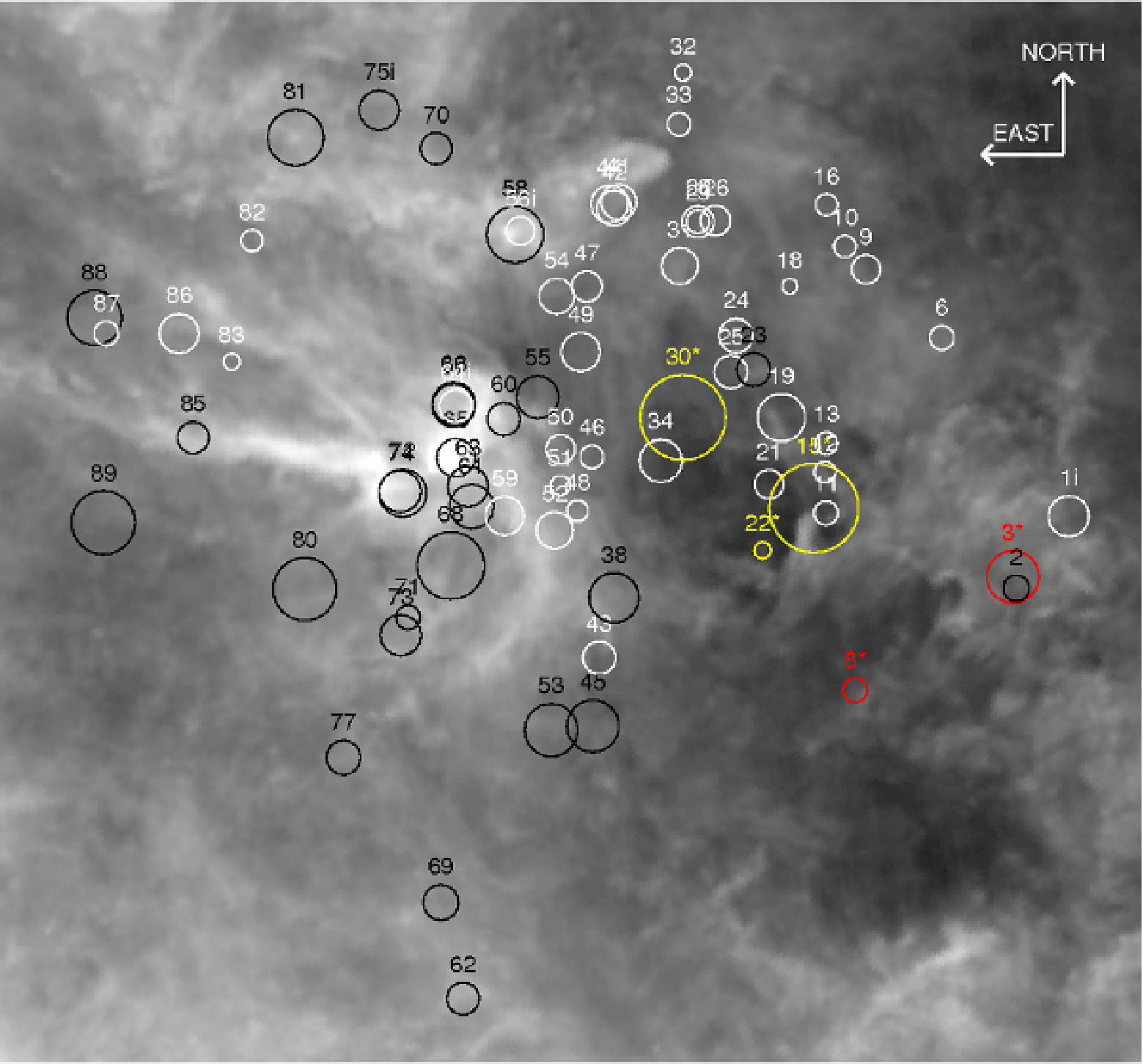}
   \caption{%%
   Circles correspond to $W_\lambda(5780)$/\Ebv\ (top) and $W_\lambda(5797)$/\Ebv\ (bottom).
   Background: IRAS 100~$\mu$m dust map.
   Targets are indicated by their target number as defined in Table~\ref{tb:basic-data}.
   White circles represent $\sigma$-type sightlines and black circles $\zeta$-type sightlines, except those 
   labeled  ``i'' which are of intermediate type. Targets labeled with ``*'' have relative errors on the ratio between 
   30 and 60\%, while errors larger than 60\% are omitted.
   The circle sizes for W(5797)/\Ebv\ are  multiplied by a factor 2 with respect to those for W(5780)/\Ebv.}
   \label{fig:EW57805797-skymap}
\end{figure}

Our dataset provides a unique opportunity to investigate the scatter on the linear relation between DIB
strength  and reddening by dust, in particular with respect to its spatial distribution.  Therefore, the
equivalent width per unit reddening is plotted on the infrared dust map (\citealt{1998ApJ...500..525S}; 
Fig.~\ref{fig:EW57805797-skymap}).  We show only the results for the 5780 and 5797~\AA\ DIBs.  The
equivalent widths for the 6196, 6379, and 6613~\AA\ DIBs behave similarly to the 5797~\AA\ DIB, but due to 
the larger relative uncertainties in the measured equivalent widths are not discussed further.

\begin{figure}[t!]
\centering
   \includegraphics[width=\columnwidth,clip]{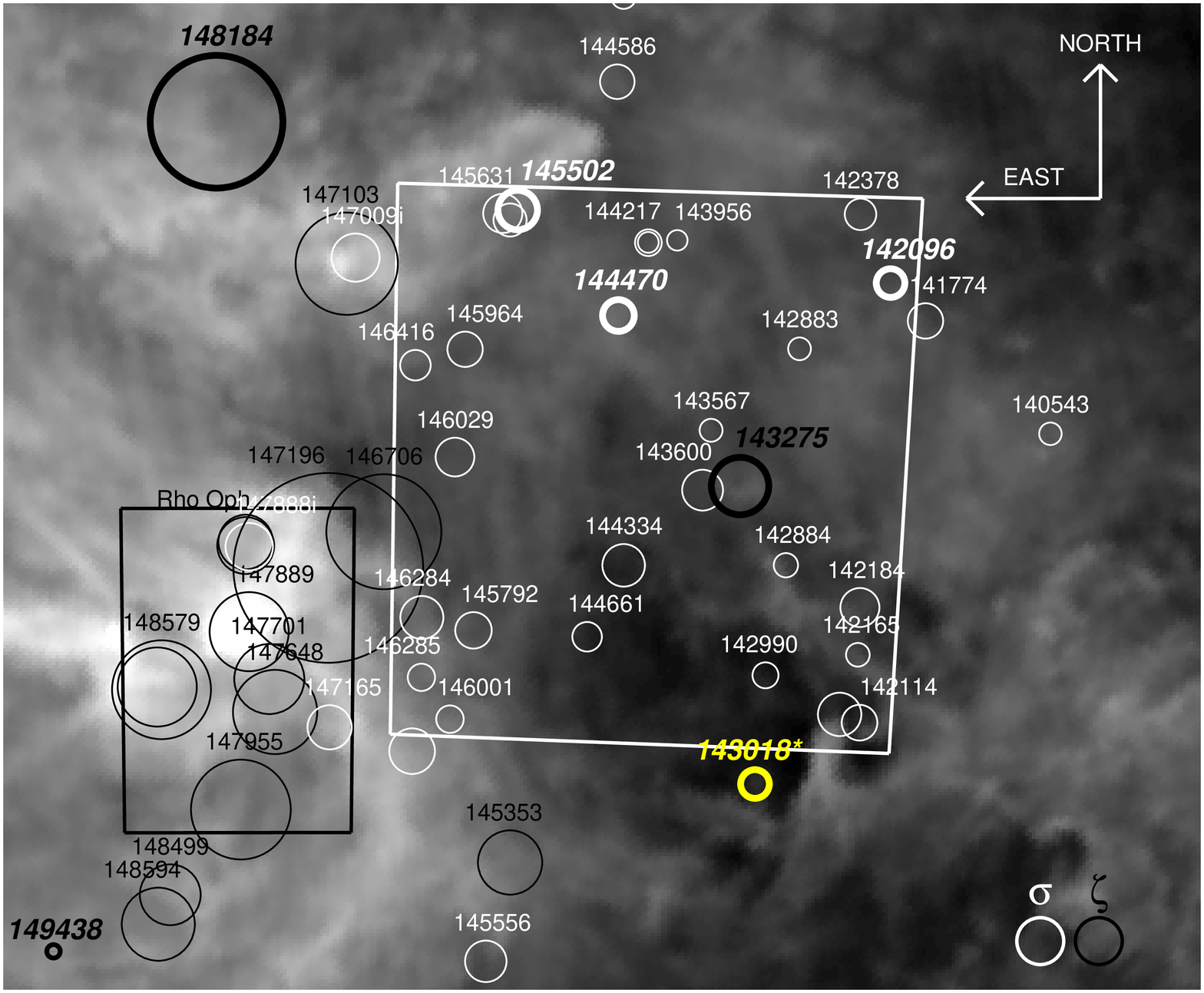} %bb=125 52 520 366,
   \caption{%%
   The circle sizes is proportional to $W_\lambda(5797)/W_\lambda(5780)$. Black and white circles 
   indicate $\zeta$ and $\sigma$ sightlines, respectively. Seven of the eight stars (one is 
   outside the map) generating 90\% of the local ISRF (\citealt{2005ApJ...633..257S}) 
   are indicated by bold circles with italic numbers. The ``high-density'' ($\rho$\,Oph
   cloud) and ``low-density'' regions selected for comparison are delineated by the
   black and white square, respectively. Intermediate classified lines-of-sight 
   are labeled with an ``i''. Note that for \object{HD149438} no DIBs were 
   detected, but it is included here for its large contribution to the local 
   ISRF. Targets labeled with ``*'' have relative errors on the ratio between 
   30 and 60\%, while errors larger than 60\% are omitted. All other targets 
   have errors smaller than 30\% (see also Table~\ref{tb:DIB-data}).}
   \label{fig:5797to5780_spatial}
\end{figure}

From Fig.~\ref{fig:EW57805797-skymap} it is apparent that the spatial  {\it behaviour} for the two DIBs is
different. In order to visualize this effect we  show the spatial distribution of the
$W_\lambda(5797)/W_\lambda(5780)$ ratio in Fig.~\ref{fig:5797to5780_spatial}. It can be seen that
$\sigma$-type sightlines are more frequently probed towards the region westward of the $\rho$\,Oph cloud
which has a low dust content, while predominantly $\zeta$-types are observed towards the high dust column
density $\rho$\,Ophiuchus cloud complex. \citet{2005ApJ...633..257S} showed that 5 of the 8 stars producing
90\% of the local ISRF are situated in this low-dust region at distances between 109 and 141 pc. Here, the
strong stellar winds have blown out most of the dust and are now impinging on the west side of the
$\rho$\,Oph cloud.

To study this difference we selected two regions, a region free of dust emission and one with strong dust
emission, respectively. The first region is centred on the $\rho$\,Oph cloud (showing a high dust column),
while the second region is centred on the region west of $\rho$\,Oph  scarce in dust emission
(Fig.~\ref{fig:5797to5780_spatial}, black and white squares, respectively). In line with the observed dust
density,  UV field strength and molecular $H_2$ fraction, 80\% of the lines-of-sight in the selected
``high-density'' (\emph{i.e}\ higher  dust column and higher $f_{\mathrm{H}_2}$) region are classified as $\zeta$
and 95\% of the sightlines in the ``low-density'' (\emph{i.e}\ low dust column density and low $f_{\mathrm{H}_2}$)
region are designated $\sigma$-type.

\begin{figure}[t!]
\centering
   \includegraphics[angle=-90,width=\columnwidth,clip]{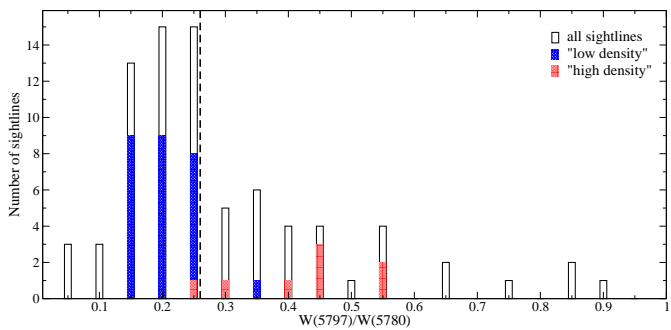}
   \caption{Histogram of the $W(5797)/W(5780)$ ratio for the lines-of-sight in the
   low-density (blue) and high-density (red) region, and for all sightlines in the sample.
   The vertical dashed line sets the division between $\sigma$ and $\zeta$-type cloud.
   (see electronic version for colour version). }
   \label{fig:DIBratioHistogram}
\end{figure}

The weighted mean and the associated error of the DIB ratio, $I_{\rm UV}$, and $f_{\mathrm{H}_2}$ are calculated for
the ``low-density'' and ``high-density'' selected regions, as well as for the total dataset  (all sightlines
with available values) and are given in Table~\ref{tb:mean-ratio}.   The two regions differ significantly
from each other, and both regions show deviations from the overall weighted mean. For the ``low-density''
region the DIB ratios peak at about 0.20 and with a distribution width of about 0.05, while the
``high-density'' DIB ratio distribution peaks at about 0.45 with a wider width of about 0.10
(Fig.~\ref{fig:DIBratioHistogram}). The mean value of 0.26 $\pm$ 0.01 for the DIB ratio has been adopted to
make the distinction between $\sigma$ and $\zeta$) type clouds.  Fig.~\ref{fig:ratio_iuv} shows that the
mean molecular hydrogen fraction, $f_{\mathrm{H}_2} = 0.34 \pm 0.21$ (Table~\ref{tb:mean-ratio}),  provides
an alternate - complementary - way to distinguish between $\sigma$ and $\zeta$ sightlines (\emph{e.g.} 
Fig.~\ref{fig:ratio_iuv}; top panel).

\begin{table}
\centering
\caption{The weighted mean and associated error of $W(5797)/W(5780)$, $I_{\rm UV}$, and $f_{\mathrm{H}_2}$
for all sightlines (for which values are available) and for the ``low-density'' and ``high-density'' regions.}
\label{tb:mean-ratio}
\begin{tabular}{llll}\hline\hline
weighted mean	  	& all sightlines 	& ``high-density''	& ``low-density''	\\ \hline
$W(5797)/W(5780)$ 	& 0.26 $\pm$ 0.01	& 0.36 $\pm$ 0.03 	& 0.16 $\pm$ 0.01   	\\
$I_{\rm UV}$	  		& 6.4 $\pm$ 2.5		& 4.2 $\pm$ 2.5 	& 8.1 $\pm$ 6.2 	\\
$f_{\mathrm{H}_2}$	& 0.34 $\pm$ 0.2	& 0.44 $\pm$ 0.2	& 0.15 $\pm$ 0.1	\\ \hline
\end{tabular}
\end{table}

\subsection{$R_V$ and $W_{\lambda}(5797)/W_{\lambda}(5780)$}\label{sec:Rv}

For Milky Way lines-of-sight a weak relation exists between the total-to-selective visual extinction ratio,
$R_V$,  and the UV extinction (\citealt{1989ApJ...345..245C}; \citealt{2007ApJ...663..320F}).  For
increasing $R_V$ the far-UV absorption decreases, suggesting that fewer small  dust particles / large
molecules absorb in the far-UV, and implying a dust size  distribution shifted towards larger grains. 
Therefore, $R_V$ is sometimes used as a tracer of high density ISM, where grain growth is more significant; 
$R_V = 3.1$ for the average diffuse ISM, while $R_V$ = 4 -- 6 (\citealt{1986ApJ...302..492C};
\citealt{1988ApJ...335..177C}) for dense ISM. \citet{1974MNRAS.168..371W} showed that sightlines penetrating
the $\rho$\,Oph cloud reveal a  higher than average $R_V$. Extinction curves with a higher $R_V$ have less
steep far-UV rise, which is associated to a lack of small particles which is often attributed to enhanced
grain growth in denser clouds. However, an obvious correlation between $R_V$ and $n_H$, N(CH), N(CN) could
not be identified.  The observed differences in $R_V$ towards Upp Sco\ showed that the dust size
distribution  is not homogeneous throughout the primordial cloud that formed the association, or that  the
dust has been processed differently (for example, due to destruction in UV exposed regions, or grain growth
in denser areas) in the various parts of the association.

In an attempt to reveal any dependence between the UV radiation field and the  dust grain size distribution
we plot $W_\lambda(5797)/W_\lambda(5780)$ versus $R_V$ for 23 targets in our sample (see
Table~\ref{tb:basic-data}). Large error bars on $R_V$ and the large differences between values obtained by 
different authors prevent us from discerning any significant trends between the DIB ratio and $R_V$. This is
consistent with the earlier work of \citet{1974ApJ...194..313S}.

\begin{figure}
\centering 
   \includegraphics[angle=-90,width=\columnwidth]{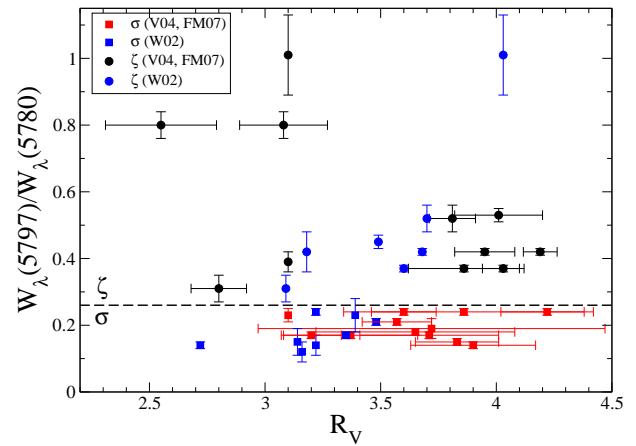}
   \caption{$W(5797)/W(5780)$ is plotted against the
   total-to-selective visual extinction ratio $R_V$ for 23 stars. 
   The low and high density sightlines are indicated by squares and circles, respectively.
   Values are from \citet{2002BaltA..11....1W} [W02]; \citet{2004ApJ...616..912V} [V04], 
   and \citet{2007ApJ...663..320F} [FM07].}
   \label{fig:ratio_vs_Rv}
\end{figure}

\section{Conclusions and summary}\label{sec:conclusion}

Lines of sight can be designated via $W(5797)/W(5780)$ as either $\zeta$-type  (sightlines penetrating cloud
cores) or $\sigma$-type (sightlines probing cloud edges) (or intermediate). We investigated the spatial
variation of the DIB strengths $W$(DIB) and the $W(5797)/W(5780)$~DIB ratio, and their dependence on
reddening in 89\ lines-of-sight within a field of 20\degr$\times$20\degr\ probing the small scale
variations in the gas and dust in the Upp Sco\ association. This represents an in-depth multi-object study
of a well-studied interstellar cloud complex, providing a valuable statistical dataset. These data cover a
wide range in dust column densities (from zero up to four magnitudes of visual extinction) to track the
sensitivity of the DIB carrier molecules in relation to their local environment.

Our results provide evidence that on average the DIB strengths in Upper Scorpius are linearly proportional
to the reddening, closely following the general relation observed for the Galactic diffuse ISM. In addition,
we showed that the scatter on these relationships, expressed for example via the $W$(5797)/$W$(5780) and
W(DIB)/\Ebv\ ratios, is significant  and can be attributed to variations in the local physical conditions,
in particular the interstellar density and the radiation field.

We found that making a distinction between $\sigma$ and $\zeta$ type sightlines clearly improved the
relation between 5780~\AA\ DIB strength and the amount of dust, \Ebv. The improvement for the other DIBs is
less pronounced although still significant, suggesting that particularly the 5780~\AA\ DIB is sensitive to
variations in the local conditions of the interstellar gas and dust.

The CH and CH$^+$ molecules are detected in 53 out of 89\ sightlines, whereas  CN and \ion{Ca}{i} are
detected in 15 and 31 sightlines, respectively. CH traces mainly cold (UV shielded) cloud material in the
line-of-sight which is confirmed by the improved regression coefficient with respect to \Ebv\ for
$\zeta$-type sightlines ($r = 0.83$) compared to its $\sigma$-type equivalent ($r = 0.53$). In line with
previous results we observe no significant velocity  difference between the CH and CH$^+$ lines in the
Upp Sco\ region, thus finding no evidence for the production of CH$^+$ in shocks
(\citealt{2011ApJ...728...36R}).

CN traces the dense cloud cores and is only detected in $\zeta$-type sightlines. The presence of (presumably
very) low column densities of CN in $\sigma$-type sightlines remains to be confirmed.

The $W(5797)/W(5780)$ DIB ratio is more strongly correlated with $W$(CH)/\Ebv\  than with $W$(CH$^+$)/\Ebv.
This result confirms that $\zeta$-type clouds (\emph{i.e.} dense and $\lambda$5797 favoured) have a higher
molecular content (\emph{e.g.} higher abundances of CH and H$_2$). The CH/CH$^+$ ratio is used to
discriminate between quiescent and turbulent regions, confirming the division between $\sigma$ and
$\zeta$-type environments. This DIB ratio is thus related to the abundance of molecular species and dust
properties inside the cloud cores.

We found a significant difference between the mean $W(5797)/W(5780)$ ratio towards the dense $\rho$\,Oph
cloud and towards the lower dust density region to the west.  The high dust density region consists
predominantly of $\zeta$-type sightlines, whereas (as expected) the low-density region contains mainly
$\sigma$-type lines of sight. This distinction was used to assign all sightlines to either $\sigma$ or
$\zeta$ type.

Radial velocities of atomic and molecular absorption lines reveal one main interstellar component  (with
possibly unresolved small-scale structure)  at approximately $-9$~km~s$^{-1}$, corresponding to the Upp Sco\ dust
sheet at a distance of 110 to 150~pc. Another, much weaker, velocity component at $\approx$-22~km~s$^{-1}$, is
linked to the tenuous foreground dust sheet at 50~pc.

Asymmetries and widths of the \ion{Na}{i} and \ion{K}{i} line profiles suggests that the main velocity
component may be resolved in narrower components at higher spectral resolution. The radial velocities for
individual sightlines reveal velocity gradient in the dust sheet associated to Upp Sco. This could be seen
as a differentially moving sheet,  with the south-west corner fixed and the north-east corner moving towards
the observer. For the sightlines with the strongest DIBs, the atomic, molecular, and DIB radial velocities
are similar within uncertainties, indicating that these species spatially co-exist with each other.

A simplified radiative transfer model of the main Upp Sco\ dust sheet was used to compute the effective
interstellar radiation field (ISRF) from CN and CH observations. Five (of the eight) stars that strongly
contribute to the local ISRF are located in the lower-density region of the Upp Sco\ cloud complex,
substantiating the inferred strong radiation field.  The most accurate values for $I_{\rm UV}$\ are derived if both
CN and CH are detected, and therefore these sightlines correspond to those probing denser $\zeta$-type
regions. From the observed peak in the CH$^+$/\Ebv\ ratio for $I_{\rm UV}$ $\approx$~5 we can infer that UV pumping
plays a role in the production of CH$^+$ in the Upp Sco\ region for diffuse clouds with sufficiently strong
UV field and significant abundance of $H_2$. The drop in CH/CH$^+$ with increasing $I_{\rm UV}$\ is probably due to
lower CH abundances for the more tenuous, UV exposed clouds. In fact the effective radiation field, $I_{\rm UV}$, is
an inverse function of the molecular hydrogen fraction, $f_{\mathrm{H}_2}$,  derived from N(CH) and
$W$(5780). And, log($f_{\mathrm{H}_2}$) is, as expected, proportional to log($n_H$/$I_{\rm UV}$). The
$W(5797)/W(5780)$ ratio versus $I_{\rm UV}$\ relation in Upp Sco\ confirms that most of the sightlines with low
$I_{\rm UV}$\ values are of $\zeta$-type for which also CN is detected, while only few $\zeta$-type sightlines have
large $I_{\rm UV}$. The average $I_{\rm UV}$\ for $\sigma$-type sightlines is higher than for $\zeta$-types, but the
standard deviation  is relatively large for the poorly determined $\sigma$-type sightlines due to limited CN
data.  Exceptionally high S/N spectra are needed to probe the CN content of these most tenuous diffuse
clouds.

Our results indicate that the relative abundance of the $\lambda$5780 DIB carrier increases with respect to
the $\lambda$5797 DIB with an increase in the effective strength of the interstellar UV radiation field. A
possible scenario is that a sufficiently large UV flux ionises the molecule whose cation has an electronic
transition at 5780~\AA.  The $\lambda$5797 DIB carrier, on the other hand, is more efficiently destroyed in
regions with high $I_{\rm UV}$\ and is only able to  survive in the deeper layers of the cloud. The behaviour of the
$\lambda\lambda$5780 and 5797 DIBs is thus strongly dependent on the local effective UV radiation field 
(and consequently $f_{\mathrm{H}_2}$) as is to be expected for molecular carriers.

The scatter on the relation between DIB strength and reddening can ultimately be translated to monitor the
life  cycle of the carriers as they get formed, excited, and destroyed throughout the complex interstellar
cloud structure associated with the $\rho$ Oph cloud and the Upp Sco\ association. The main driver for this
cycle is the UV radiation field. More information is needed on the exact conditions of these diffuse clouds
to disentangle additional (secondary) drivers. Measurements of additional atomic and molecular lines, such
as C$_2$, CN, H$_2$, and CO, can improve the constraints on the chemical networks and physical conditions of
individual clouds and thus improve our insight in the effects of these parameters on the abundance of DIB
carriers. In addition to the two strong narrow DIBs at 5780 and 5797~\AA\ discussed in  this work, accurate
- high S/N - observations of additional narrow and broad diffuse bands are essential to fully probe the
effects of changing environmental conditions on the ensemble of carriers giving rise to the DIB spectrum.
Moreover, additional detailed studies of the Upp Sco\ region need to be complemented with similar studies of
other specific  Galactic and extra-galactic regions to probe a variety and range of different conditions
that affect formation and destruction of the DIB carriers.

\begin{acknowledgements} 

We thank Jos de Bruijne for obtaining the FEROS spectra. DAIV thanks Paul Groot and Gijs Nelemans for
stimulating discussions. PE is supported by the NASA Astrobiology Institute. Furthermore, we thank the
referee for a very thorough reading of the manuscript and many interesting and helpful suggestions that
helped to improve this paper.

\end{acknowledgements}

\bibliographystyle{aa}
\bibliography{/lhome/nick/Desktop/ReadingMaterial/Astronomy/Bibtex/bibtex}  % .bib

\Online

\begin{appendix}
\section{Error on equivalent width}\label{sec:appendix-ew}.

\begin{figure}[h!]
\centering
\includegraphics[bb=0 0 624 431,width=0.7\columnwidth]{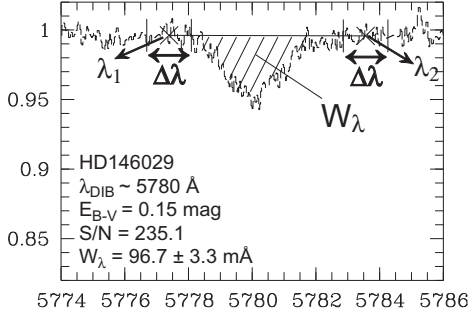}
   \caption{%%
   Typical example illustrating the determination of the DIB equivalent width.
   }
   \label{fig:EW-measure}
\end{figure}

For the equivalent width measurement ($W = \int_{\lambda_1}^{\lambda_2} \left( \frac{F_c - F_\lambda}{F_c}
\right)$),  two integration borders $\lambda_1$ and $\lambda_2$ are selected for which the average intensity
in the continuum is derived in a window of width $\Delta\lambda_c = 1.4$~\AA\ centred on the borders
(Fig.~\ref{fig:EW-measure}).

The broader DIBs at 5778~and 5796~\AA\ underlying the strong 5780 and 5797~\AA\ DIBs are largely removed
from the spectra in the local continuum normalisation. They are very weak, even for high \Ebv, and do not
significantly contaminate  the $W$ measurements of the 5780 and 5797~\AA\ DIBs.

Taking into account both the noise of the measured flux ($\sigma_{\rm noise}$) and the uncertainty in
estimating the continuum ($\sigma_{c}$), the error on the equivalent width is given according to
\citet{1983A&A...127..279C} [CM83], in our notation:

\begin{eqnarray}
\sigma_{W}^2 & = & \sigma_{\rm noise}^2 + \sigma_{\rm continuum}^2 \\\smallskip
~	     &   & {\rm via\ equation\ A.9\ from\ CM83:} \nonumber\\ 
~	     & = & M \frac{\delta\lambda}{SNR}^2 (\frac{F_j}{F_c})^2 + \frac{\sigma(F_c)}{F_c}^2 (\Delta\lambda - W)^2   \\
~	     & = & \Delta\lambda_{\rm line} \frac{\delta\lambda}{SNR^2} (F_j)^2 + \frac{\delta\lambda}{\Delta\lambda_c}\ \frac{1}{SNR^2} (\Delta\lambda_{\rm line} - W)^2 \\
~	     & = & \frac{\delta\lambda}{SNR^2} \left( \Delta\lambda_{\rm line} F_j^2 + \frac{(\Delta\lambda_{\rm line} - W)^2}{\Delta\lambda_c}\right) \\
~	     & = & \frac{\Delta\lambda\ \delta\lambda}{SNR^2}\ \left( F_j + \frac{(\Delta\lambda - W)}{\Delta\lambda^2} \right) \\ \label{eq:step5}
~	     & \approx & \frac{\Delta\lambda\ \delta\lambda}{SNR^2}\ \left( 1 + \frac{(\Delta\lambda - W)}{\Delta\lambda^2} \right) \\ \label{eq:step6}
%\sigma_W     & \approx & \frac{\sqrt{\Delta\lambda\ \delta\lambda}}{SNR}\ \sqrt{2}
\sigma_W     & \approx & \frac{\sqrt{\Delta\lambda\ \delta\lambda}}{SNR}\ \sqrt{2}
\end{eqnarray}
where we used the integration range $\lambda_2 - \lambda_1 = M\ \delta\lambda = \Delta\lambda_{\rm line}$,
$SNR$ is the signal-to-noise ratio (per pixel), $\frac{\sigma^{2}_{F_c}}{F_c^2} = \frac{1}{N_c\ SNR^2}$,
$N_c = \Delta\lambda / \delta\lambda$,  $\delta\lambda$ is the spectral dispersion in \AA/pix, $M$ the
number of pixels in the integration range, and we take $\Delta\lambda_c \approx \Delta\lambda_{\rm line} =
\Delta\lambda$. In Eq.~\ref{eq:step5} the last term is approximately 2 as for shallow-weak lines $W \ll
\Delta\lambda = 3$~\AA. We have $\delta\lambda$ = 0.03~\AA, FWHM $\approx$ 1~\AA\ (typical for a narrow
DIB), so that $\Delta\lambda \approx\ 3$~FWHM. This error estimate yields similar results as when applying
the error estimate in the weak line limit by \citet{2006AN....327..862V} (Eq. A.X): $\sigma_W = \sqrt{2}
\frac{( \Delta\lambda - W )}{SNR} = \frac{\sqrt{2}}{SNR}\ \sqrt{\Delta\lambda \delta\lambda}$.

For the strong narrow \ion{K}{i} line the approximation in Eq.~\ref{eq:step5} is not valid and the given
$\sigma_W$  will be an upper limit to the true $\sigma_W$.

\end{appendix}

\begin{appendix}
\section{Contamination with the photospheric spectrum}\label{sec:stellar}

In order to evaluate possible contamination from stellar lines to the measured equivalent widths of the
interstellar lines (both DIBs and molecules) we have checked synthetic spectra of B main-sequence [A0V, B5V,
B2V, B0V, O9V] stars  for the presence of atmospheric lines in the regions of interest in this study (see
\citealt{1996HotNews..22..16G}  for details on the models).  A brief overview of the expected contamination
is given in Table~\ref{tb:contamination}.

For the 5797, 6196~\AA\ DIBs no contamination above the 2~m\AA\ level is expected for these B-type
main-sequence stars. The \ion{N}{ii}	line at 6610.6~\AA\ present in spectra of O9V to B2V stars is
clearly separated from the 6613~\AA\ DIB at the  spectral resolving power of FEROS. Both the \ion{O}{ii}
line at 4303.6~\AA\ and the \ion{Fe}{ii} line at 4303.2~\AA\  are also well resolved from the CH 4300.3~\AA\
line (the rest wavelength velocity difference is more than 200~km~s$^{-1}$). For CH (4300.3~\AA) the contamination
in A0V spectra may be severe due to the presence of \ion{Ti}{ii}, however in that case also the \ion{Fe}{ii}
line at 4303.2~\AA\ should be clearly present (with $W$ $\sim$200~m\AA). Strong \ion{Fe}{i} lines appear in
A0V spectra at 3872.5~\AA\ and 3878.6~\AA, however, these are also clearly separated from the interstellar
CN line (3874.6~\AA).  The velocity profile plots cover a region of $\pm$0.6~\AA\ around the CN line. Also
the weak to strong ($W_{\rm total} \sim 34$ to $\sim 185$~m\AA) \ion{C}{ii} lines around 3876.0 --
3876.7~\AA\  fall outside the plotted window. Only the \ion{Ne}{ii} line at 3875.3~\AA\ is relatively close
to the CN line (with rest wavelength difference of 54~km~s$^{-1}$).

There are several ways to investigate the potential problem of contamination. For example, the line widths
of all stellar lines are set by the rotational velocity ($v$~sin\,$i$), whereas the di-atomic lines have
intrinsically small widths, and can thus be distinguished from each other. Also, the radial velocities of
the stellar and interstellar lines are (in virtually all cases) different.  Therefore, any (di)atomic line
with a velocity significantly different from \emph{e.g.} that of the \ion{K}{i}  doublet (no contamination
expected) is suspect. Furthermore, in the case of relatively strong CH$^+$ line, the CH$^+$ transition at
3957.7~\AA\ can be used to verify and confirm the abundance of CH$^+$. For the relatively broad DIBs this
differentiation due to velocity differences is less secure.  Fortunately, the contamination expected at the
DIB wavelengths is very low.

\begin{table}[ht!]
\caption{Overview of possible significant contamination from stellar lines to the interstellar line equivalent width measurements}
\label{tb:contamination}
\resizebox{\columnwidth}{!}{
\begin{tabular}{llll}\hline\hline
measured line	&	contaminating line	& Spectral types 	&  contamination $W$	\\\hline
5780~\AA	&   \ion{Fe}{ii} (5780.1~\AA)	& B5V-A0V		& 13~m\AA		\\
		&	"			& $<$ B5V		& $<$ 4~m\AA		\\
		&   \ion{He}{i} (5780.5~\AA)	& O9V			& $\sim$2~m\AA		\\
6379~\AA	&   \ion{Ne}{ii} (6379.6~\AA)	& B0V, B1V, B2V		& 8, 22, 8~m\AA		\\
CH (4300.3~\AA) &   \ion{Ti}{ii} (4300.0~\AA)	& B5V			& 4~m\AA		\\		
		&       "                      	& A0V			& 161~m\AA		\\
%		&   \ion{Ti}{ii} (4301.9~\AA)	& A0V			& 87~m\AA		\\
CH$^+$ (4232.5~\AA)&  \ion{Fe}{ii} (4232.8~\AA)	& B2V			& $<$ 12~m\AA		\\
		& 	"			& B5V			& $\sim$98~m\AA	\\
		&				& A0V			& $\sim$300~m\AA	\\
		&   \ion{Si}{ii} (4233.2~\AA)	& B2V			& $<$ 12~m\AA		\\
\hline
\end{tabular}
}
\end{table}

\end{appendix}

\newpage

\begin{appendix}
\section{Correlation plots}

\begin{figure}[h!]
   \includegraphics[angle=-90,width=\columnwidth,clip]{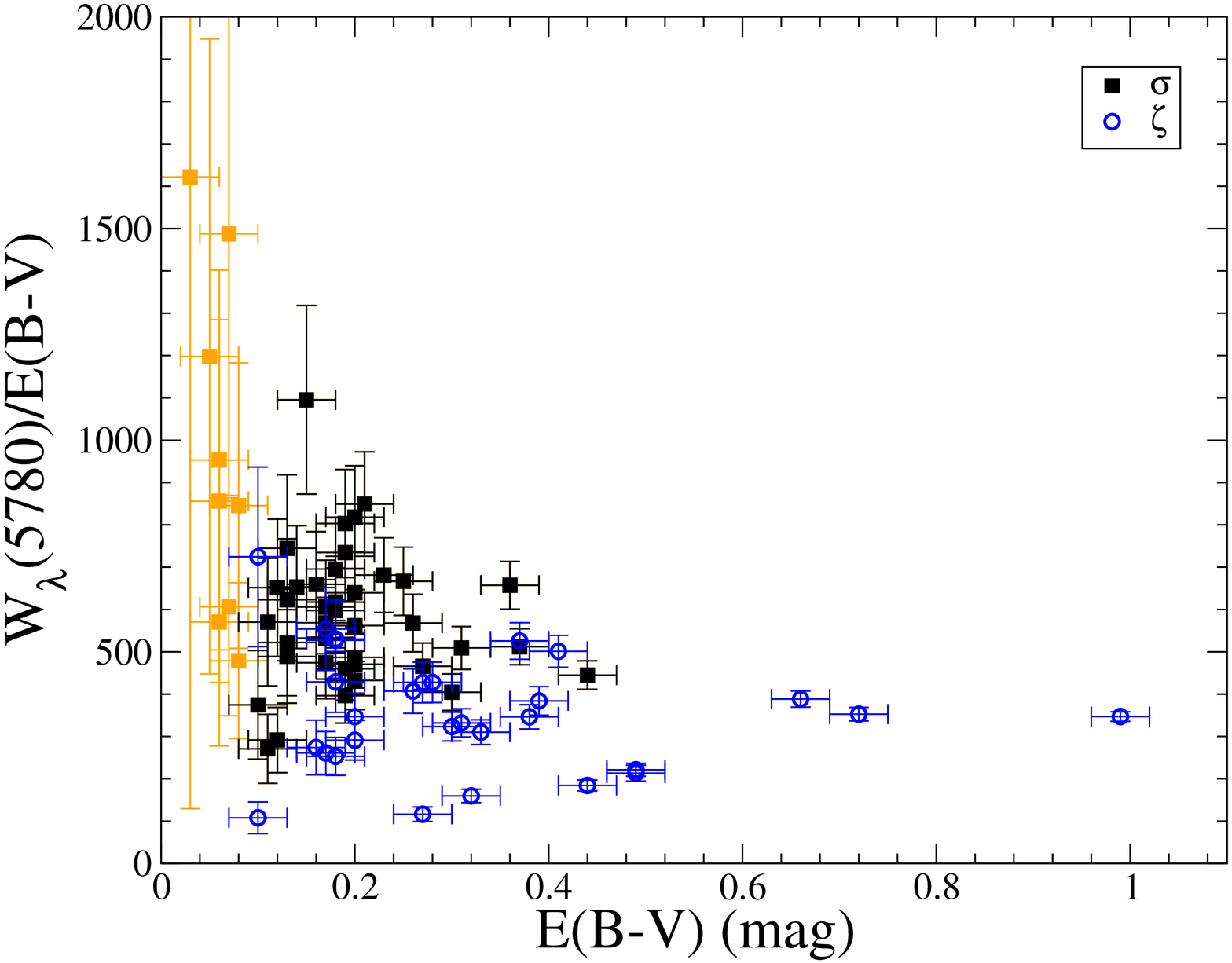}
   \includegraphics[angle=-90,width=\columnwidth,clip]{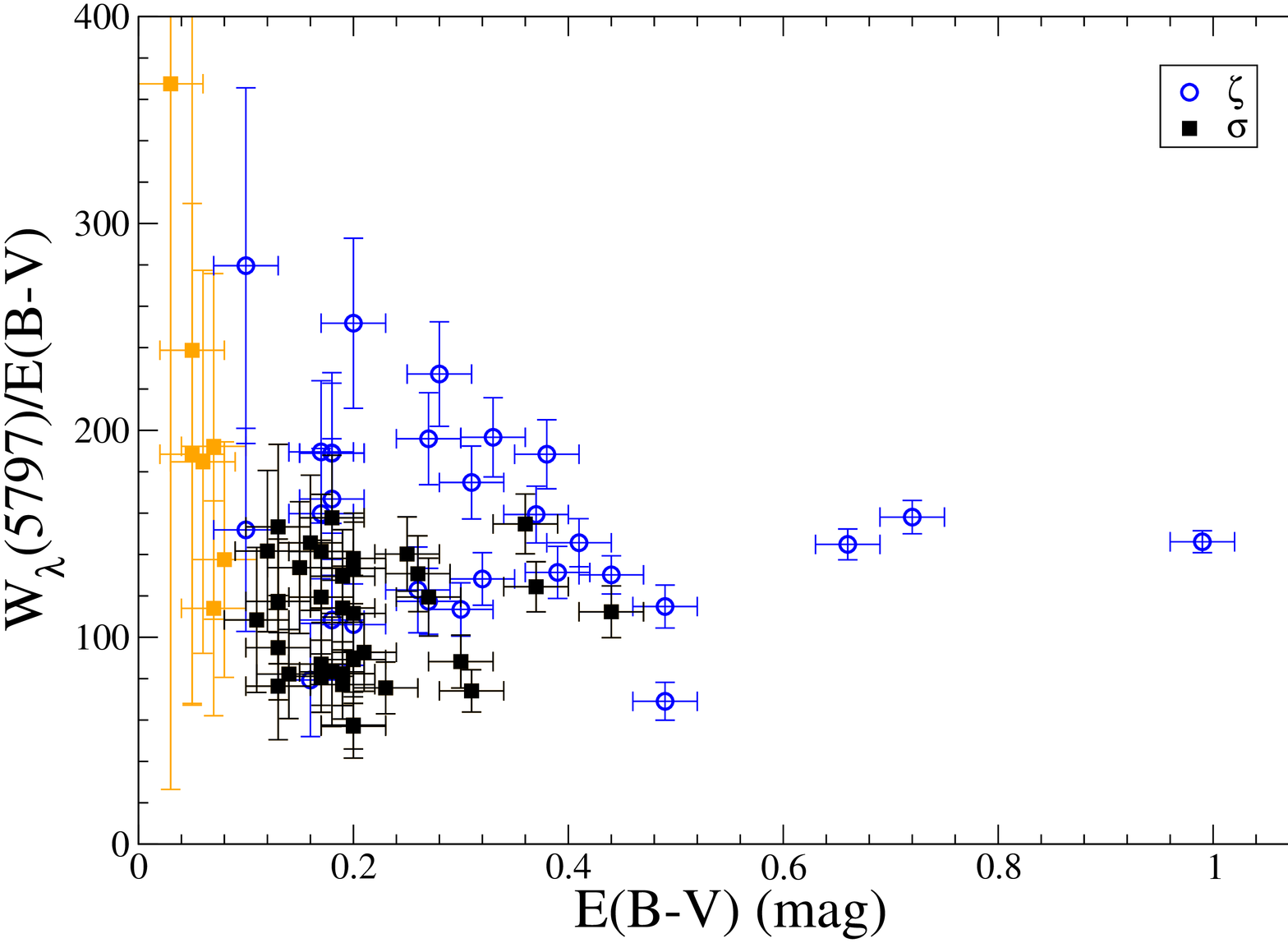}
   \caption{Equivalent width per unit reddening
   versus \Ebv\ for the 5780 and 5797~\AA\ DIBs.
   Reduced chi-squared for linear fits are indicated in the respective panels.
   This plot reveals an intrinsic scatter on the relationship between the DIB strength and amount of dust
   in Upper Scorpius. Sightlines with \Ebv\ $<0.1$~mag (orange/grey squares) are omitted for calculating the correlation coefficient;
   $r_{5780} = -0.32$ and $r_{5797} = 0.07$.}
   \label{fig:EWDIBvsEbv2}
\end{figure}

\begin{figure}[h!]
   \includegraphics[angle=-90,width=\columnwidth,clip]{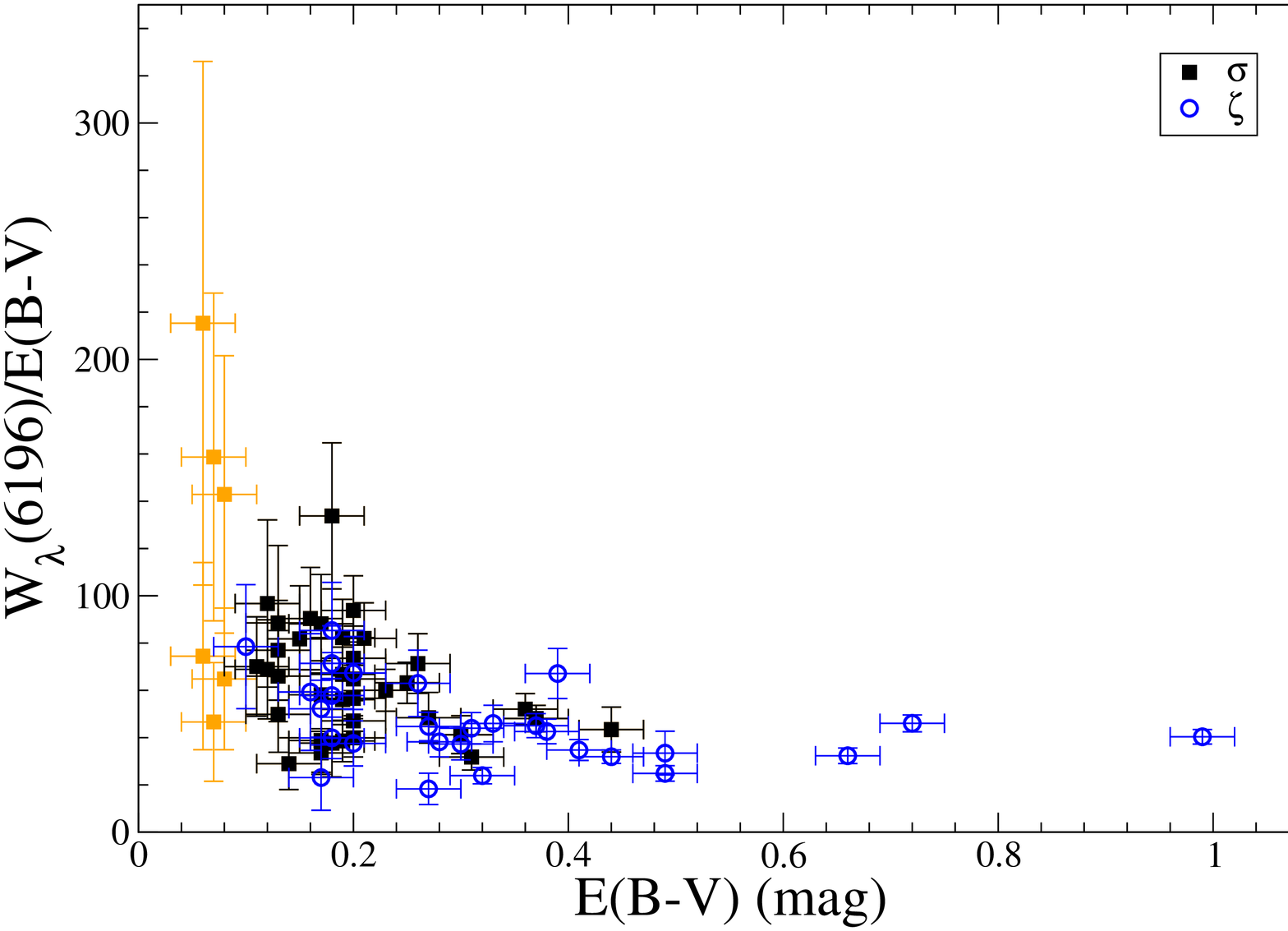}
   \includegraphics[angle=-90,width=\columnwidth,clip]{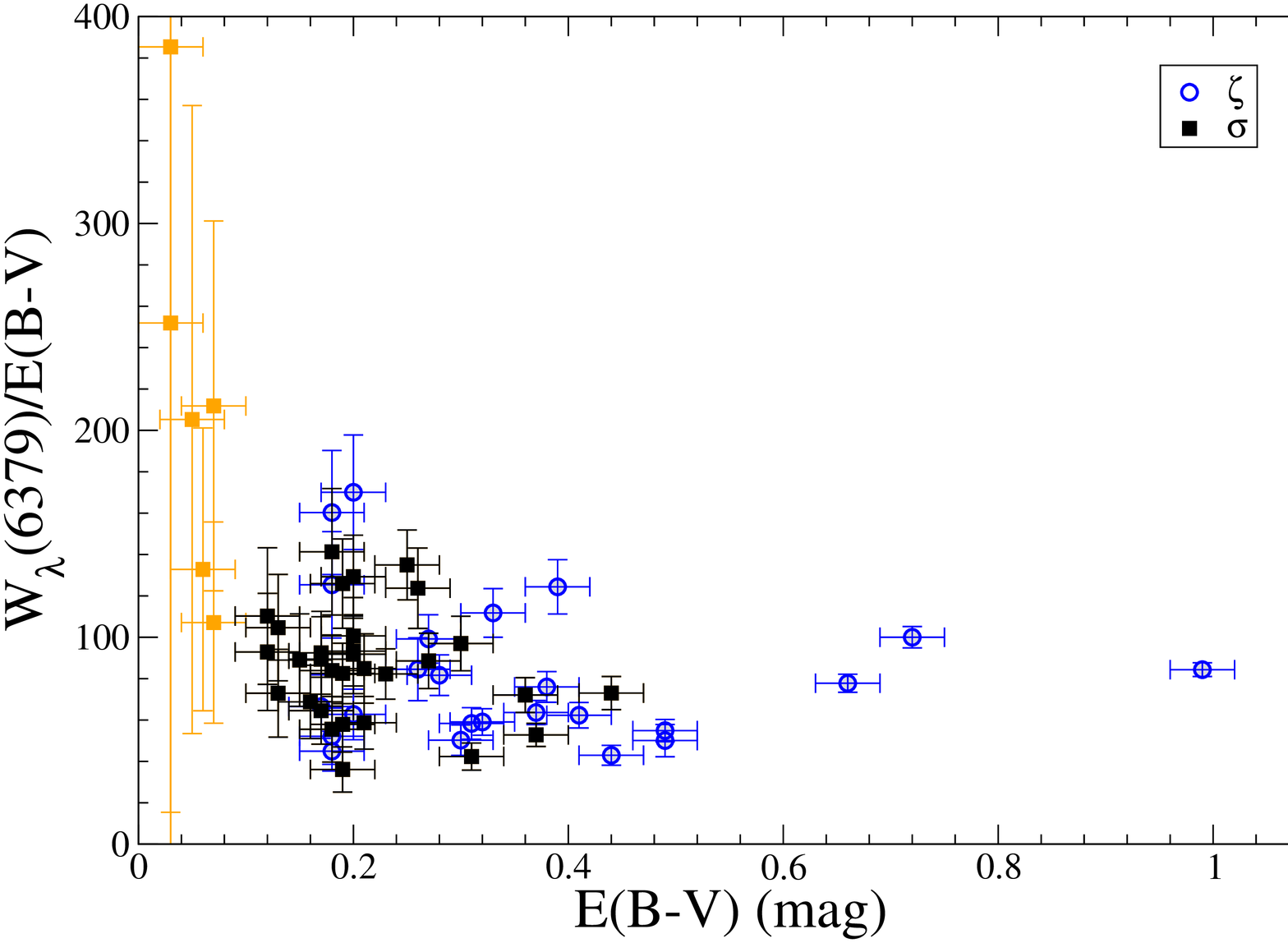}
   \includegraphics[angle=-90,width=\columnwidth,clip]{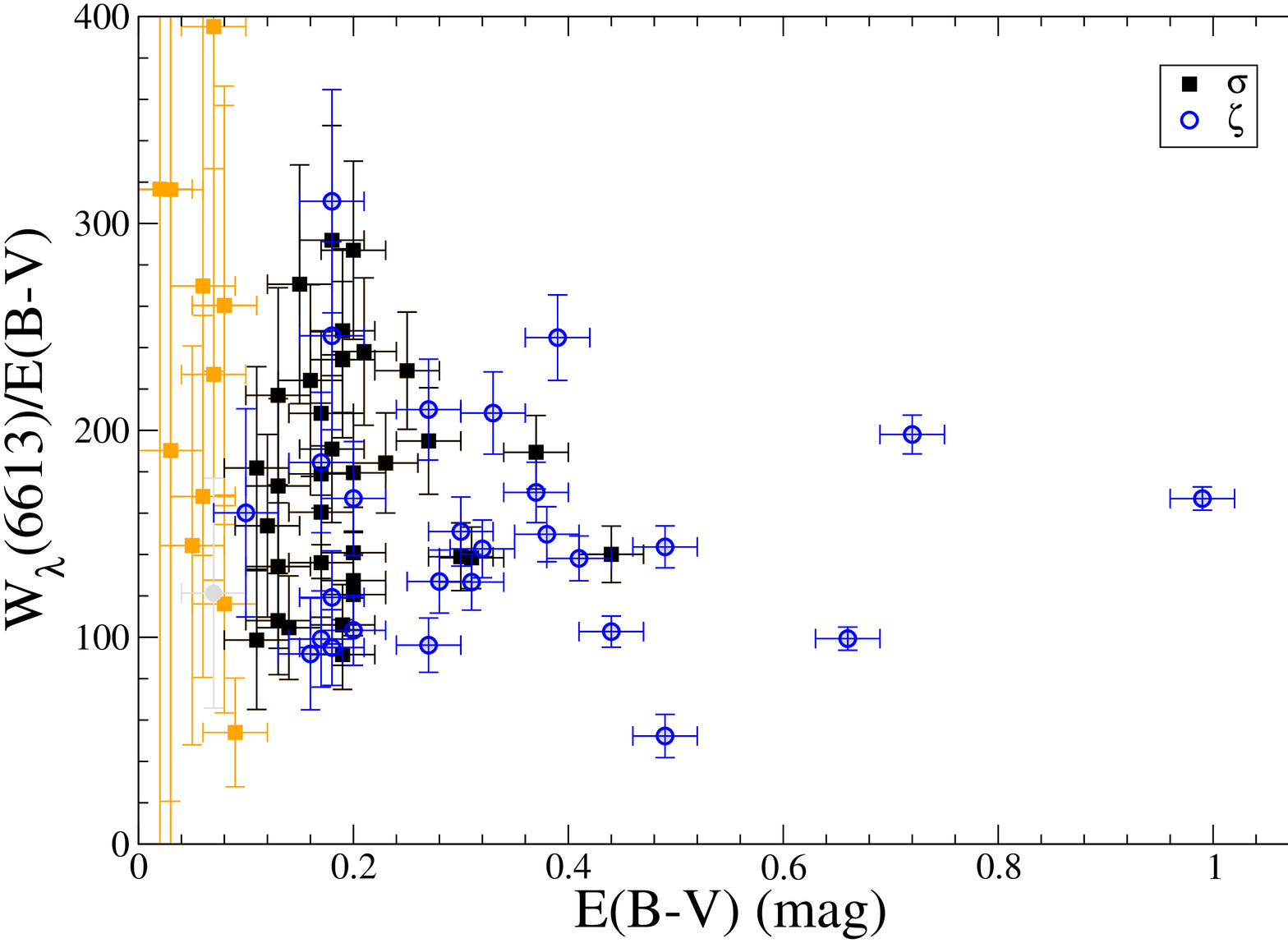}
   \caption{Equivalent widths per unit reddening
   versus \Ebv\ for the 6196, 6379, and 6613~\AA\ DIBs.
   Reduced chi-square for linear fits are indicated in the respective panels.
   This plot reveals an intrinsic scatter on the relationship between the DIB strength and amount of dust
   in Upper Scorpius. Sightlines with \Ebv\ $<0.1$~mag (orange/grey squares) are omitted for calculating the correlation coefficient;
   $r_{6196} = -0.41$, $r_{6379} = -0.17$, and $r_{6613} = -0.12$.}
   %Intercepts and slopes for the linear regressions are given in Table~\ref{tb:linearfit}.
   \label{fig:EWDIBvsEbv3}
\end{figure}

\begin{figure}[h!]
\centering
   \includegraphics[angle=-90,width=\columnwidth,clip]{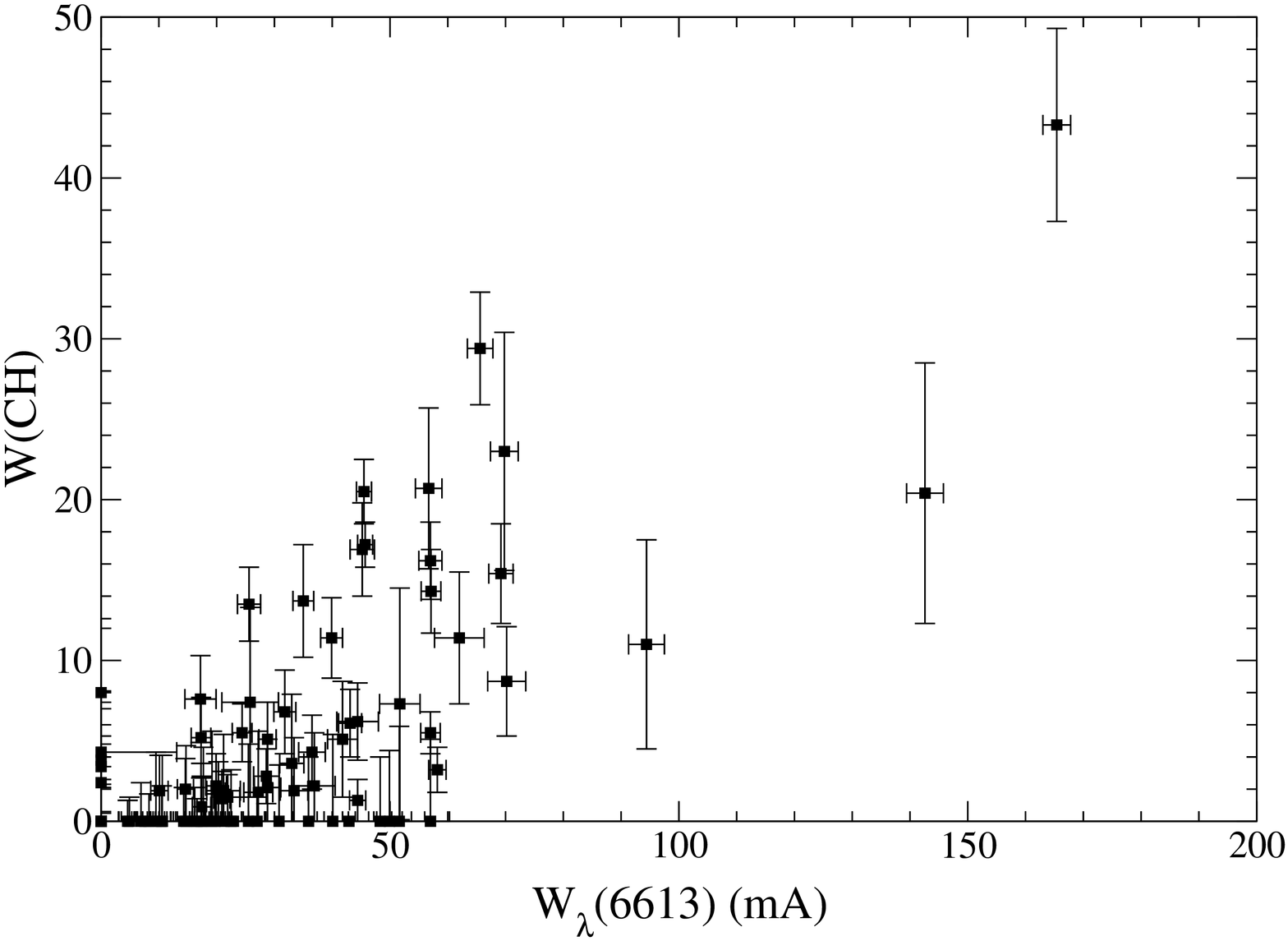}
   \includegraphics[angle=-90,width=\columnwidth,clip]{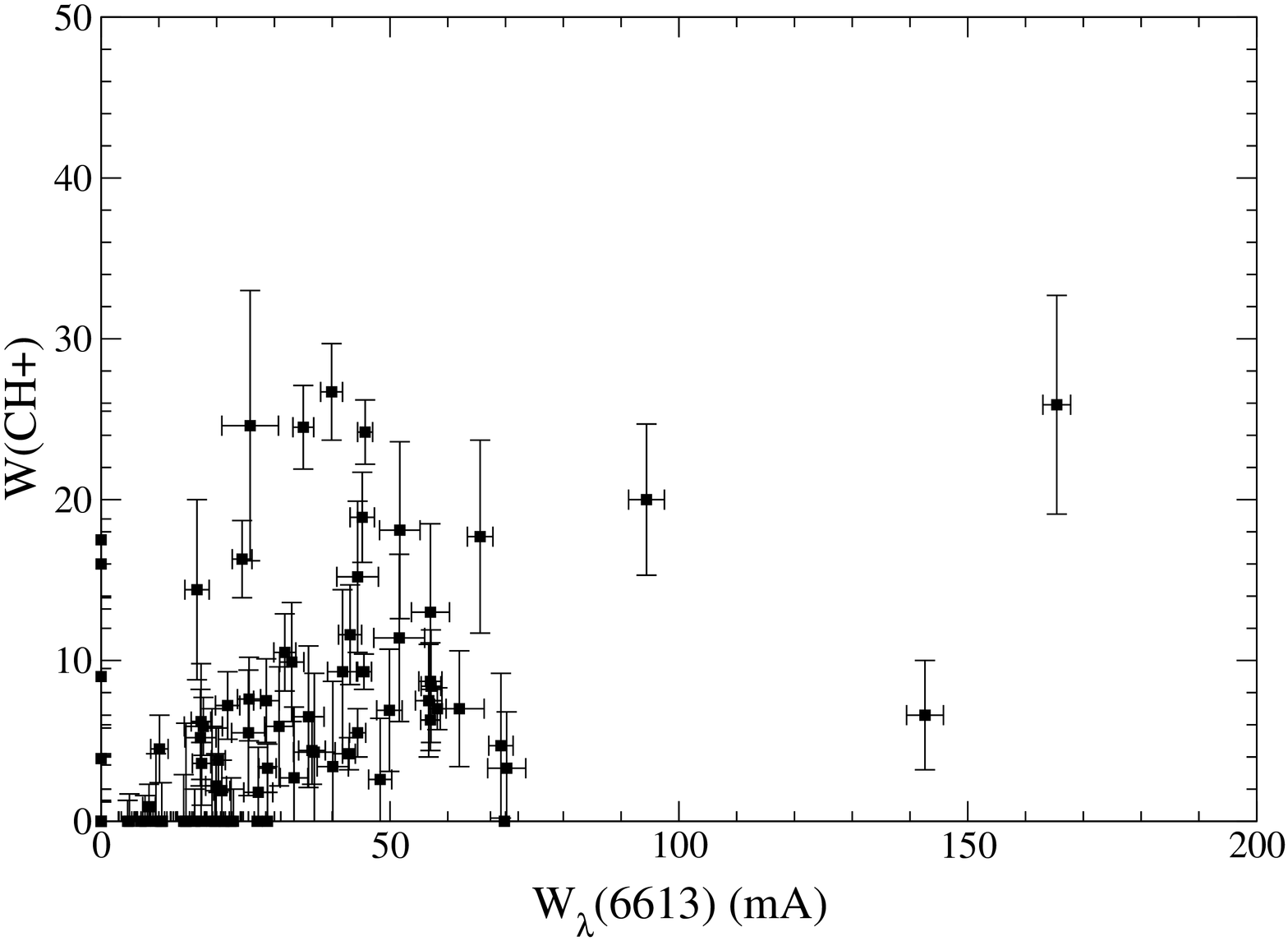}
   \caption{%
   Equivalent widths for CH (top) and CH$^+$ (bottom) are plotted versus W(6613),
   the strength of the 6613~\AA\ DIB. See Sect.~\ref{sec:dibs-diatomics}.
   }     
   \label{fig:DIBvsCH3}
\end{figure}

\begin{figure}[h!]
\centering 
   \includegraphics[angle=-90,width=1.1\columnwidth,clip]{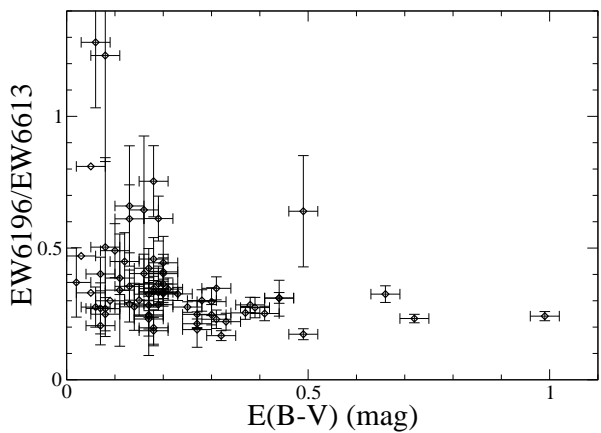}
   \caption{%%
   The $W(6196)/W(6613)$ ratio plotted against \Ebv.
   Contrary to the $(5797)/W(5780)$ ratio discussed in Sect.~\ref{subsec:DIBratios} this ratio is less
   sensitive to reddening.}
   \label{fig:ratio_vs_ebv2}
\end{figure}

\end{appendix}

\newpage
~
\newpage
\onecolumn

\begin{appendix}
\section{Line profiles}\label{sec:profiles}

{
\begin{figure*}[bth!] 
	\includegraphics[bb=100 40 565 300, angle=-90, width=6cm,clip]{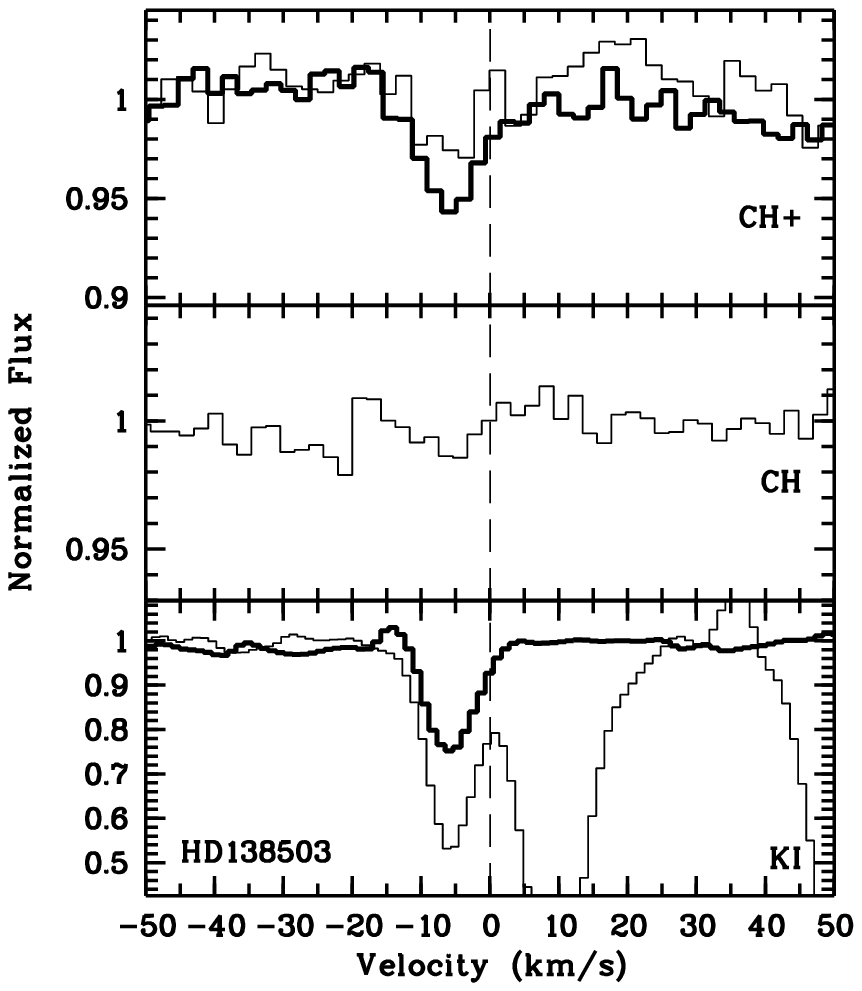} 
	\includegraphics[bb=100 40 565 300, angle=-90, width=6cm,clip]{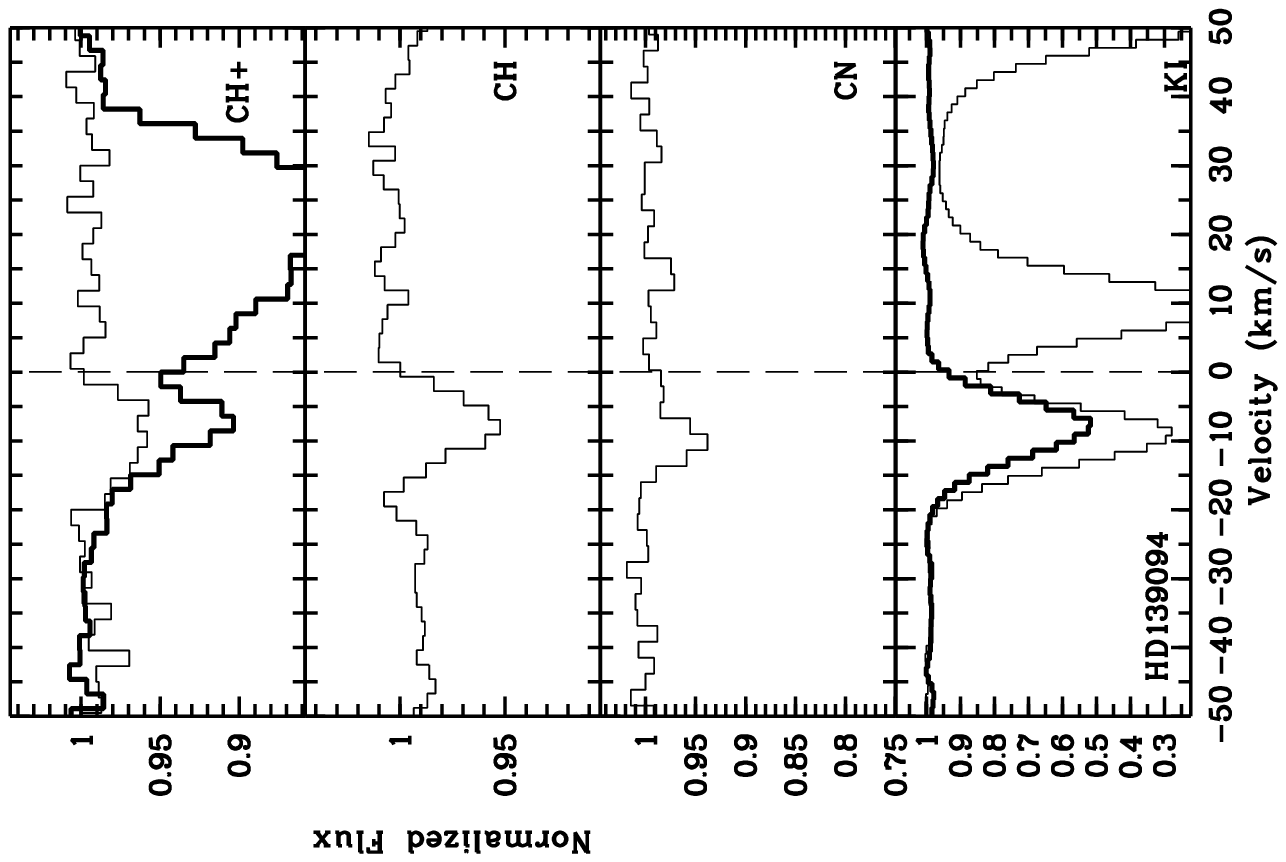} 
	\includegraphics[bb=100 40 565 300, angle=-90, width=6cm,clip]{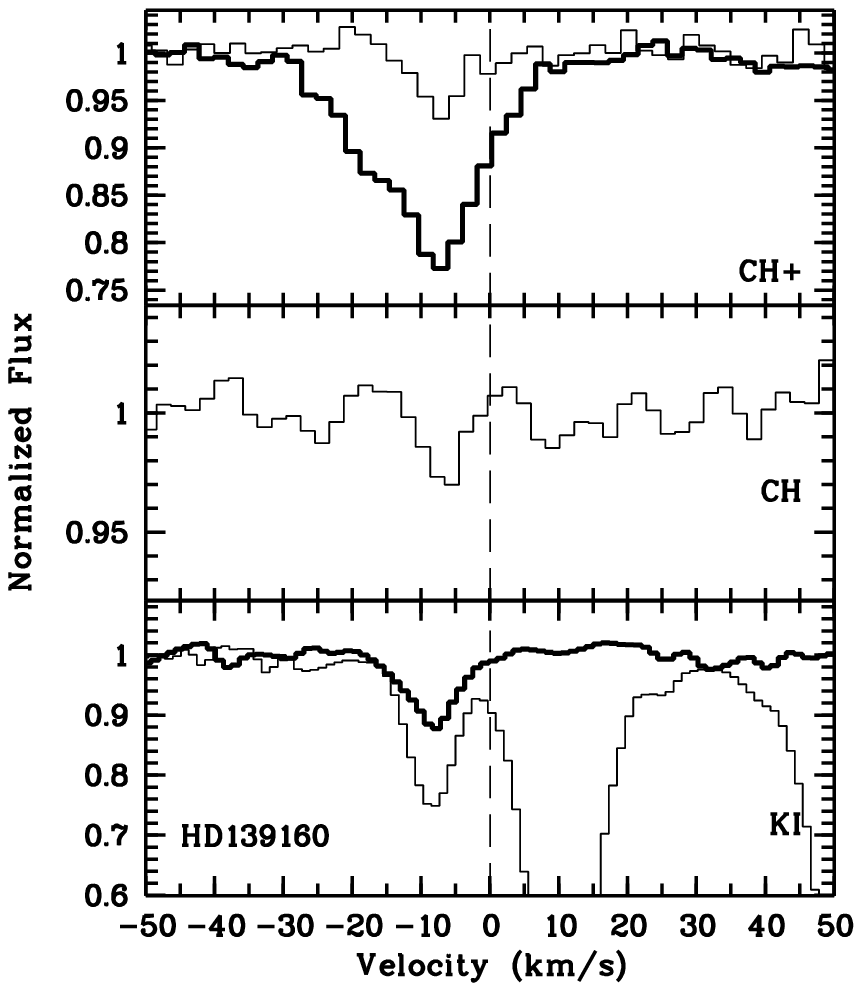} 
	\includegraphics[bb=100 40 565 300, angle=-90, width=6cm,clip]{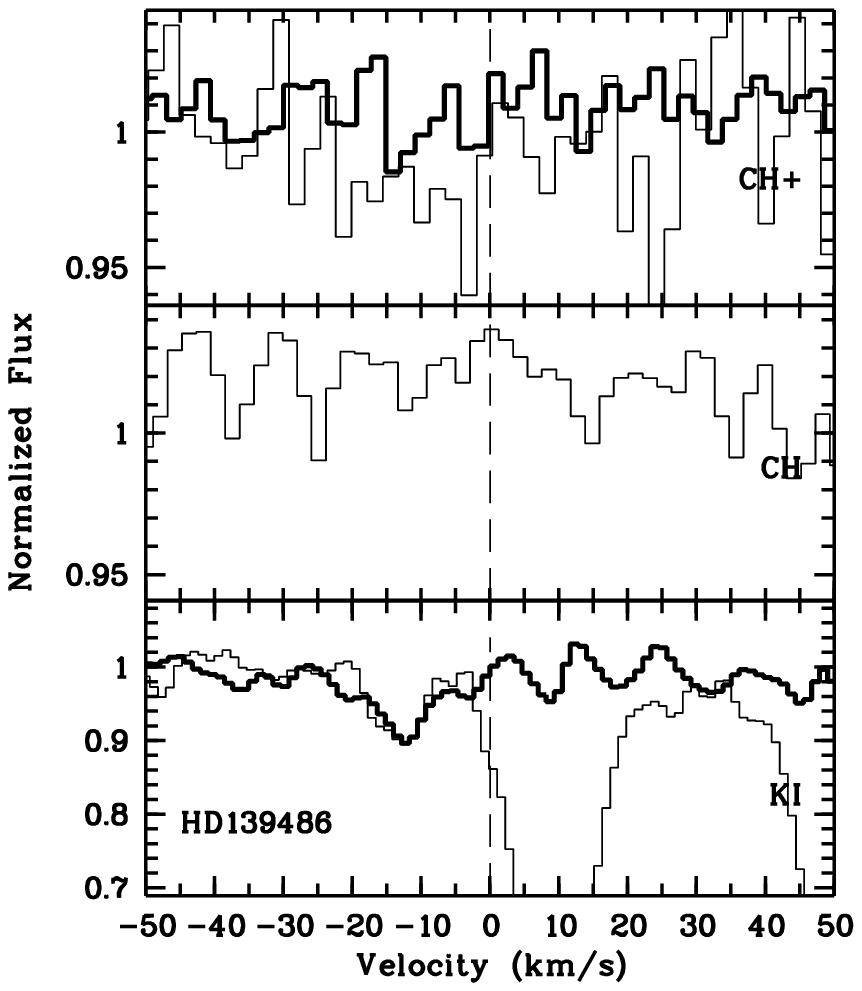} 
	\includegraphics[bb=100 40 565 300, angle=-90, width=6cm,clip]{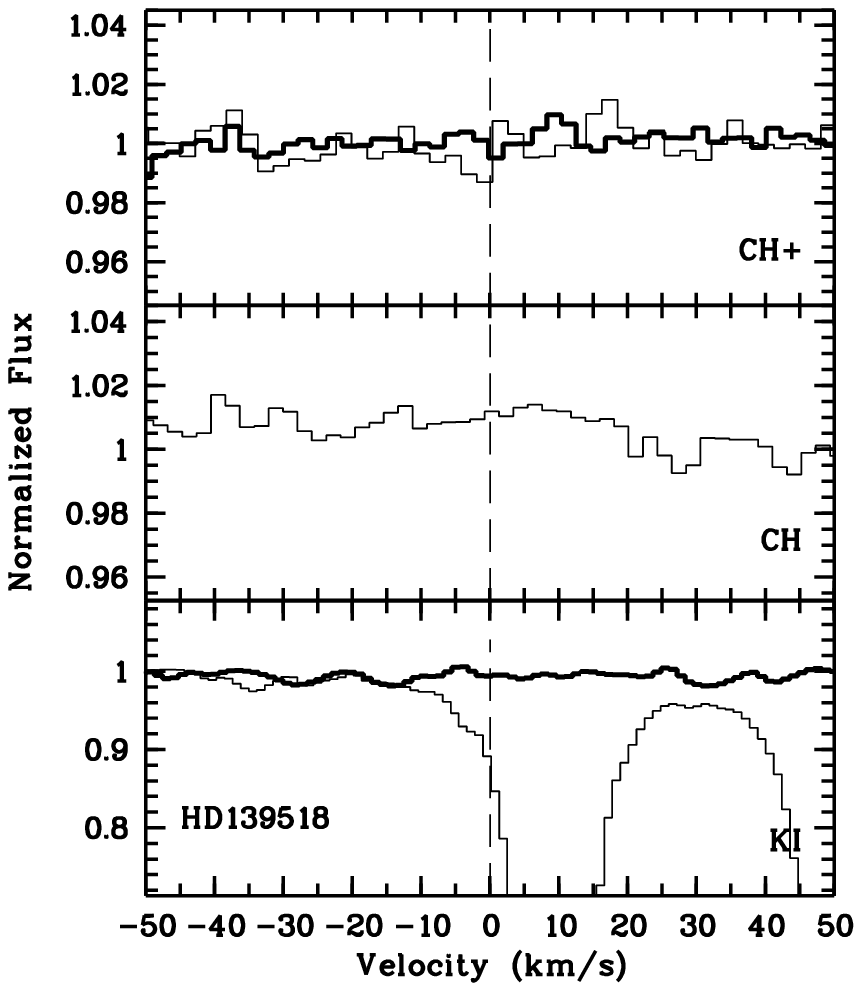} 
	\includegraphics[bb=100 40 565 300, angle=-90, width=6cm,clip]{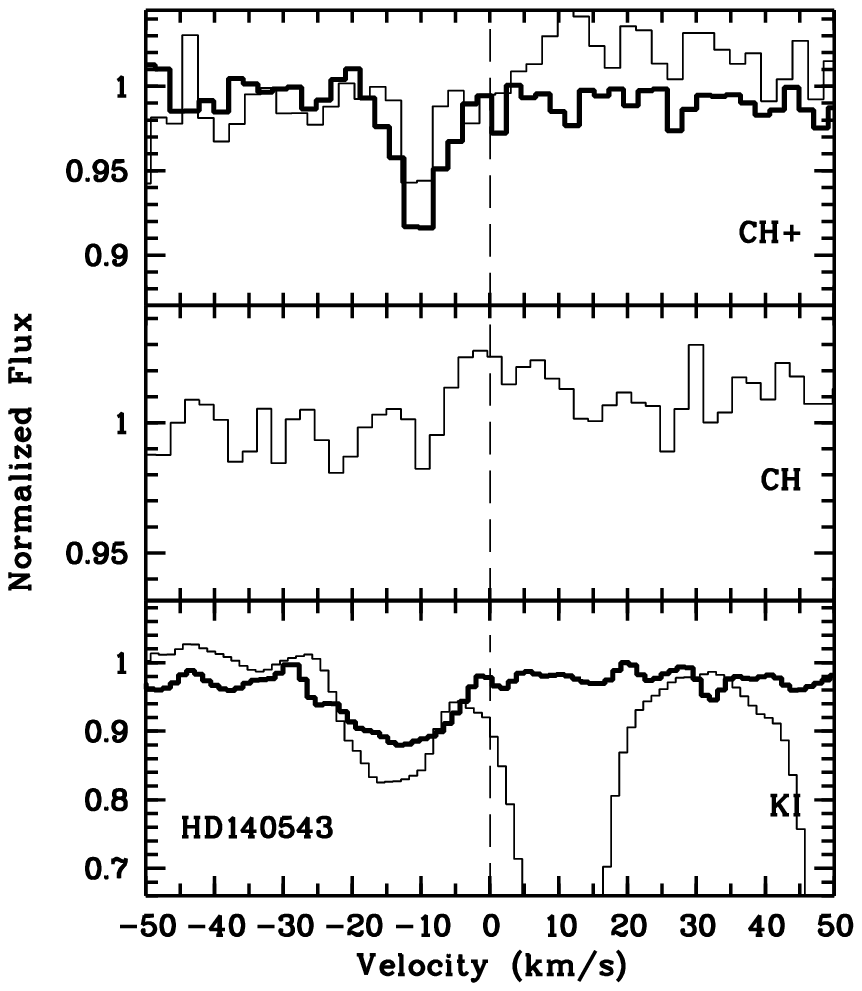} 
	\caption{Observed profiles for the interstellar lines of CH$^+$, CH, and \ion{K}{i} in sightlines
	probing the Upp Sco\ region. HD numbers correspond to those given in Table~\ref{tb:basic-data}.
	In the bottom panel the weaker \ion{K}{i} doublet line at 7699~\AA\ is over-plotted (thick solid line) 
	on top of the stronger (telluric contaminated) 	\ion{K}{i} doublet line at 7665~\AA.
	The \ion{Ca}{i} and CN lines are included only if potentially detected.
	The CH and CH$^+$ lines are shown for all lines-of-sight. 
	The stronger CH$^+$ line at 4232~\AA\ is over-plotted (thick solid line) on top of the 3957~\AA\ line.
	}
	\label{fig:lineprofiles}\end{figure*}}\clearpage	

\addtocounter{figure}{-1}

{
\begin{figure*}[h!]  
	\includegraphics[bb=100 40 565 300, angle=-90, width=6cm,clip]{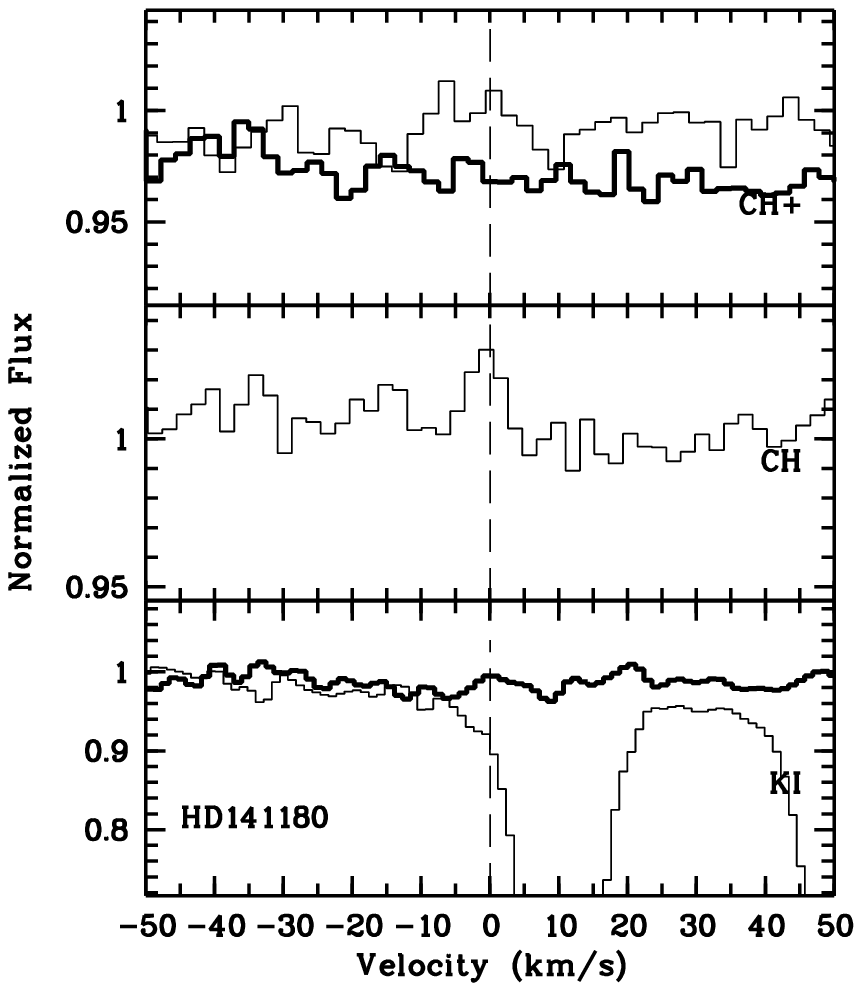}    
	\includegraphics[bb=100 40 565 300, angle=-90, width=6cm,clip]{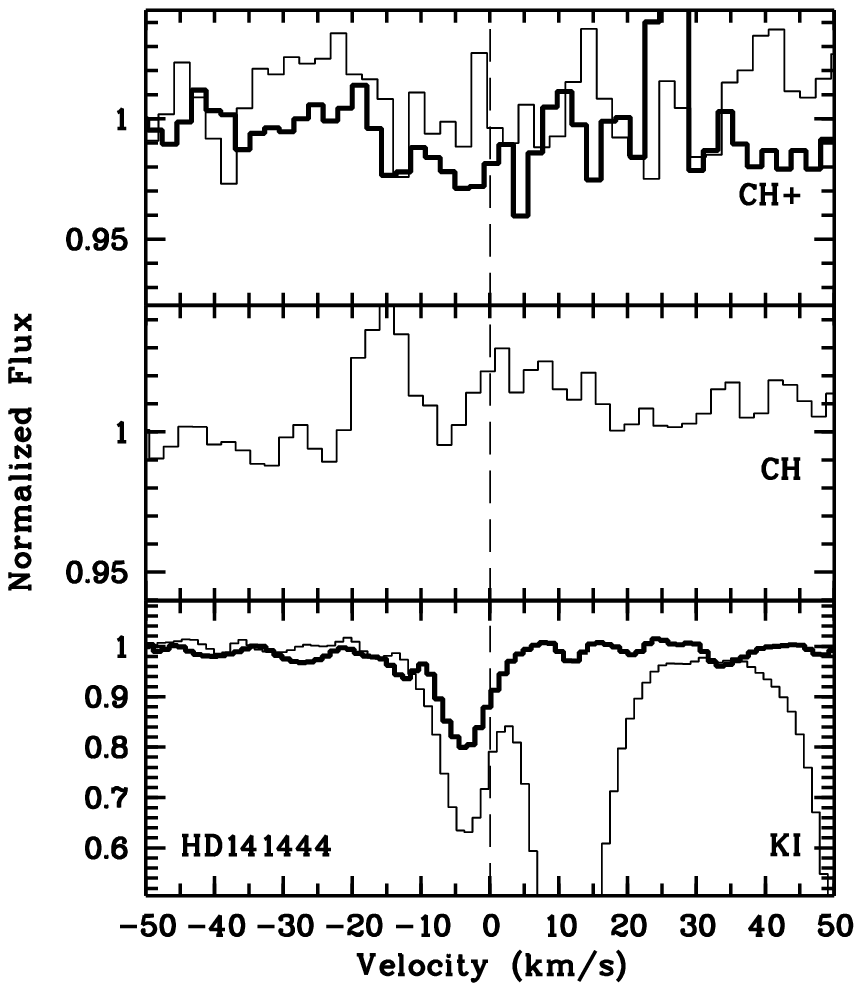}    
	\includegraphics[bb=100 40 565 300, angle=-90, width=6cm,clip]{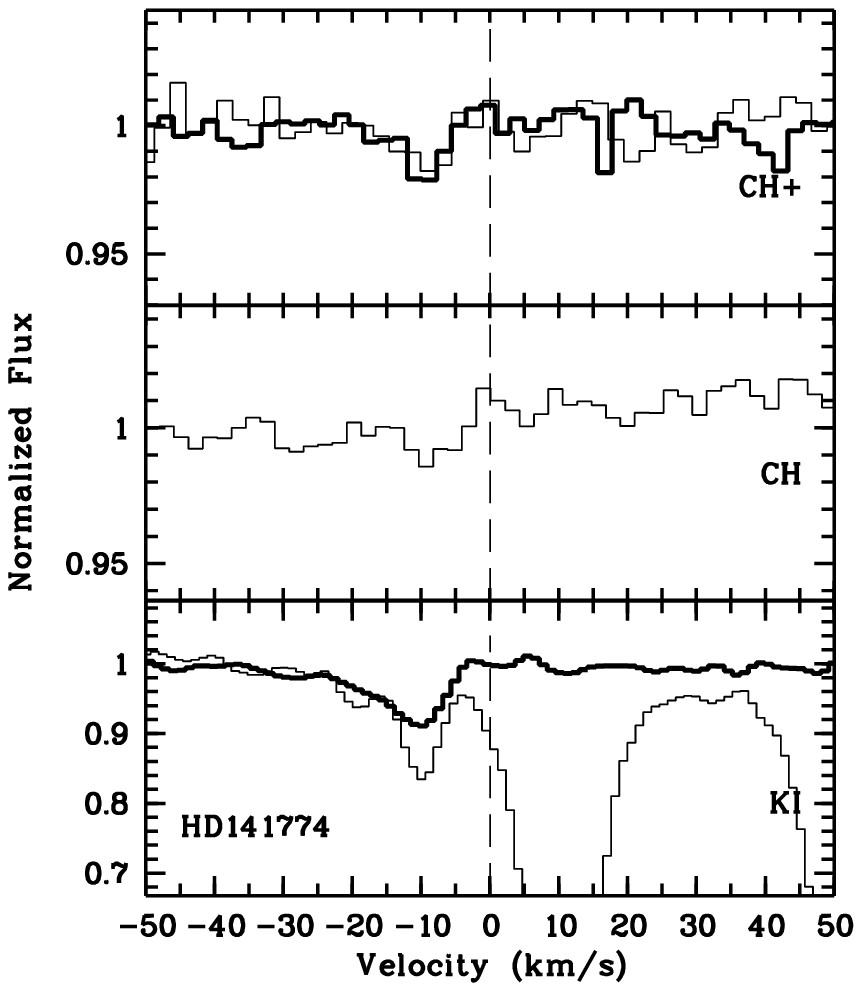}    
	\includegraphics[bb=100 40 565 300, angle=-90, width=6cm,clip]{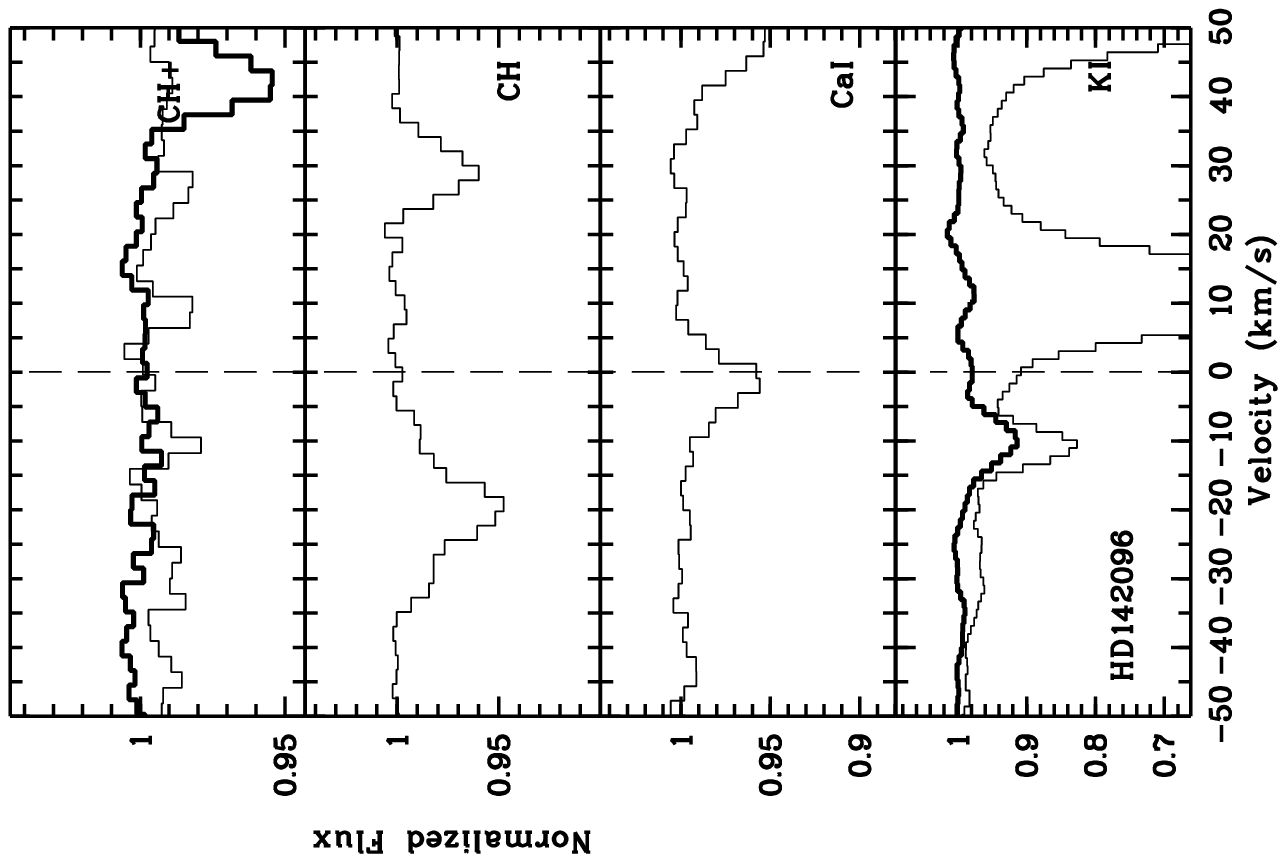}
	\includegraphics[bb=100 40 565 300, angle=-90, width=6cm,clip]{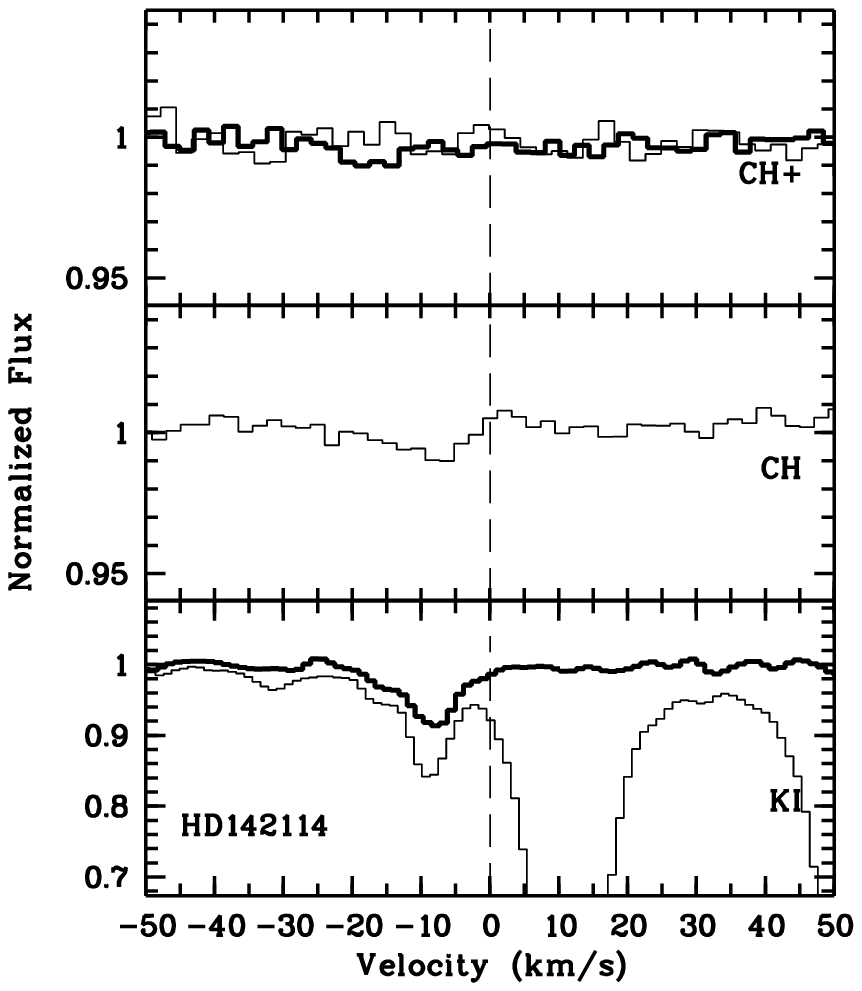}    
	\includegraphics[bb=100 40 565 300, angle=-90, width=6cm,clip]{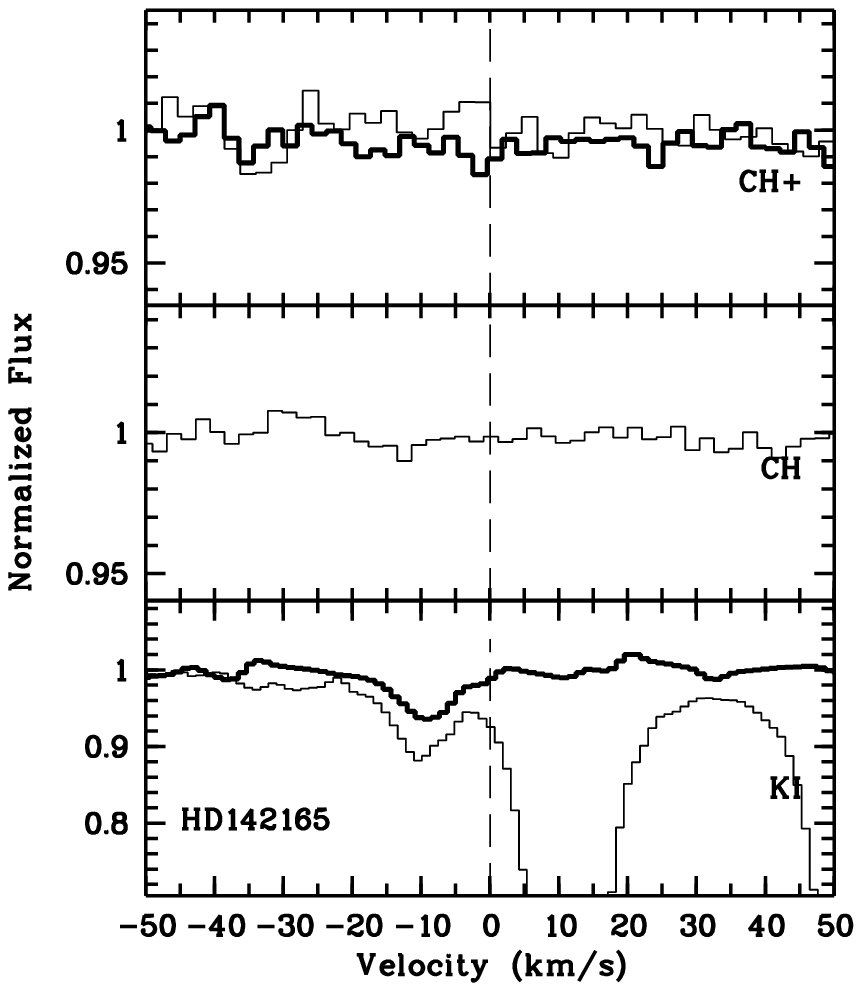}  \caption{Continued.} \end{figure*}}\clearpage    
\addtocounter{figure}{-1}

{
\begin{figure*}[h!] 
	\includegraphics[bb=100 40 565 300, angle=-90, width=6cm,clip]{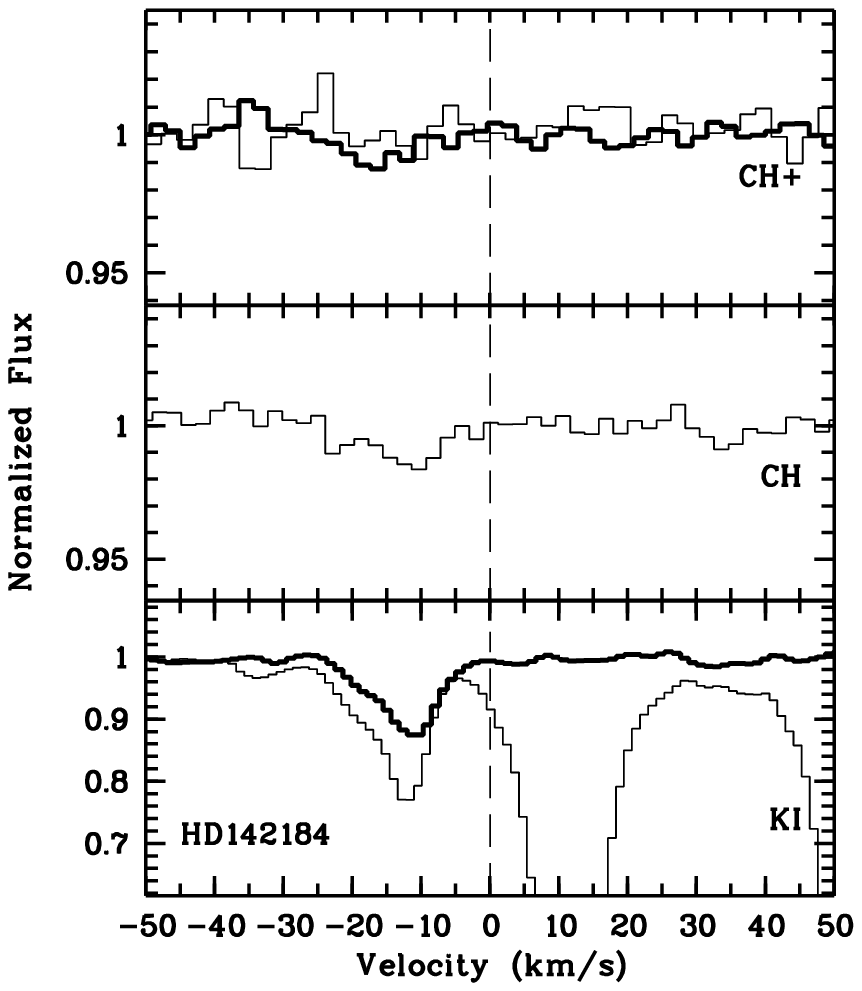}   
	\includegraphics[bb=100 40 565 300, angle=-90, width=6cm,clip]{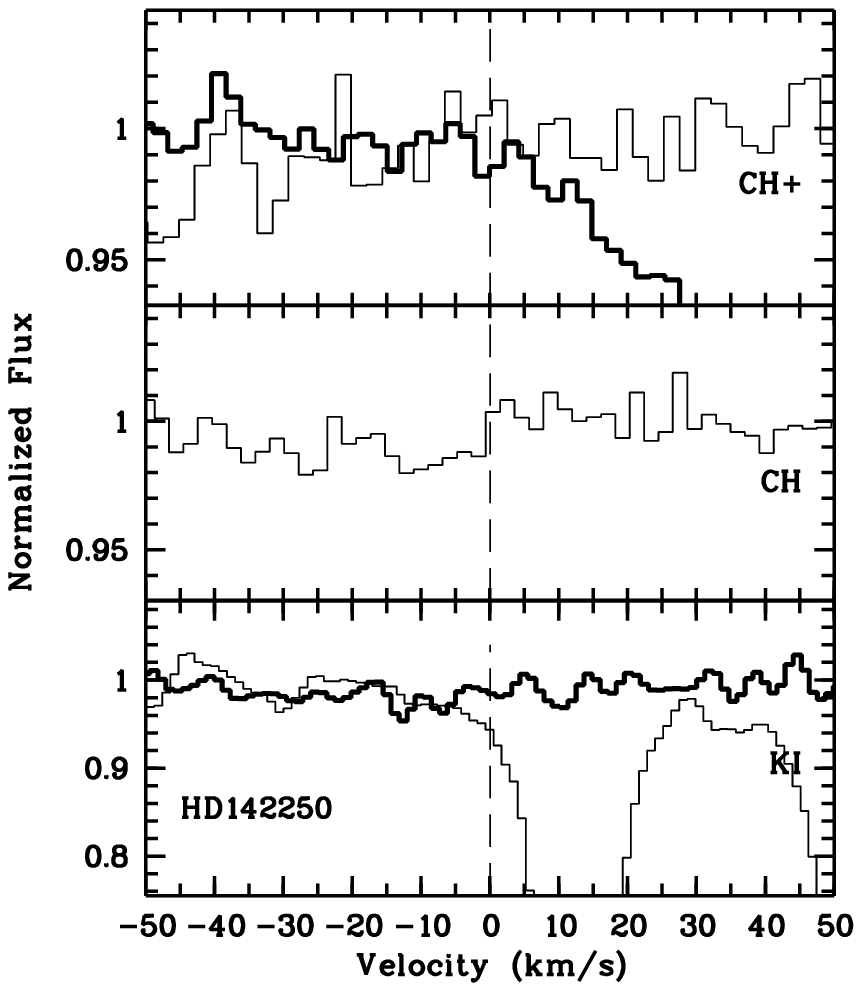}   
	\includegraphics[bb=100 40 565 300, angle=-90, width=6cm,clip]{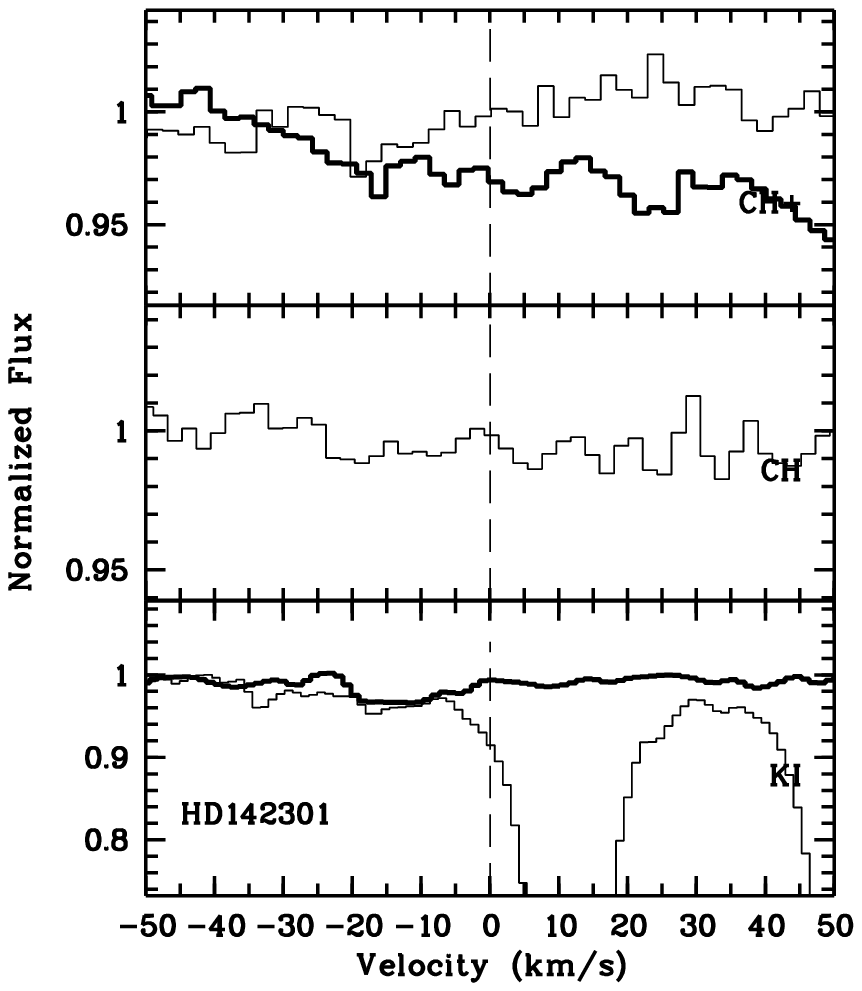}
	\includegraphics[bb=100 40 565 300, angle=-90, width=6cm,clip]{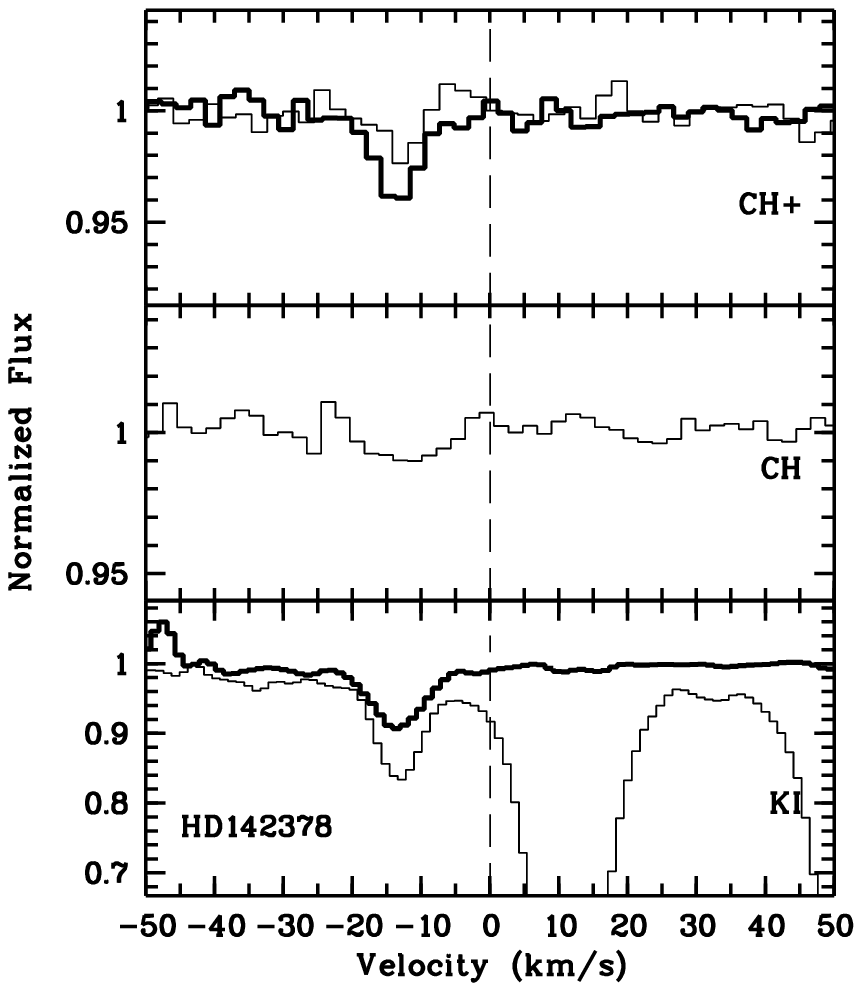}    
	\includegraphics[bb=100 40 565 300, angle=-90, width=6cm,clip]{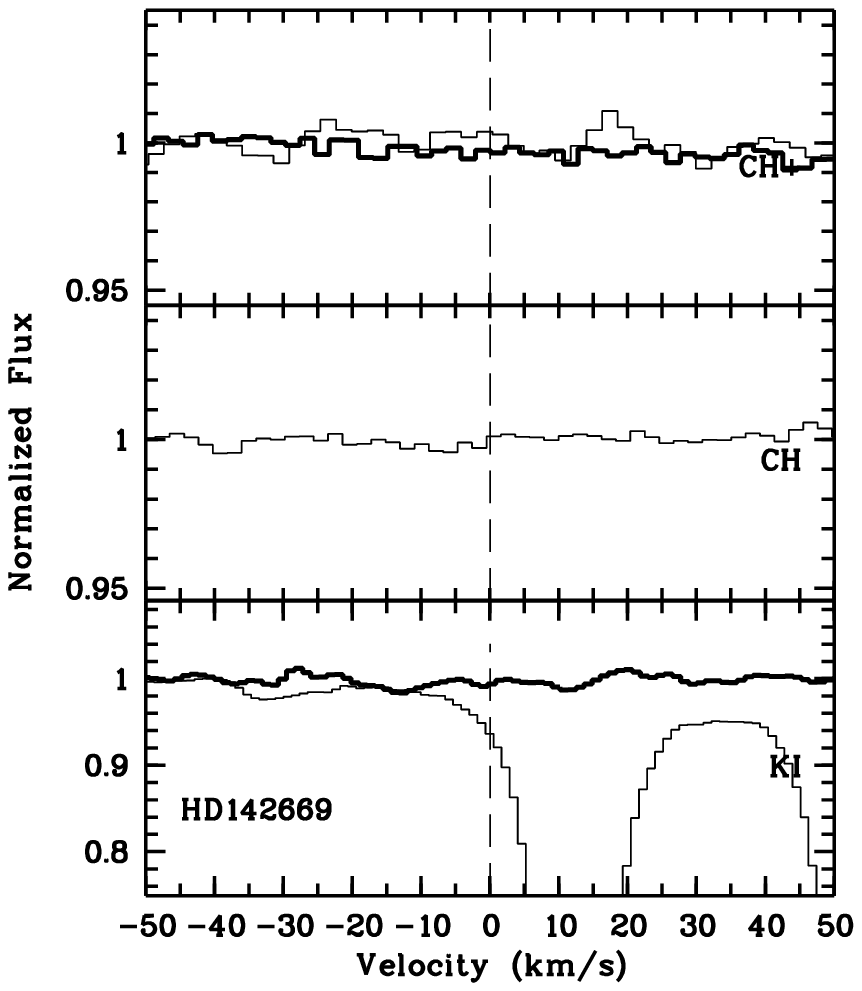}    
	\includegraphics[bb=100 40 565 300, angle=-90, width=6cm,clip]{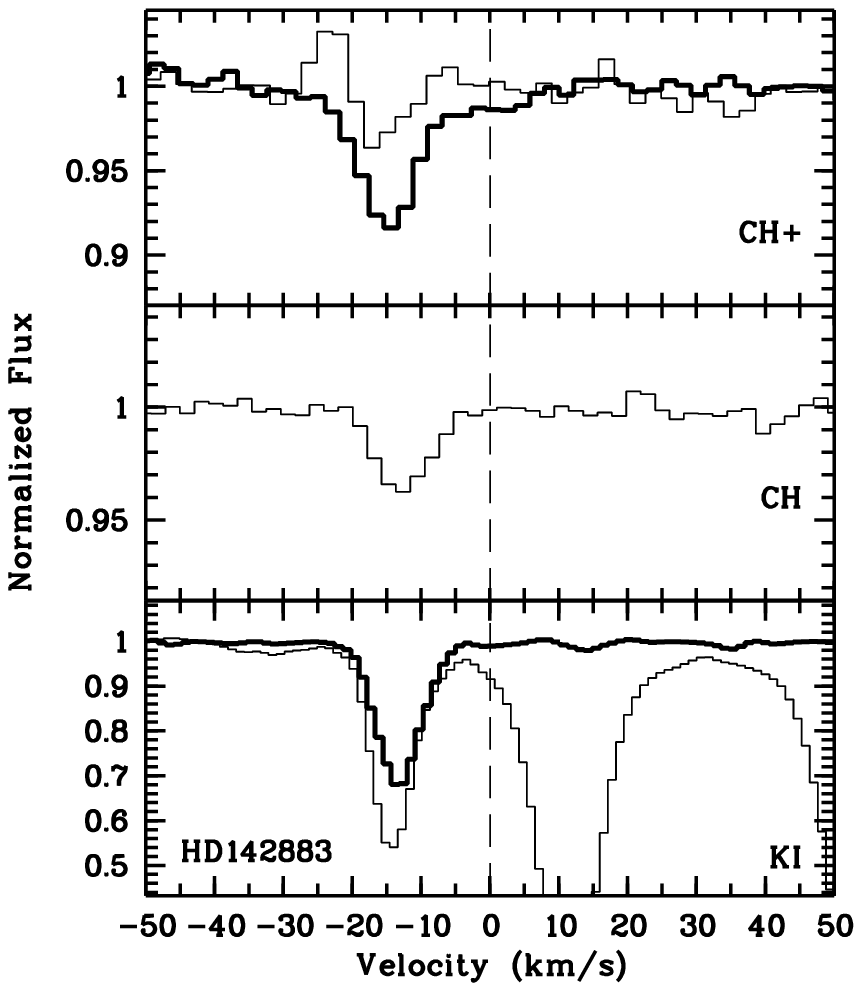}  \caption{Continued.} \end{figure*}}\clearpage      
\addtocounter{figure}{-1}

{
\begin{figure*}[h!]   
	\includegraphics[bb=100 40 565 300, angle=-90, width=6cm,clip]{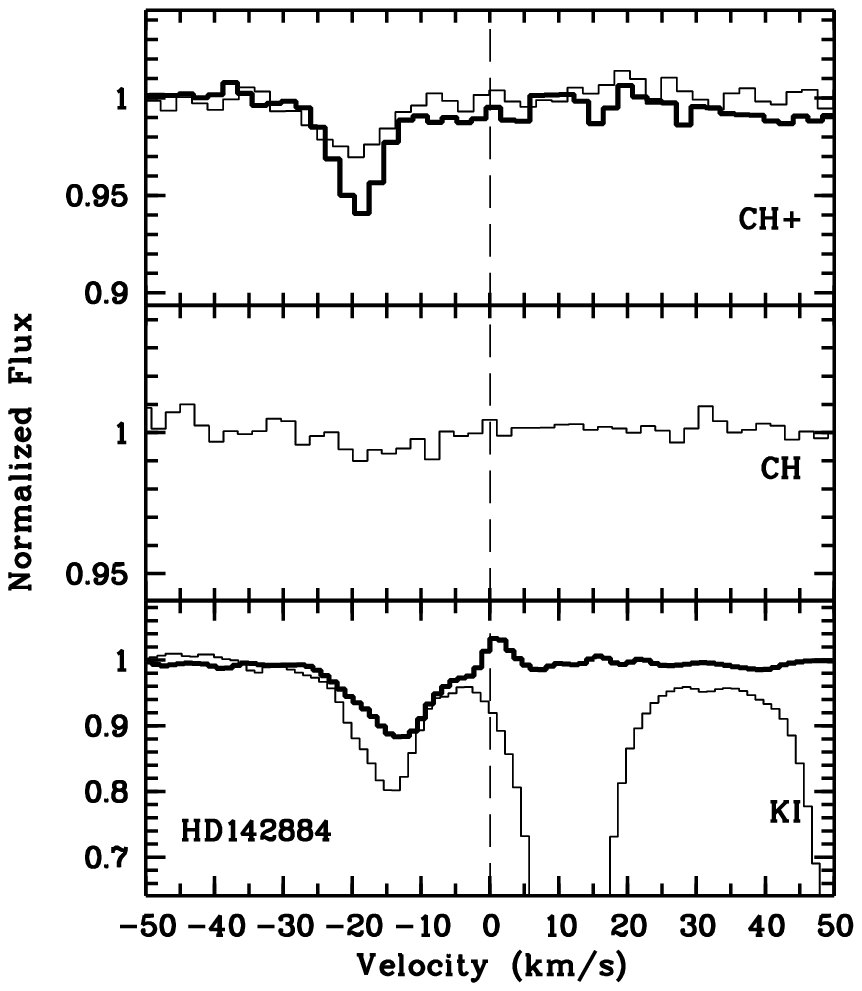}    
	\includegraphics[bb=100 40 565 300, angle=-90, width=6cm,clip]{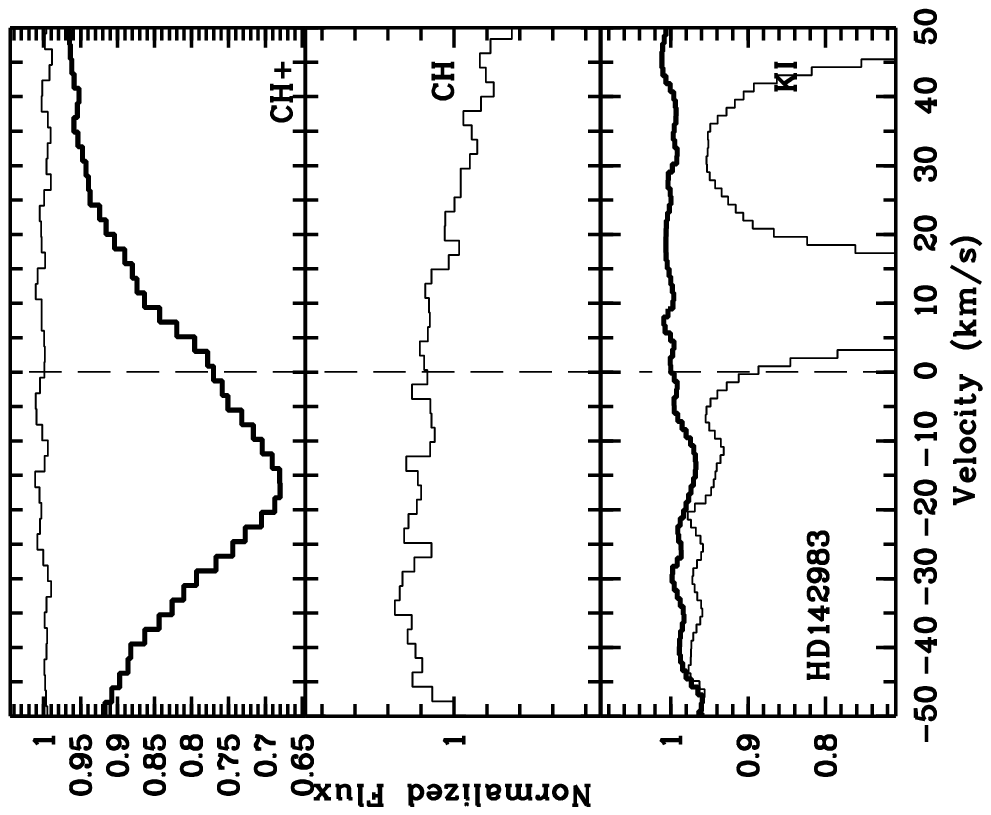}
	\includegraphics[bb=100 40 565 300, angle=-90, width=6cm,clip]{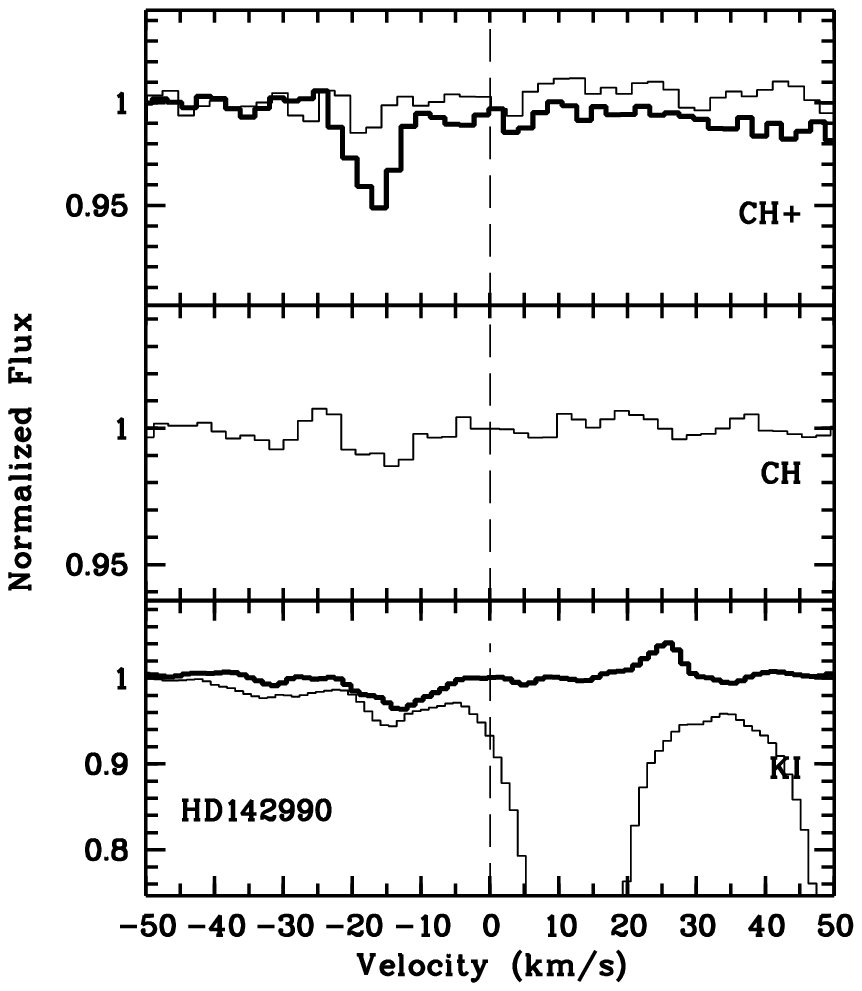}   
	\includegraphics[bb=100 40 565 300, angle=-90, width=6cm,clip]{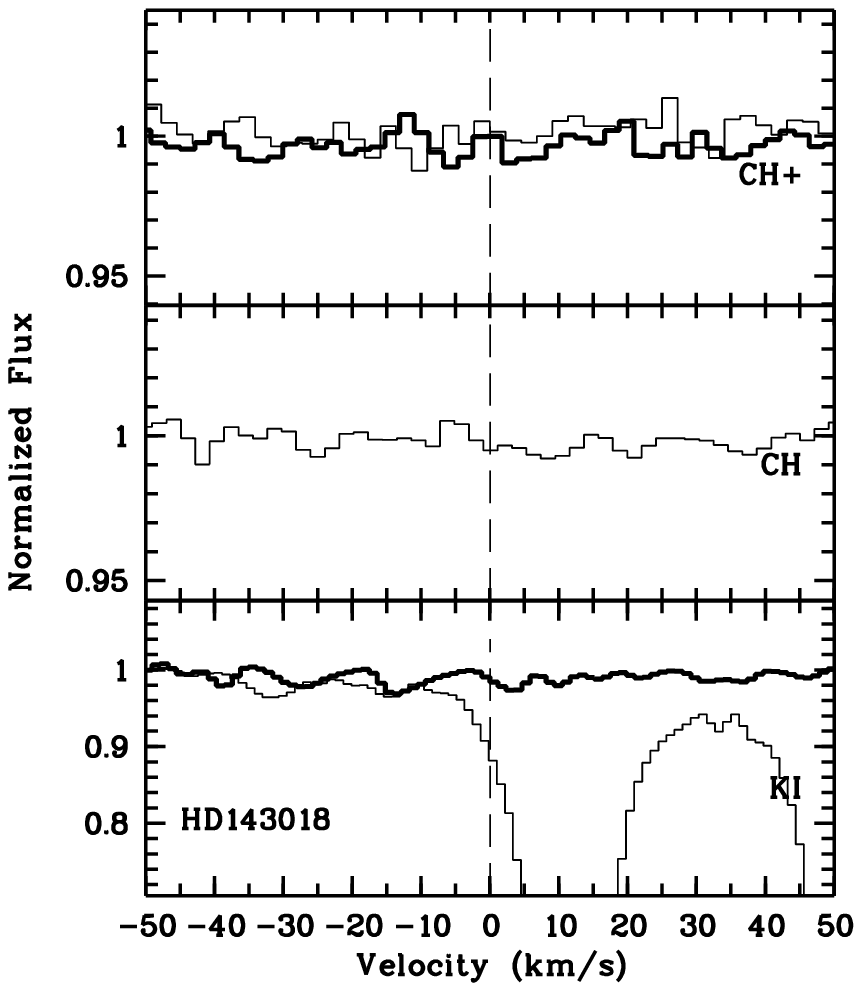}    
	\includegraphics[bb=100 40 565 300, angle=-90, width=6cm,clip]{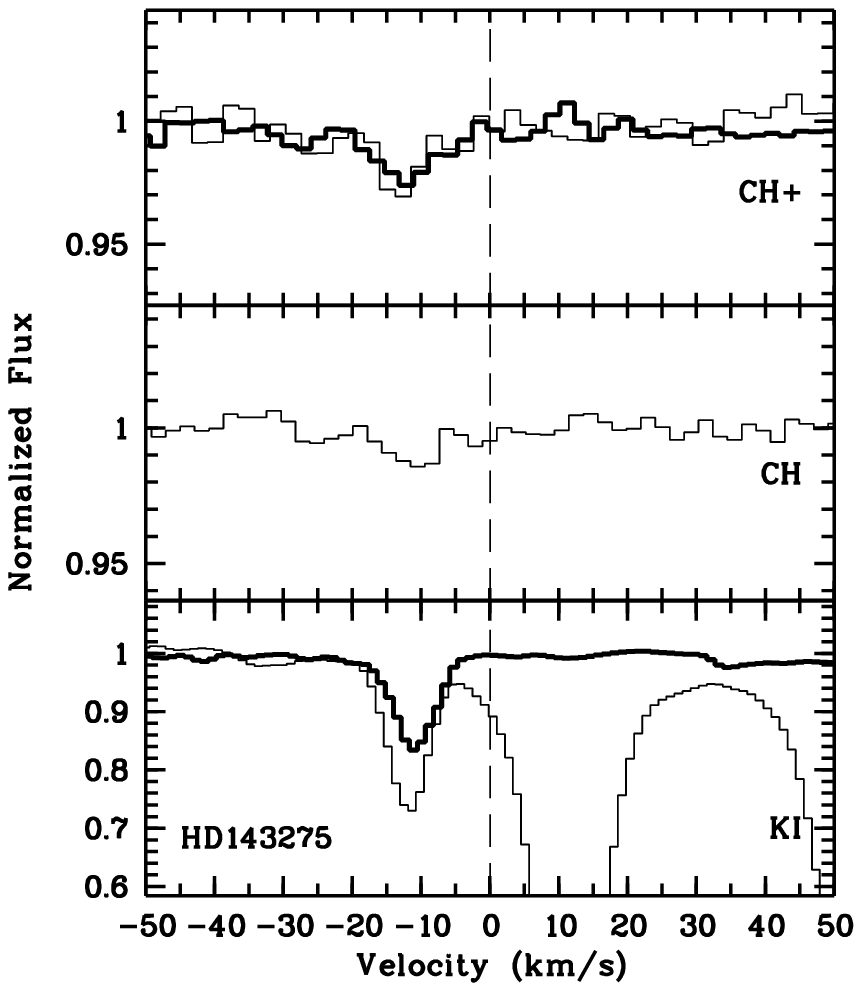}    
	\includegraphics[bb=100 40 565 300, angle=-90, width=6cm,clip]{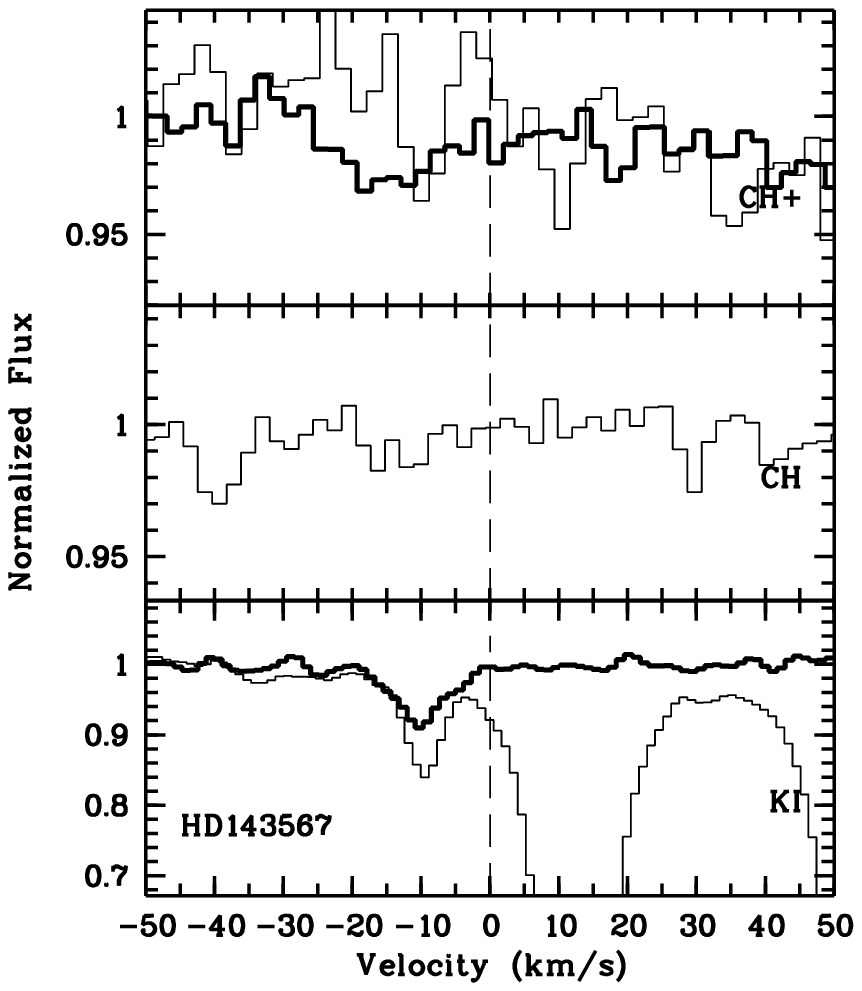}  \caption{Continued.} \end{figure*}}\clearpage      
\addtocounter{figure}{-1}

{
\begin{figure*}[h!]   
	\includegraphics[bb=100 40 565 300, angle=-90, width=6cm,clip]{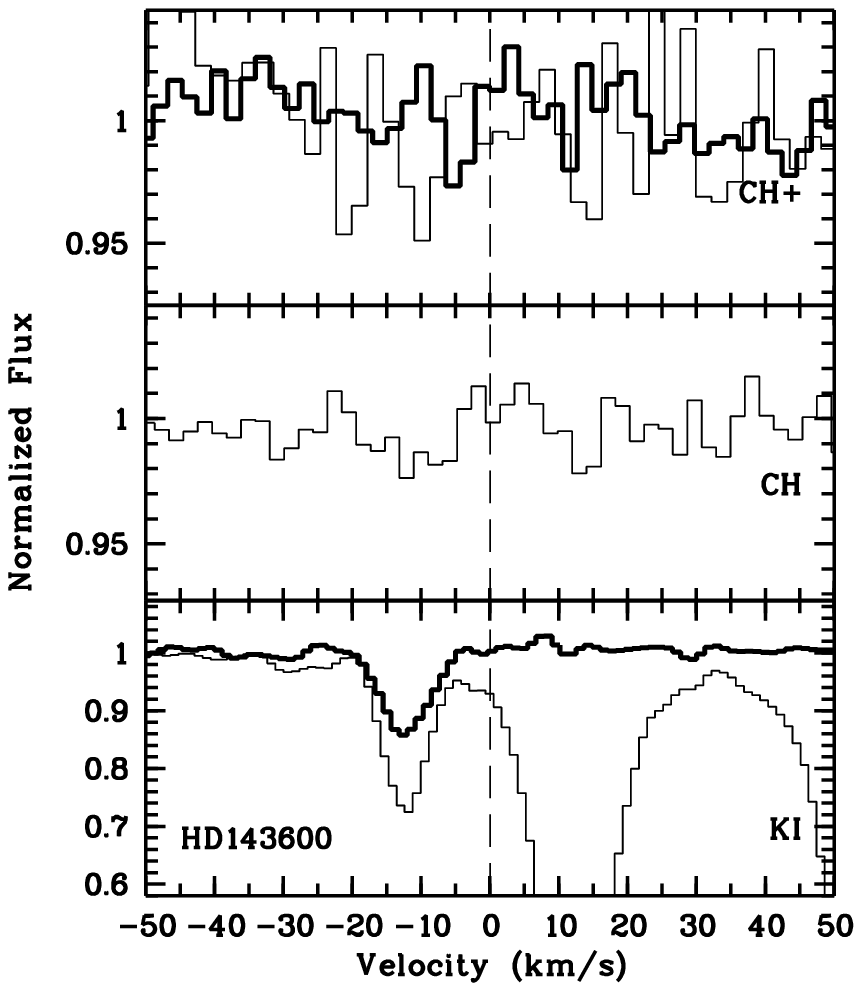}
	\includegraphics[bb=100 40 565 300, angle=-90, width=6cm,clip]{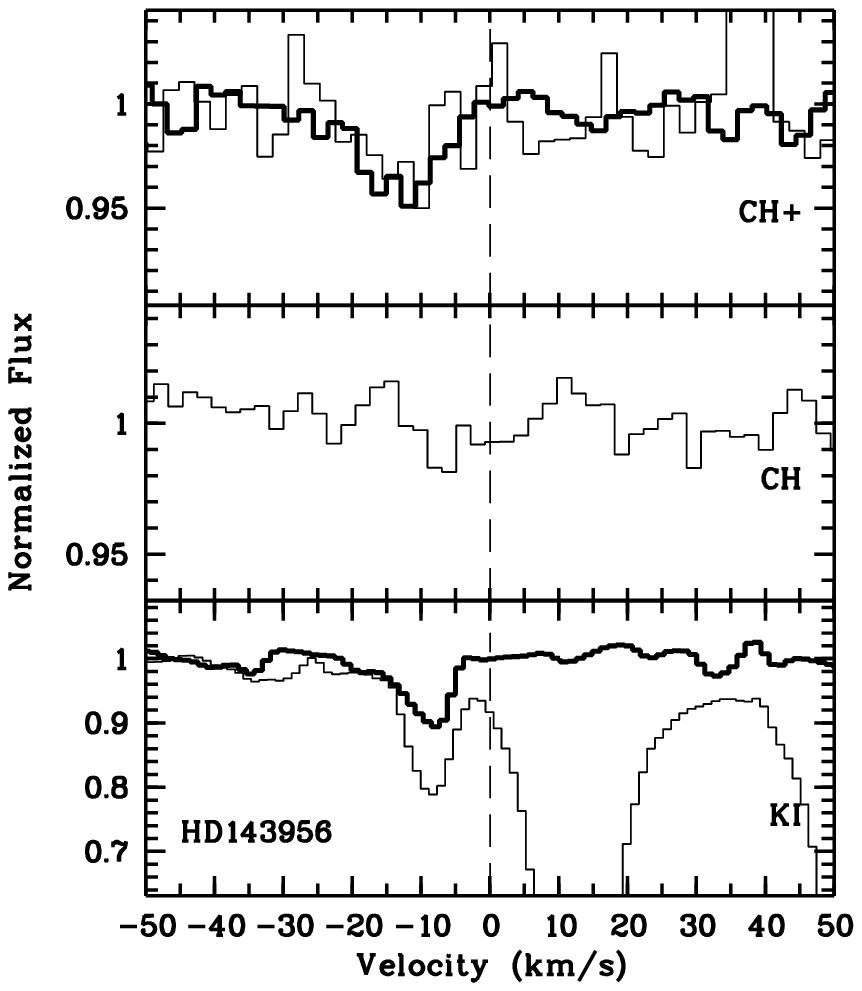} 
	\includegraphics[bb=100 40 565 300, angle=-90, width=6cm,clip]{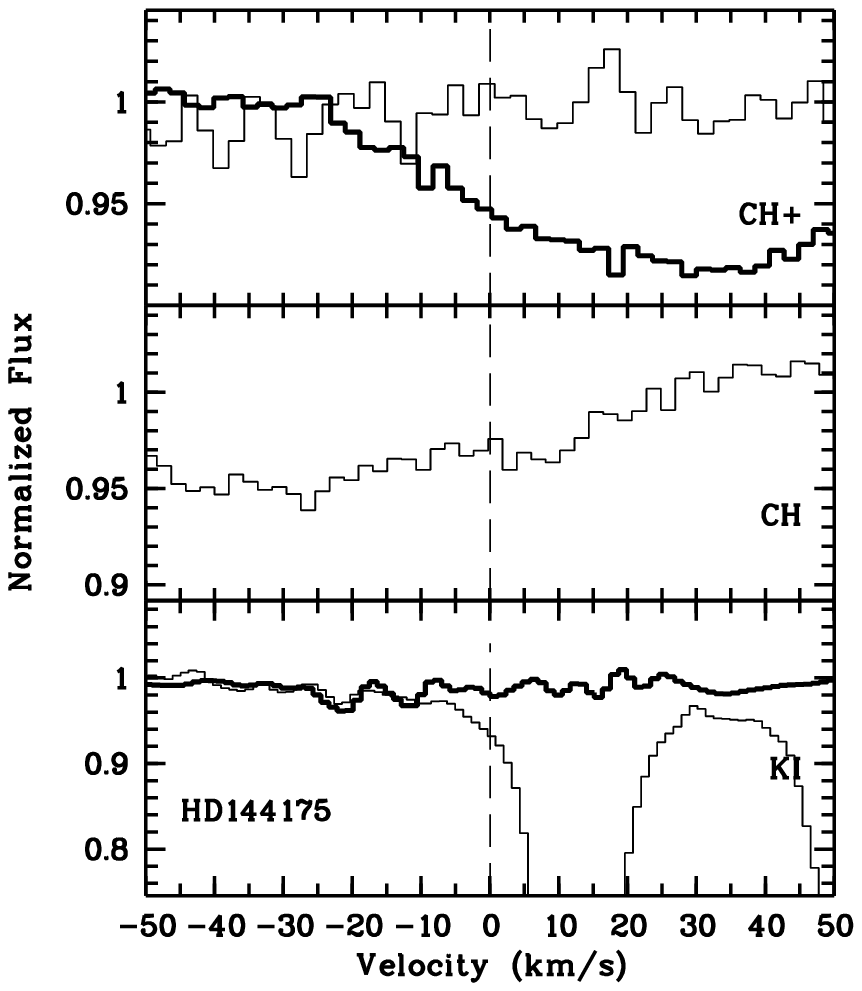} 
	\includegraphics[bb=100 40 565 300, angle=-90, width=6cm,clip]{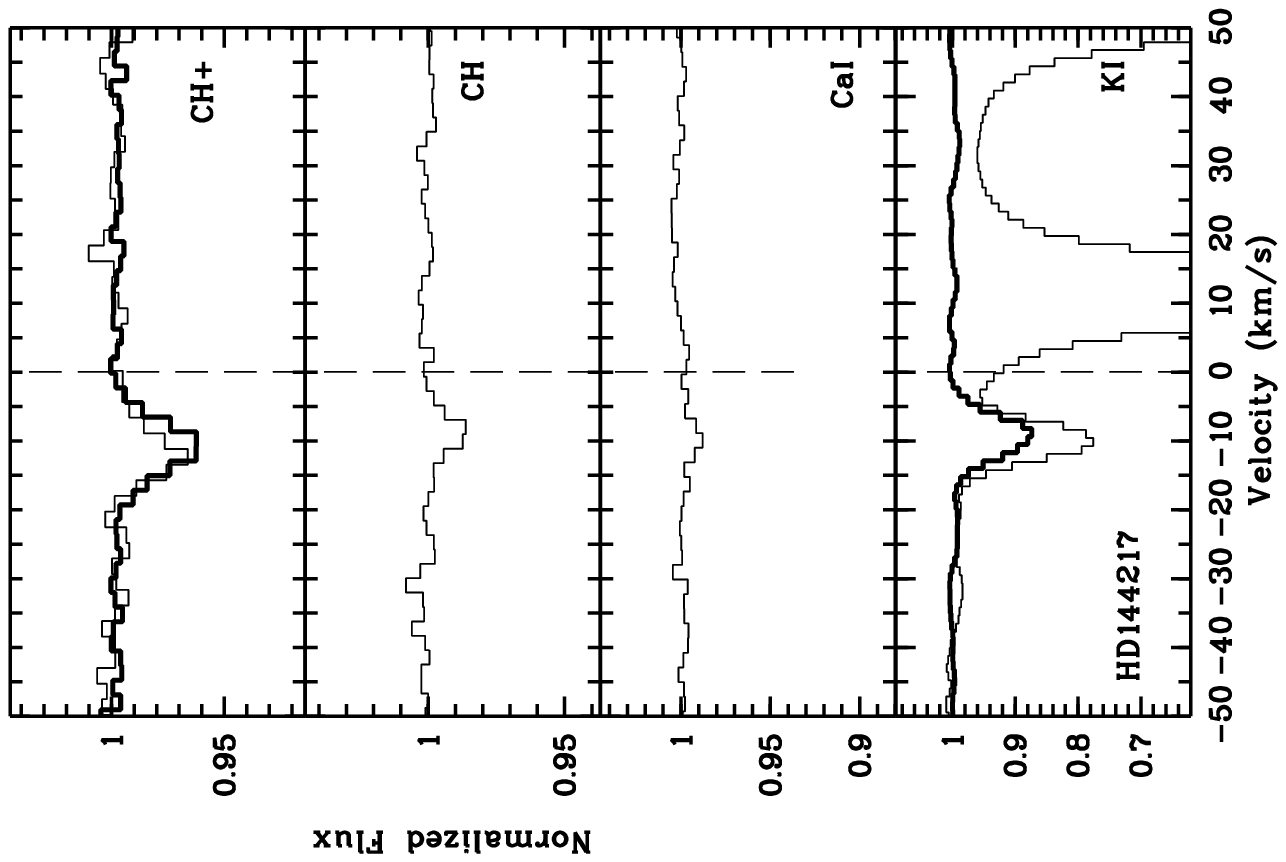} 
	\includegraphics[bb=100 40 565 300, angle=-90, width=6cm,clip]{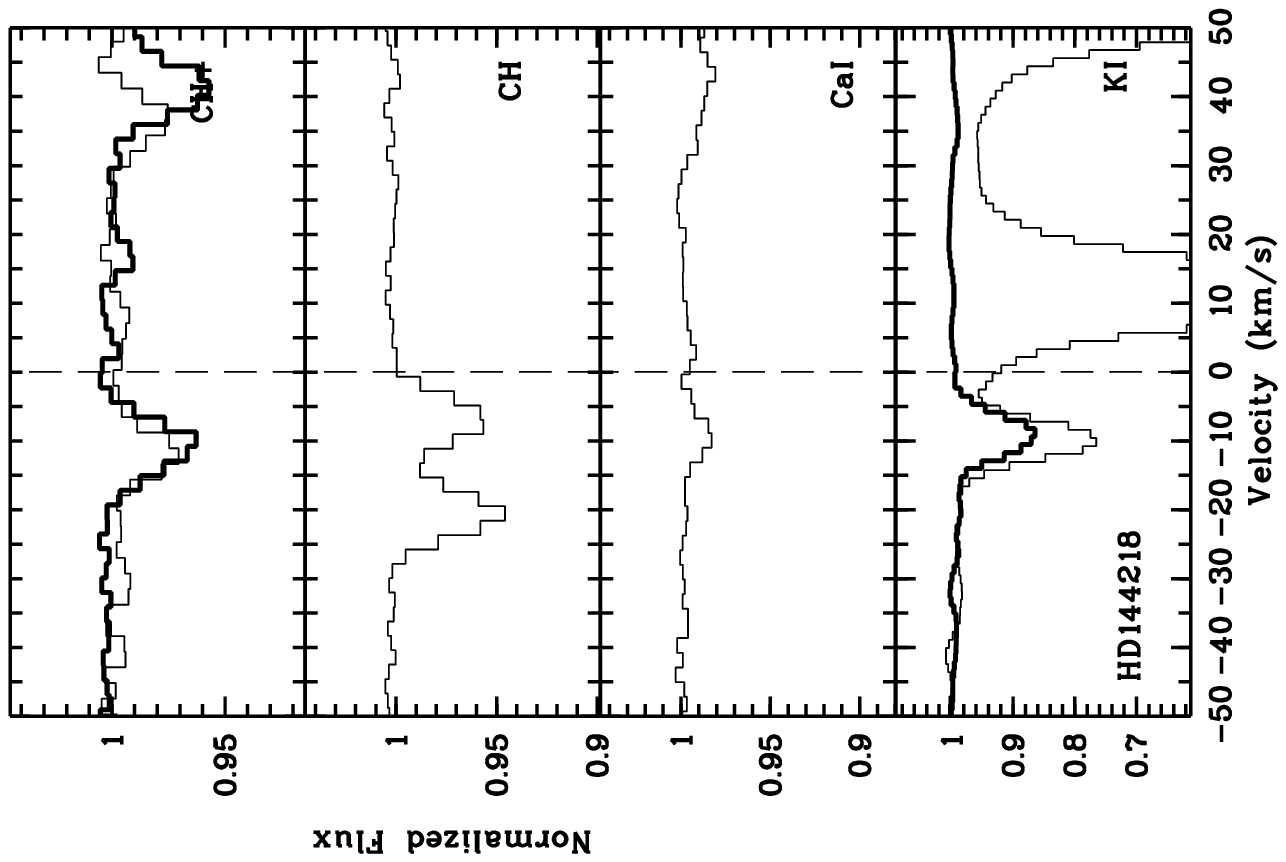} 
	\includegraphics[bb=100 40 565 300, angle=-90, width=6cm,clip]{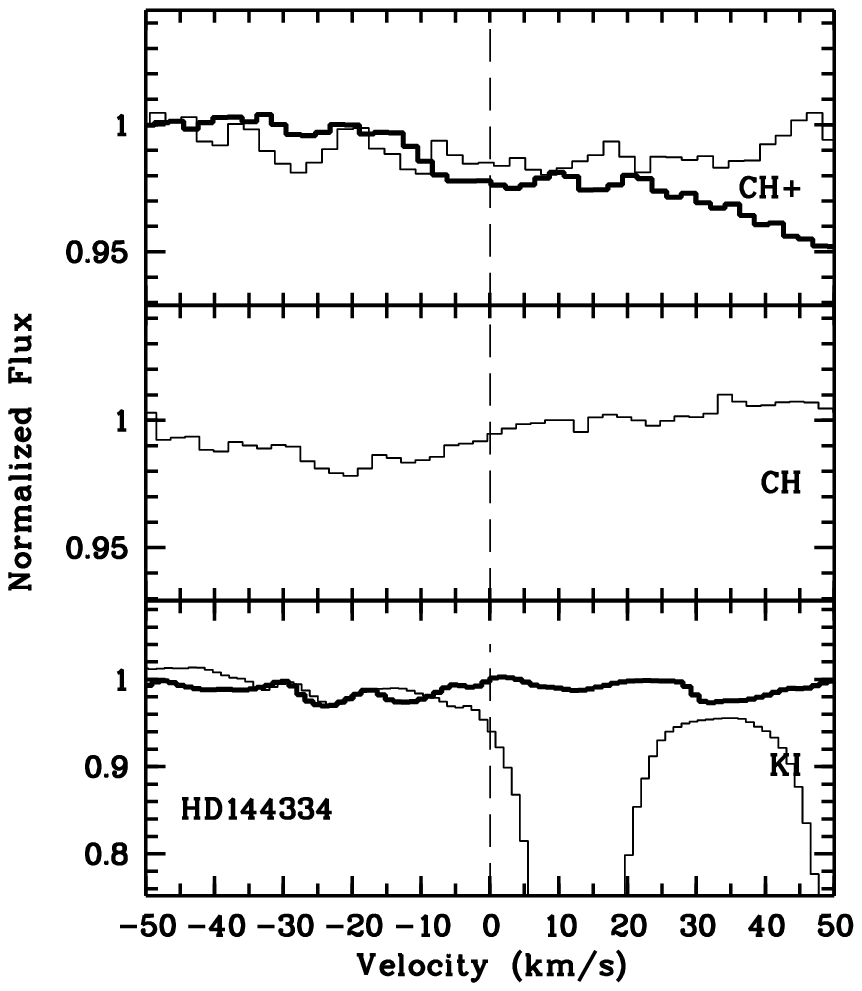} \caption{Continued.} \end{figure*}}\clearpage      
\addtocounter{figure}{-1}

{
\begin{figure*}[h!]
        \includegraphics[bb=100 40 565 300, angle=-90, width=6cm,clip]{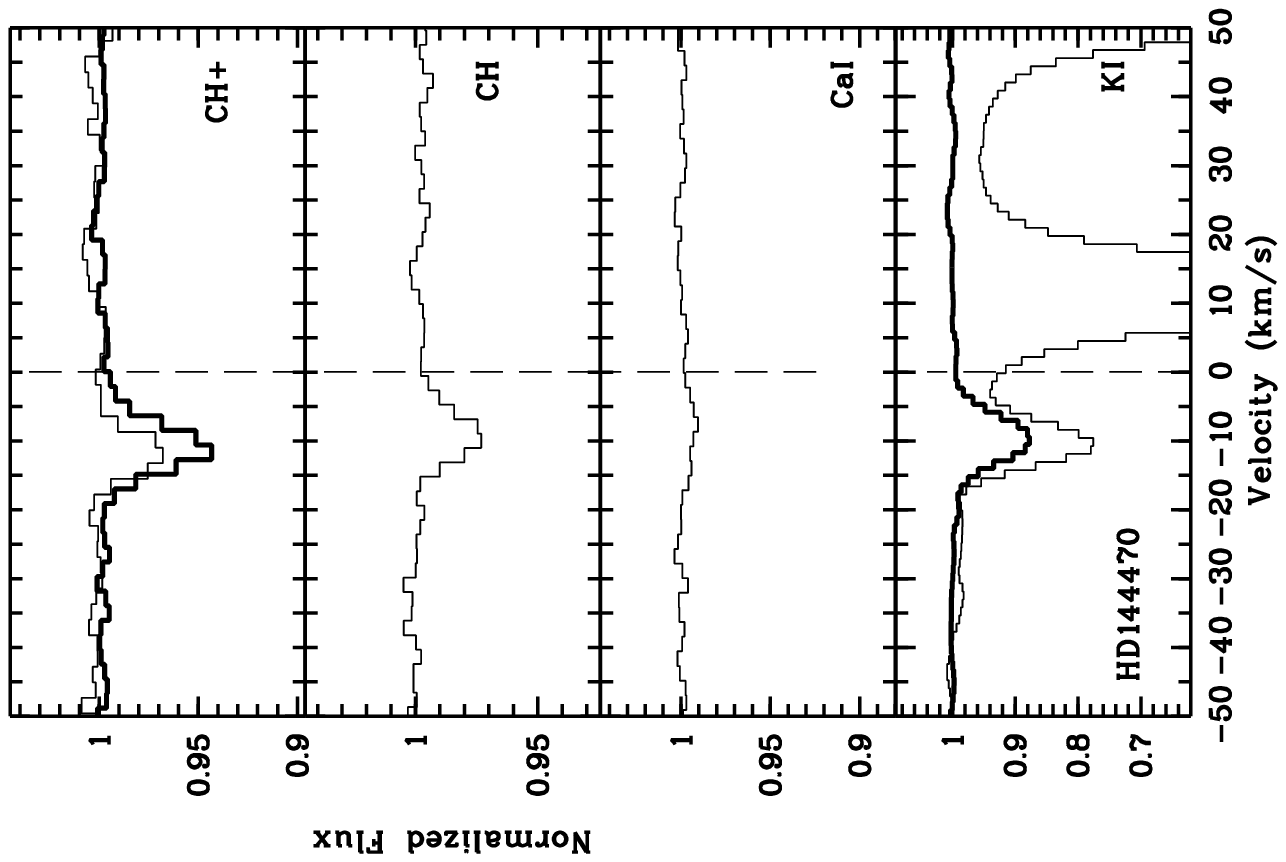}    
	\includegraphics[bb=100 40 565 300, angle=-90, width=6cm,clip]{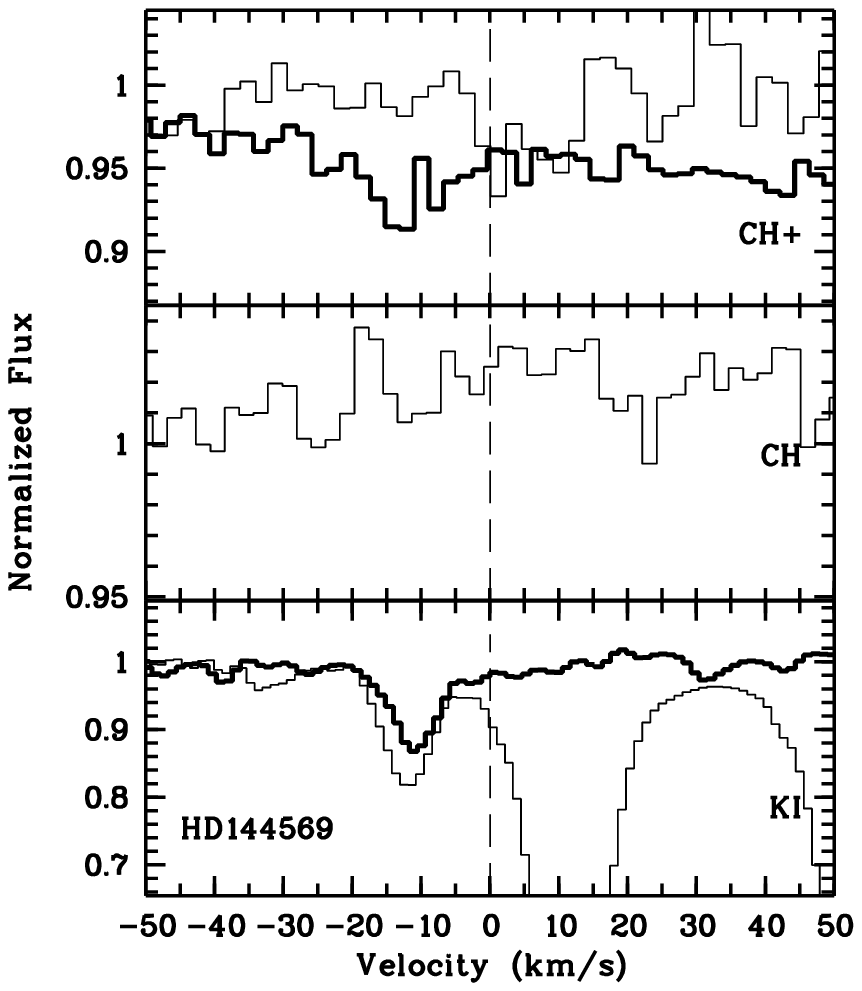}    
	\includegraphics[bb=100 40 565 300, angle=-90, width=6cm,clip]{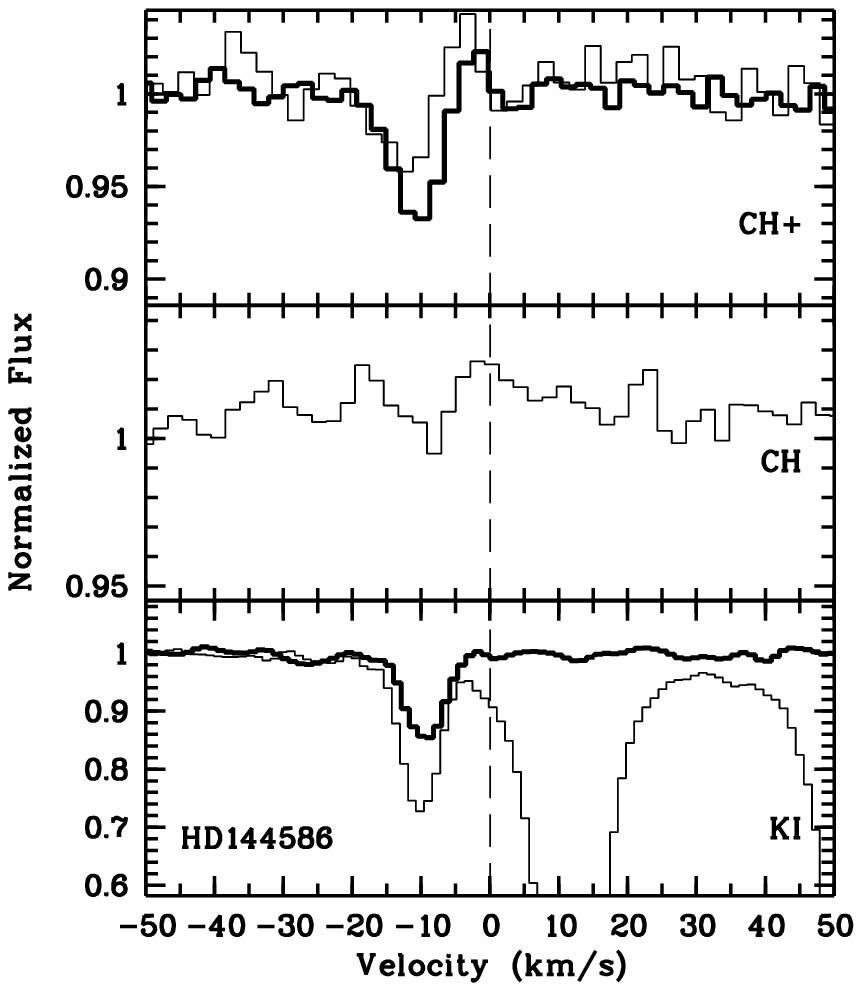}    
	\includegraphics[bb=100 40 565 300, angle=-90, width=6cm,clip]{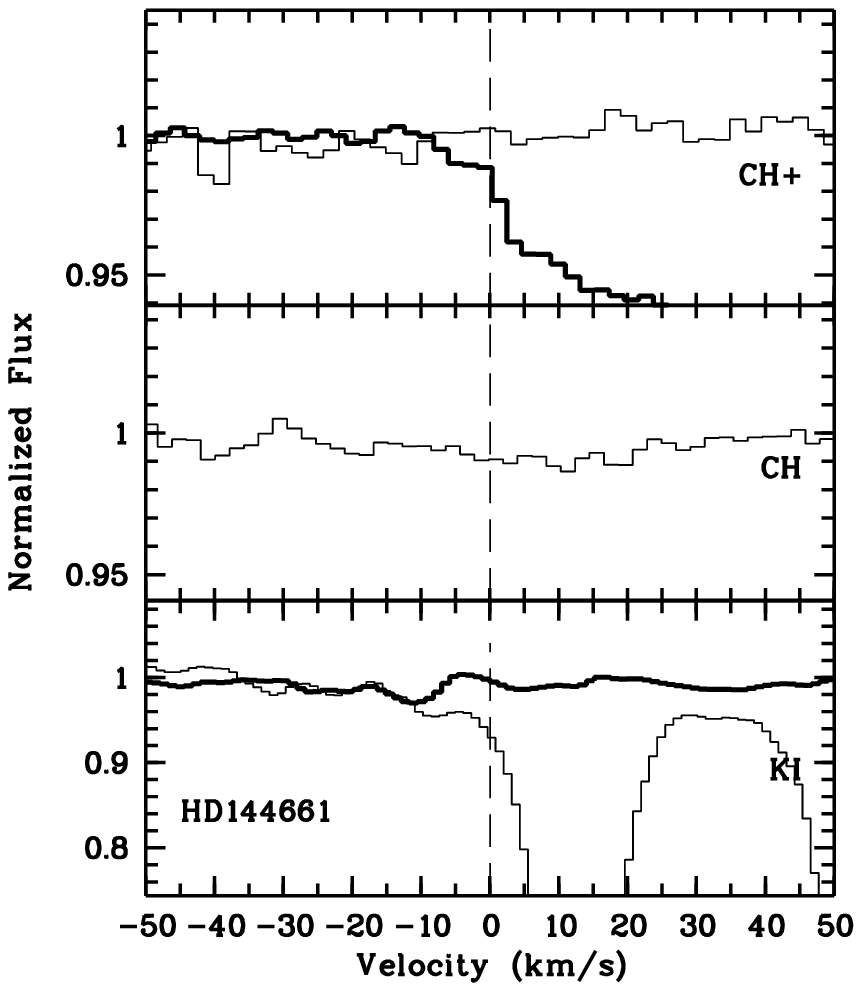}    
	\includegraphics[bb=100 40 565 300, angle=-90, width=6cm,clip]{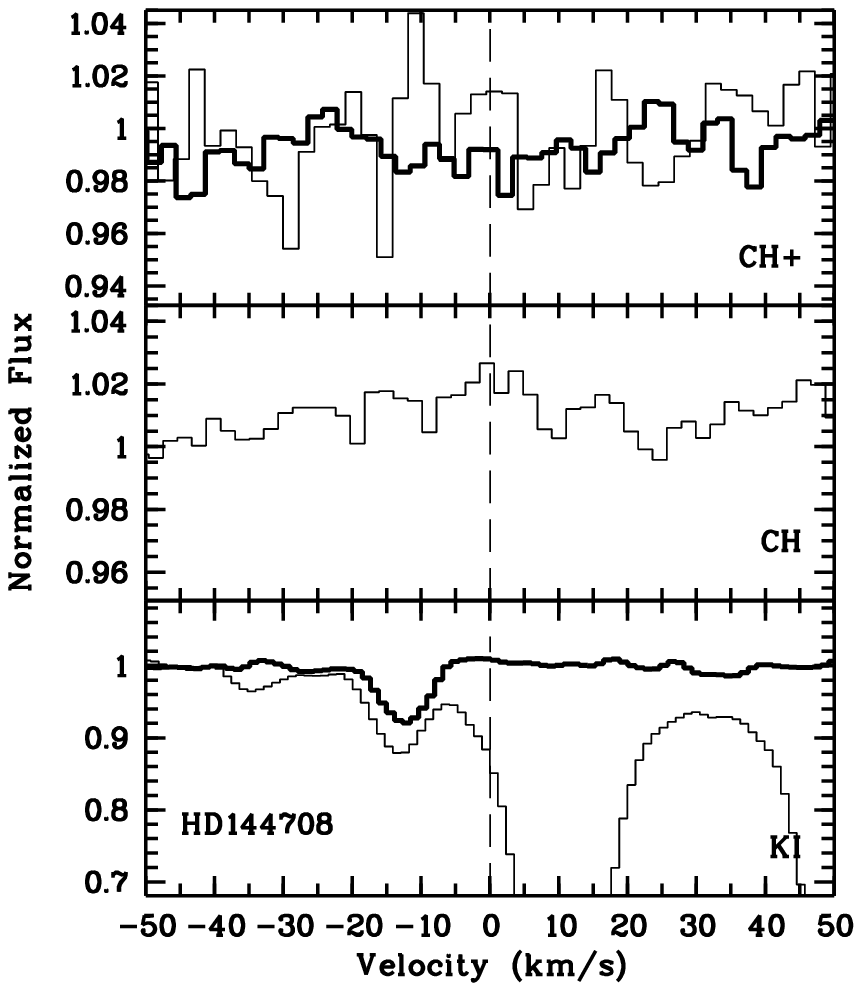}
	\includegraphics[bb=100 40 565 300, angle=-90, width=6cm,clip]{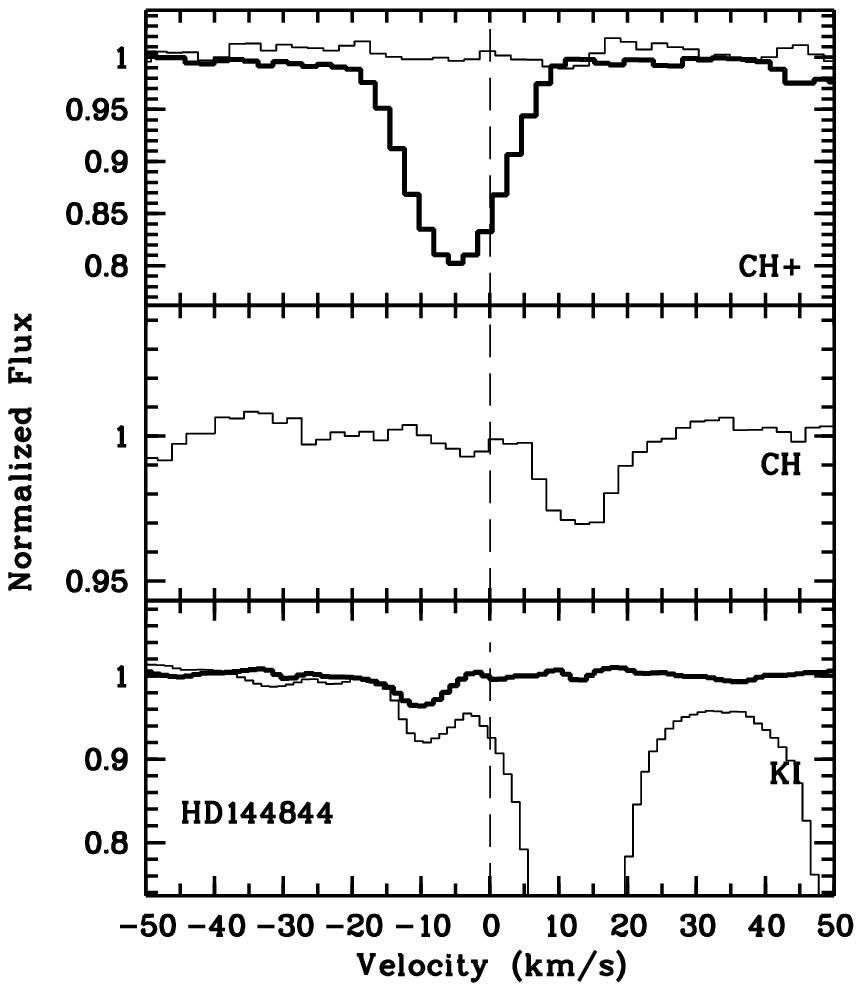} \caption{Continued.}  \end{figure*}}\clearpage      
\addtocounter{figure}{-1}

{
\begin{figure*}[h!]  
	\includegraphics[bb=100 40 565 300, angle=-90, width=6cm,clip]{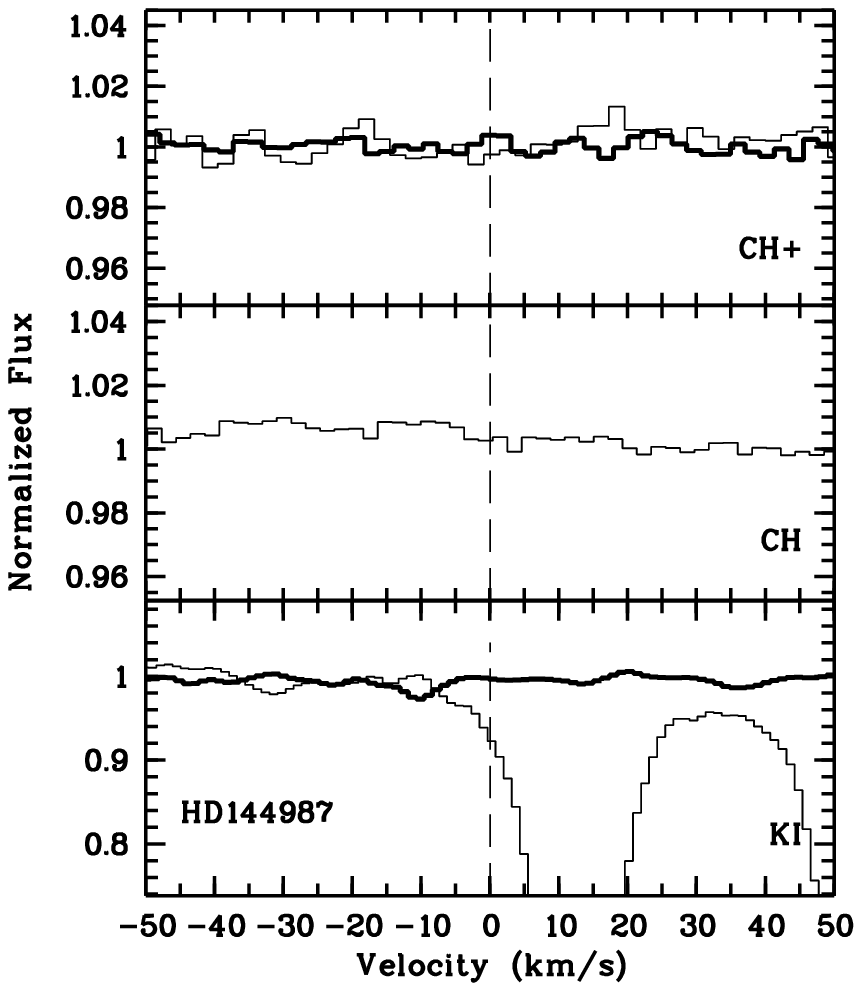}    
	\includegraphics[bb=100 40 565 300, angle=-90, width=6cm,clip]{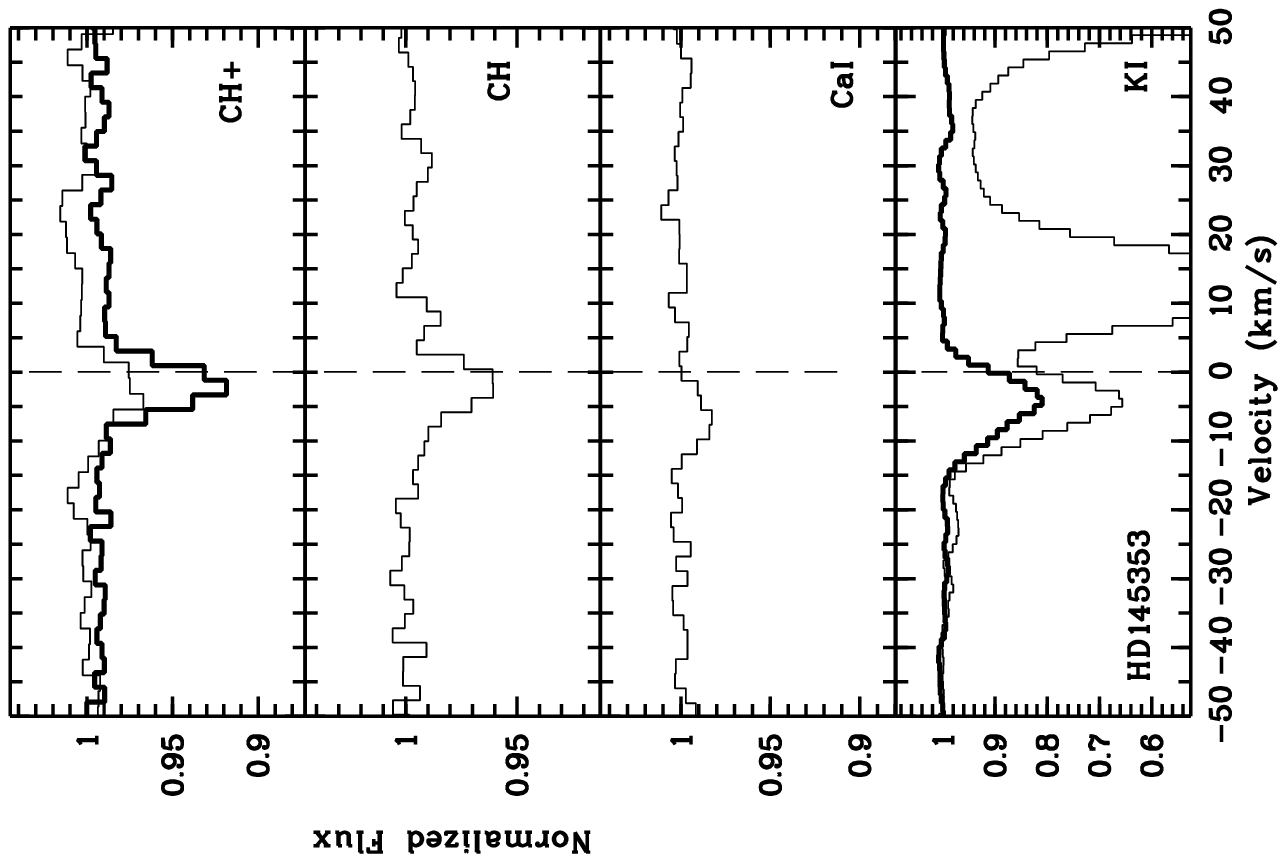}    
	\includegraphics[bb=100 40 565 300, angle=-90, width=6cm,clip]{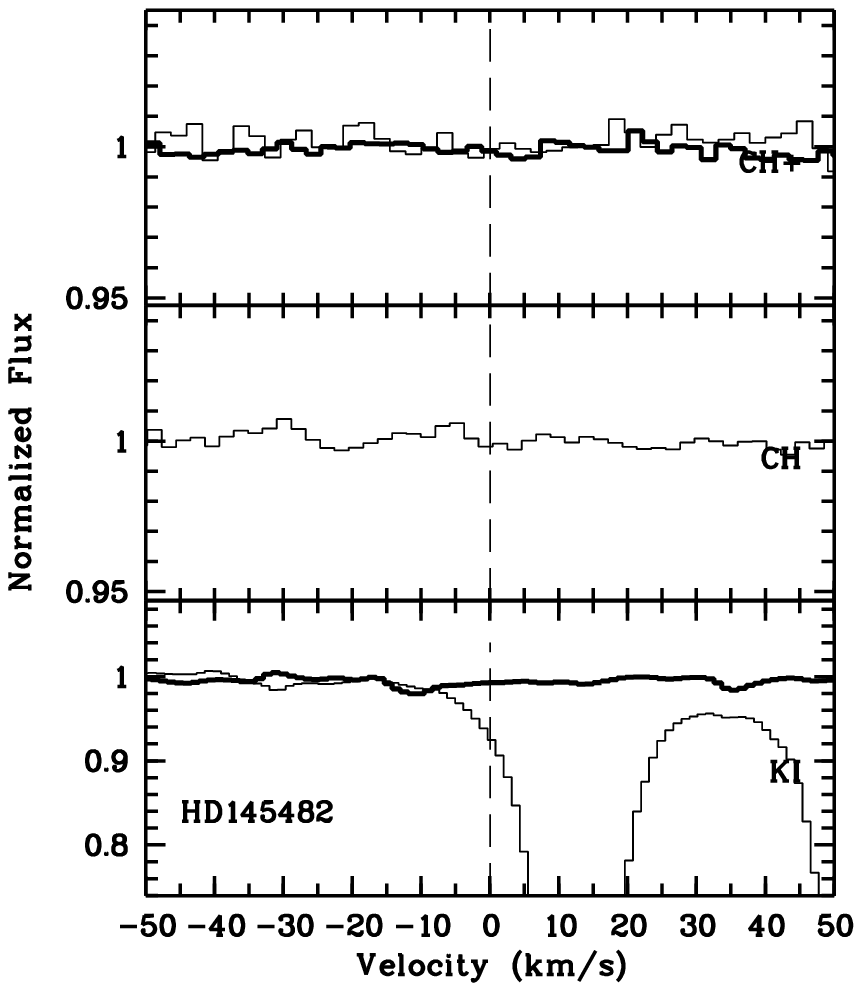}    
	\includegraphics[bb=100 40 565 300, angle=-90, width=6cm,clip]{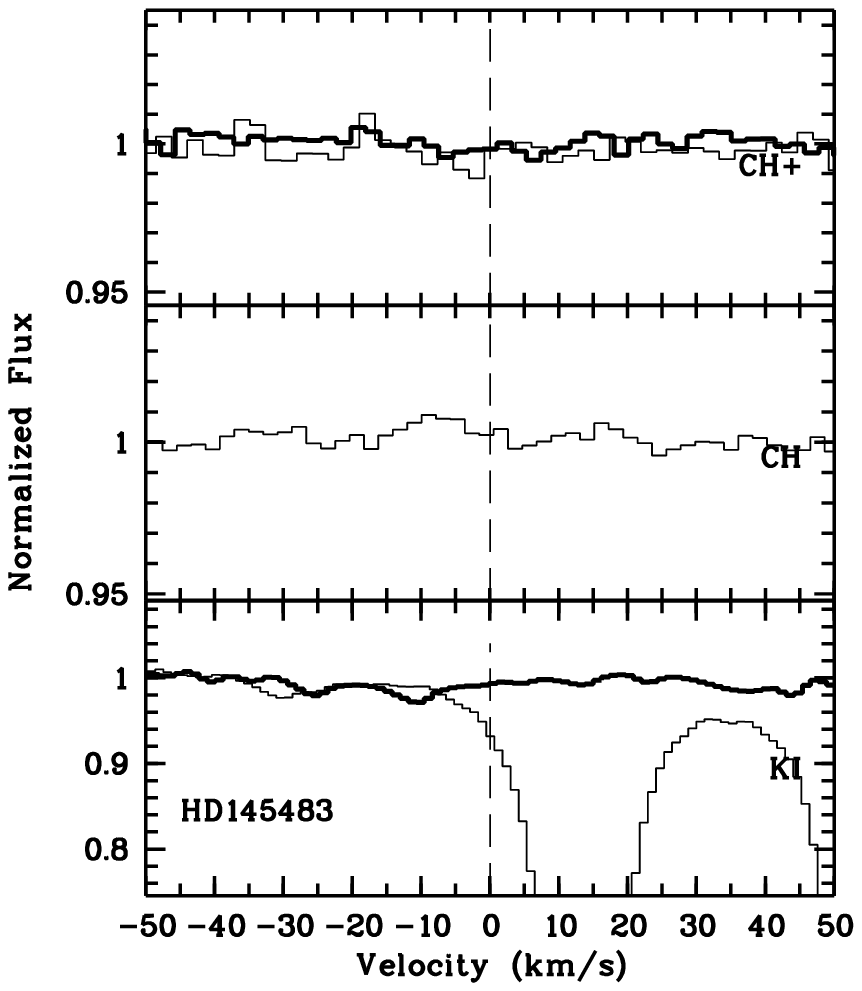}
	\includegraphics[bb=100 40 565 300, angle=-90, width=6cm,clip]{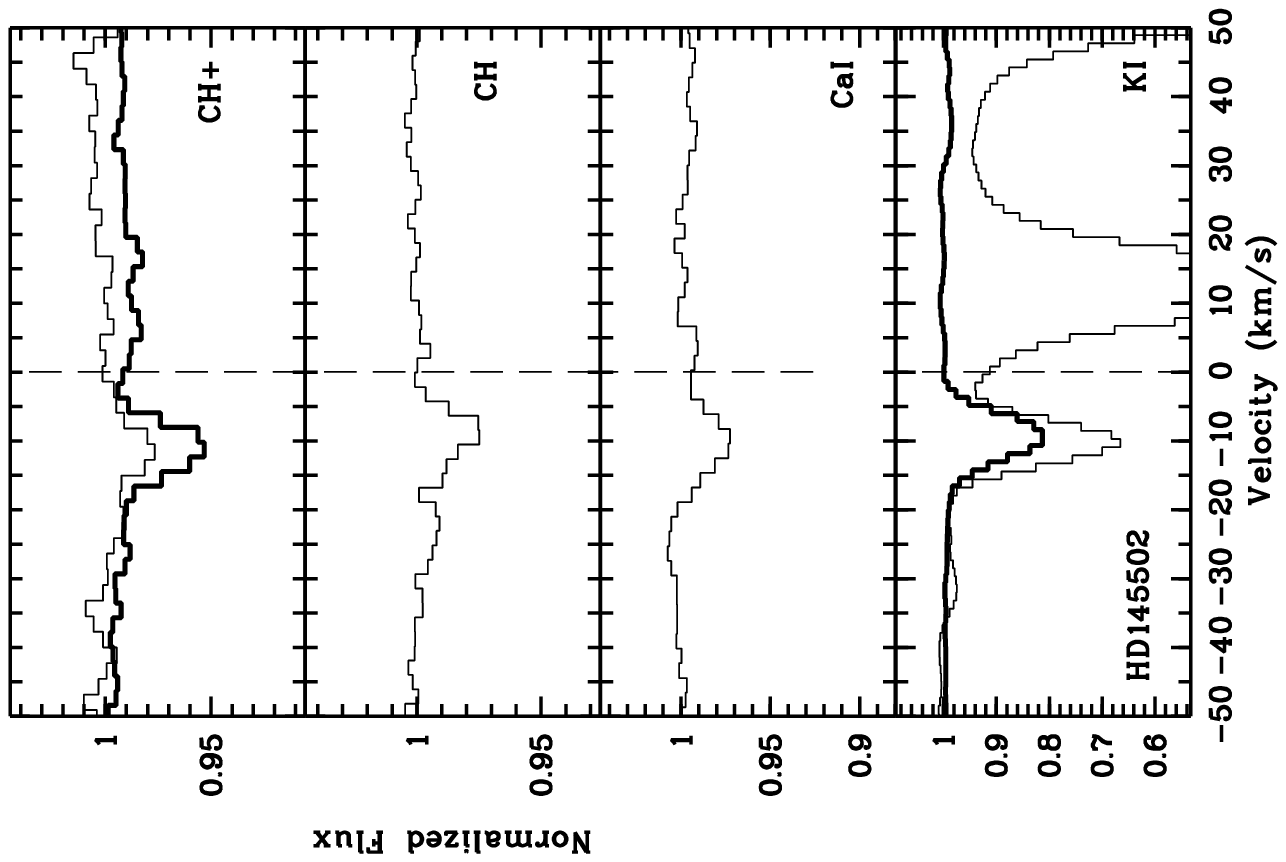}    
	\includegraphics[bb=100 40 565 300, angle=-90, width=6cm,clip]{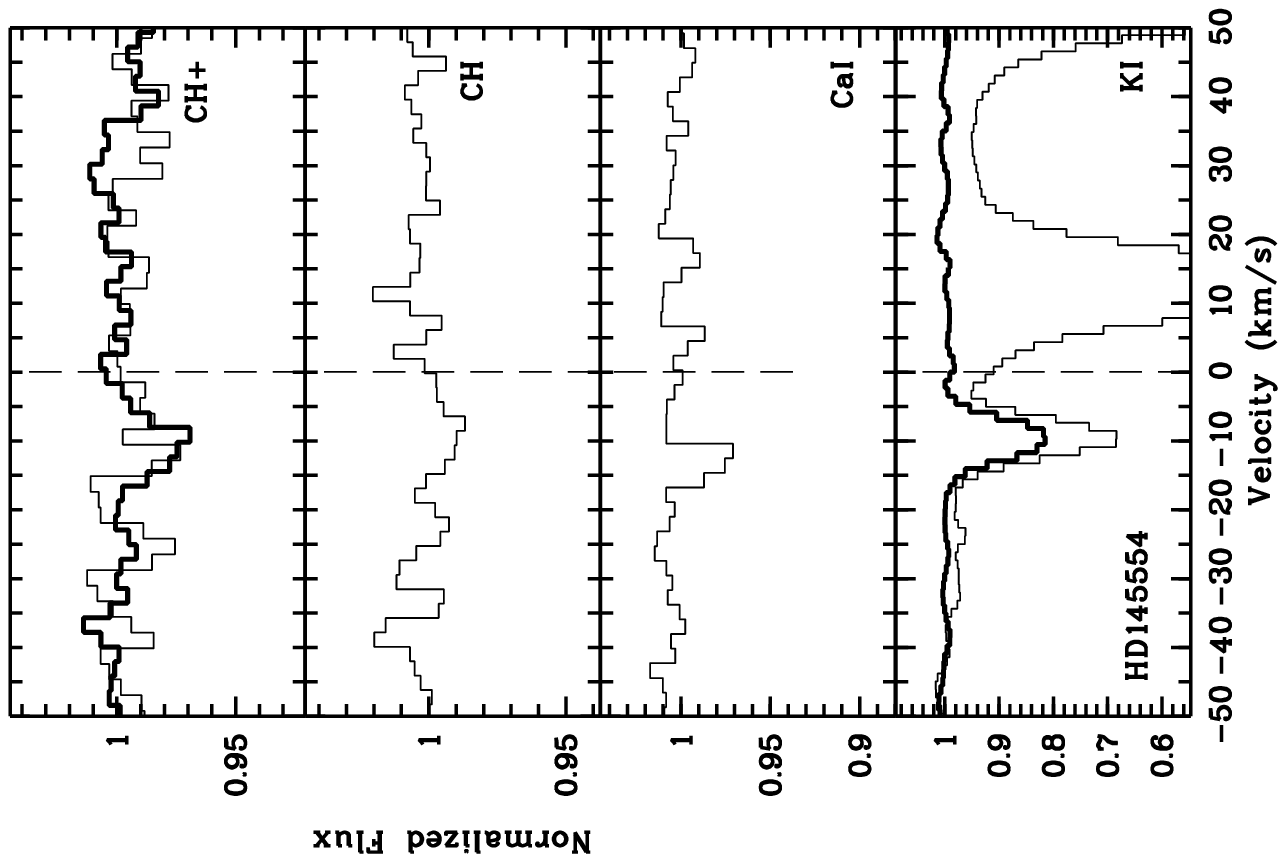} \caption{Continued.} \end{figure*}}\clearpage      
\addtocounter{figure}{-1}

{
\begin{figure*}[h!]    
	\includegraphics[bb=100 40 565 300, angle=-90, width=6cm,clip]{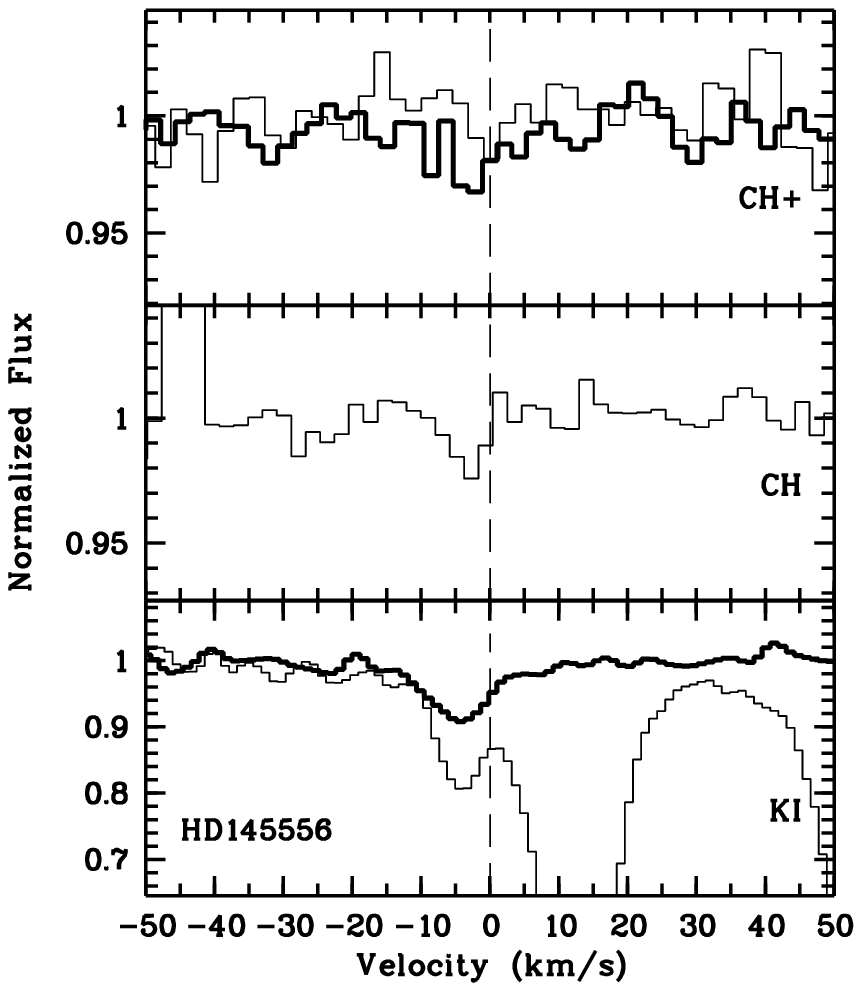}    
	\includegraphics[bb=100 40 565 300, angle=-90, width=6cm,clip]{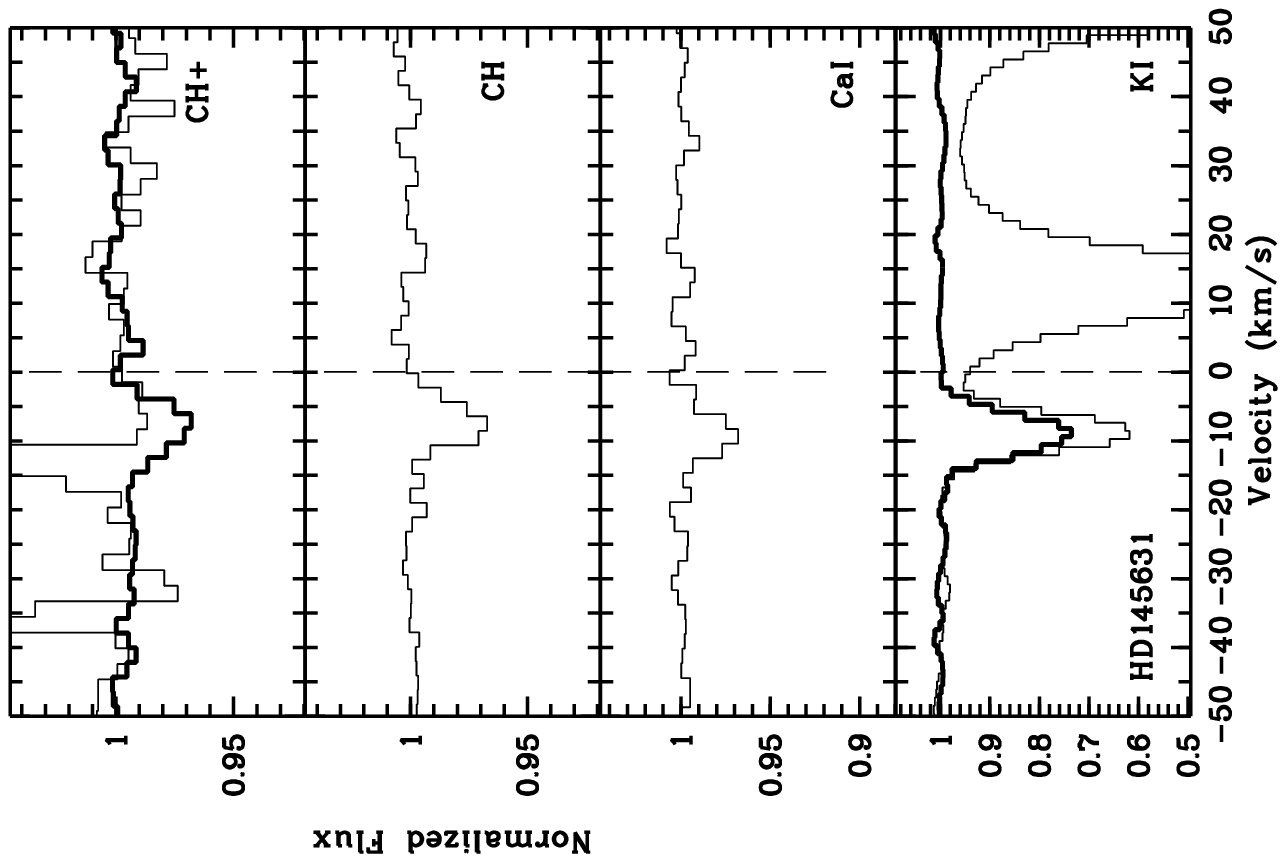}    
	\includegraphics[bb=100 40 565 300, angle=-90, width=6cm,clip]{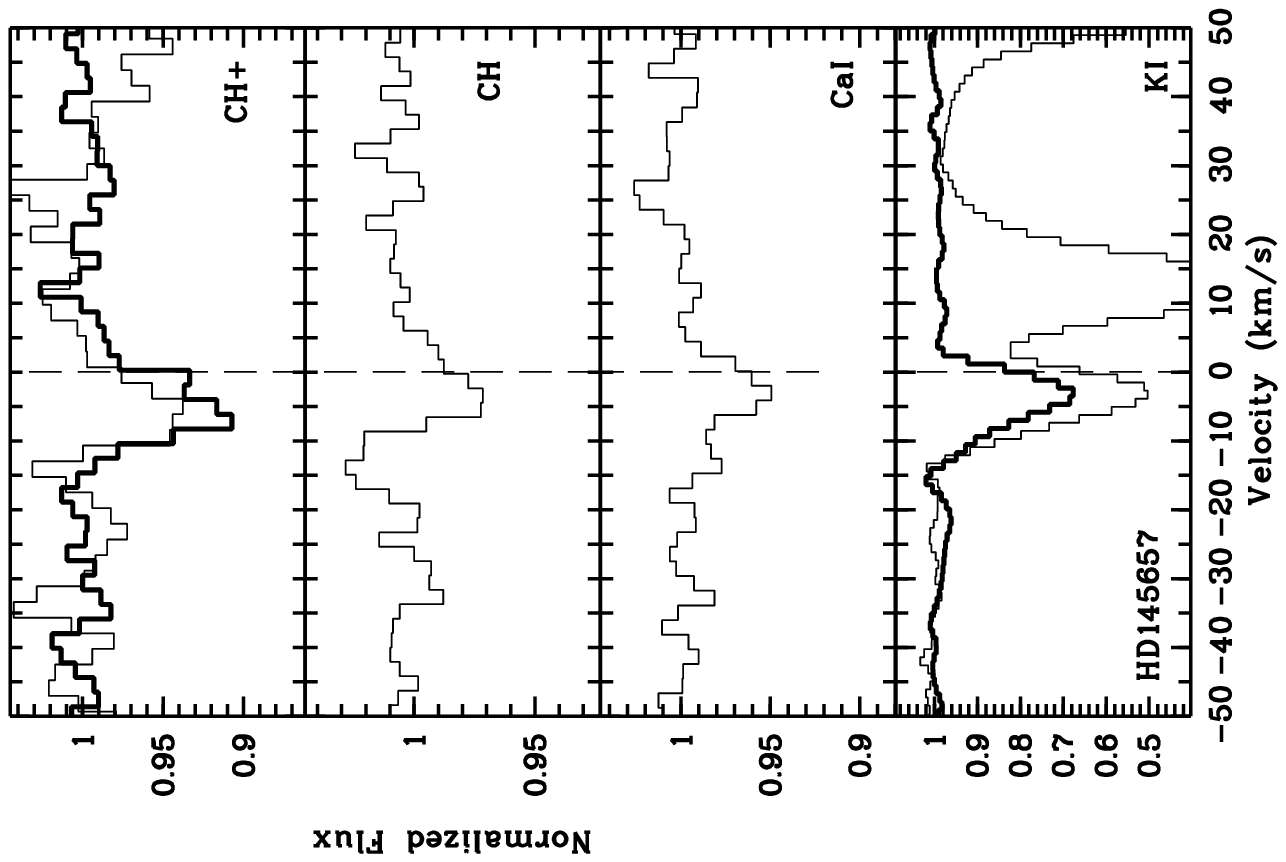}
	\includegraphics[bb=100 40 565 300, angle=-90, width=6cm,clip]{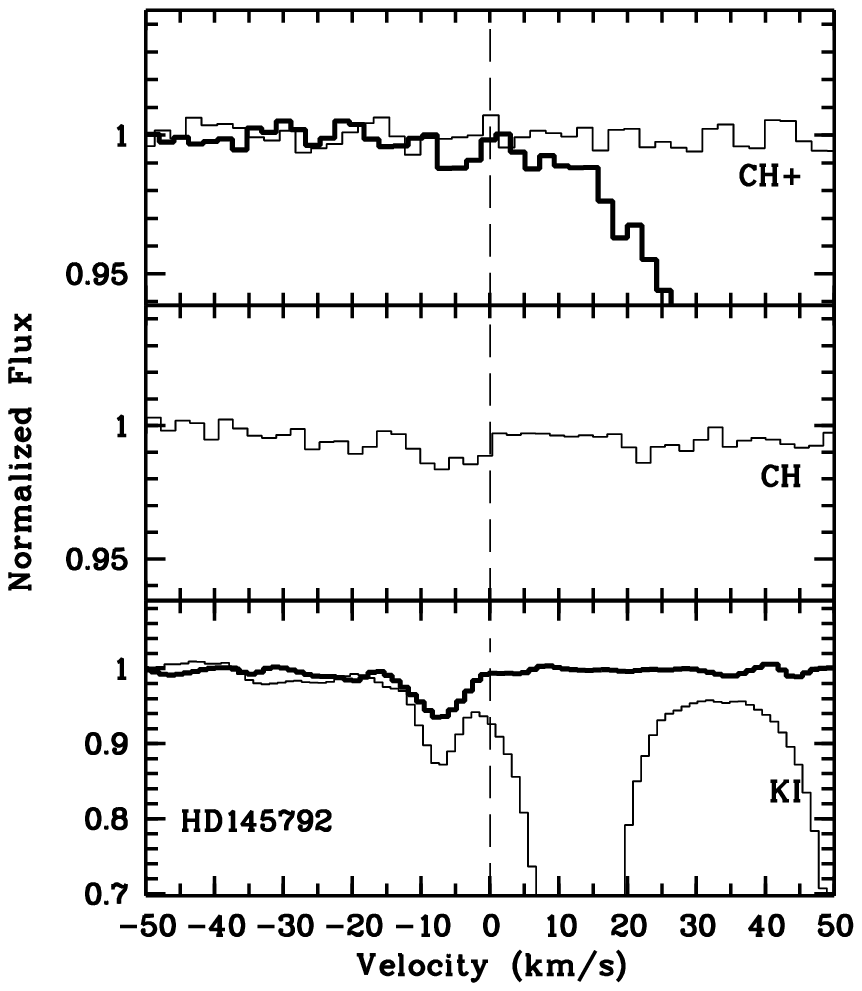}    
	\includegraphics[bb=100 40 565 300, angle=-90, width=6cm,clip]{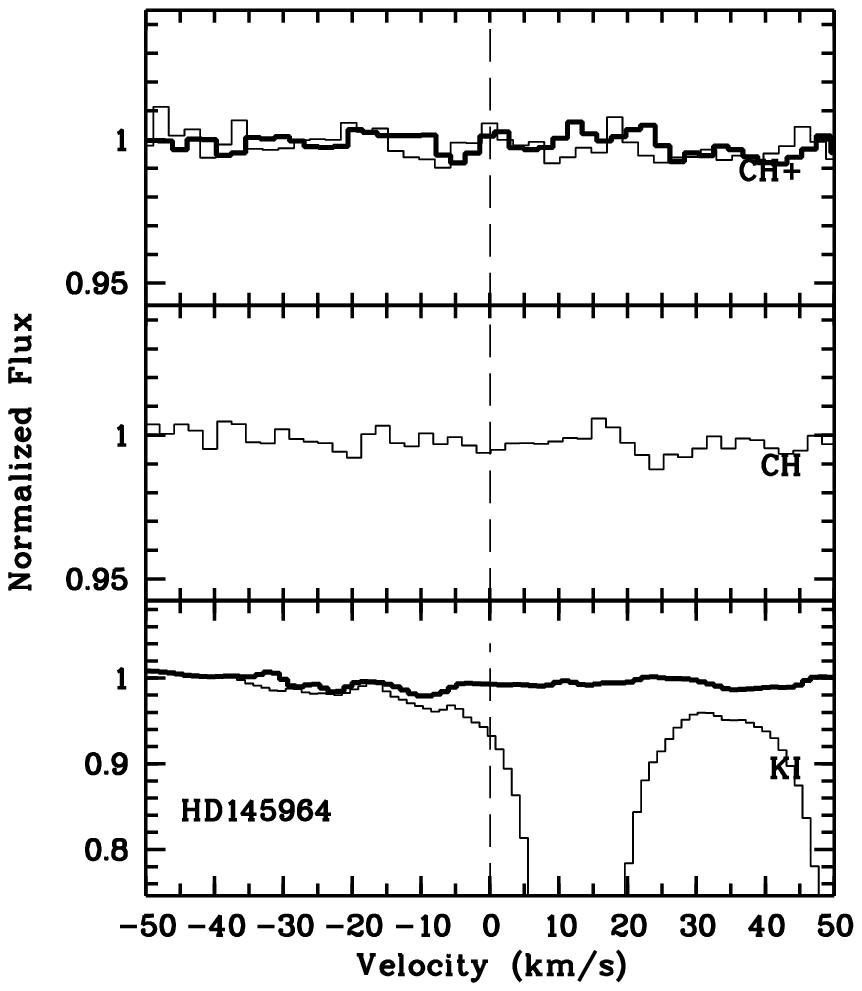}    
	\includegraphics[bb=100 40 565 300, angle=-90, width=6cm,clip]{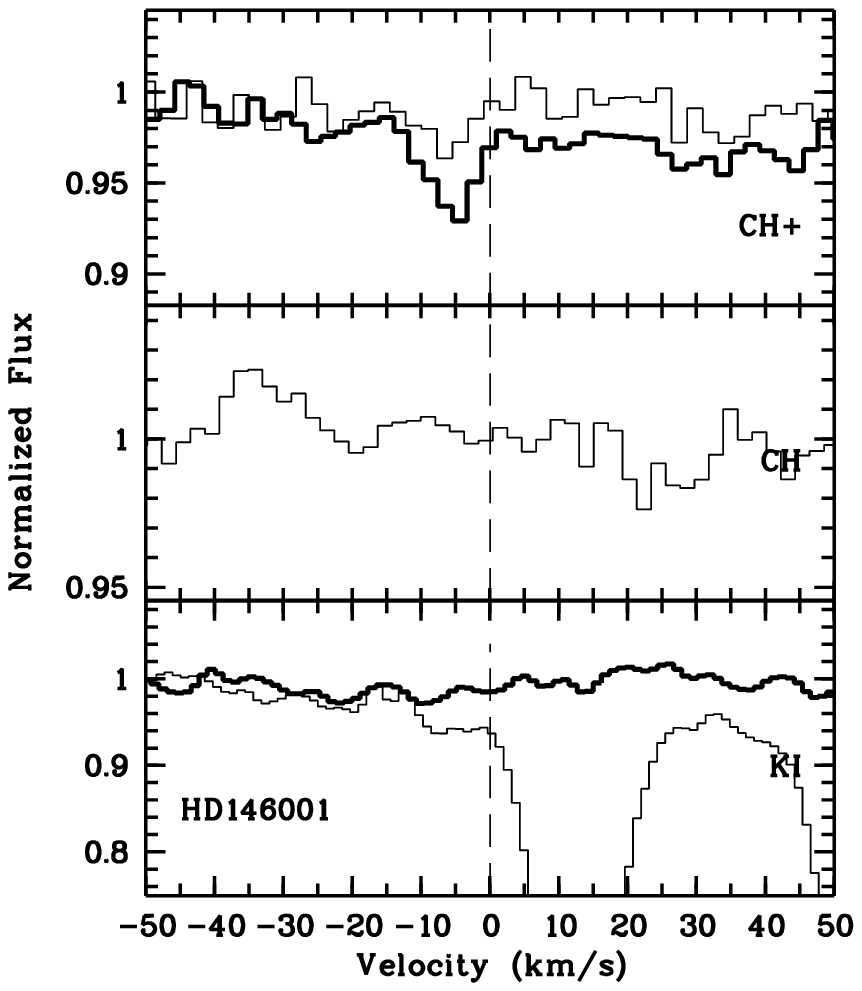} \caption{Continued.} \end{figure*}}\clearpage      
\addtocounter{figure}{-1}

{
\begin{figure*}[h!]
	\includegraphics[bb=100 40 565 300, angle=-90, width=6cm,clip]{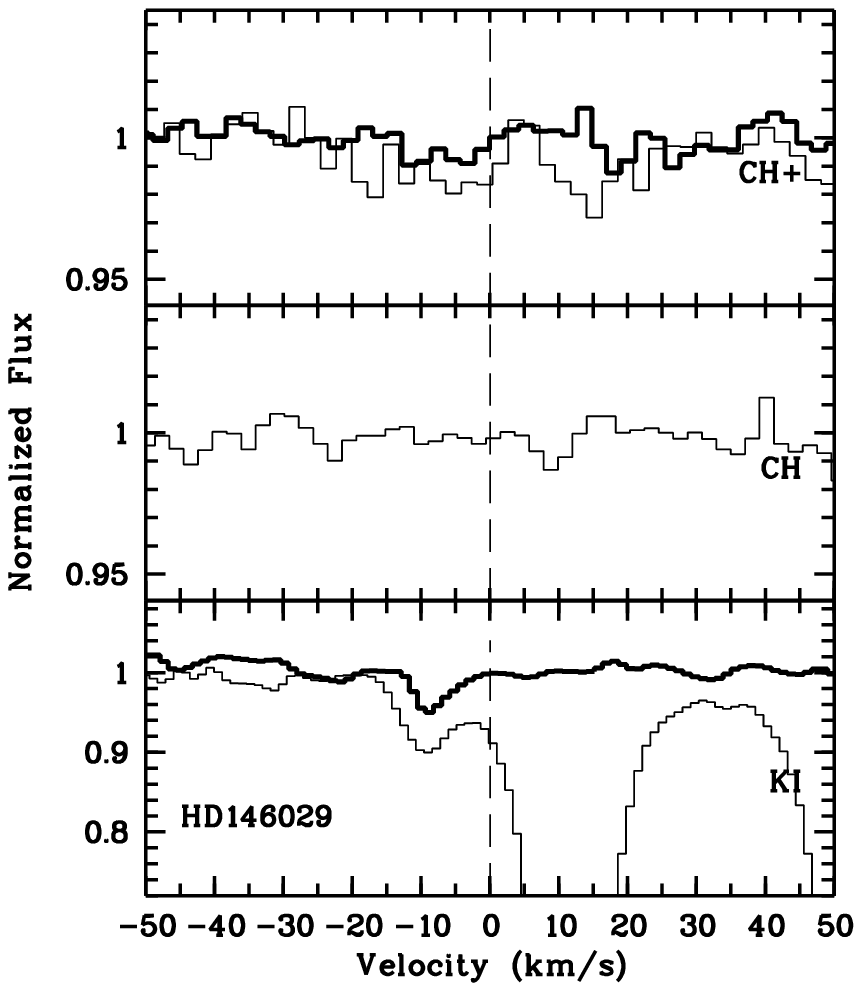}    
	\includegraphics[bb=100 40 565 300, angle=-90, width=6cm,clip]{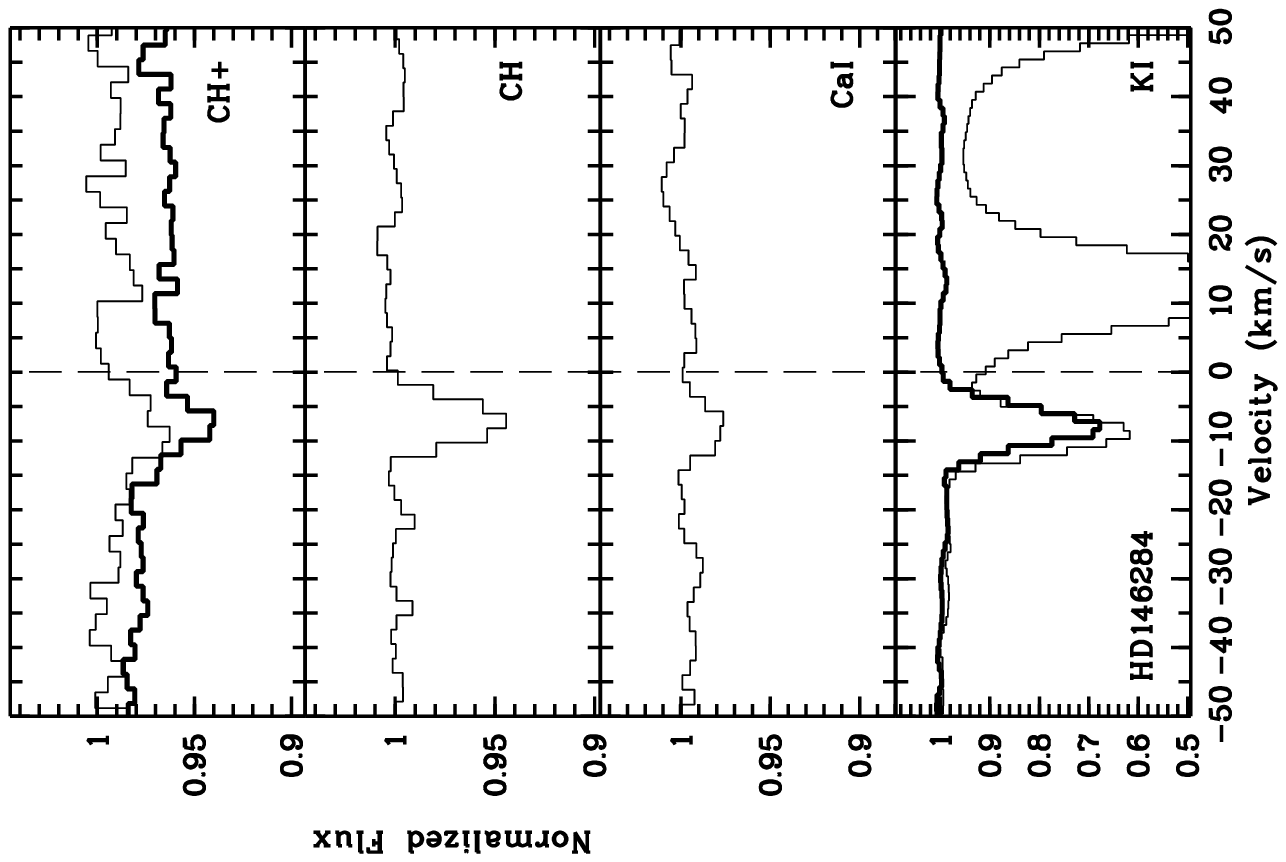}
	\includegraphics[bb=100 40 565 300, angle=-90, width=6cm,clip]{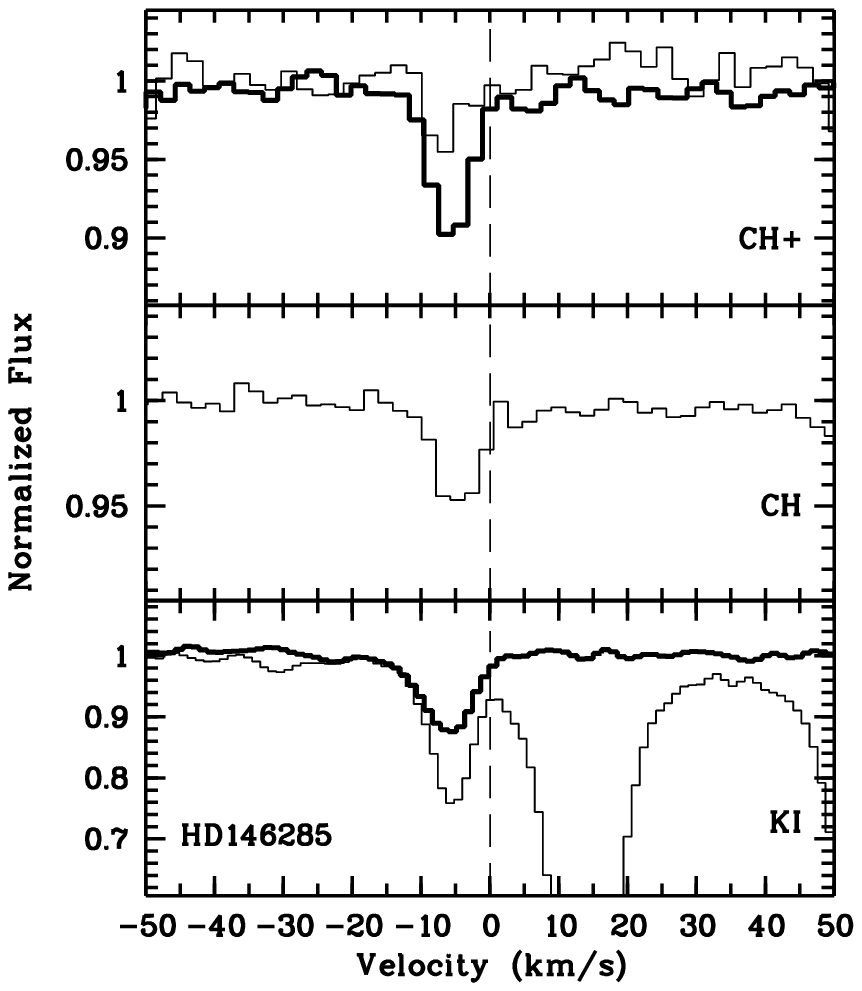}     
	\includegraphics[bb=100 40 565 300, angle=-90, width=6cm,clip]{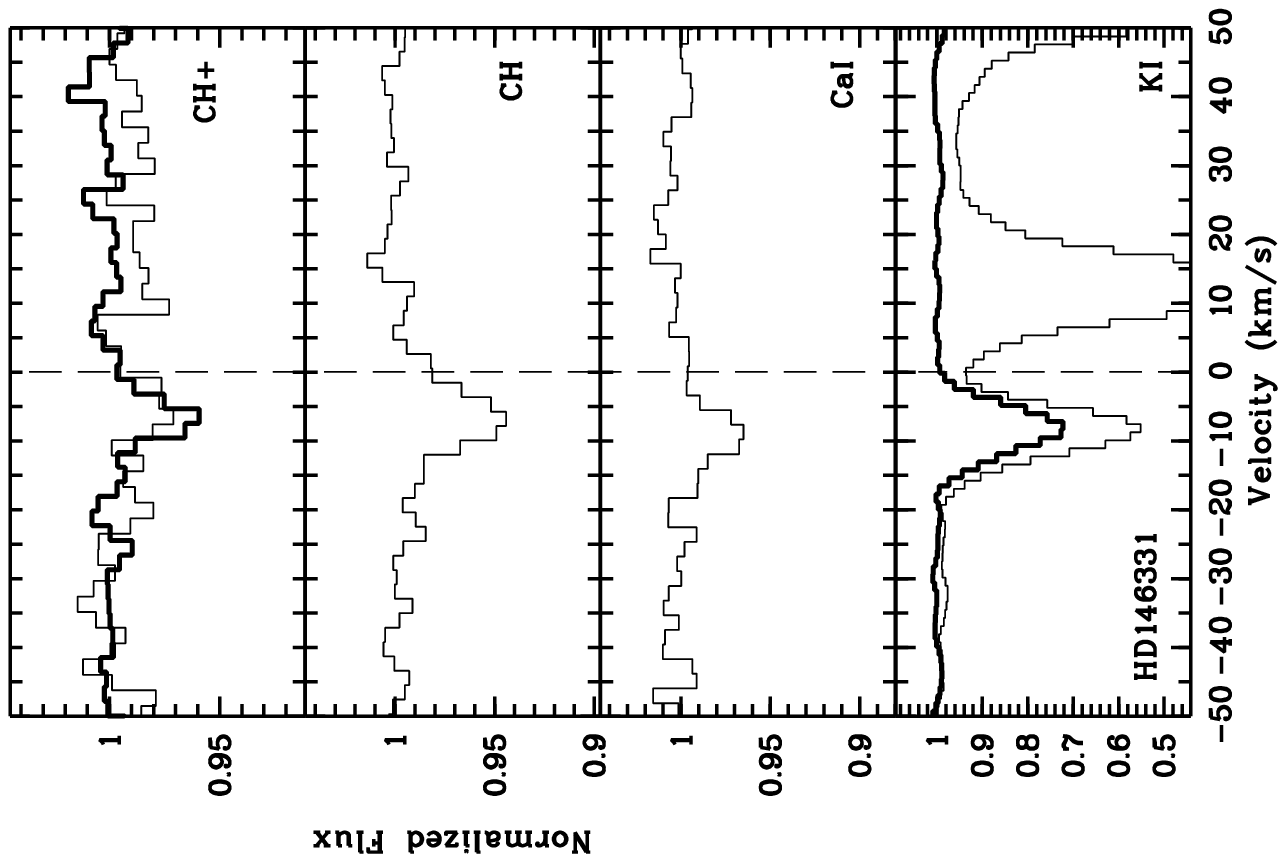}   
	\includegraphics[bb=100 40 565 300, angle=-90, width=6cm,clip]{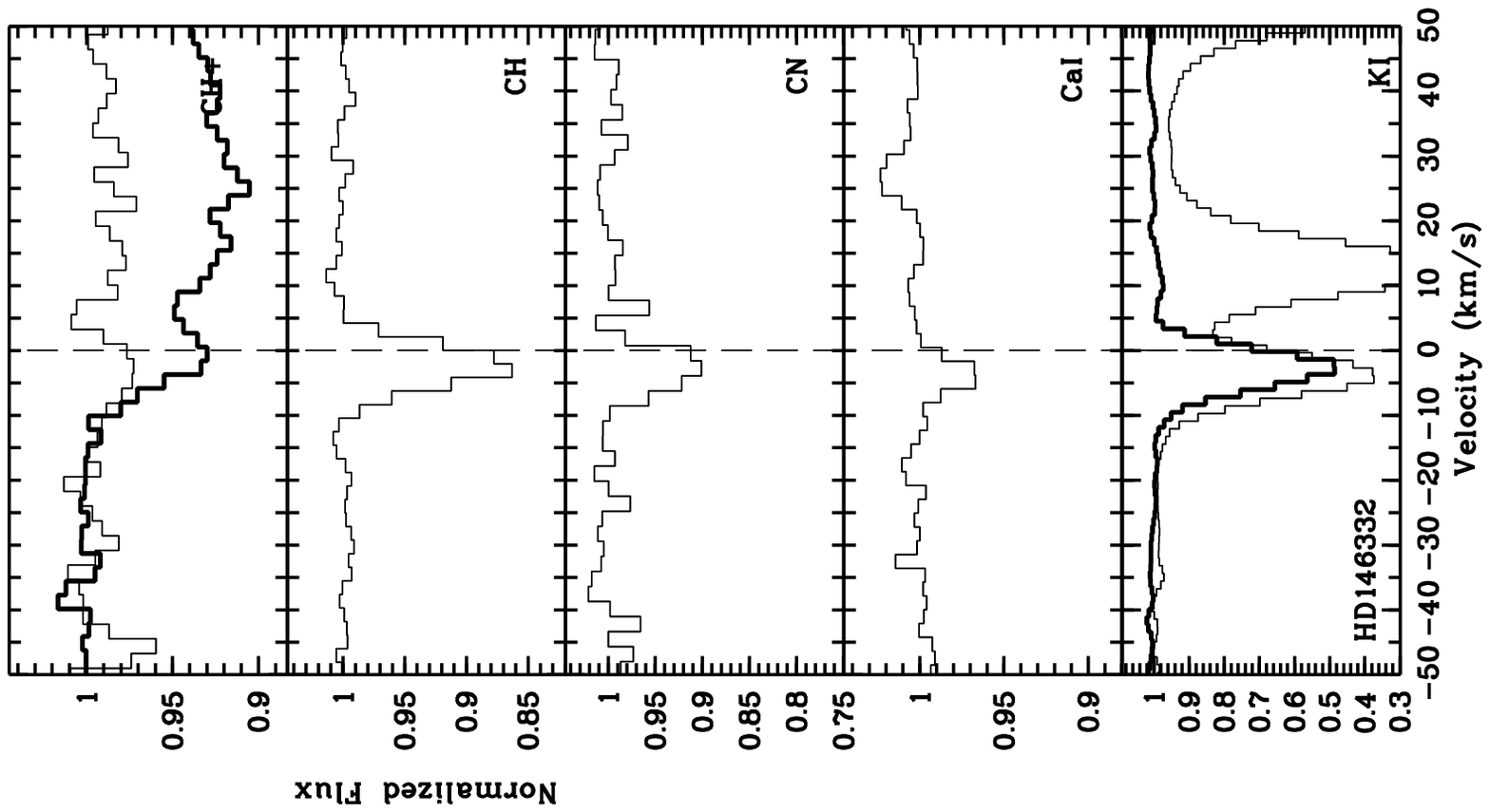}   
	\includegraphics[bb=100 40 565 300, angle=-90, width=6cm,clip]{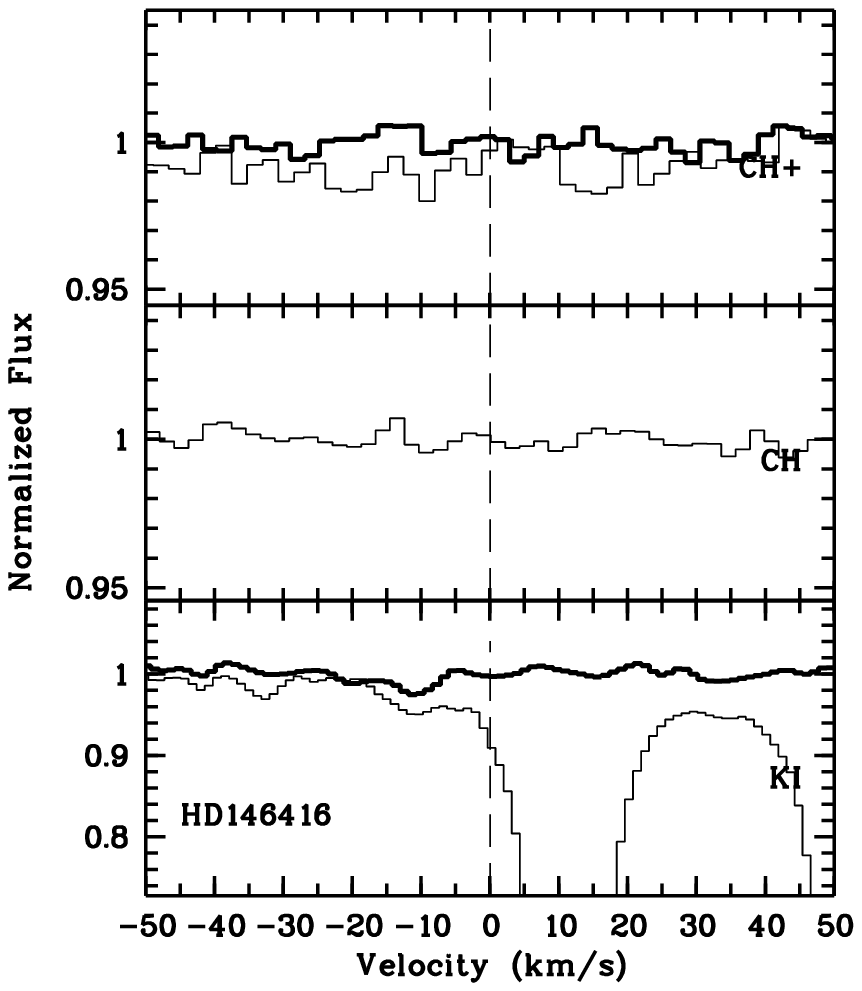}\caption{Continued.} \end{figure*}}\clearpage    

\addtocounter{figure}{-1}

{
\begin{figure*}[h!]
	\includegraphics[bb=100 40 565 300, angle=-90, width=6cm,clip]{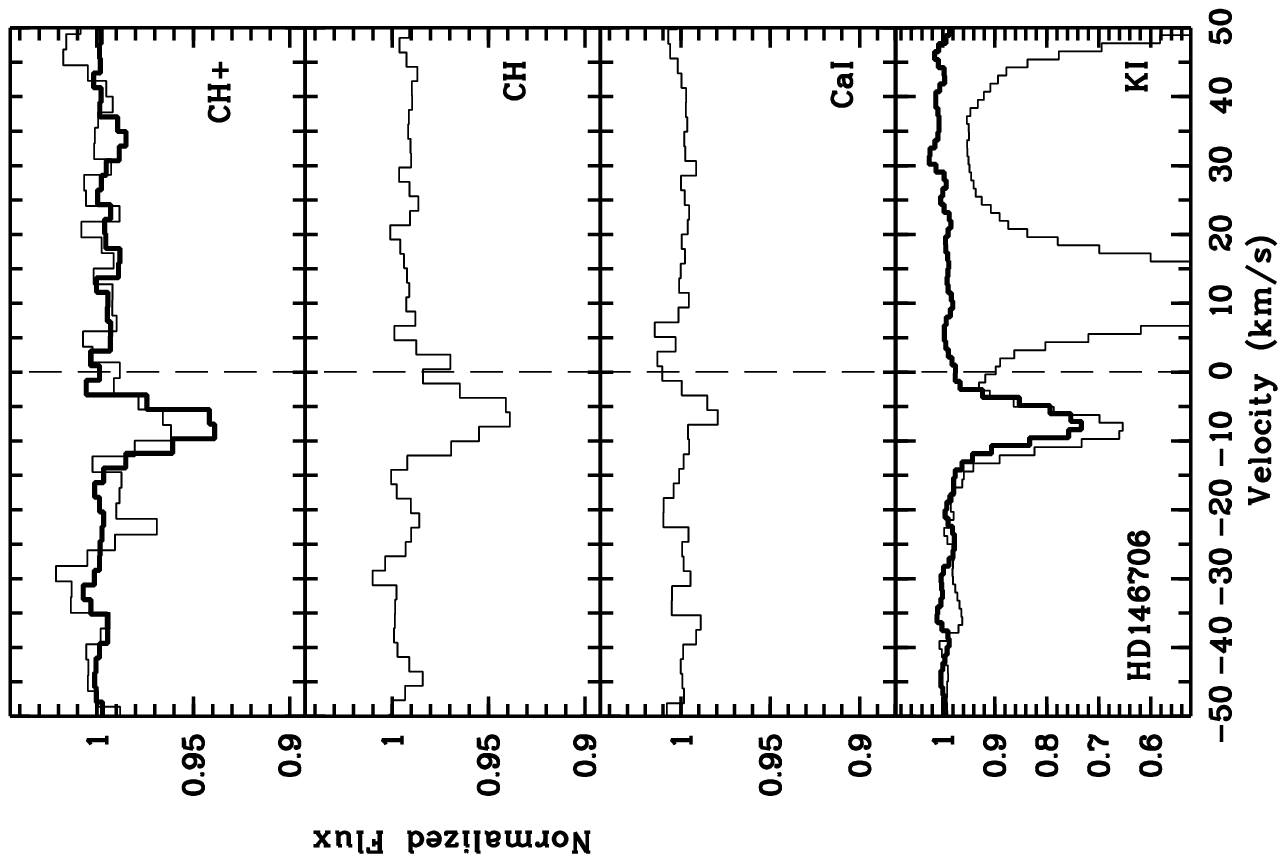}
	\includegraphics[bb=100 40 565 300, angle=-90, width=6cm,clip]{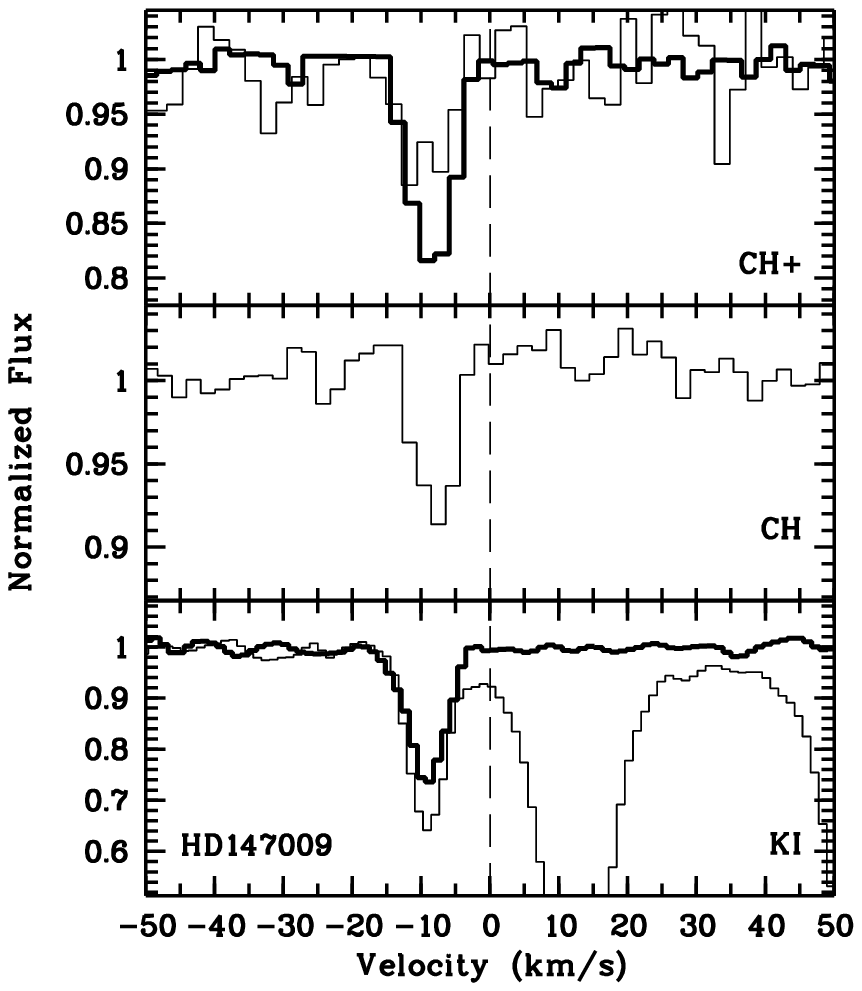}    
	\includegraphics[bb=100 40 565 300, angle=-90, width=6cm,clip]{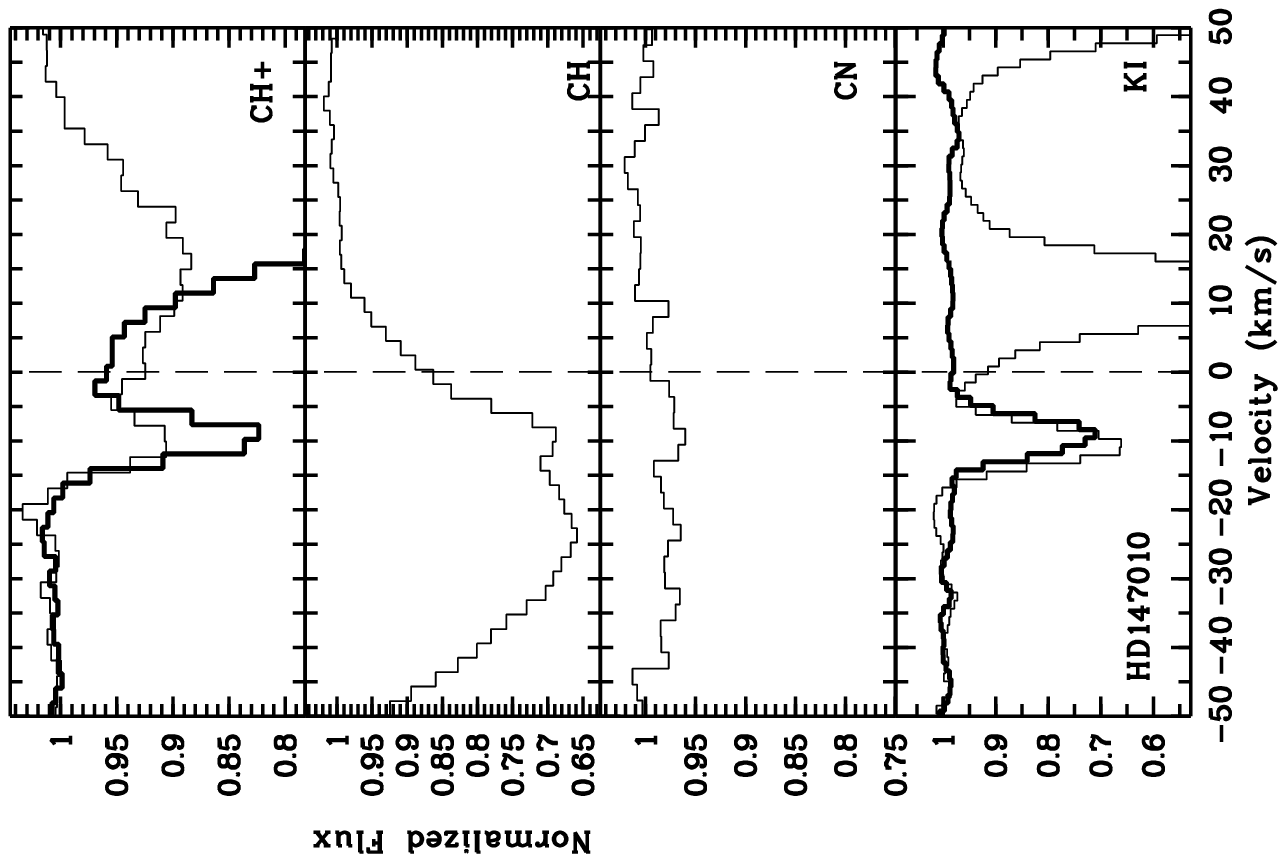}    
	\includegraphics[bb=100 40 565 300, angle=-90, width=6cm,clip]{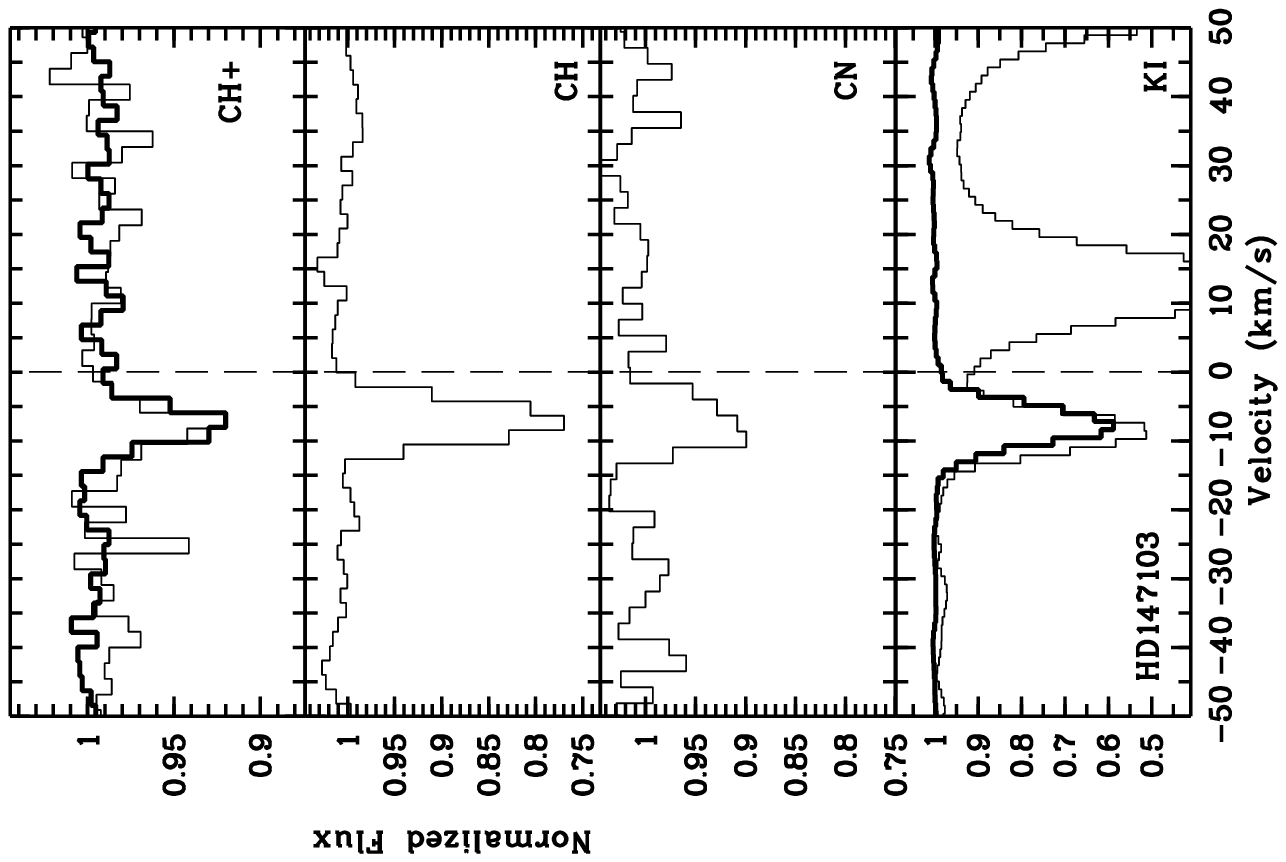}    
	\includegraphics[bb=100 40 565 300, angle=-90, width=6cm,clip]{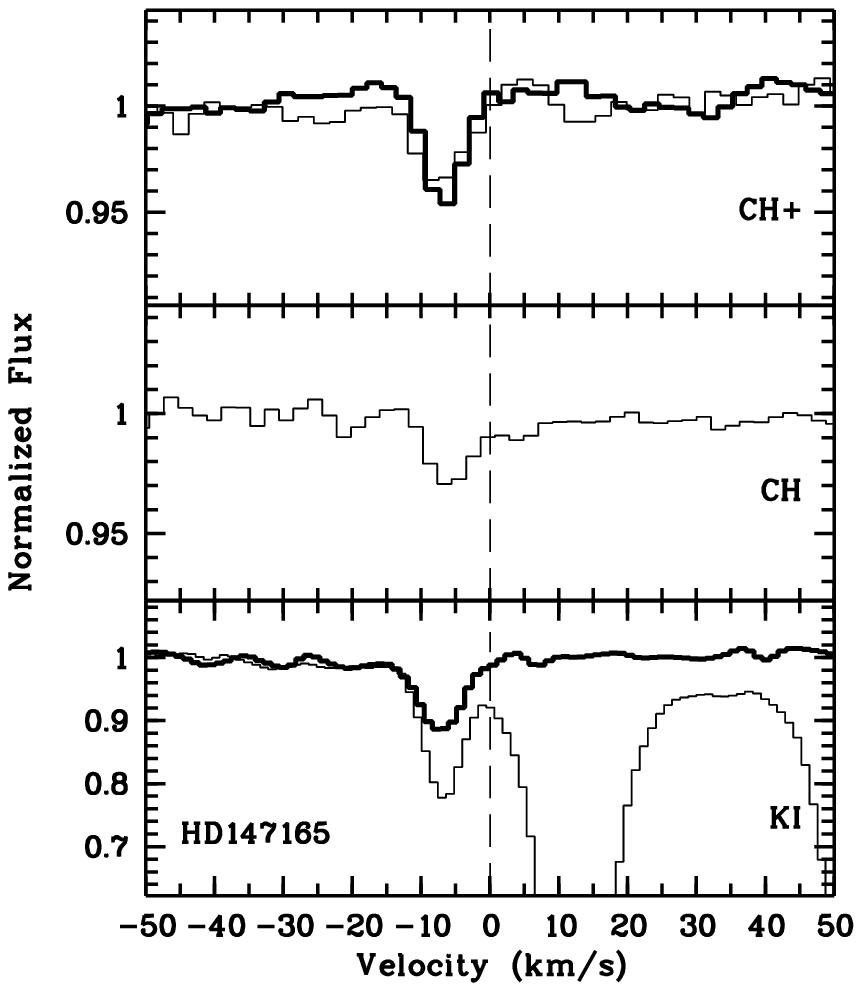}    
	\includegraphics[bb=100 40 565 300, angle=-90, width=6cm,clip]{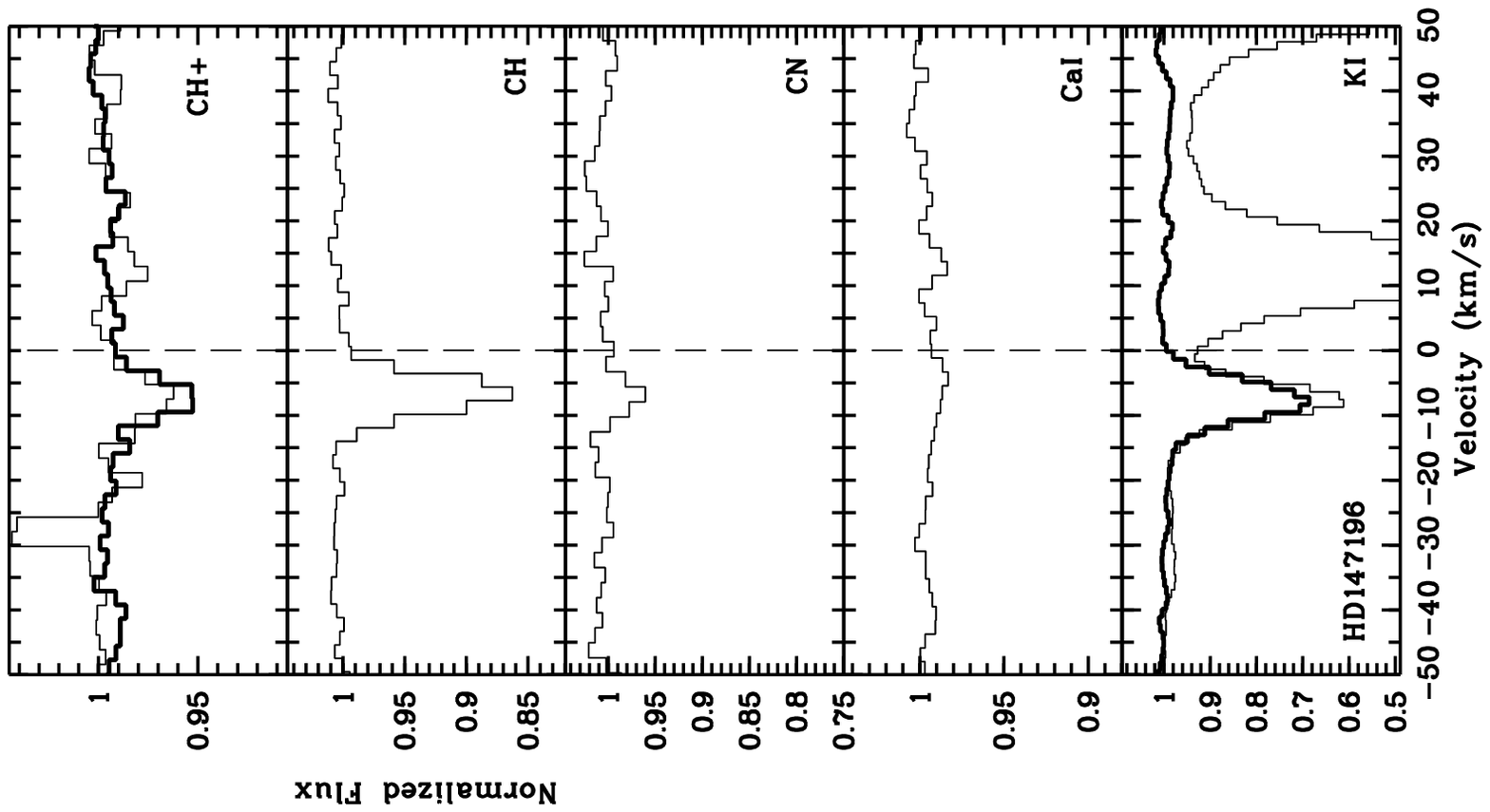}\caption{Continued.} \end{figure*}}\clearpage       
\addtocounter{figure}{-1}

{
\begin{figure*}[h!]
	\includegraphics[bb=100 40 565 300, angle=-90, width=6cm,clip]{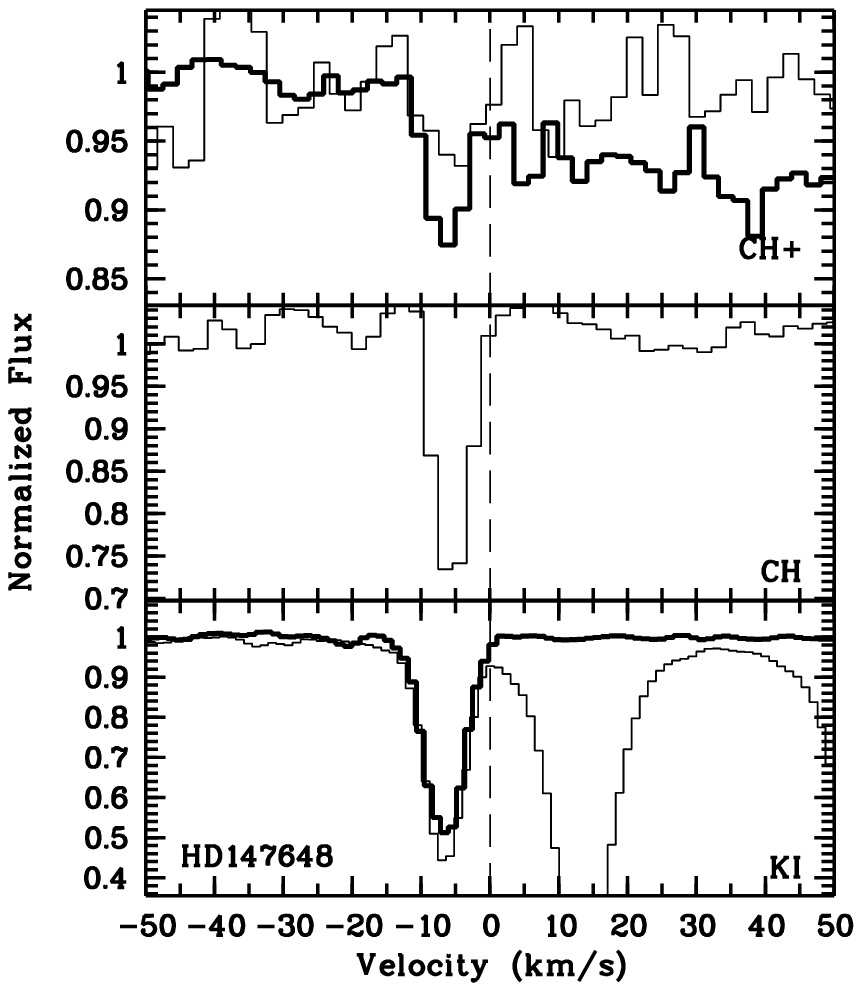}   
	\includegraphics[bb=100 40 565 300, angle=-90, width=6cm,clip]{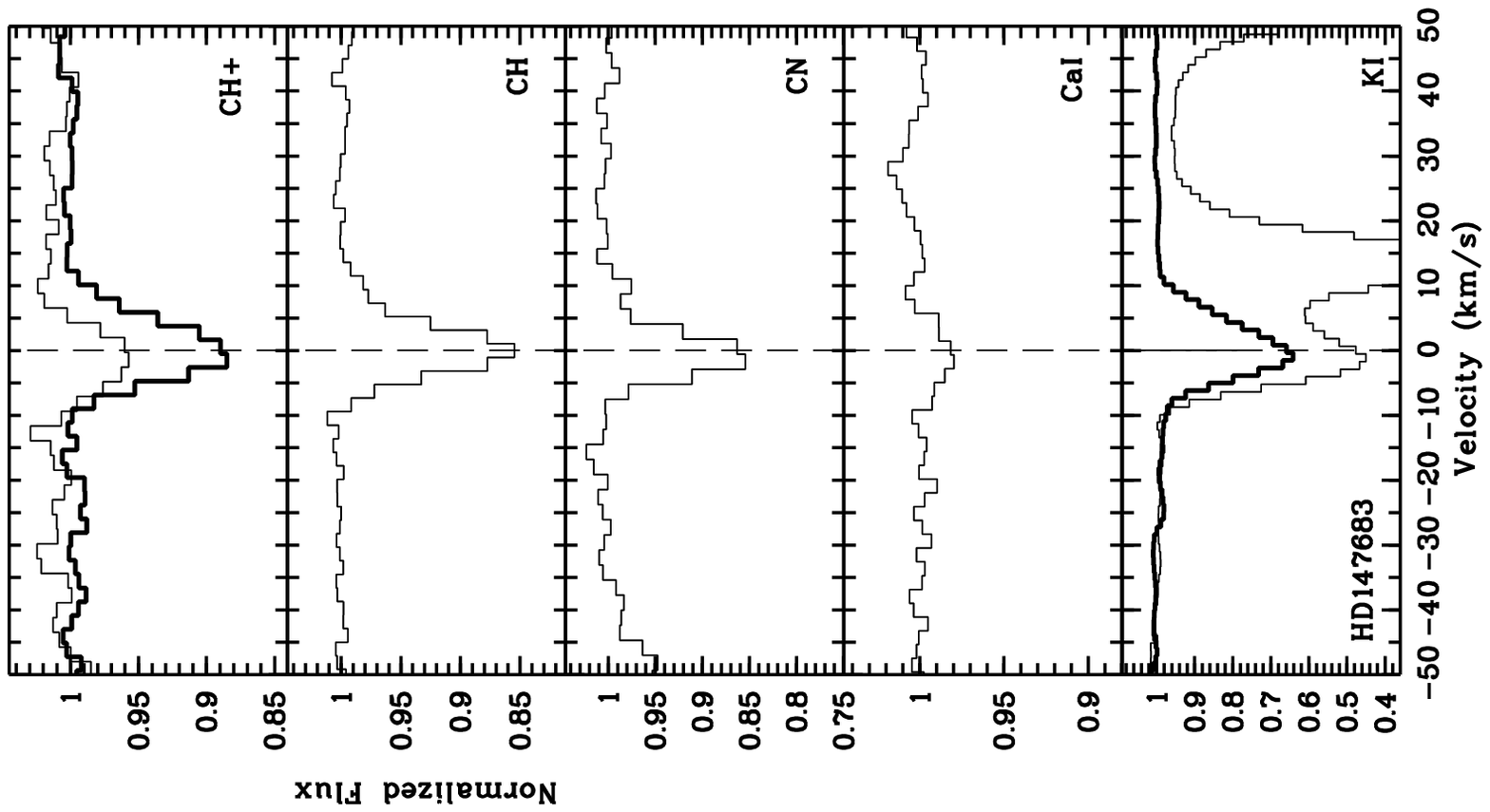}    
	\includegraphics[bb=100 40 565 300, angle=-90, width=6cm,clip]{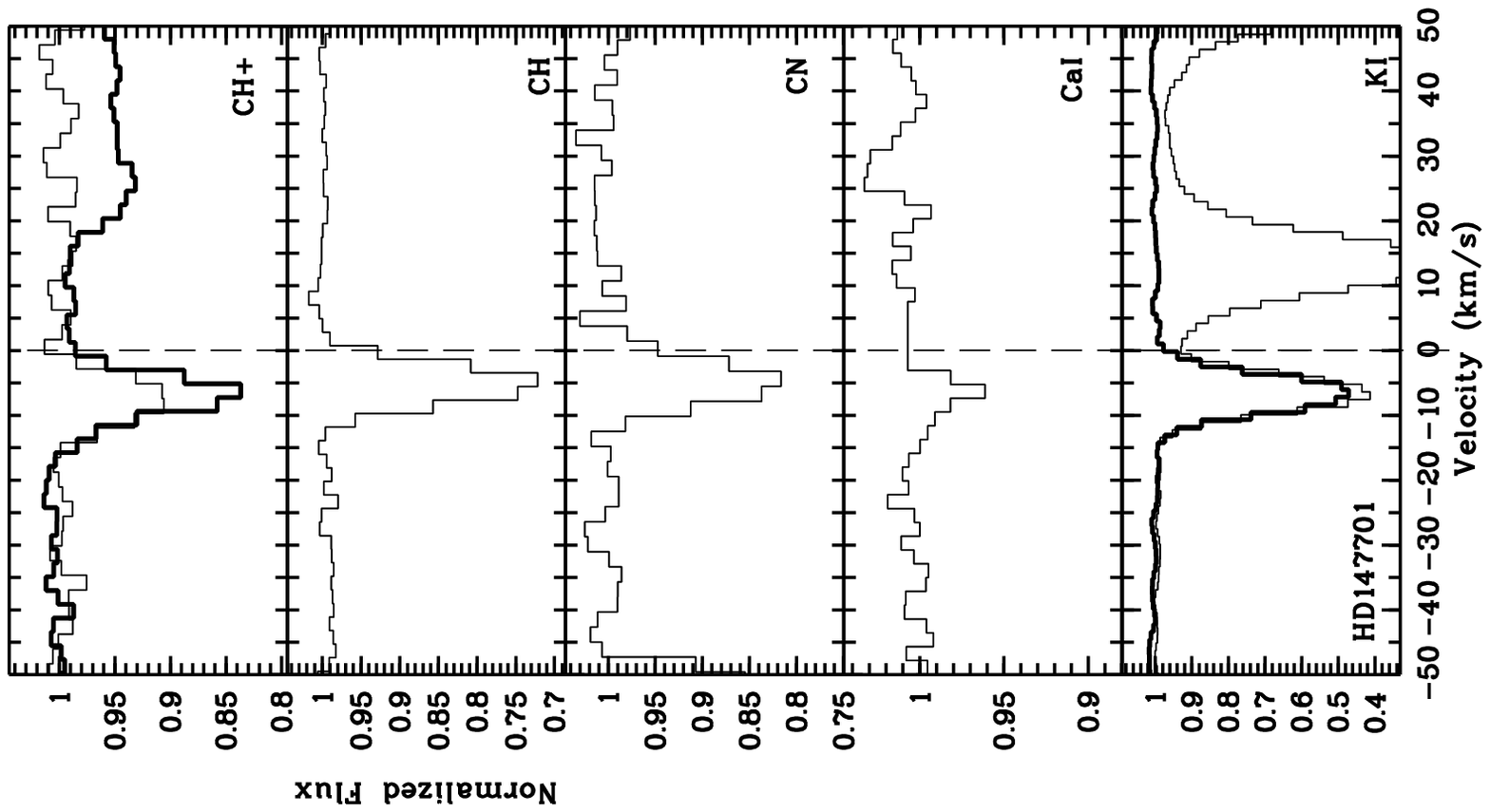}    
	\includegraphics[bb=100 40 565 300, angle=-90, width=6cm,clip]{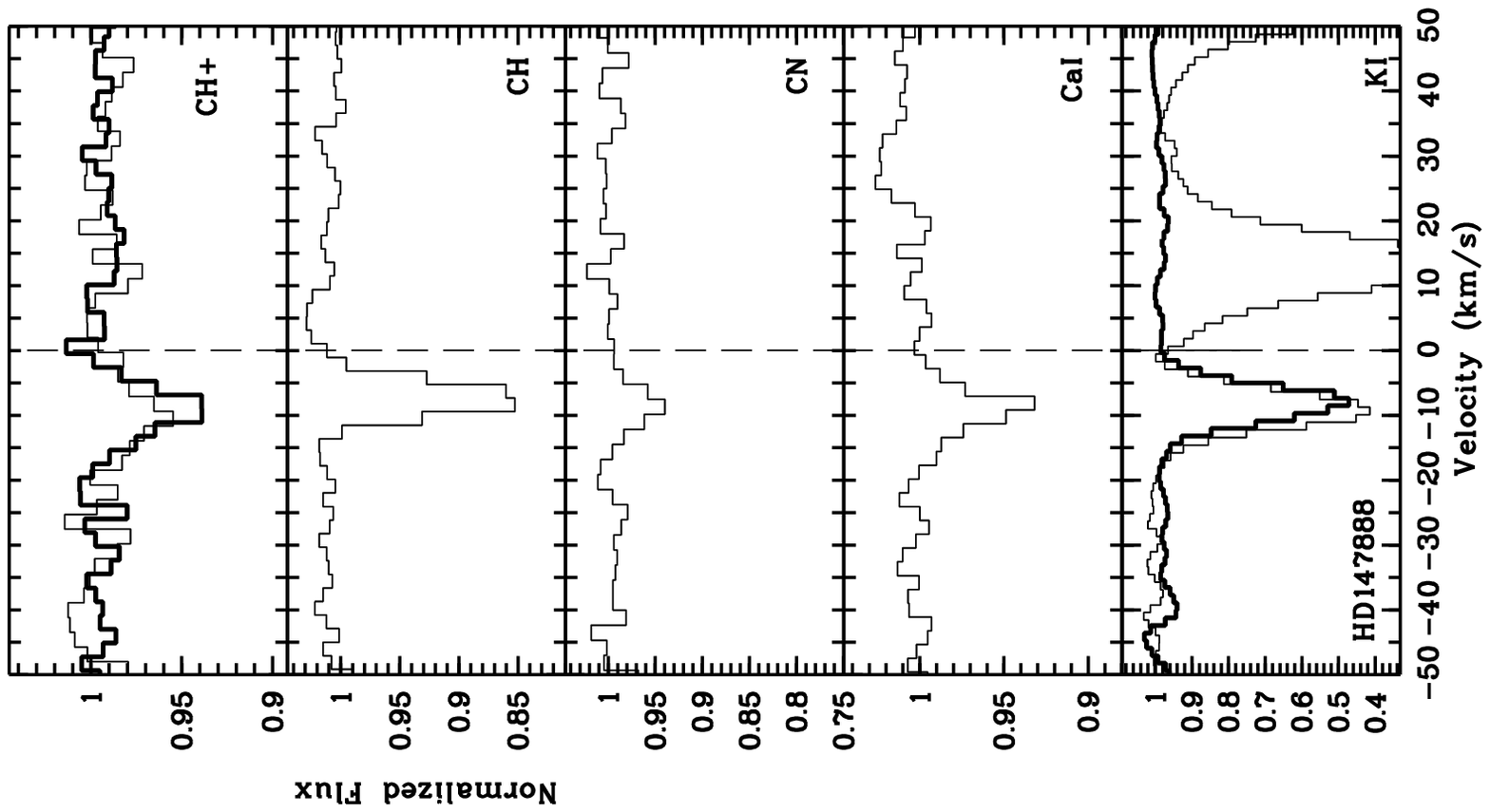}    
	\includegraphics[bb=100 40 565 300, angle=-90, width=6cm,clip]{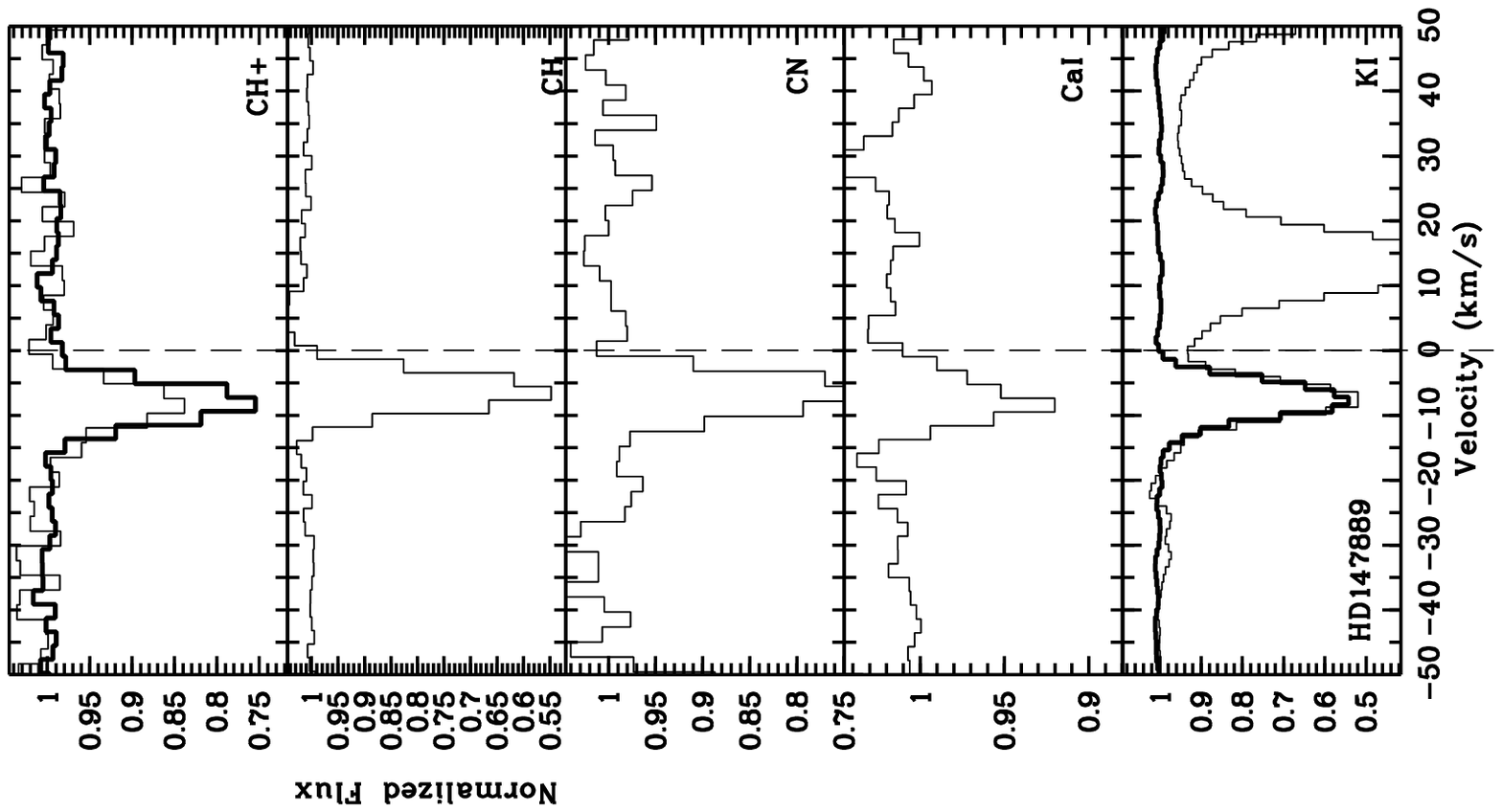}
	\includegraphics[bb=100 40 565 300, angle=-90, width=6cm,clip]{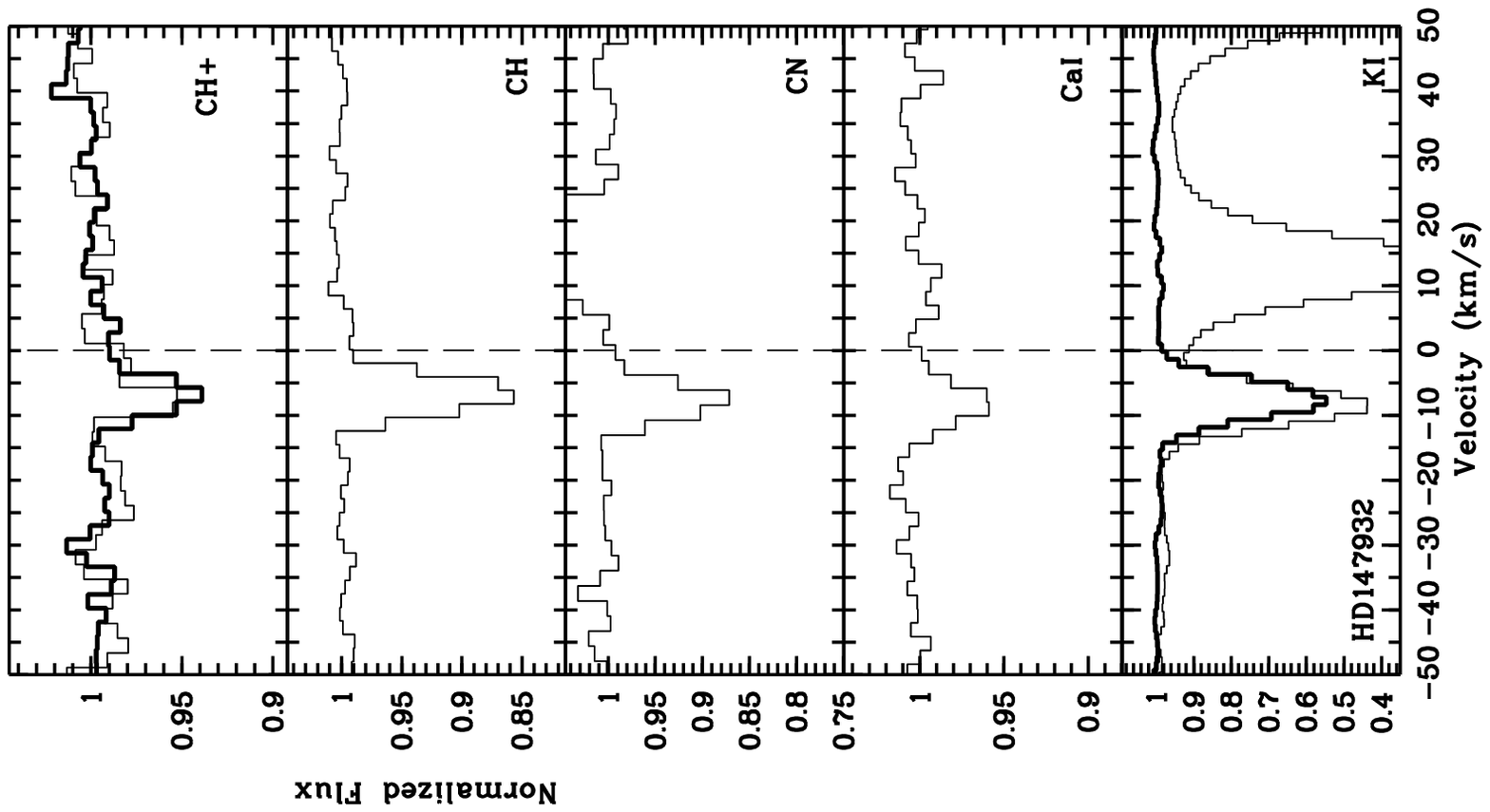}\caption{Continued.} \end{figure*}}\clearpage    
\addtocounter{figure}{-1}

{
\begin{figure*}[h!]	
        \includegraphics[bb=100 40 565 300, angle=-90, width=6cm,clip]{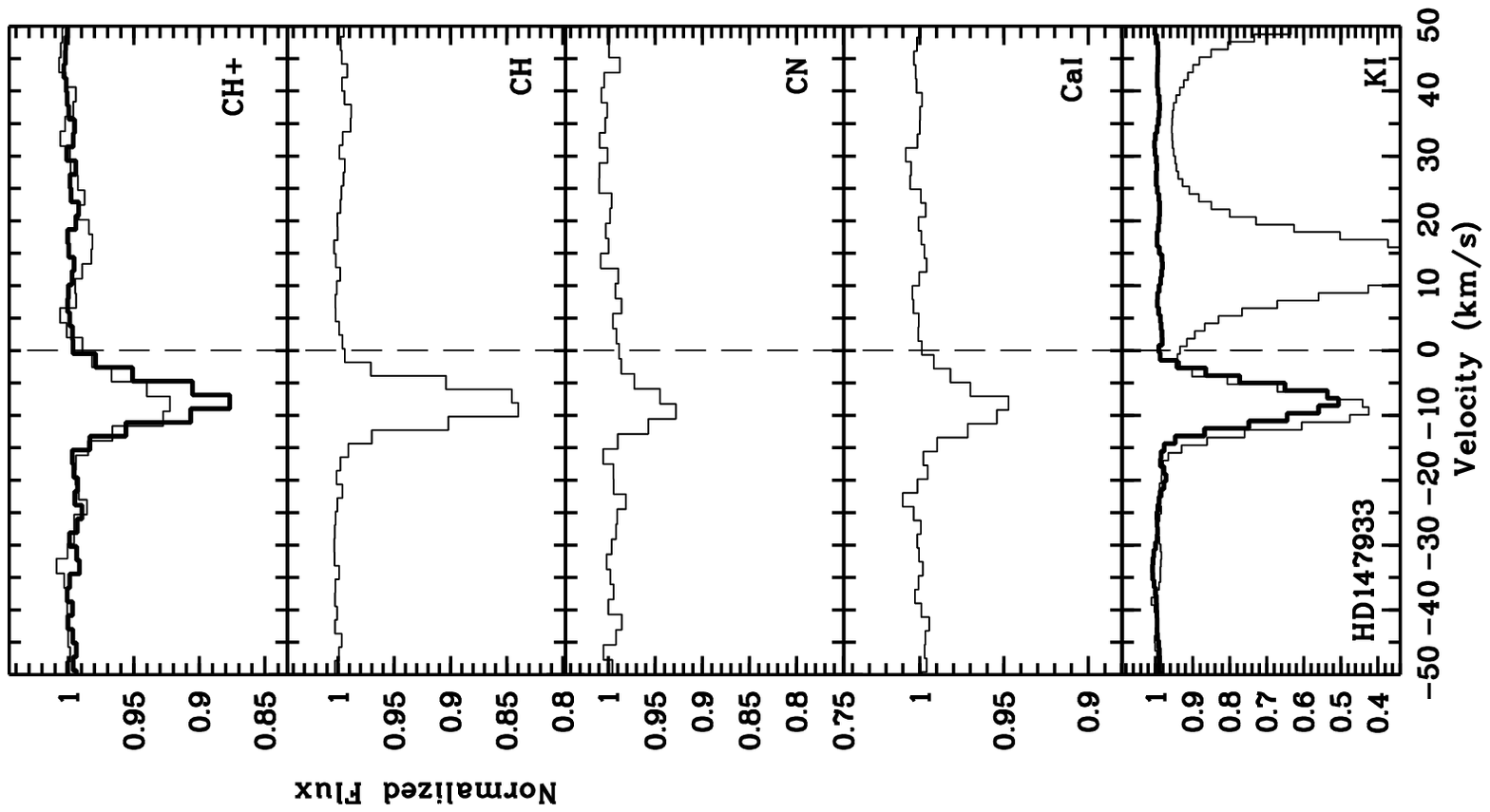}
	\includegraphics[bb=100 40 565 300, angle=-90, width=6cm,clip]{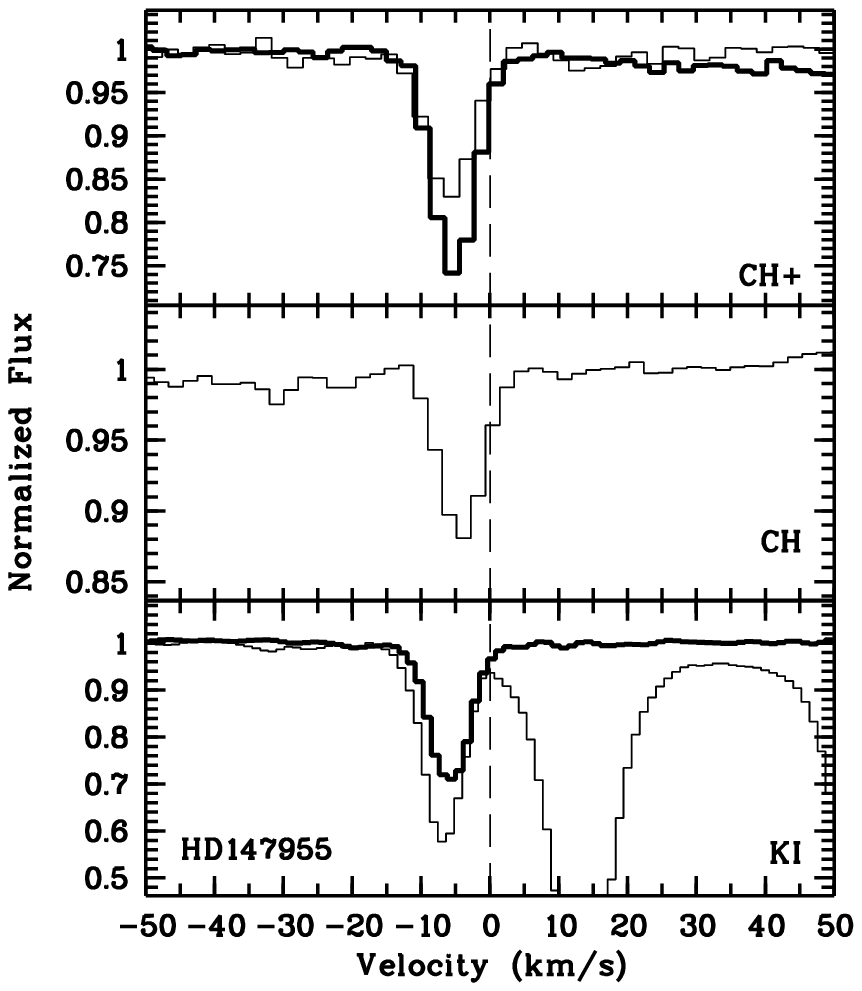}    
	\includegraphics[bb=100 40 565 300, angle=-90, width=6cm,clip]{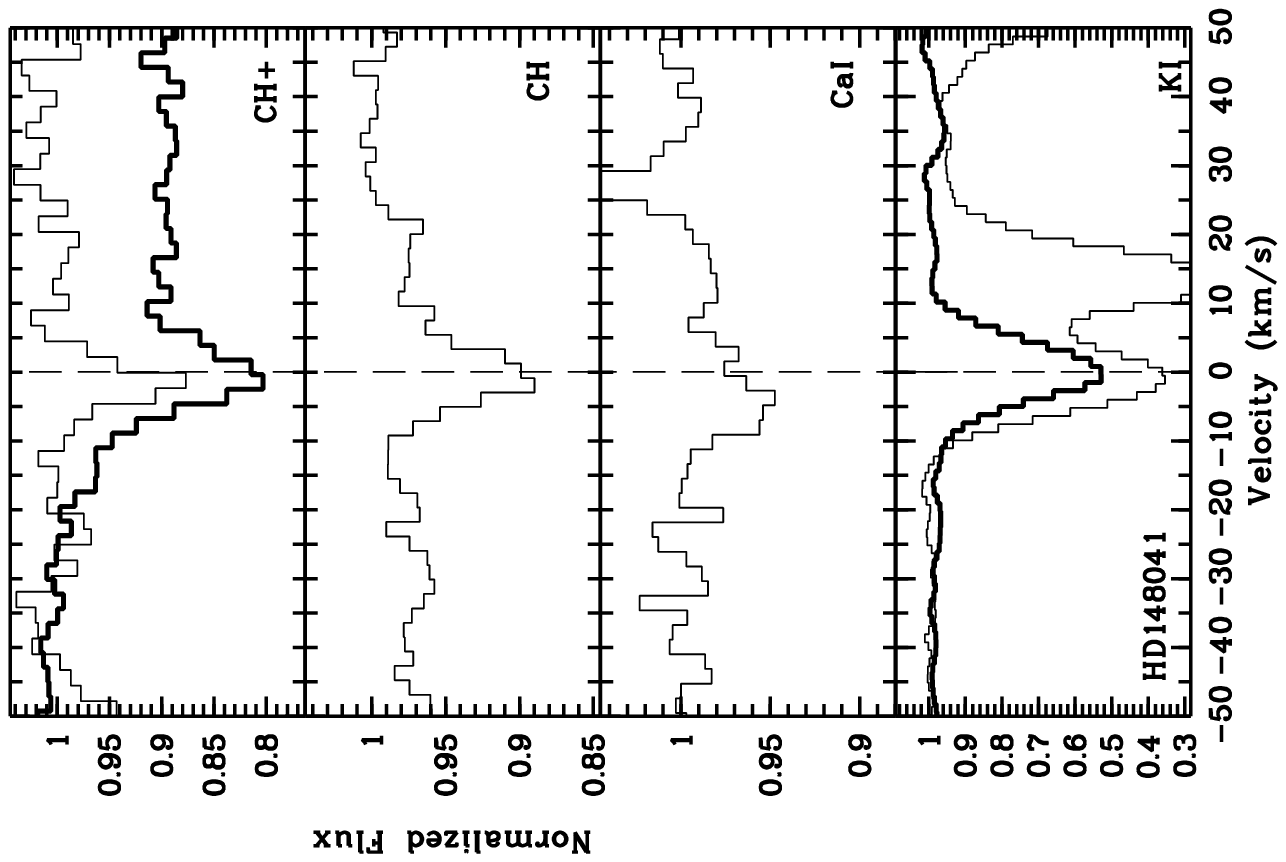}    
	\includegraphics[bb=100 40 565 300, angle=-90, width=6cm,clip]{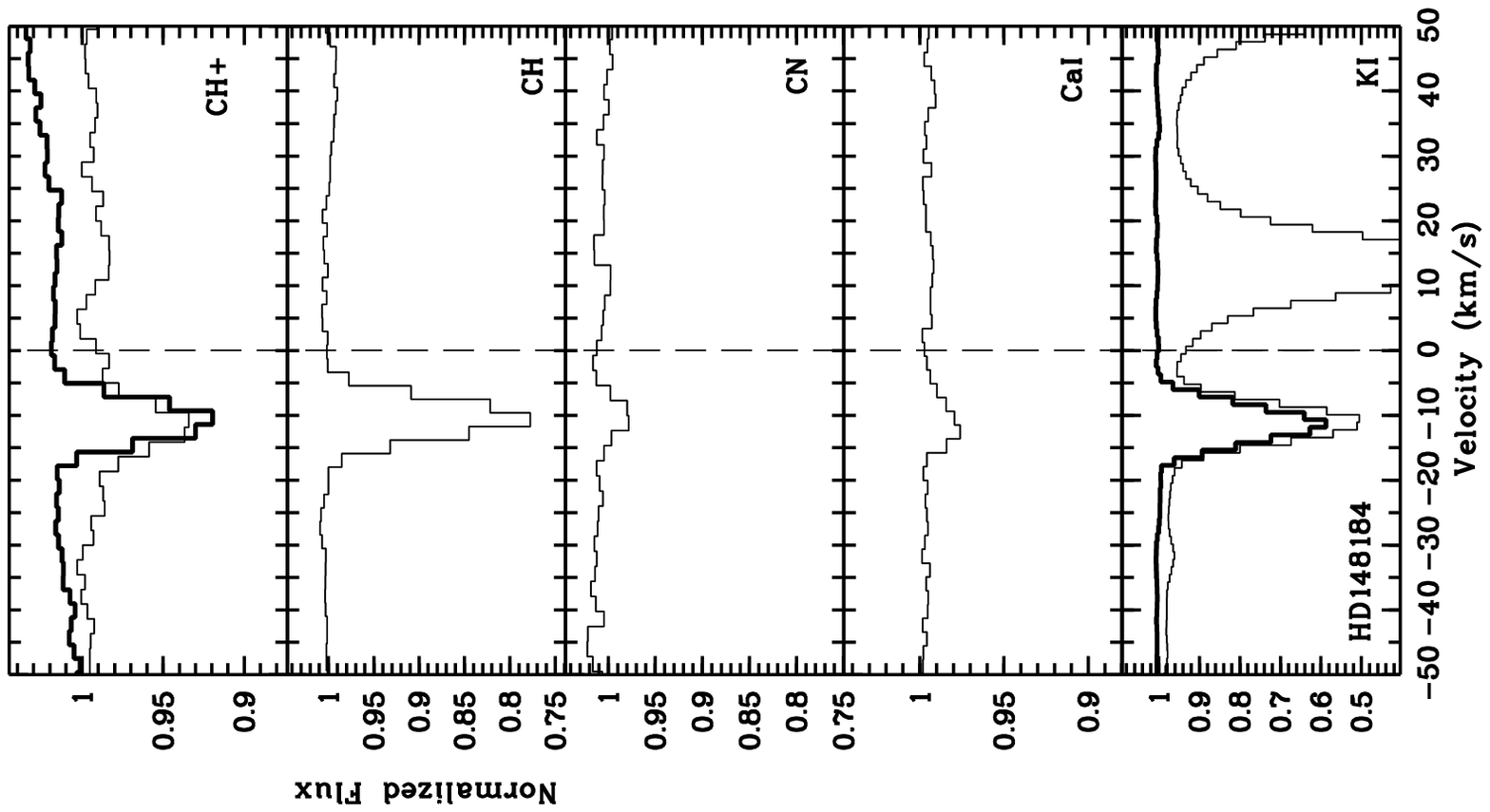}    
	\includegraphics[bb=100 40 565 300, angle=-90, width=6cm,clip]{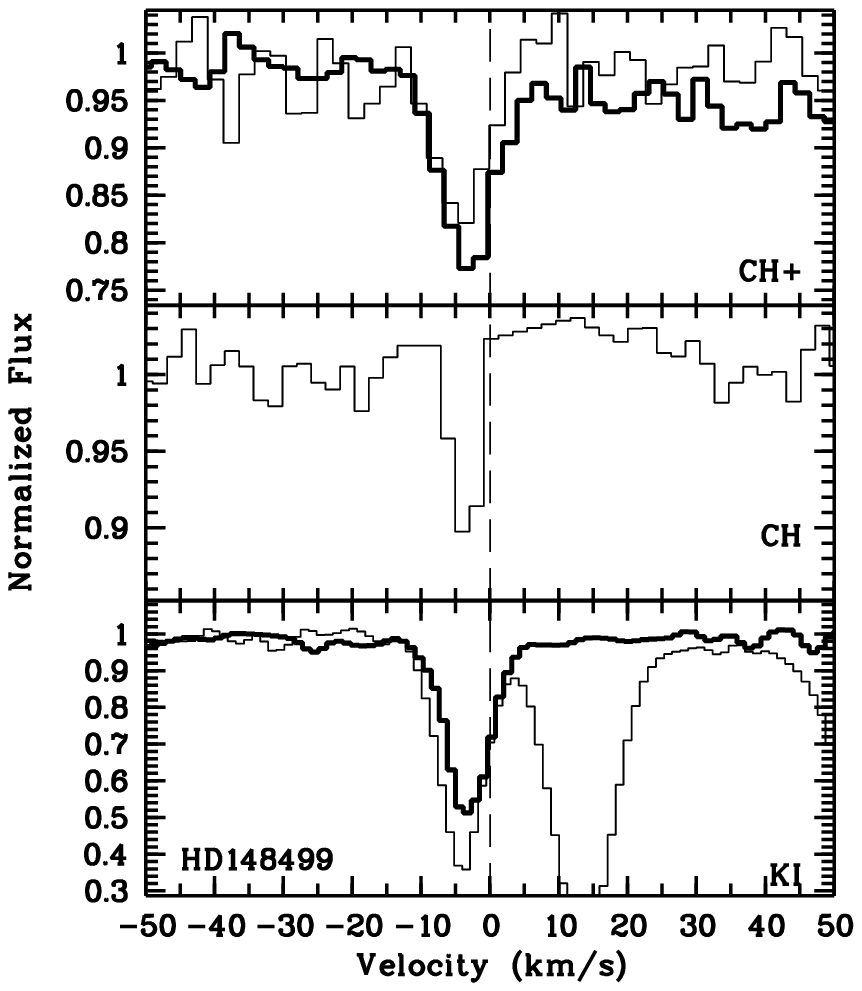}
	\includegraphics[bb=100 40 565 300, angle=-90, width=6cm,clip]{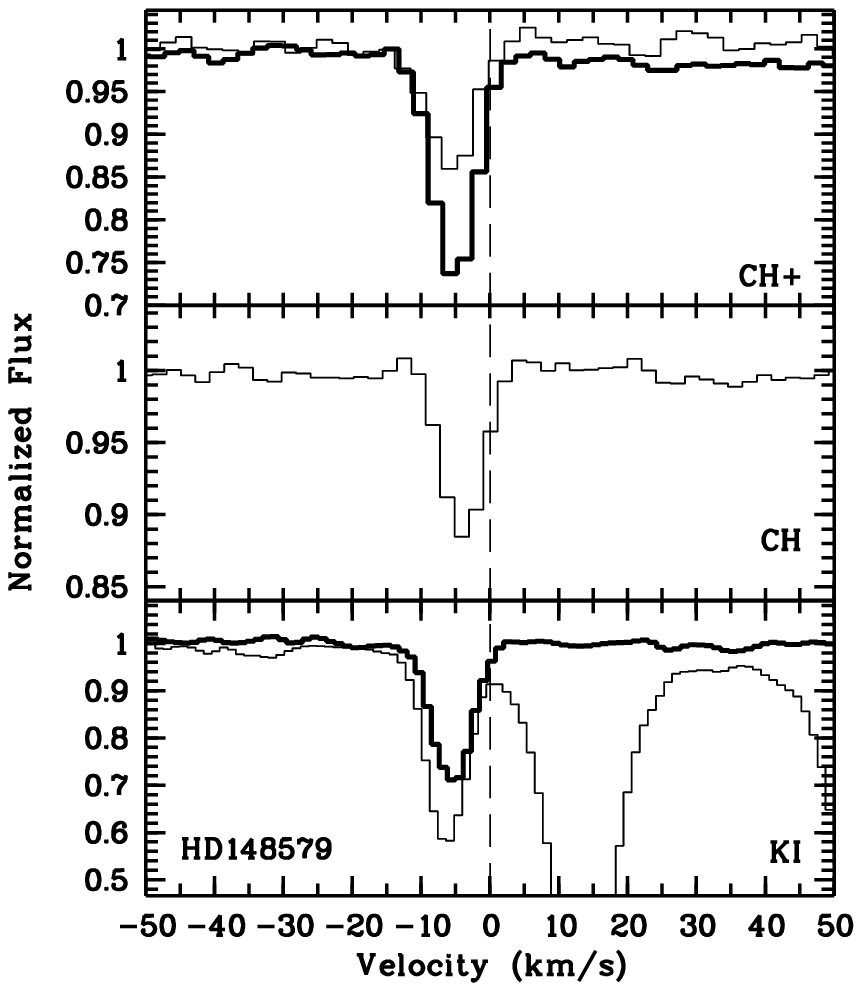}\caption{Continued.}  \end{figure*}}\clearpage    
\addtocounter{figure}{-1}

{
\begin{figure*}[h!]	 
	\includegraphics[bb=100 40 565 300, angle=-90, width=6cm,clip]{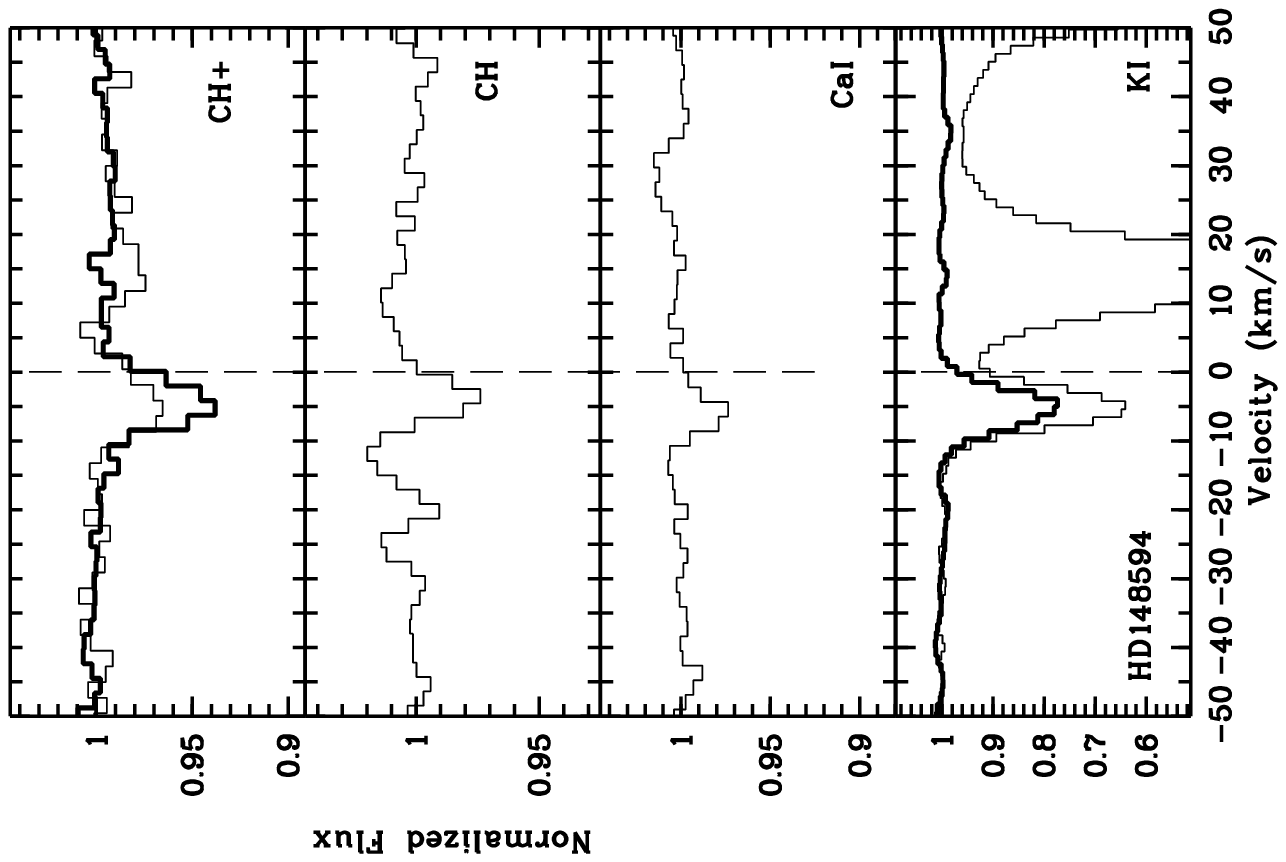}
	\includegraphics[bb=100 40 565 300, angle=-90, width=6cm,clip]{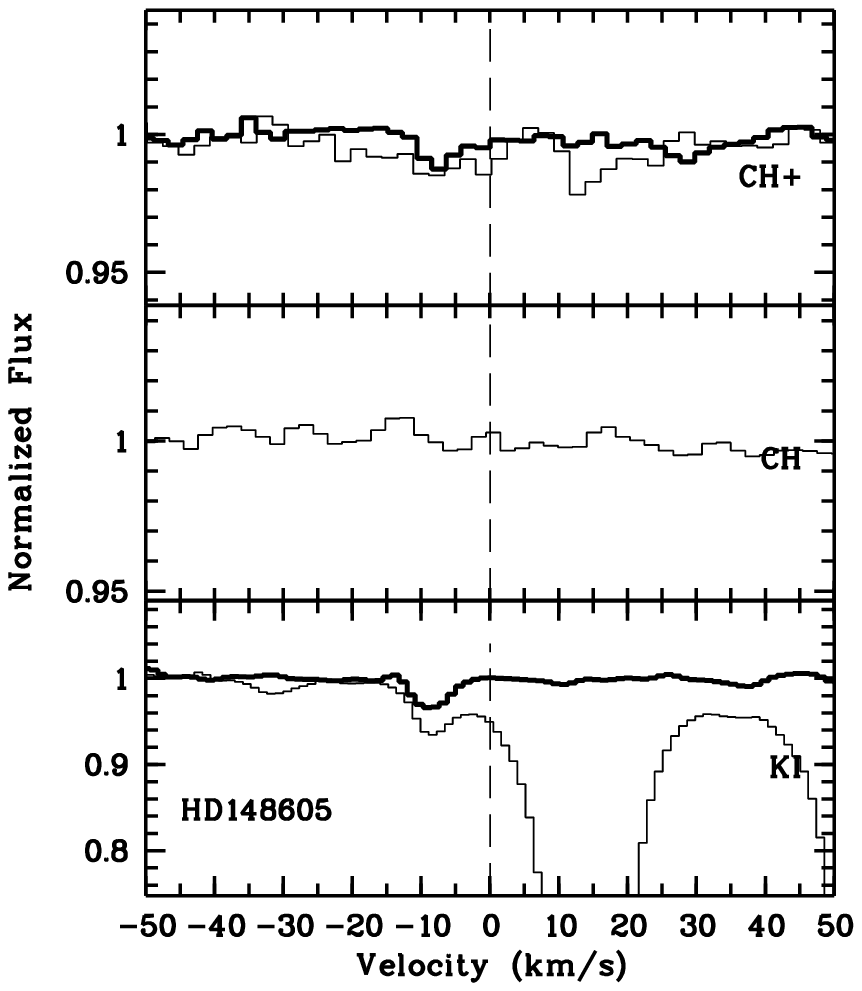}    
	\includegraphics[bb=100 40 565 300, angle=-90, width=6cm,clip]{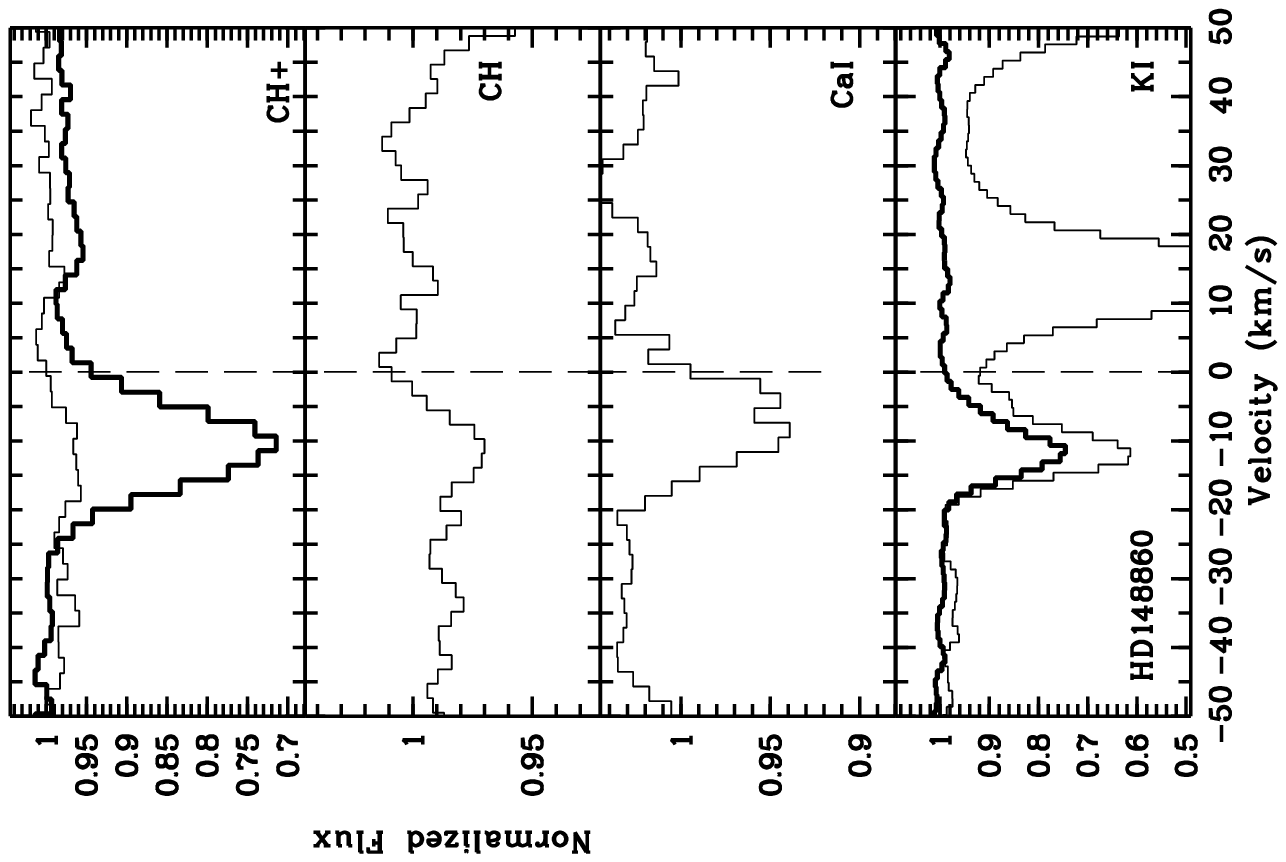}    
	\includegraphics[bb=100 40 565 300, angle=-90, width=6cm,clip]{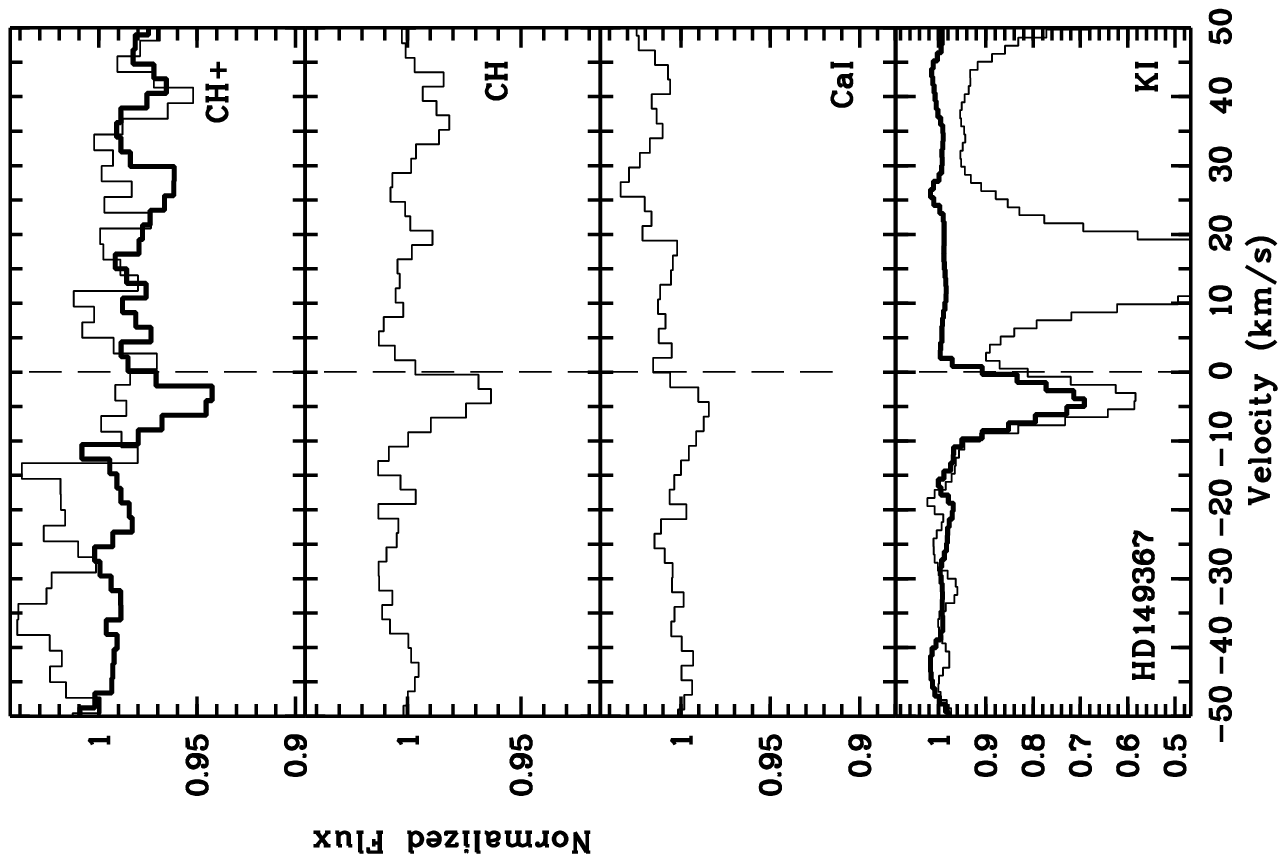}
	\includegraphics[bb=100 40 565 300, angle=-90, width=6cm,clip]{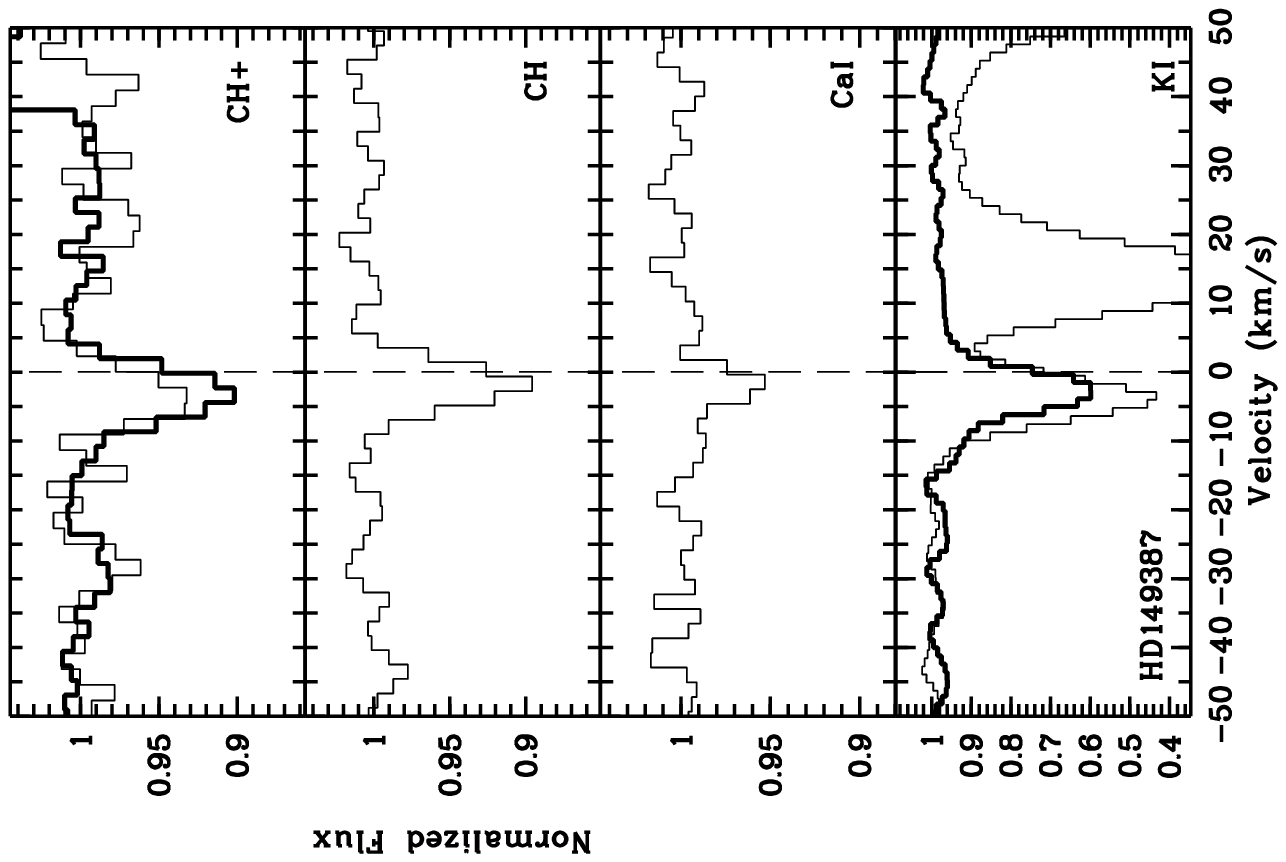}    
	\includegraphics[bb=100 40 565 300, angle=-90, width=6cm,clip]{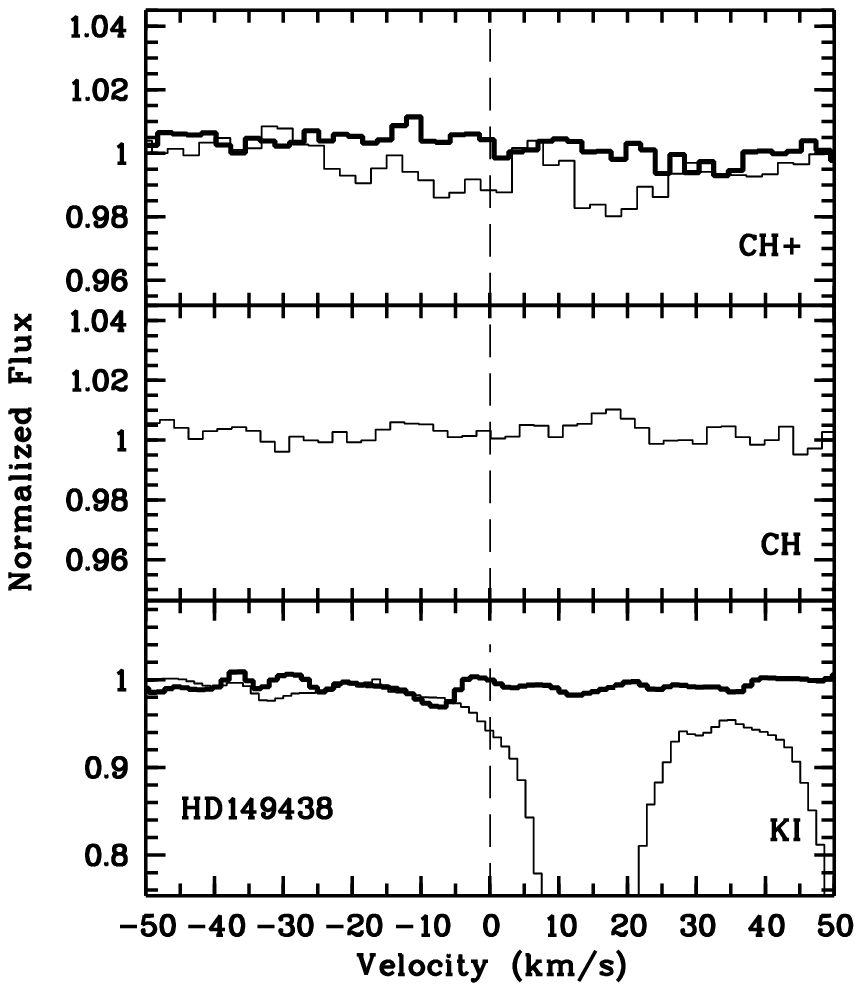}\caption{Continued.} \end{figure*}}\clearpage    
\addtocounter{figure}{-1}

{
\begin{figure*}[h!]	
	\includegraphics[bb=100 40 565 300, angle=-90, width=6cm,clip]{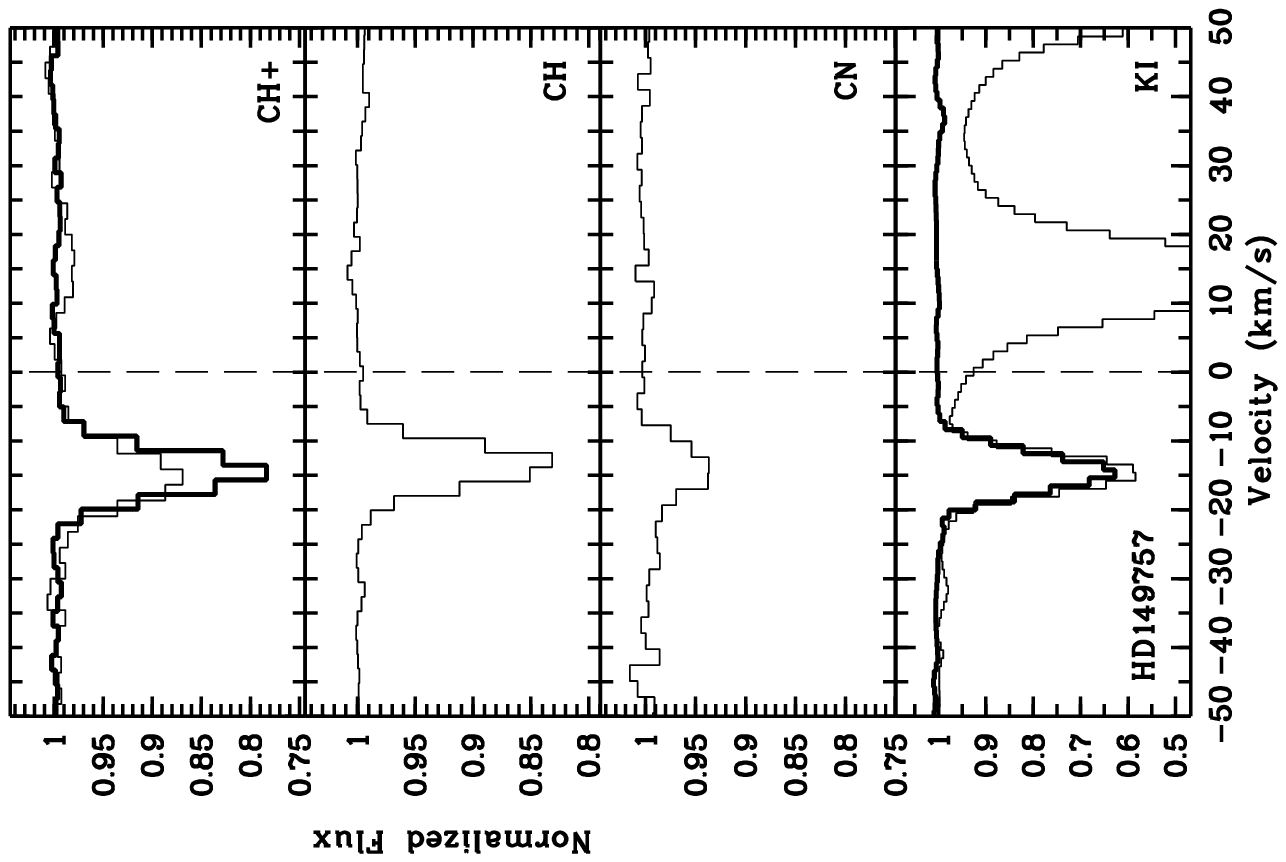}
	\includegraphics[bb=100 40 565 300, angle=-90, width=6cm,clip]{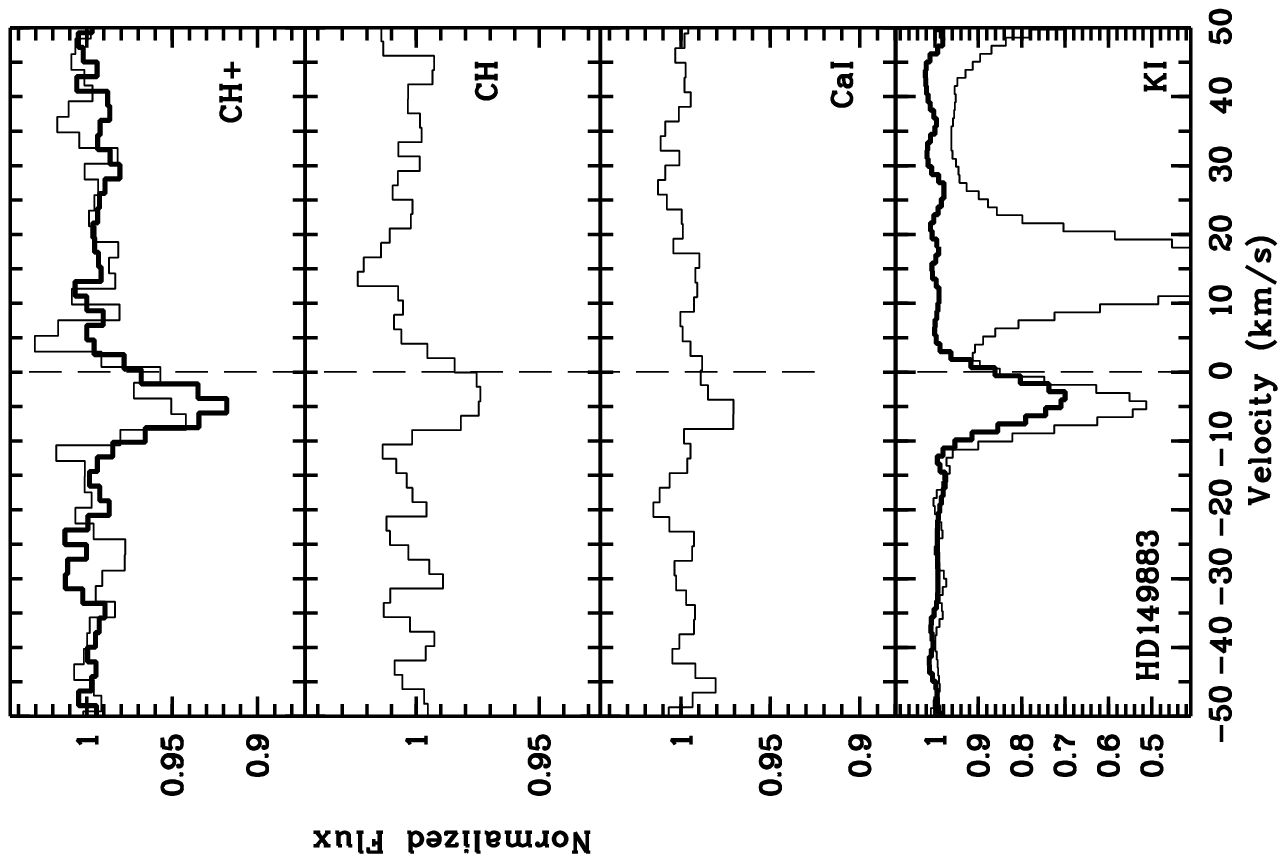}    
	\includegraphics[bb=100 40 565 300, angle=-90, width=6cm,clip]{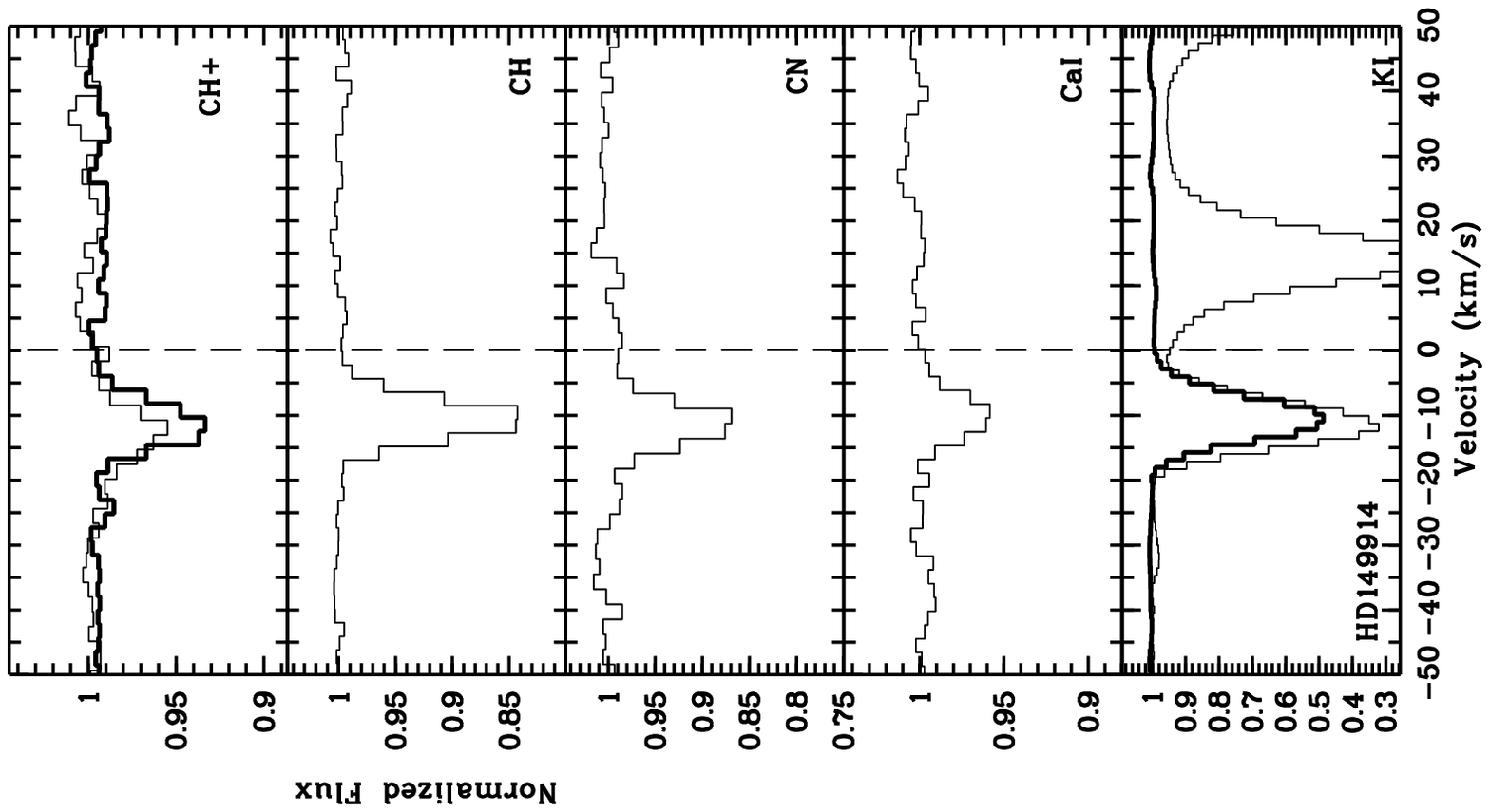}
	\includegraphics[bb=100 40 565 300, angle=-90, width=6cm,clip]{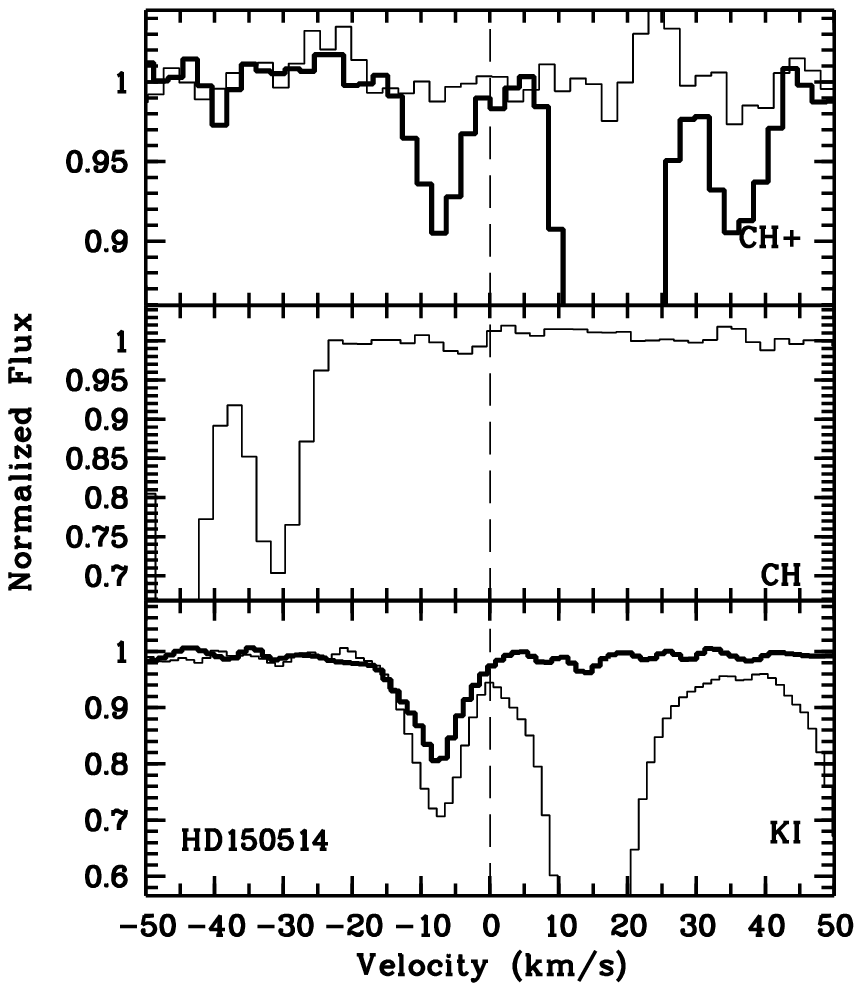}    
	\includegraphics[bb=100 40 565 300, angle=-90, width=6cm,clip]{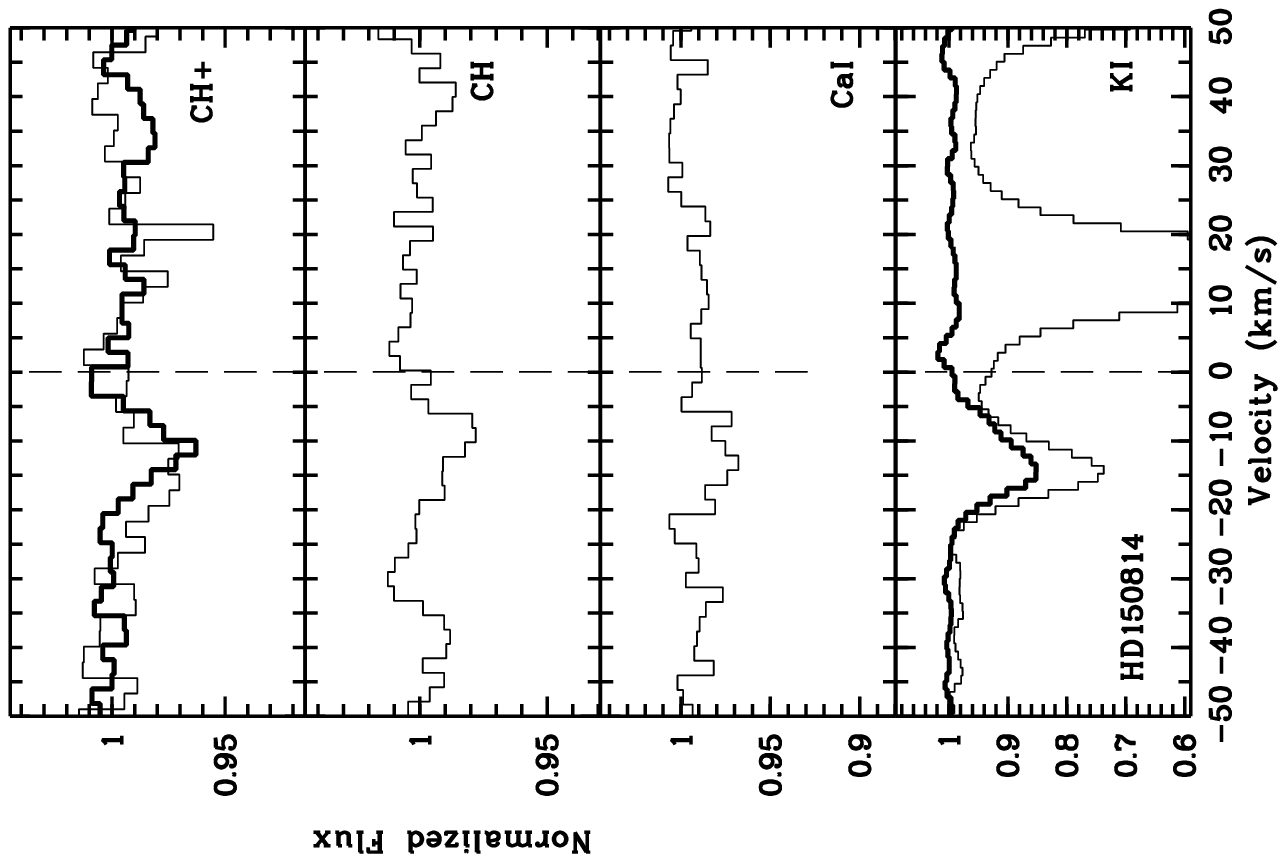}     
	\includegraphics[bb=100 40 565 300, angle=-90, width=6cm,clip]{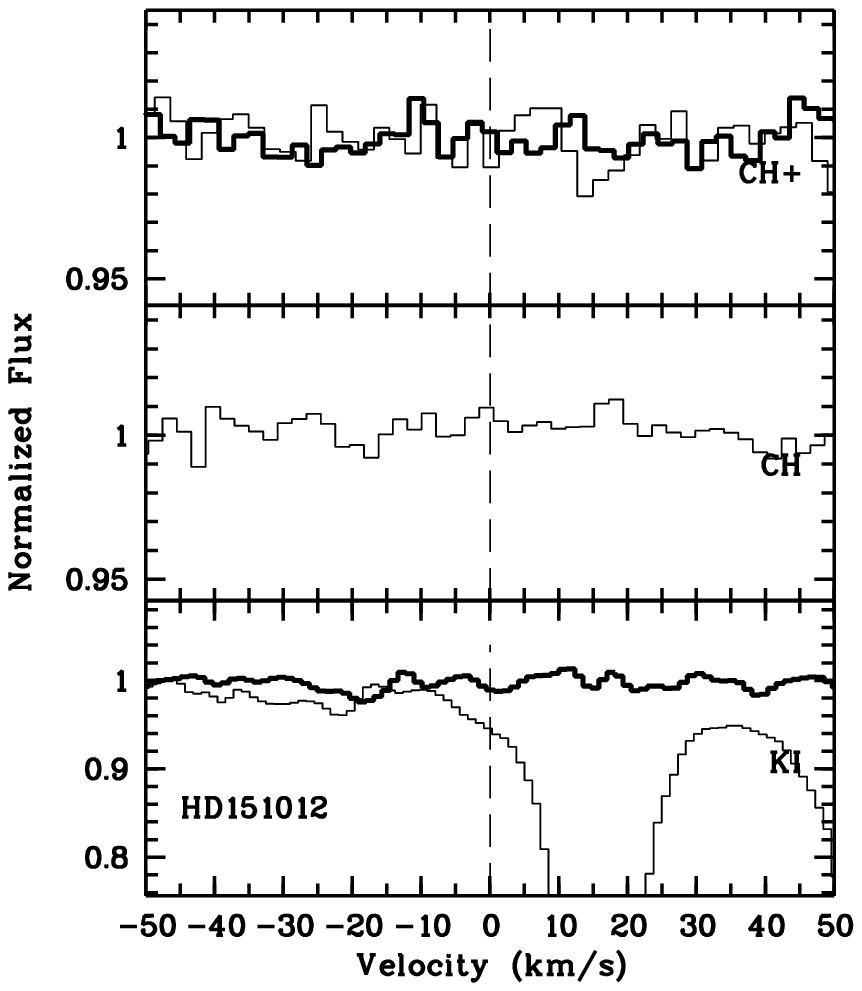} \caption{Continued.} \end{figure*}}\clearpage      
\addtocounter{figure}{-1}

{
\begin{figure*}[h!]    	 
	\includegraphics[bb=100 40 565 300, angle=-90, width=6cm,clip]{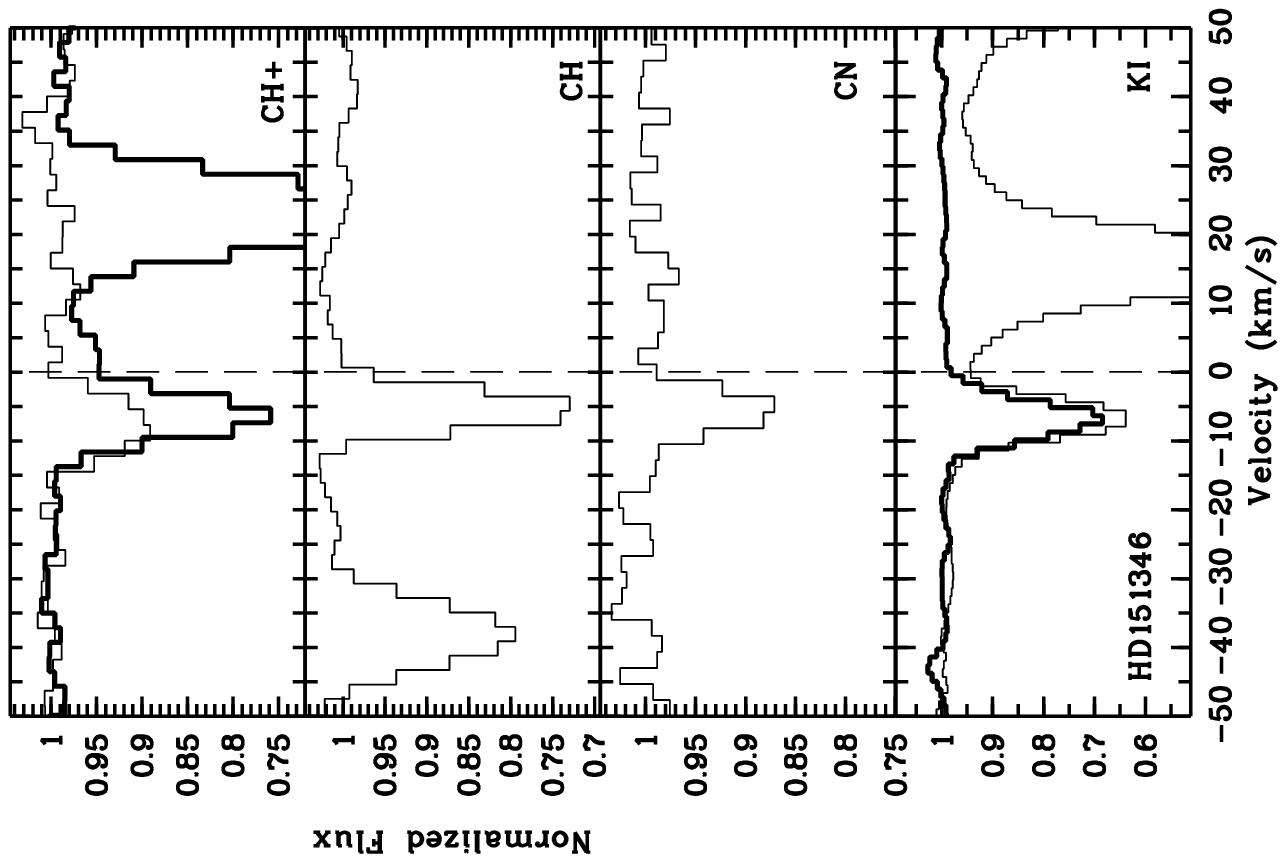}
	\includegraphics[bb=100 40 565 300, angle=-90, width=6cm,clip]{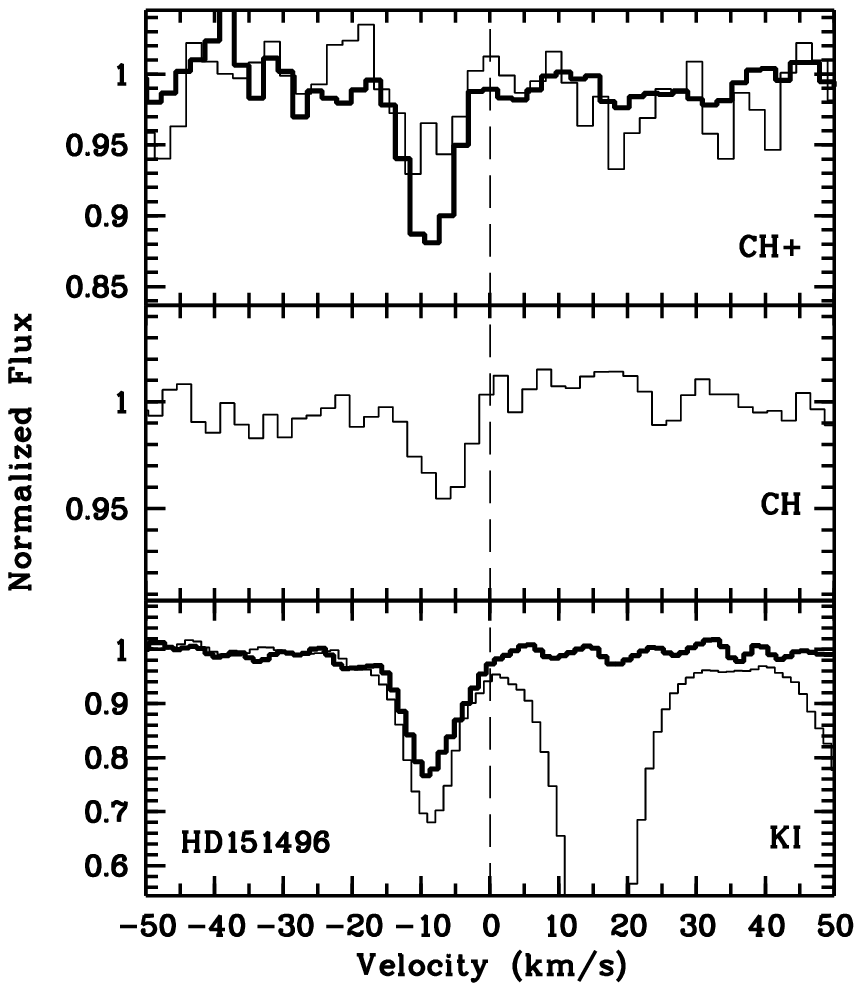}
	\includegraphics[bb=100 40 565 300, angle=-90, width=6cm,clip]{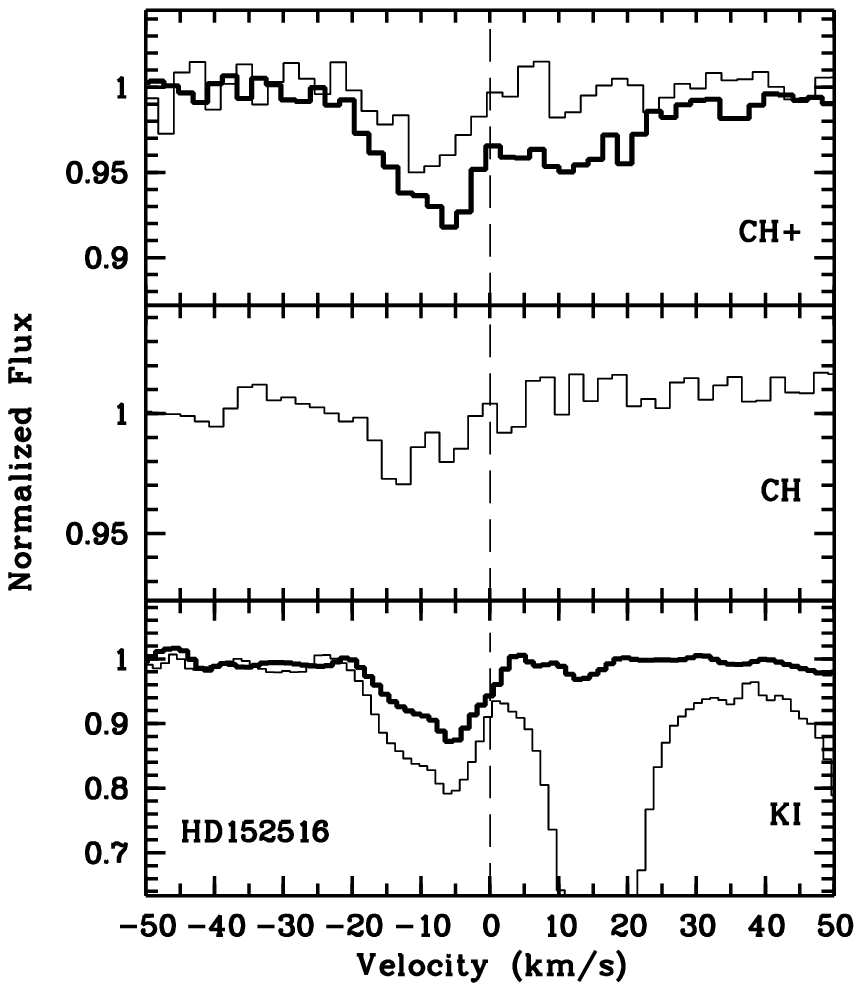}   
	\includegraphics[bb=100 40 565 300, angle=-90, width=6cm,clip]{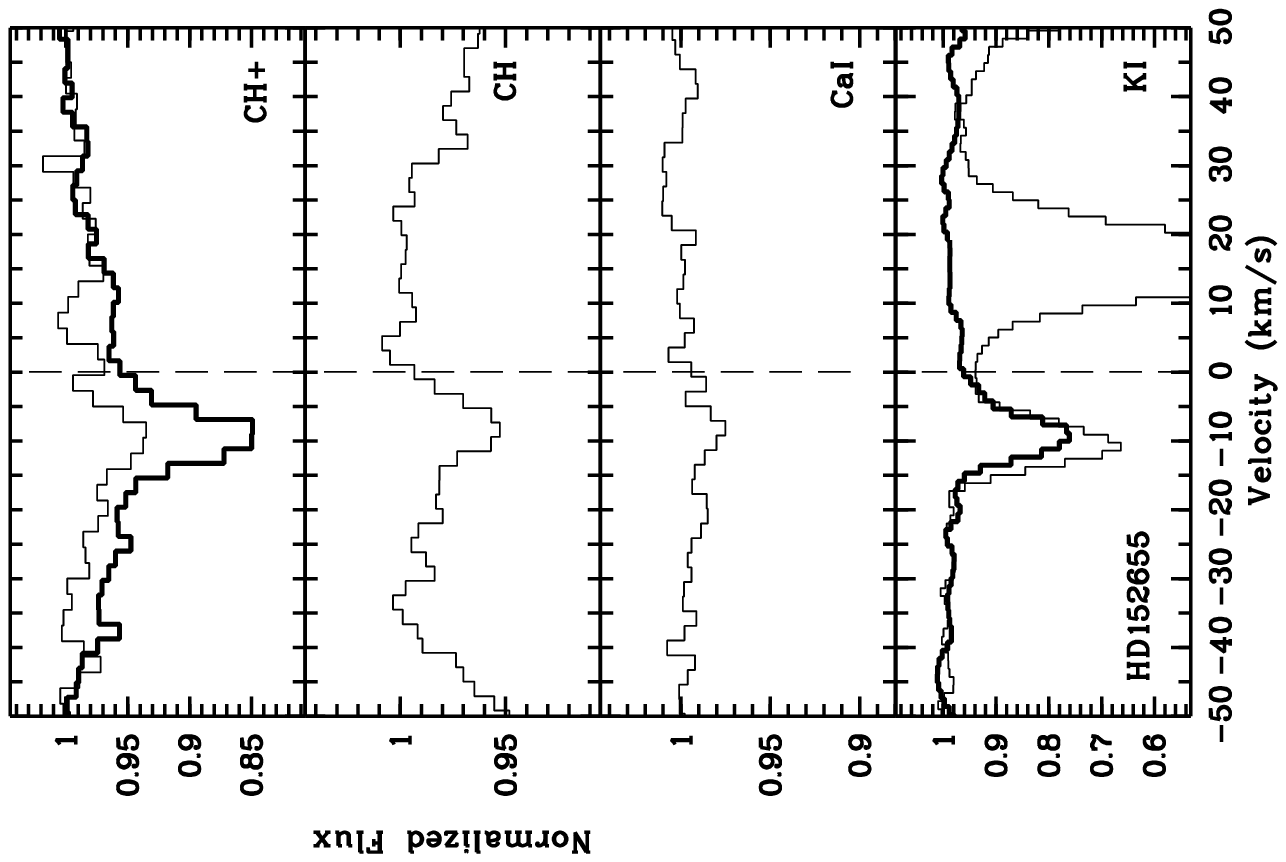}   
	\includegraphics[bb=100 40 565 300, angle=-90, width=6cm,clip]{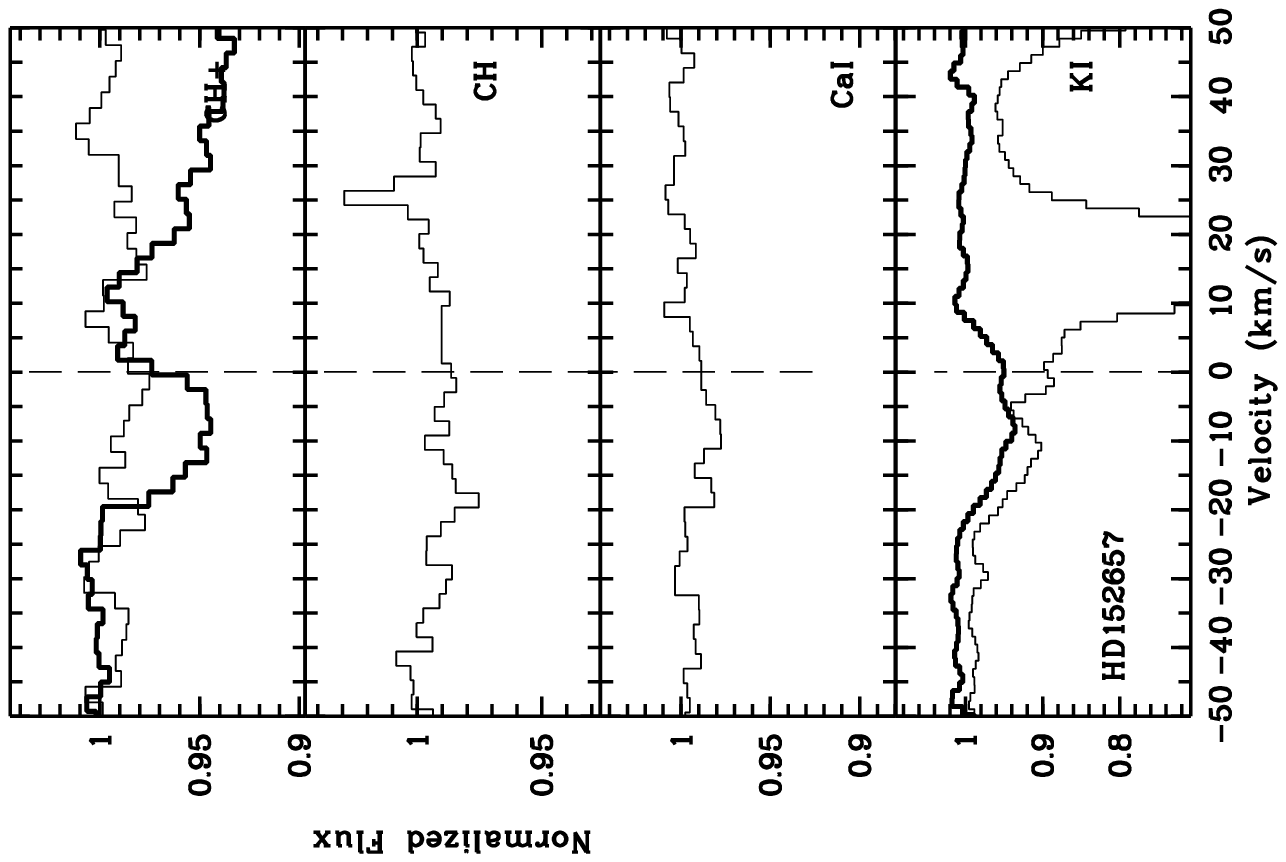} \caption{Continued.} \end{figure*}}\clearpage  
								       
\pagebreak
\end{appendix}

\twocolumn
\end{document}